\documentclass[prx,epsf,showpacs,twocolumn,superscriptaddress,footinbib]{revtex4-2}

\usepackage[pdftex]{graphicx}
\usepackage{dcolumn}
\usepackage{bm}
\usepackage{epsfig}
\usepackage{latexsym}
\usepackage{amsmath}
\usepackage{amsfonts}
\usepackage[position=top,caption=false,singlelinecheck=off, justification=raggedright]{subfig}
\usepackage{color}
\usepackage{array}
\usepackage{braket}
\usepackage{bbold}
\usepackage{dsfont}
\usepackage{mathrsfs}
\usepackage{hyperref}
\usepackage{soul}
\hypersetup{
    colorlinks,
    citecolor=blue,
    filecolor=red,
    linkcolor=magenta,
    urlcolor=blue
}
\renewcommand{\>}{\rangle}

\makeatletter
\DeclareFontFamily{OMX}{MnSymbolE}{}
\DeclareSymbolFont{MnLargeSymbols}{OMX}{MnSymbolE}{m}{n}
\SetSymbolFont{MnLargeSymbols}{bold}{OMX}{MnSymbolE}{b}{n}
\DeclareFontShape{OMX}{MnSymbolE}{m}{n}{
    <-6>  MnSymbolE5
   <6-7>  MnSymbolE6
   <7-8>  MnSymbolE7
   <8-9>  MnSymbolE8
   <9-10> MnSymbolE9
  <10-12> MnSymbolE10
  <12->   MnSymbolE12
}{}
\DeclareFontShape{OMX}{MnSymbolE}{b}{n}{
    <-6>  MnSymbolE-Bold5
   <6-7>  MnSymbolE-Bold6
   <7-8>  MnSymbolE-Bold7
   <8-9>  MnSymbolE-Bold8
   <9-10> MnSymbolE-Bold9
  <10-12> MnSymbolE-Bold10
  <12->   MnSymbolE-Bold12
}{}

\let\llangle\@undefined
\let\rrangle\@undefined
\DeclareMathDelimiter{\llangle}{\mathopen}%
                     {MnLargeSymbols}{'164}{MnLargeSymbols}{'164}
\DeclareMathDelimiter{\rrangle}{\mathclose}%
                     {MnLargeSymbols}{'171}{MnLargeSymbols}{'171}

\newcommand{\fref}[1]{Fig.~\ref{#1}}
\newcommand{\eref}[1]{Eq.~\eqref{#1}}
\newcommand{\sref}[1]{Sec.~\ref{#1}}

\newcommand{\diagramt}[1]{\vcenter{\hbox{\includegraphics[width=0.36\textwidth]{#1.pdf}}}}

\newcommand{\diagram}[1]{\vcenter{\hbox{\includegraphics[width=0.42\textwidth]{#1.pdf}}}}

\newcommand{\diagramb}[1]{\vcenter{\hbox{\includegraphics[width=0.4\textwidth]{#1.pdf}}}}

\newcommand{\diagrams}[1]{\vcenter{\hbox{\includegraphics[width=0.5\textwidth]{#1.pdf}}}}

\newcommand{\diagramf}[1]{\vcenter{\hbox{\includegraphics[width=0.32\textwidth]{#1.pdf}}}}

\newcommand{\Z}{\mathbb{Z}}

\newcommand{\E}{\mathcal{E}}
\newcommand{\tildeE}{\tilde{\mathcal{E}}}

\newcommand{\diff}[1]{\textcolor{black}{#1}}
\newcommand{\ddiff}[1]{\textcolor{black}{#1}}

\begin{document}

\title{Digital quantum magnetism on a trapped-ion quantum computer}

%%%%%%%%% Front %%%%%%%%%%

\author{R. Haghshenas}
\thanks{These authors contributed equally.}
\affiliation{Quantinuum, 303 S. Technology Ct., Broomfield, Colorado 80021, USA}

\author{E. Chertkov}
\thanks{These authors contributed equally.}
\affiliation{Quantinuum, 303 S. Technology Ct., Broomfield, Colorado 80021, USA}

%%%%%%%%%%%%%%%%%%%%%%%%%%

\author{M. Mills} 
\affiliation{Quantinuum, 303 S. Technology Ct., Broomfield, Colorado 80021, USA}

\author{W. Kadow}
\affiliation{Technical University of Munich, TUM School of Natural Sciences, Physics Department, 85748 Garching, Germany}
\affiliation{Munich Center for Quantum Science and Technology (MCQST), Schellingstr. 4, 80799 M\"unchen, Germany}

\author{S.-H. Lin}
\affiliation{Quantinuum, Leopoldstrasse 180, 80804 Munich, Germany}
\affiliation{Technical University of Munich, TUM School of Natural Sciences, Physics Department, 85748 Garching, Germany}
\affiliation{Munich Center for Quantum Science and Technology (MCQST), Schellingstr. 4, 80799 M\"unchen, Germany}

\author{Y. H. Chen}
\affiliation{Quantinuum, 303 S. Technology Ct., Broomfield, Colorado 80021, USA}

\author{C. Cade}
\author{I. Niesen}
\affiliation{Fermioniq, Science Park 408, 1098 XH Amsterdam, The Netherlands}

\author{T. Begu\v{s}i\'c}
\affiliation{Division of Chemistry and Chemical Engineering, California Institute of Technology, Pasadena, California 91125, USA}
\affiliation{Current affiliation: University of W\"urzburg, Institute of Physical and Theoretical Chemistry, 97074 Würzburg, Germany}

\author{M. S. Rudolph}
\affiliation{Institute of Physics, Ecole Polytechnique Fédérale de Lausanne (EPFL), Lausanne CH-1015, Switzerland}

\author{C. Cirstoiu}
\affiliation{Quantinuum, Terrington House, 13-15 Hills Road, Cambridge CB2 1NL, UK}

\author{K. Hemery}
\affiliation{Quantinuum, Leopoldstrasse 180, 80804 Munich, Germany}

\author{C. Mc Keever}
\affiliation{Quantinuum, Partnership House, Carlisle Place, London SW1P 1BX, UK}

\author{M. Lubasch}
\affiliation{Quantinuum, Partnership House, Carlisle Place, London SW1P 1BX, UK}

\author{E. Granet}
\affiliation{Quantinuum, Leopoldstrasse 180, 80804 Munich, Germany}

%%%%%%%%%%%%% Alphabetical %%%%%%%%%%%%%%%

\author{C. H. Baldwin}
\affiliation{Quantinuum, 303 S. Technology Ct., Broomfield, Colorado 80021, USA}

\author{J. P. Bartolotta}
\affiliation{Quantinuum, 303 S. Technology Ct., Broomfield, Colorado 80021, USA}

\author{M. Bohn}
\affiliation{Quantinuum, 303 S. Technology Ct., Broomfield, Colorado 80021, USA}

\author{J. J. Burau}
\affiliation{Quantinuum, 303 S. Technology Ct., Broomfield, Colorado 80021, USA}

\author{J. Cline}
\affiliation{Quantinuum, 303 S. Technology Ct., Broomfield, Colorado 80021, USA}

\author{M. DeCross}
\affiliation{Quantinuum, 303 S. Technology Ct., Broomfield, Colorado 80021, USA}

\author{J. M. Dreiling}
\affiliation{Quantinuum, 303 S. Technology Ct., Broomfield, Colorado 80021, USA}

\author{C. Foltz}
\affiliation{Quantinuum, 303 S. Technology Ct., Broomfield, Colorado 80021, USA}

\author{D. Francois}
\affiliation{Quantinuum, 303 S. Technology Ct., Broomfield, Colorado 80021, USA}

\author{J. P. Gaebler}
\affiliation{Quantinuum, 303 S. Technology Ct., Broomfield, Colorado 80021, USA}

\author{C. N. Gilbreth}
\affiliation{Quantinuum, 303 S. Technology Ct., Broomfield, Colorado 80021, USA}

\author{J. Gray}
\affiliation{Division of Chemistry and Chemical Engineering, California Institute of Technology, Pasadena, California 91125, USA}

\author{D. Gresh}
\affiliation{Quantinuum, 303 S. Technology Ct., Broomfield, Colorado 80021, USA}

\author{A. Hall}
\affiliation{Quantinuum, 303 S. Technology Ct., Broomfield, Colorado 80021, USA}

\author{A. Hankin}
\affiliation{Quantinuum, 303 S. Technology Ct., Broomfield, Colorado 80021, USA}

\author{A. Hansen}
\affiliation{Quantinuum, 303 S. Technology Ct., Broomfield, Colorado 80021, USA}

\author{N. Hewitt} 
\affiliation{Quantinuum, 303 S. Technology Ct., Broomfield, Colorado 80021, USA}

\author{C. A. Holliman}
\affiliation{Quantinuum K.K., Otemachi Financial City Grand Cube 3F, 1-9-2 Otemachi, Chiyoda-ku, Tokyo, Japan}

\author{R. B. Hutson}
\affiliation{Quantinuum, 303 S. Technology Ct., Broomfield, Colorado 80021, USA}

\author{M. Iqbal}
\affiliation{Quantinuum, Leopoldstrasse 180, 80804 Munich, Germany}

\author{N. Kotibhaskar}
\affiliation{Quantinuum, 303 S. Technology Ct., Broomfield, Colorado 80021, USA}

\author{E. Lehman}
\affiliation{Quantinuum, 303 S. Technology Ct., Broomfield, Colorado 80021, USA}

\author{D. Lucchetti}
\affiliation{Quantinuum, 303 S. Technology Ct., Broomfield, Colorado 80021, USA}

\author{I. S. Madjarov}
\affiliation{Quantinuum, 303 S. Technology Ct., Broomfield, Colorado 80021, USA}

\author{K. Mayer}
\affiliation{Quantinuum, 303 S. Technology Ct., Broomfield, Colorado 80021, USA}

\author{A. R. Milne}
\affiliation{Quantinuum, Partnership House, Carlisle Place, London SW1P 1BX, UK}

\author{S. A. Moses}
\affiliation{Quantinuum, 303 S. Technology Ct., Broomfield, Colorado 80021, USA}
\affiliation{Current affiliation: AWS Center for Quantum Computing, Pasadena, CA 91125.}

\author{B. Neyenhuis}
\affiliation{Quantinuum, 303 S. Technology Ct., Broomfield, Colorado 80021, USA}

\author{G. Park}
\affiliation{Division of Engineering and Applied Science, California Institute of Technology, Pasadena, California 91125, USA}

\author{A. R. Perry}
\affiliation{Quantinuum, 303 S. Technology Ct., Broomfield, Colorado 80021, USA}

\author{B. Ponsioen}
\affiliation{Fermioniq, Science Park 408, 1098 XH Amsterdam, The Netherlands}

\author{M. Schecter}
\affiliation{Quantinuum, 303 S. Technology Ct., Broomfield, Colorado 80021, USA}

\author{P. E. Siegfried}
\affiliation{Quantinuum, 303 S. Technology Ct., Broomfield, Colorado 80021, USA}

\author{D. T. Stephen}
\affiliation{Quantinuum, 303 S. Technology Ct., Broomfield, Colorado 80021, USA}

\author{B. G. Tiemann}
\affiliation{Quantinuum, 303 S. Technology Ct., Broomfield, Colorado 80021, USA}

\author{M. D. Urmey}
\affiliation{Quantinuum, 303 S. Technology Ct., Broomfield, Colorado 80021, USA}

\author{J. Walker}
\affiliation{Quantinuum, 303 S. Technology Ct., Broomfield, Colorado 80021, USA}

%%%%%%%%% Back %%%%%%%%%%

\author{A. C. Potter}
\affiliation{Quantinuum, 303 S. Technology Ct., Broomfield, Colorado 80021, USA}

\author{D. Hayes}
\affiliation{Quantinuum, 303 S. Technology Ct., Broomfield, Colorado 80021, USA}

\author{G. K.-L. Chan}
\affiliation{Division of Chemistry and Chemical Engineering, California Institute of Technology, Pasadena, California 91125, USA}
\affiliation{Institute for Quantum Information and Matter, California Institute of Technology, Pasadena, California 91125, USA}

\author{F. Pollmann}
\affiliation{Technical University of Munich, TUM School of Natural Sciences, Physics Department, 85748 Garching, Germany}
\affiliation{Munich Center for Quantum Science and Technology (MCQST), Schellingstr. 4, 80799 M\"unchen, Germany}

\author{M. Knap}
\affiliation{Technical University of Munich, TUM School of Natural Sciences, Physics Department, 85748 Garching, Germany}
\affiliation{Munich Center for Quantum Science and Technology (MCQST), Schellingstr. 4, 80799 M\"unchen, Germany}

\author{H. Dreyer}
\affiliation{Quantinuum, Leopoldstrasse 180, 80804 Munich, Germany}

\author{M. Foss-Feig}
\email{michael.feig@quantinuum.com}
\affiliation{Quantinuum, 303 S. Technology Ct., Broomfield, Colorado 80021, USA}

\begin{abstract}

Digital quantum matter---realized when discrete quantum gates approximate continuous time evolution---is susceptible to heating into chaotic, structureless states \cite{PhysRevX.4.041048}. \ddiff{If digitization errors are adequately suppressed, a long-lived transient regime of approximately energy-conserving dynamics \cite{Bukov2015, KUWAHARA201696,PhysRevB.95.014112,doi:10.1126/sciadv.aau8342,Weidinger2017,HO2023169297} can be observed on gate-based quantum computers. Conservation of energy, in turn, enables the exploration of a wide variety of complex behaviors observed in equilibrium systems, ranging from the nontrivial microscopic origins of thermalization itself \cite{dallesio_ETH} to the stabilization of effective models hosting exotic emergent properties.} Here, we use Quantinuum's system model H2 quantum computer \cite{Moses2023,decross2024computational} to simulate digitized dynamics of the quantum Ising model, suppressing digitization errors well enough to observe thermalization on timescales that severely challenge classical simulation methods. Relaxation of an inhomogeneous state reveals an emergent hydrodynamics due to approximate energy conservation, and we compute the associated diffusion constant. By reprogramming our simulations to take place on a triangular lattice with periodic boundary conditions, we observe thermalization consistent with emergent gauge and topological constraints resulting from lattice frustration \cite{PhysRevLett.84.4457,PhysRevB.63.224401,savary2016quantum}. Our results were enabled by continued advances in two-qubit gate quality (native partial entangler fidelities of $99.94(1)\%$), and establish digital quantum computers as powerful tools for studying (effectively) continuous-time dynamics.

\end{abstract}

\maketitle

\section{Introduction}

Quantum computers can efficiently simulate complex quantum many-body systems that are challenging to describe with classical computers \cite{doi:10.1126/science.273.5278.1073}. However, prior to the development of large-scale fault-tolerant quantum computers, the presence of noise severely limits the scope of feasible simulations: It is difficult to simultaneously suppress hardware noise (requiring shallower circuits) and digitization errors incurred when approximating time evolution with discrete gates (requiring deeper circuits). While analog quantum simulators circumvent this challenge to some extent \cite{daley_qa_review,PhysRevLett.113.147205,bernien_2017,zhang_2017,Joshi_2022_hydro,andersen2024thermalization,Wienand_2024_hydro,king2024computational}, they are specialized machines that cannot be easily adapted to a variety of problems. If both digitization errors and noise are adequately controlled, the flexibility of quantum computers will expand the scope of feasible quantum simulations~\cite{ibm_utility,cochran2024visualizing,will2025probing}.

%Figure 1
%
\begin{figure*}[!!t]
\centering
\includegraphics[width=2.04\columnwidth]{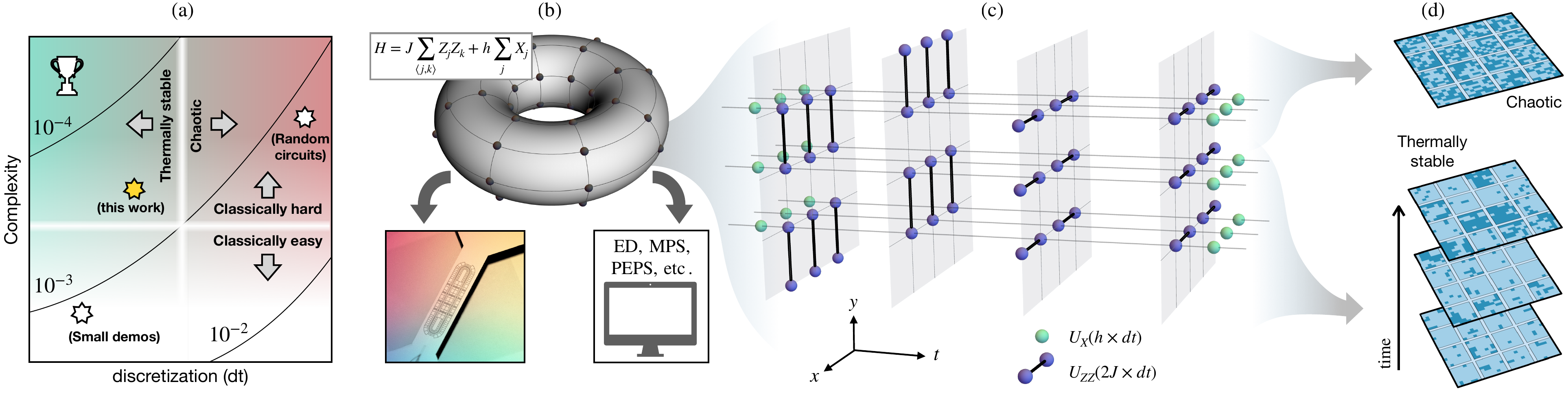}
\caption{{\bf Digitizing an Ising quantum magnet:} (a) Schematic tradeoff between discretization error and achievable complexity, with contours indicating various two-qubit gate fidelities. (b) We simulate digitized dynamics of the transverse-field Ising model on a torus, comparing results from Quantinuum's H2 quantum computer to classical simulation methods. (c) Digitization of the dynamics via a second-order Trotterization of the time evolution operator. (d) Depending on the Trotter step size two outcomes are possible: (Top) For step sizes that are too large the system swiftly heats into a chaotic, infinite-temperatures state. (Bottom) For step sizes small enough, the system evolves into a long-lived Floquet-prethermalized state (here evidenced in raw shots output by H2 showing persistent ferromagnetic correlations).
\label{fig:setup}}
\end{figure*}

Trotterization is a standard way to digitize time evolution, ensuring accurate simulation by using sufficiently small time steps \cite{PhysRevX.11.011020}. While there have been many demonstrations of this approach, quantitative accuracy comes at a cost: Small time steps yield large circuit depths, so far restricting accurate simulations to small scales and low simulation complexity (lower left, \fref{fig:setup}a). To achieve computationally complex dynamics, quantum computers have thus resorted to heavily discretized dynamics bearing no resemblance to the time evolution of physical systems, as epitomized by random circuit sampling \cite{arute_2019,PhysRevLett.127.180501,decross2024computational} (upper right, \fref{fig:setup}a) or Floquet dynamics at very large time steps \cite{ibm_utility,rosenberg2024}. How small digitization errors must be to capture the essential \emph{qualitative} consequences of time being continuous is less clear \cite{doi:10.1126/sciadv.aau8342}. For example, consider the relaxation of an isolated quantum system. When time is continuous, energy is conserved, and local observables relax to non-trivial thermal values \cite{dallesio_ETH}. When time is digitized and energy conservation is lost, the system instead heats into a structureless, infinite-temperature state. However, recent studies \cite{Bukov2015, KUWAHARA201696,PhysRevB.95.014112,doi:10.1126/sciadv.aau8342,Weidinger2017,HO2023169297} show that while Floquet heating dominates at long times, time-discretized systems can exhibit \emph{Floquet prethermalization} at intermediate times, thermalizing with respect to an effective Hamiltonian resembling the targeted one. Here, we show that quantum computers can now access the \emph{qualitatively} important features of continuous-time dynamics in regimes of high classical simulation complexity (upper left, \fref{fig:setup}a).

Specifically, we use Quantinuum's H2 trapped-ion quantum computer~\cite{Moses2023,decross2024computational} to simulate the digitized quantum Ising model. Other recent works targeting quantum advantage via Ising model dynamics have worked in the limit of strong Floquet heating \cite{ibm_utility} (with all local observables converging quickly to infinite-temperature values) or employed analog quantum simulations with native (non-programmable) interactions \cite{king2024computational}. We use a gate-based quantum computer to simulate \emph{effectively} continuous-time TFIM dynamics. First, we show that we can discretize time finely enough to stabilize Floquet prethermal behavior. We assess various classical methods to simulate the dynamics accessible to H2, including tensor network \cite{ayral2023density,Tomislav:2024,LuCiBa14a}, neural network \cite{Schmitt2020}, and operator truncation methods \cite{beguvsic2024real,rudolph2023classical,bermejo2024quantum,angrisani2024classically, cirstoiu2024fourier}. H2 provides trustworthy data in a regime of scale and complexity where no known
classical methods are both efficient and trustworthy. 
The accurate simulation of effectively continuous time evolution allows us to probe distinct emergent dynamical phenomena. First, we probe characteristic signatures of thermalization in the form of \emph{emergent hydrodynamics}, and use H2 to compute the diffusion constant of the associated thermal transport. Second, by utilizing the flexibility of the digital quantum architecture, we simulate antiferromagnetic interactions on a frustrated triangular lattice with periodic boundary conditions and observe dynamics consistent with an \emph{emergent gauge-theoretic description} in terms of a quantum dimer model.

\section{Digitized Ising dynamics}

We investigate digitized non-equilibrium dynamics of the nearest-neighbor transverse-field Ising model (TFIM) on two-dimensional (2D) lattices, with Hamiltonian
\begin{align}
\label{eq:tfim}
H&=J\sum_{\langle j,k\rangle} Z_{j}Z_{k}+h\sum_{j}X_j
\equiv H_{ZZ}+H_X.
\end{align}
All simulations use periodic boundary conditions (PBC, \fref{fig:setup}b) to mitigate boundary effects. Furthermore, PBC facilitates the extraction of momentum-resolved energy relaxation in \sref{sec:hydro}, and modifies thermalization through induced topological constraints in \sref{sec:triangular}.

The TFIM describes a collection of spin-1/2 magnetic moments with Pauli operators $X_j,Y_j,Z_j$ for each site $j$. Their tendency to magnetically order (via nearest-neighbor spin exchange coupling, $J$) competes with the disordering effect of quantum fluctuations (driven by the transverse field, $h$). Despite the simplicity of the TFIM, dynamics induced by $H$ are generally chaotic in $D>1$ dimensions. Consequentially, while all of our simulations evolve pure states, the system can act as its own bath to produce \textit{effectively} thermal local observables \cite{dallesio_ETH}. Provided Trotterization is performed accurately enough that energy is approximately conserved, the initial state energy density controls the effective temperature of these thermal observables. In this setting, the initial state and Hamiltonian parameters can be tuned to reveal (in late-time thermal observables) the equilibrium phase diagram of the TFIM. For ferromagnetic ($J<0$) interactions, the system exhibits an ordered phase (breaking the $\Z_2$ symmetry of $H$) at small temperature and transverse field values, becoming disordered (by way of a continuous phase transition) at increasing temperature and/or transverse field.

We digitize the time evolution operator $U(t)=\exp(-i H t)$ with a second-order Trotter expansion,
\begin{align}
\label{eq:Trotter}
U(dt)\approx e^{-i dt H_X/2}e^{-i dt H_{ZZ}}e^{-i dt H_{X}/2}\equiv \mathscr{U}.
\end{align}
The Trotter step unitary $\mathscr{U}$ is implemented as a quantum circuit with two layers of single-qubit gates $U_{X}(h dt)=\exp[-i (h dt/2) X]$ enclosing a fixed number of layers of two-qubit gates $U_{ZZ}(2 J dt) = \exp[-i (J dt) Z\otimes Z]$, as in \fref{fig:setup}c.  The digitized dynamics that results from repeated applications of this quantum circuit can be viewed as continuous time evolution under a Floquet Hamiltonian $\mathscr{H}=i \log [\mathscr{U}]/dt$. While this description appears to immediately imply energy-conserving dynamics, it is important to note that $\mathscr{H}$ need not look anything like $H$. Indeed, it is generally expected that local observables will ultimately approach infinite-temperature values at late times, as energy is continuously pumped into the system by the drive (\fref{fig:setup}d, top). However, recent work has shown that this infinite-temperature long-time limit can be preceded by \textit{Floquet prethermalization} surviving up to the characteristic Floquet heating timescale $\tau_H\sim \exp(1/|J dt|)$ \cite{KUWAHARA201696,PhysRevB.95.014112,doi:10.1126/sciadv.aau8342}. Until this time, local observables equilibrate with respect to a dressed Hamiltonian that is very similar to $H$. Hence thermal physics of the TFIM can be accessed up to an exponentially long time scale $\tau_H$ (\fref{fig:setup}d, bottom). Any failure of the dynamics induced by repeated applications of $\mathscr{U}$ to reproduce the exact dynamics induced by $U$ is a direct consequence of the Trotter error in \eref{eq:Trotter}. The exponentially long window of prethermal behavior suggests that as the time-step becomes small, the stability of many qualitative features of the desired dynamics can be much more robust than known rigorous bounds on Trotter error suggest \cite{doi:10.1126/sciadv.aau8342}.

%Figure 2
%
\begin{figure*}[!!t]
\centering
\includegraphics[width=2.06\columnwidth]{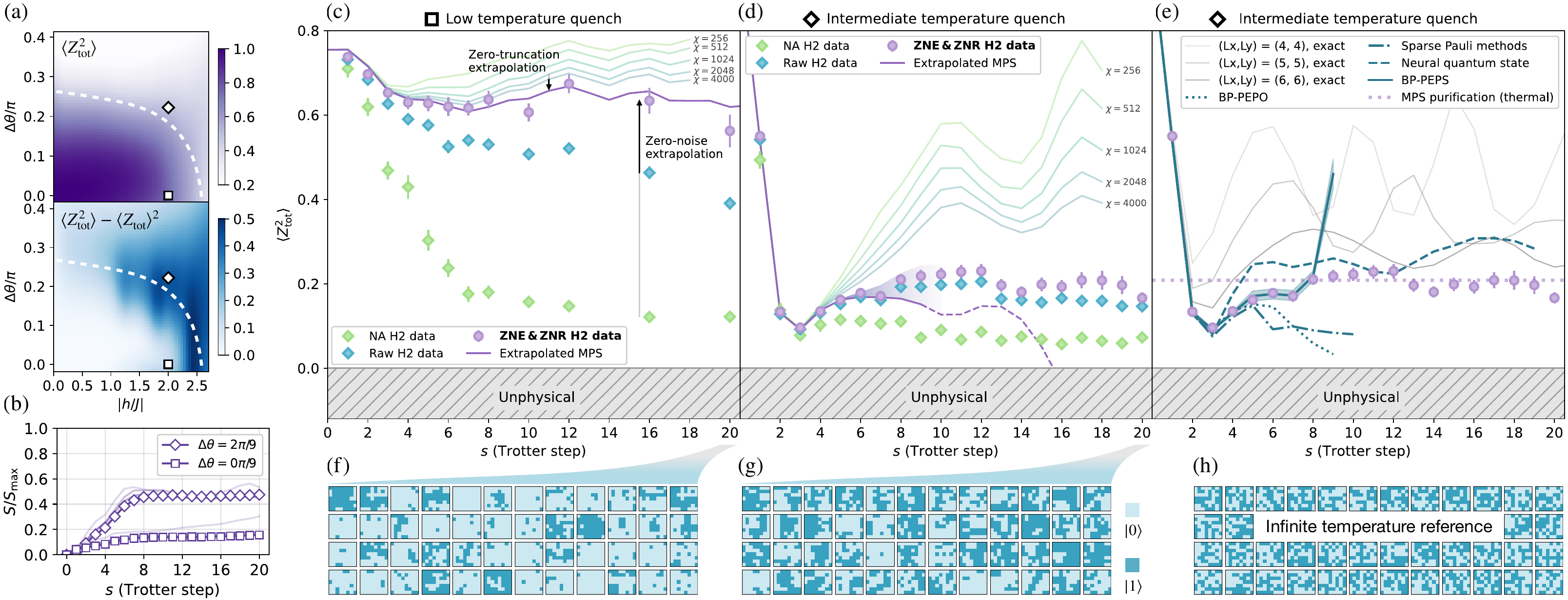}
\caption{{\bf Prethermalization dynamics with $56$ qubits.} (a) Small-scale numerical simulations reveal that a time step $dt |J| =0.25$ is small enough to observe Ising thermal physics at a temperature controlled by $\Delta \theta$. Both $\langle Z_{\rm tot}^2 \rangle$ (top) and $\langle Z_{\rm tot}^2 \rangle-\langle Z_{\rm tot}\rangle^2$ (bottom) are averaged over 30 Trotter steps. \ddiff{White dashed-lines are guides to
the eye denoting the ordered region.} (b) Entanglement entropy for the low-temperature and intermediate-temperature quenches. (c) Low-temperature quench, $(h/|J|,dt|J|,\Delta\theta)=(2,0.25,0)$: Accurate results can be obtained classically using MPS methods and their extrapolation to zero truncation error (ZTE).  We use this quench to benchmark the efficacy of our error mitigation methods (ZNE \& ZNR) on the quantum data. ZNR removes the effect of qubit leakage errors from the raw (blue) and noise amplified (NA, green) data, which is then extrapolated to the limit of zero two-qubit gate noise to produce the ZNE \& ZNR (purple) estimates for $\langle Z_{\rm tot}^2\rangle$. (d) Intermediate-temperature quench $(h/|J|,dt |J|,\Delta\theta)=(2,0.25,2\pi/9)$: ZTE of the MPS results is only controlled at early times. For $s\leq9$ (purple solid line) the $\chi=4000$ MPS fidelity is high enough that, while the extrapolation itself is not necessarily accurate, we can have reasonable confidence about where the exact observable lies (purple shaded region). For $s>9$ (purple dashed line) the $\chi=4000$ MPS fidelity is too low to even quantify how wrong the extrapolation might be. (e) Various classical simulation methods compared to the quantum data. (f,g) Samples of raw data taken at $s=20$ demonstrate the late-time stability of (at least short-range) ferromagnetic order, while (h) shows (randomly generated) infinite temperature samples for visual reference. All error bars are standard errors of the mean obtained by bootstrap resampling 100 resamples of the data (see Methods).
\label{fig:main_results}}
\end{figure*}

\section{Quantum thermalization}

We begin by simulating ferromagnetic ($J<0$) dynamics on a $L_x\times L_y=7\times 8$ square lattice. For all simulations, the system is initiated in a product state of the form
\begin{align}
\label{eq:initial_state}
\ket{\Psi}=\bigotimes_{j}\Big(\cos(\theta_j/2)\ket{0}_j+\sin(\theta_j/2)\ket{1}_j\Big).
\end{align}
In this section we restrict our attention to quenches from a uniform state, denoted $\ket{\Psi(\theta)}$, and Trotterized time evolution is carried out with a time step $dt = 0.25/|J|$. We evolve the system for $s$ Trotter steps and then make $Z$-basis \ddiff{($\ket{0},\,\ket{1}$)} measurements, giving us access to correlation functions of the Ising order parameter \footnote{For the intermediate-temperature quench discussed in this work, the Ising order parameter itself is small ($\langle Z_{\rm tot}\rangle^2\sim 0$) except at very early times, and $\langle Z_{\rm tot}^2\rangle$ serves as a direct measure of order parameter fluctuations.}
\begin{align}
\label{eq:corr}
\langle  Z_{\rm tot}^2(s)\rangle =\frac{1}{N^2}\sum_{j,k}\langle \Psi(\theta)|(\mathscr{U}^{\dagger})^s Z_{j}Z_{k}(\mathscr{U})^s|\Psi(\theta)\rangle.
\end{align}

To situate the quench dynamics explored here in the context of the TFIM thermal phase diagram, it is helpful to define the mean-field ground state $\ket{\Psi(\theta_{\rm min})}$ with $\theta_{\rm min}=\sin^{-1}[h/(zJ)]$. The initial state angle measured relative to the mean-field ground state, $\Delta\theta = \theta-\theta_{\rm min}$, sets the initial energy density of the state, and plays a role analogous to temperature in determining late-time (equilibrated) expectation values. At zero temperature, the TFIM on a 2D square lattice is ordered for $|h|<h_c\approx 3|J|$, with the order persisting to finite temperatures. We perform all quenches in or near the boundary of the ferromagnetic phase by choosing $J<0$, $h=2|J|$, for which the mean-field ground state (at $\theta_{\rm min}=-\pi/6$) lies in the ordered phase. Small-scale ($L_x\times L_y = 4\times 4$)(\fref{fig:main_results}a) indicate that time-averaged correlation functions reveal the expected thermal phase diagram of the TFIM, and that we can cross the finite-temperature phase transition of the TFIM by increasing $\Delta\theta$. Figure \ref{fig:main_results}b shows that as $\Delta\theta$ increases, the late time entanglement entropy saturates to progressively larger values as the system moves from the (low-temperature) ordered phase into the (high-temperature) disordered phase. For sufficiently small amounts of saturated entanglement entropy we expect matrix-product-state (MPS) based methods to work well; conversely, increasing $\Delta\theta$ enables us to controllably tune the quantum quench into a regime for which classical simulations are potentially challenging.  We use this flexibility to identify both a \textit{low temperature quench} ($\Delta\theta =0$) and an \textit{intermediate temperature quench} ($\Delta\theta = 2\pi/9$) close to the thermal phase transition. The former can be simulated with high accuracy via MPS methods, while the latter appears to require an extremely large amount of classical resources to simulate accurately (see supplemental material, Sec. V).

Quantum and classical simulation results for the low temperature quench, using up to $s=20$ Trotter steps on a $7\times 8$ lattice, are shown in \fref{fig:main_results}c.  Semi-transparent solid lines are obtained using MPS-based quantum circuit simulations \cite{fermioniq} employing a method similar to that described in Ref.\,\cite{ayral2023density}. We performed these simulations up to a maximum MPS bond dimension of $\chi=4000$, for which MPS-based truncation errors were small enough that we could reliably extrapolate to the limit of unity state fidelity (purple line, zero-truncation extrapolation or ZTE, see supplemental material Sec.\,V). Raw (completely unprocessed) data from the H2 quantum computer are shown as blue symbols. Imperfect two qubit gates, which have an average fidelity of $99.94(1)\%$ (see supplemental material, Sec. IIB) are the dominant source of error in this data. We also show data from noise amplified (NA) circuits, in which two-qubit Pauli operators are stochastically inserted (based on a learned noise channel of our two-qubit gates) in order to artificially reduce the quality of the quantum computation \cite {ibm_utility}. From the raw and NA data, noiseless quantum estimates are extracted via zero-noise extrapolation (ZNE) \cite{PhysRevLett.119.180509}. Further details of the ZNE methods (including the calculation of bootstrapped error bars for ZNE data), along with a novel strategy to remove the impact of detectable leakage errors without post-selection (zero-noise regression, or ZNR), can be found in the methods and supplemental material. The good agreement of the processed data with the extrapolated MPS results gives us confidence in the error-mitigation strategies. While somewhat longer times (deeper circuits) are likely accessible at current error rates, eventually bias introduced by the error mitigation will impact the results.

Figure \ref{fig:main_results}d shows results of the intermediate temperature quench, for which ZTE for classical MPS is ineffective (except at very short times) due to much larger late-time entanglement entropies. The shaded purple region above the MPS-extrapolation curve indicates our best estimate of the likely region in which the exact answer sits after the point where the extrapolation becomes inaccurate, but before the MPS fidelity becomes so low that extrapolation is completely uncontrolled (as estimated from an analysis of extrapolation errors in simulations of smaller systems, see Sec.\,\ref{methods:ZTE}). After about $s=9$ (dashed purple line) the extrapolation should be viewed as completely unreliable, and indeed at $s=16$ it becomes negative (which is unphysical, since $Z_{\rm tot}^2$ is a positive operator). Estimates from simulations of system sizes up to $36$ qubits suggest (see Sec.\,\ref{methods:ZTE}) that on the $7\times8$ lattice, extrapolating $\langle Z_{\rm tot}^2\rangle$ to roughly $5\%$ relative accuracy for all steps $s\leq 20$ may be possible with extremely large bond dimensions \cite{PRXQuantum.4.010317}. However, even a modest increase in system size would frustrate this approach to simulation. The quantum and classical calculations are complementary in that the ZTE extrapolates by much less than (and therefore provides more accurate results than) the ZNE at small $\Delta \theta$ (small temperature), while the opposite is true at intermediate $\Delta \theta$ (intermediate temperature). For the deepest circuits, which have over $2000$ two-qubit gates, we estimate that only about 10\% of the data has no errors in it. We attribute the quality of the raw data in general, and in particular the improvements at higher temperatures, to recent arguments regarding the dilution of stochastic errors in thermalizing systems Refs.\,\cite{PRXQuantum.4.030320,PRXQuantum.6.010333,chertkov2024robustness}.

Figure \ref{fig:main_results}e shows the same error-mitigated quantum data for the intermediate temperature quench compared to exact diagonalization results for smaller systems (gray curves), along with further approximate classical methods for the $7\times 8$ lattice, providing a comprehensive comparison of the quantum data with a variety of classical approaches at the frontier of numerical simulations. We explored two different classes of methods: Schr\"odinger picture methods that evolve a compressed representation of the wave function, and Heisenberg picture methods that evolve a compressed representation of the observable $Z_{\rm tot}^2$. For the former, we attempted to evolve the wave function using a projected entangled pair state (PEPS) ansatz and a variety of compression schemes. We achieved the best accuracy by evolving a PEPS using belief propagation (BP-PEPS) and then computing $\langle Z_{\rm tot}^2\rangle$ by sampling, though we were still not able to obtain converged results beyond about 8 Trotter steps (mainly due to GPU memory limitations).  Simulations in the Heisenberg picture were carried out using both projected entangled pair operator (PEPO) and sparse-Pauli methods \cite{Tomislav:2024,beguvsic2024real,rudolph2023classical,bermejo2024quantum,angrisani2024classically, cirstoiu2024fourier}. In general, we found PEPO evolution to be less accurate than PEPS simulations with comparable resources.  Sparse Pauli dynamics simulations retained about 4 billion Pauli operators and used roughly $1{\rm TB}$ of memory, and while they are in principle systematically improvable with more memory they could not be converged with available resources. We also performed neural network quantum state (NQS) simulations using a convolutional neural network (CNN) architecture similar to Ref.~\cite{Schmitt2020}, evolved with time-dependent variational Monte Carlo (tVMC)~\cite{carleo2012localization} using the NetKet library \cite{netket3:2022}. By systematically increasing the number of parameters of the ansatz, we obtained good results for smaller systems, but we were not able to obtain quantitatively accurate results except at very short times for a $7 \times 8$ lattice. \ddiff{Further details on the PEPS, PEPO, SPD, and NQS simulations described in this section can be found in the supplemental material.}

 The dominant noise sources in our quantum simulations are unital, and would (at much longer times than those explored here) drive the system to an infinite-temperature state with $\langle Z_{\rm tot}^2\rangle= 1/N$. The quantum data is consistent with weak persistent oscillations of $\langle Z_{\rm tot}^2\rangle$ around a much larger prethermal value. \ddiff{Figure \ref{fig:main_results}(e)} shows the expected thermal value of $\langle Z_{\rm tot}^2\rangle$ for the Floquet Hamiltonian (to order $dt^2$) computed in the canonical ensemble using an MPS purification ansatz \cite{Verstraete2004_mpdo, Barthel2009} (see supplemental material Sec.\,IV). We note that exact agreement between (even time-averaged) quantum data and the thermal result is not necessarily expected, as both finite-size differences between the canonical and diagonal ensemble (filtered by the time-evolved state), as well as $O(dt^3)$ corrections to the Floquet Hamiltonian, are not guaranteed to be small. Figures \fref{fig:main_results}(f,g) show snapshots of the raw quantum data at the latest times explored (20 Trotter steps) and show clearly the formation of correlated spin domains expected for a state thermalizing near the phase transition of the underlying TFIM [cf.\,the simulated infinite-temperature snapshots in \fref{fig:main_results}(h)].

\

\section{Emergent hydrodynamics\label{sec:hydro}}

%Figure 3
%
\begin{figure}[!t]
\centering
\includegraphics[width=1\columnwidth]{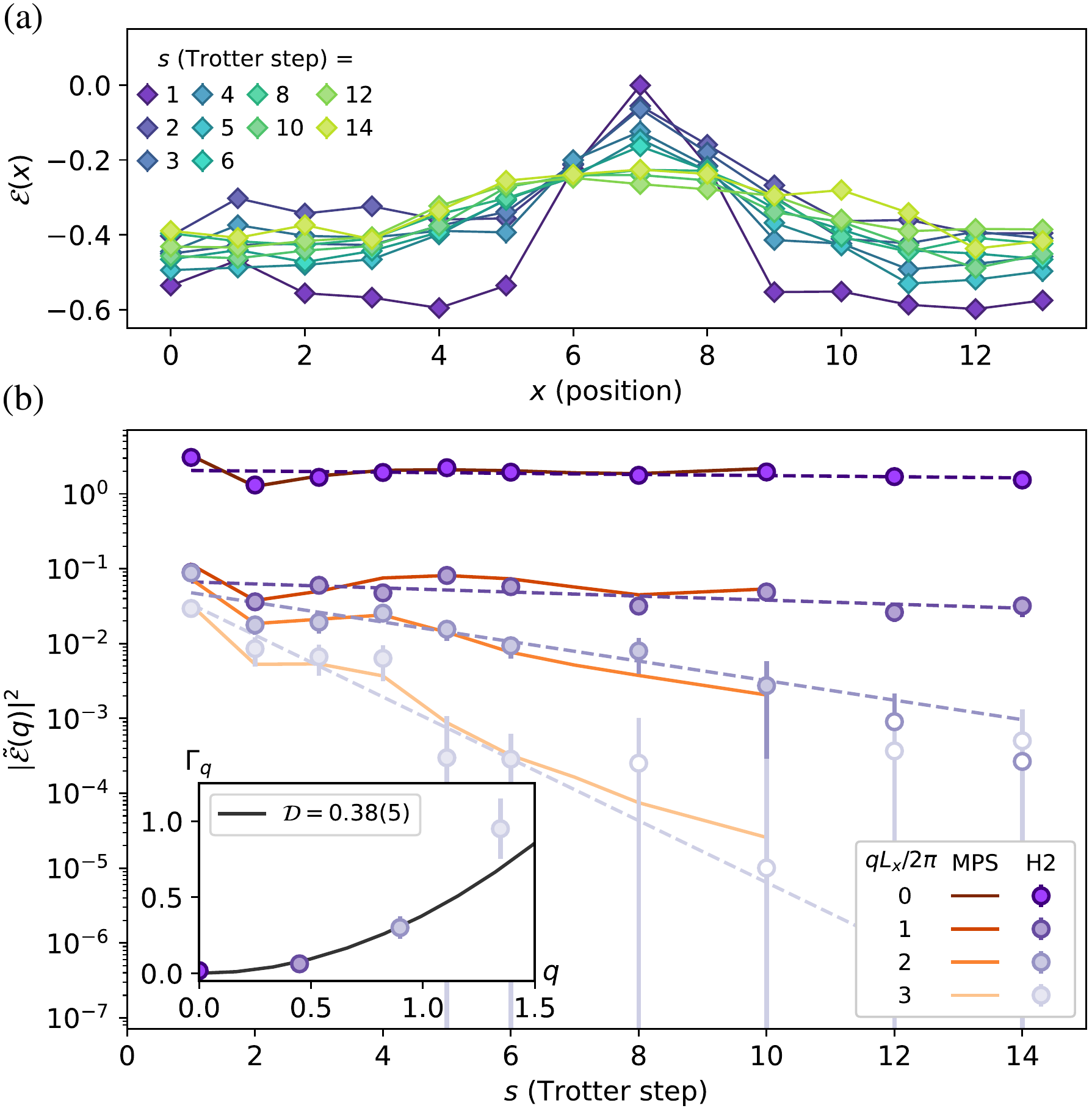}
\caption{{\bf Emergent hydrodynamics.} Hydrodynamic relaxation of the spin-exchange contribution to energy  density
for a quench in an $14\times 4$ strip with an inhomogeneous initial state (described in text). (a) Evolution of the y-averaged exchange energy density profiles, $\E(x)$. (b) $\log|\tildeE(q)|^2$ for $q = 2\pi n/L_x$, with $n=0,1,2,3$, both from H2 (circles) and MPS simulations (up to to $s=10$, solid lines). Dashed lines show a fit to $\log(a e^{-\Gamma_q s}+c\delta_{q,0})$, excluding data points for which the statistical error exceeds the mean (unfilled symbols).  Extracted decay rates are consistent with the expected diffusive scaling of the heat equation, $\Gamma_q \approx \mathcal{D} q^2$ (inset) with diffusion constant $\mathcal{D}=0.38(5)$. All error bars are standard errors of the mean obtained by bootstrap resampling 100 resamples of the data (see \sref{methods:EM}).}
\label{fig:hydro}
\end{figure}
%
%

%Figure 4
%
\begin{figure*}[!t]
\centering
\includegraphics[width=2.06\columnwidth]{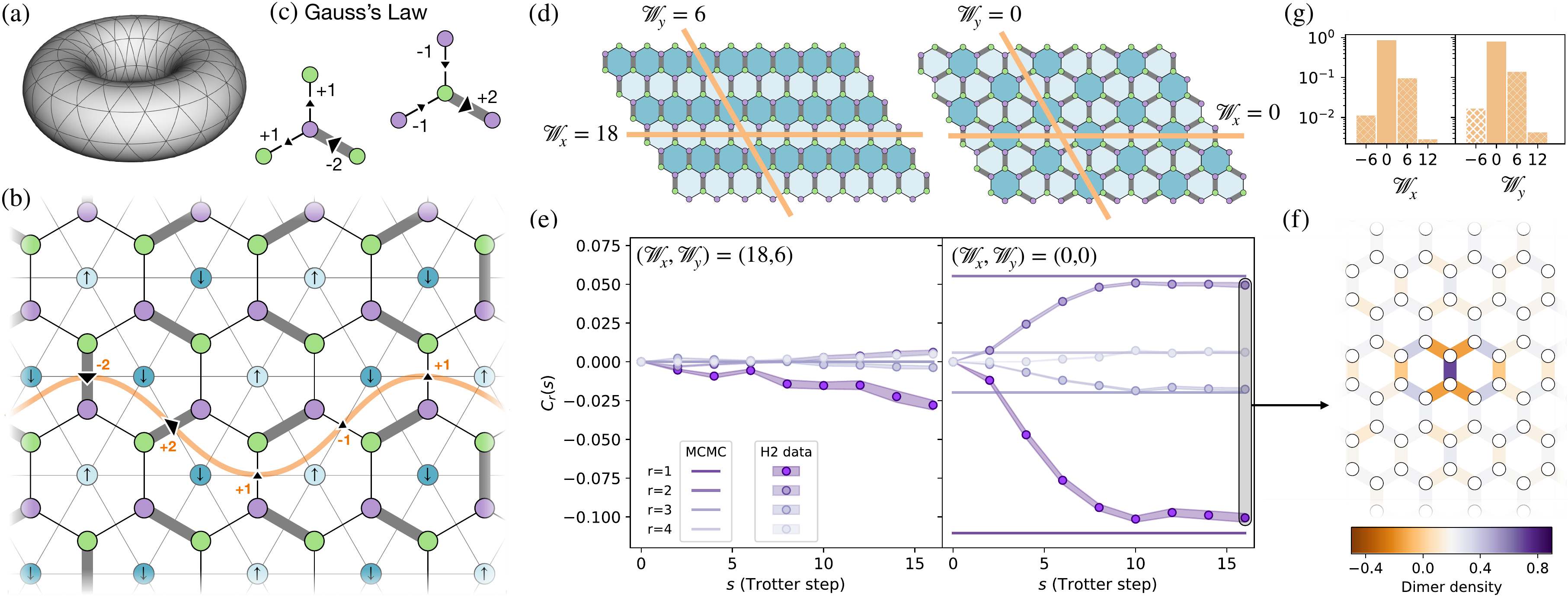}
\caption{{\bf Emergent gauge theory on a frustrated lattice.} (a) Triangular lattice on a torus. (b) The lowest-energy classical configurations can be identified with dimer coverings of a dual hexagonal lattice (an edge having a dimer when AFM order of adjacent spins is frustrated, as indicated by the thick gray link). In the low-energy subspace only one dimer can touch any vertex; by defining two sub-lattices (green/purple) and assigning electric field values of $\pm1(2)$ to the absence (presence) of a dimer as in (c), the low-energy subspace obeys a discrete Gauss's law (zero flux at each vertex).  On a torus, the total flux through a topologically non-trivial loop (orange line) is conserved to low-order in perturbation theory. (d) Two different initial states in the maximal [$(\mathscr{W}_x,\mathscr{W}_y)=(18,6)$] and minimal [$(\mathscr{W}_x,\mathscr{W}_y)=(0,0)$] winding number sectors. (e) All data taken at $(h/|J|,dt|J|) = (0.3,0.25)$. The maximal winding number sector has no dynamics under the effective Hamiltonian $H_{\rm QDM}$, Eq.~\eqref{eq:dimer_ham}. To the contrary, a state in the minimal winding number sector quickly relaxes, as evidenced by the dimer-dimer correlations $C_{r}(s)$ approaching their ergodic values obtained by Markov-chain Monte Carlo (MCMC). The full spatial structure of the dimer-dimer correlations (f), as well as the full counting statistics of the nominally conserved topological invariants (g), are both shown for the mobile state at the latest time $s=16$. All error bars are standard errors of the mean obtained by bootstrap resampling 200 resamples of the data (see \sref{methods:EM}).
\label{fig:triangular}}
\end{figure*}

An essential aspect of prethermalization in Floquet systems is the emergence of an approximately conserved energy, whose violation only becomes apparent at exponentially late times, $t\gtrsim \tau_H$.
In chaotic (pre)thermalizing systems, local observables that overlap with the conserved energy density are expected to exhibit a short-time transient relaxation to a local thermal quasi-equilibrium followed by a slow, hydrodynamic relaxation of long-wavelength energy inhomogeneities at late times. 
To probe the expected emergence of hydrodynamics, we simulate a quench from an inhomogeneous state in which a thin vertical domain of down ($\ket{1}$) spins is embedded in an otherwise uniform background, corresponding to Eq.~\eqref{eq:initial_state} with site dependent $\theta_j =\pi$ if $x_j=L_x/2$ or otherwise $\theta_j=\theta_{\rm min}+2\pi/9$ ($x_j$ being the $x$-coordinate of qubit $j$). 
To create an appropriate separation of time-scales between the local thermalization time and late-time hydrodynamics, we consider an elongated rectangular geometry with dimensions $(L_x ,L_y) = (14, 4)$.
We measure the local density of $H_{ZZ}$: $\E_i = \frac{J}{4}\sum_{j\in \langle i\rangle} \langle Z_iZ_j\rangle$, average this quantity over the $y$ coordinate, $\E(x) = \frac{1}{L_y} \sum_{i:x_i=x}\E_i$ (\fref{fig:hydro}a), and compute its 1D Fourier transform $\tildeE(q) =  \frac{1}{\sqrt{L_x}}\sum_x e^{iqx} \E(x)$ (\fref{fig:hydro}b). 
The $q=0$ mode quickly saturates to a steady-state value, indicating its overlap with an emergent conserved energy.
The smallest few non-zero ($q_n = 2\pi n/L_x$, $n=1,2,3$) wavevectors relax with characteristic rates $\Gamma_q$, consistent with the expected behavior for the heat equation, $\Gamma_q \approx \mathcal{D}q^2$ (valid asymptotically for long wavelengths, $|{q}|\ll 1$), with diffusion constant $\mathcal{D}=0.38(5)$. Components with larger $q$ decay rapidly within the initial local thermalization and do not survive into the hydrodynamic regime.

These results provide direct evidence for the emergence of (approximately) conserved energy current within the Floquet prethermalization regime. The computation of hydrodynamic transport coefficients such as $\mathcal{D}$ is an important task for materials simulation, which is expected to be difficult for classical methods (though we do not claim that the particular hydrodynamic transport simulation described in this section is beyond the limits of classical simulation). Existing quantum simulations of hydrodynamics in quantum magnets~\cite{Zu_2021_hydro, Joshi_2022_hydro, Wienand_2024_hydro} have been conducted exclusively in analog quantum simulators, and observing this physics in programmable digital quantum computers has previously been challenging due to limited accessible circuit depths or system sizes.

\section{Emergent gauge theory on a frustrated lattice \label{sec:triangular}}

The previous sections focused on a ferromagnetic Ising model on a square lattice. However, the all-to-all effective connectivity of H2 enables flexible exploration of essentially arbitrary lattice geometries and spin-exchange couplings. To demonstrate this capability, we next implement a triangular lattice TFIM with antiferromagnetic (AFM)  $J>0$ couplings, which exhibits geometric frustration. As we will show below, the approximate conservation of energy---achieved by using sufficiently small Trotter step size---manifests in this context as emergent gauge and topological constraints that substantially alter the thermalization dynamics.

When placing classical spins on the triangular lattice, each triangular plaquette must have at least one edge with a ferromagnetic (FM) spin configuration. The classical ($h=0$) ground-space is extensively degenerate, consisting of states with exactly one (FM) bond per triangle. These configurations are equivalently represented by dimer coverings on the dual hexagonal lattice \cite{PhysRev.79.357}, with a dimer assigned to each link of the dual lattice separating ferromagnetically aligned spins in the original lattice (Fig.~\ref{fig:triangular}b). For weak transverse field ($0<h\ll J$), the classical ground-state degeneracy is lifted by quantum fluctuations of dimer patterns, described to first order by an effective quantum dimer model (QDM):
\begin{align} 
\label{eq:dimer_ham}
\diagramt{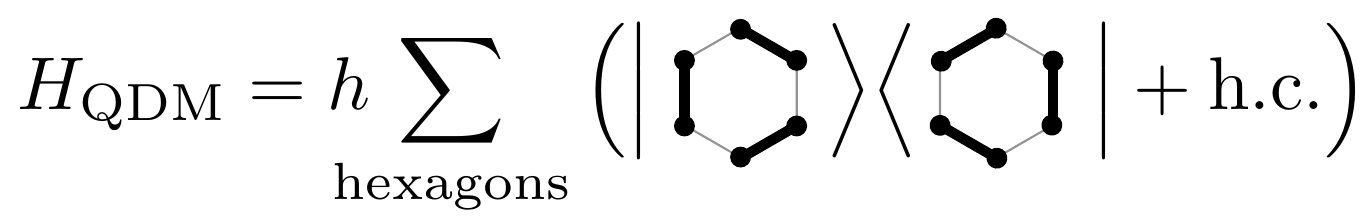}.
\end{align}
This and related QDMs have been extensively studied in the context of spin-liquids with emergent electromagnetic gauge fields \cite{PhysRevLett.84.4457,PhysRevB.63.224401,savary2016quantum}, and resonating valence bond (RVB) theory of high-temperature superconductivity~\cite{anderson1987resonating}. Computational basis states within the low-energy dimer subspace satisfy a local Gauss-law constraint [\fref{fig:triangular}(c)] that is conserved by dynamics under $H_{\rm QDM}$, but violated in evolution under $H_{\rm TFIM}$ by processes that are higher-order in $h/J$. 
Consequently, with periodic boundary conditions, the QDM dynamics also conserves topological winding numbers, $\mathscr{W}_{x}$  and $\mathscr{W}_{y}$, respectively defined by the total electric flux through the two non-contractible loops wrapping around the torus in the $x$ and $y$ directions [see Fig.~\ref{fig:triangular}(b) for an illustration of one of the non-contractible loops].

We initialize the dynamics with configurations in both the maximal $(\mathscr{W}_{x},\mathscr{W}_y)=(18,6)$ and minimal $(\mathscr{W}_{x},\mathscr{W}_y)=(0,0)$ winding number sectors of a $9\times6$ lattice, respectively [Figure \ref{fig:triangular}(d)]. Subsequently, we time evolve these two initial states using H$2$ for up to $16$ Trotter steps, taking measurements in the computational basis.  The maximal winding-number state is (ideally) completely frozen under dynamics induced by $H_{\rm QDM}$, while the minimal winding number state undergoes a quantum walk within the $(\mathscr{W}_{x},\mathscr{W}_y)=(0,0)$ sector and quickly relaxes.  At late times this state becomes ergodic while fulfilling both the Gauss-law and winding-number constraints, which impose non-trivial late time correlations between the dimers.  Figure \ref{fig:triangular}(e) shows the temporal relaxation of connected dimer-dimer correlation functions 
\begin{align}
C_{r}(s) = \sum_{|a-b|=r}\langle n_a(s)n_b(s)\rangle-\langle n_a(s)\rangle \langle n_b(s)\rangle.
\end{align}
Here $n_a$ is the density operator for dimers on edge $a$ of the dual lattice, and the distance $r=|a-b|$ is defined as the number of vertices traversed along the shortest path connecting (and including) those edges. While the maximal-winding number sector (left) is indeed effectively frozen, the minimal winding sector state (right) exhibits correlations that quickly relax to the ergodic values obtained from classical Markov-chain Monte Carlo (MCMC) calculations.  The late time ($s=16$) spatially-resolved correlations in the minimal winding sector are structured due to the gauge constraints [\fref{fig:triangular}(f)], and strong persistence in the initialized topological sector is verified in the full-counting statistics of winding numbers [\fref{fig:triangular}(g), histogram shown on logarithmic scale].
We attribute the small departures from zero correlations in the minimal winding number sector, and from the MCMC values at late times in the maximal winding number sector, to a combination of Trotter errors and non-infinitesimal transverse field resulting in higher order in $h/J$ corrections not captured by \eref{eq:dimer_ham}.

\

\section{Outlook}

Early demonstrations of quantum advantage with noisy quantum computers have relied on carefully orchestrated quantum circuits in which the contribution of every gate to the classical simulation difficulty is optimized, ultimately with the goal of ensuring that classical difficulty sets in before the impact of gate noise is too large.  As digital quantum computers continue to grow in scale and improve in fidelity, so will the prospects for obtaining quantum advantage in more useful algorithms for which numerous constraints---such as the need to suppress digitization errors in Hamiltonian simulation---inevitably bias the competition towards classical methods.  In this work, we have provided the first evidence for the emergence of stable equilibrium behavior from digitized quantum dynamics at a scale for which numerically exact classical simulations appear to be difficult.

\ddiff{Moving forward, it is likely that improvements in both scale and gate fidelities will be required to accurately simulate more general and realistic models, e.g., with more complex spin-spin interactions or involving fermionic degrees of freedom (considerable progress in this direction is already being made \cite{granet2025,alam2025fermionic,alam2025programmable}). In the meantime, data available from current quantum computers may prove useful for benchmarking emerging classical simulation heuristics.} While we cannot predict the future capabilities of such heuristics, a careful analysis of the limitations of numerous approximate methods indicates that the quantum data provided by the H2 quantum computer is arguably the most convincing standard to which they should be compared.

\section{Methods}

\subsection{Quantum hardware and data acquisition}

In this work, we perform four sets of quench experiments on Quantinuum's H2-1 and H2-2 trapped-ion 56-qubit quantum computers: a low-temperature quench on a $7 \times 8$ square lattice, an intermediate-temperature quench on a $7 \times 8$ square lattice, a hydrodynamics quench on a $14 \times 4$ square lattice, and a $9 \times 6$ triangular-lattice quench. Immediately before and after each quench, we learn an error model for our two-qubit gates by running cycle benchmarking circuits (see supplemental material Sec.\,IIA) on the quantum computers used in the quench. For each quench, we randomly interleave benchmarking circuits with the quench circuits, which allows us to estimate the average two-qubit gate infidelity at each Trotter step, which can drift over time [see supplement Fig.~7(c)]. The low-temperature quench experiments are run on H2-1 and H2-2; the intermediate-temperature quench are run only on H2-2; the hydrodynamics quench are run only on H2-1; and the triangular-lattice quench are run only on H2-2 (see supplemental material Sec.\,III for more details).

\subsection{\label{methods:EM}Error mitigation and data processing}

In our quench experiment circuits, we use multiple circuit-level strategies to suppress the effect of hardware errors. To suppress coherent memory errors that build up on qubits as they idle in the device, we use dynamical decoupling -- where we apply $X$ pulses on our qubits periodically to cause coherent cancellation of errors (see supplemental material, Sec.\,IA). To convert coherent (memory and other) errors into incoherent errors, we use randomized-compiling (or Pauli twirling) -- where we randomly conjugate every two Trotter layers of our circuit with random single qubit Paulis (see supplemental material Sec.\,IB). To suppress leakage errors, we utilize a circuit-level leakage detection gadget available on our trapped-ion hardware (see supplemental material, Sec.\,IC). To suppress two-qubit gate errors, we use zero-noise extrapolation -- a standard error mitigation technique that allows us to extrapolate to the zero-noise limit by intentionally amplifying errors, which we do by randomly inserting two-qubit Pauli operators before our two-qubit gates according to the error model learned from cycle benchmarking (see supplemental material, Sec.\,ID). All of these circuit-level protocols are combined together in the circuits we execute on hardware, with additional care taken to prevent the accumulation of coherent single-qubit gate errors [see supplement Sec.~IE]. To implement the Pauli insertion for zero-noise extrapolation, each circuit is run in two modes: with no errors inserted or with errors inserted to amplify the two-qubit error rate to a particular value; circuits of these two modes are randomly interleaved.

Additionally, we use two post-processing techniques to mitigate leakage and two-qubit gate errors. Rather than post-selecting on data with no detected leakage errors, we instead use a technique that we call ``zero-noise regression'' to estimate the zero-noise limit of the detected errors, similar in spirit to zero-noise extrapolation though in the context of detectable errors (see supplemental material, Sec.\,IIC). This technique produces error bars on the zero-noise estimate that are improved relative to standard post-selection (at the cost of a potential heuristic bias). After zero-noise regression on the leakage errors, zero-noise extrapolation is then applied to the two-qubit gate errors. We perform this extrapolation using only two points and use an exponential fit (see supplemental material, Sec.\,IIB). The results of the benchmarking circuits that are interleaved with the quench circuits are used to estimate the error amplification rate, as that can change during the experiment as the gate infidelity drifts (see supplemental material Fig.\,7). Error bars reported in this work are standard errors of the mean obtained by bootstrap resampling, either applied to raw observables or observables obtained after post-processing (see supplemental material, Sec.\,III).

\subsection{\label{methods:ZTE}Zero truncation extrapolation of MPS data}

Accurate numerical results for the low-temperature quench are obtained by extrapolating imperfect MPS simulations to the limit of zero infidelity.  Extensive details of the numerical simulations justifying the efficacy and limitations of ZTE are provided in Sec.\,V of the supplemental material, but here we give a brief overview of the methodology and our conclusions.

ZTE attempts to estimate the value of an observable $O$ by performing MPS simulations for various bond dimensions $\chi$, and tabulating $\langle O\rangle(\chi)$ for all bond dimensions. For each bond-dimension, an estimate of the associated simulation fidelity $F(\chi)$ can be made by multiplying together the individual fidelities associated with every compression step in the simulation.  These two sets of results enable us to make an implicit estimate of $\langle O\rangle(F)$, which we can then attempt to extrapolate (over the available dataset) to the limit $F\rightarrow1$. We call this procedure zero-truncation extrapolation (ZTE). Absent an a-priori correct functional form for $\langle O \rangle (F)$, we perform a linear extrapolation using the three largest bond dimensions available. By analyzing the efficacy of this procedure for a various evolution times and systems sizes (all chosen small enough to determine the inaccuracy of the procedure by comparing ZTE observables to their exact values), we have verified that high global state fidelities (in excess of $\sim 80\%$) lead to reliable extrapolations. For the low-temperature quench, adequate simulation fidelities ($\gtrsim 80\%$) can be maintained for all Trotter steps $s\leq 20$ using a maximum bond dimension of $\chi = 4000$, enabling us to report $\langle Z_{\rm tot}^2\rangle$ to a level of accuracy that is considerably smaller than the statistical error bars of our data.

For the intermediate quench data, except at very short times we were not able to achieve high enough state fidelities from MPS simulations with $\chi\leq4000$ to report accurate observables. Numerics from simulations on system sizes up to $6\times 6$ can be used to estimate what bond-dimension much be reached (as a function of system size) to enable ZTE to produce results with a given level of accuracy. This estimation is described in detail in Sec.\,VB2 of the supplemental material, and the results are reported in Fig.\,16 therein. It appears that achieving $5\%$ relative accuracy on $\langle Z_{\rm tot}^2\rangle$ for the intermediate quench (up to all times $s\leq 20$) may be achievable with MPS bond dimensions $\sim2^{16}$, comparable to the largest ever utilized in classical simulations.

\section*{Data availability}
All experimental data presented in this manuscript are available online at \url{https://doi.org/10.5281/zenodo.18487089}.

\section*{Acknowledgments}
We acknowledge the entire Quantinuum team for their many contributions towards successful operation of the H2 quantum computer, and Honeywell for fabricating the trap used in this experiment. We thank Konstantinos Meichanetzidis, Chris Langer, Grahame Vittorini, Stephen Erickson, James Hostetter, and David Liefer for helpful comments on the manuscript and thank Dan Mills for useful discussions. S.A.M. contributed to the construction of the experimental apparatus while at Quantinuum. Fermioniq was supported by the Dutch National Growth Fund (NGF), as part of the Quantum Delta NL Programme. W.K., M.K., and F.P. acknowledge support from the Deutsche Forschungsgemeinschaft (DFG, German Research Foundation) under Germany’s Excellence Strategy–EXC–2111–390814868, TRR 360 – 492547816 and DFG grants No. KN1254/1-2, KN1254/2-1, FOR 5522
(project-id 499180199), the European Research Council (ERC) under the European Union’s Horizon 2020 research and innovation programme (grant agreement No 851161), the European Union (grant agreement No 101169765), as well as the Munich Quantum Valley, which is supported by the Bavarian state government with funds from the Hightech Agenda Bayern Plus. E.G. acknowledges funding by the Bavarian Ministry of Economic Affairs, Regional Development and Energy (StMWi) under project Bench-QC
(DIK0425/01). M.S.R. acknowledges funding from the 2024 Google PhD Fellowship, the NCCR
MARVEL, a National Centre of Competence in Research, funded by the Swiss National Science Foundation (grant
number 205602), and the the Swiss National Science Foundation (grant number 200021-219329). SPD simulation work at Caltech was supported by the US Department of Energy, Office of Science, 
 National Quantum Information Science Research Centers, Quantum Systems
Accelerator. TN-BP algorithm development at Caltech was supported by the US Department of Energy, Office of Science, Basic Energy Sciences and Advanced Scientific Computing Research, Quantum Utility through Advanced Computational Quantum Algorithms.

\section*{Author contributions}
M.M, J.P.B., M.B., J.J.B., J.C., J.M.D., C.F., D.F., J.P.G., C.N.G., D.G., A. Hall, A. Hankin, A. Hansen, N.H., C.A.H., R.B.H., N.K., E.L., D.L., I.S.M., A.R.M., S.A.M., B.N., A.R.P., M.S., P.E.S., B.G.T., M.D.U., J.W., and D.H. contributed to designing, building, optimizing, or characterizing, the H2 quantum computers used in this work. R.H., W.K., S.-H.L., Y.H.C, C.C, I.N., T.B., M.S.R., C.C., K.H., C.M., M.L., J.G., M.I., G.P., B.P., G.K.-L.C., F.P., M.K., H.D., and M.F.-F. performed and/or analyzed the numerical simulations of quantum dynamics. E.C., M.M., Y.H.C., E.G., C.H.B, M.D., J.P.G., K.M., D.T.S. A.C.P., D.H., H.D., and M.F.-F. studied the impact and mitigation of errors. A.C.P., D.H., G.K.-L.C, F.P., M.K., H.D., and M.F.-F. supervised the theoretical aspects of this work. All authors participated in writing, reviewing, and revising the manuscript.

\clearpage

\section*{Supplemental Material}

\setcounter{figure}{0}
\setcounter{equation}{0}
\setcounter{section}{0}
\renewcommand{\thefigure}{S\arabic{figure}}
\renewcommand{\theequation}{S\arabic{equation}}
\renewcommand{\thesection}{S\arabic{section}}

% \tableofcontents

\section{Implementation of time evolution on H2} \label{sec:implementation}

In this work, we perform quantum simulations of the second-order Trotter circuit performed by iterating the unitary $\mathscr{U}$ in Eq.~(2) of the main text, and study the dynamics of local observables. The direct implementation of the Trotter circuit, which amounts to repeated applications of four layers of $e^{-i\theta Z\otimes Z/2}$ gates surrounded by $e^{-i\phi X/2}$ gates, would have its measured observables degraded primarily by coherent memory errors, coherent and incoherent gate errors, and leakage errors present in the H2 device. Instead, we implement a modified version of this circuit that incorporates multiple error suppression and error mitigation protocols---dynamical decoupling, randomized compiling, leakage detection, and zero noise extrapolation (ZNE)---to mitigate these error sources and obtain high-accuracy results. In this section, we describe how we physically implement each individual error mitigation protocol and how we combine them together into the final circuits executed on H2. How the data are post-processed and used to produce improved estimates of noisy observables is discussed in Sec.~\ref{sec:postprocess}.

\subsection{Dynamical decoupling} \label{sec:dd_implementation}

In H2, idling qubits experience quantum errors that are collectively referred to as memory errors. For a fixed idling time, these errors are mostly a combination of coherent dephasing (apply $e^{-i\phi_j Z_j}$ to qubit $j$), incoherent dephasing (randomly apply $Z_j$ to qubit $j$), and leakage errors (qubit $j$ leaks to a non-computational basis state). In general, these errors could have complicated spatial and temporal dependence. From numerical simulations, we find that coherent errors have particularly large effects on measured observables. For this reason, we first focus on mitigating coherent memory errors using a form of dynamical decoupling \cite{viola1999}.

Our dynamical decoupling (DD) scheme involves physically applying single-qubit $X$ gates periodically in time. We apply the $X$ gates to each qubit before each of the four layers of $U_{ZZ}(\theta)$ gates in a Trotter step. Because $X\otimes X$ commutes with $U_{ZZ}(\theta)$ and $X^2=I$, the DD circuit is logically equivalent to the original circuit. However, it significantly reduces the build-up of coherent memory errors. Roughly the same amount of time elapses between each layer of $U_{ZZ}(\theta)$ gates. Assuming coherent memory errors are time-independent, each qubit's memory error is $e^{-i\phi Z_j}$ between each gate layer (note that this error commutes with all of the $U_{ZZ}(\theta)$ gates). For the DD circuit, the $X$ gates affect the sign of the error operators, resulting in cancellation for the four layers in a Trotter step,
\begin{align}
&Xe^{-i\phi Z_j}X e^{-i\phi Z_j}X e^{-i\phi Z_j}Xe^{-i\phi Z_j}\\
=&e^{+i\phi Z_j}e^{-i\phi Z_j}e^{+i\phi Z_j}e^{-i\phi Z_j}=1.
\end{align}

In reality, DD will not work perfectly because dephasing-type memory errors are not completely coherent over the duration of a single Trotter step, and because the spacing in time between $X$ gates is not perfectly uniform (due to the ion transport performed in between the layers of gates). However, we find that the $X$ pulses in our circuit on H2 are to a good approximation uniformly spaced in time.

An important consideration is that DD significantly increases the number of single-qubit (1Q) gates implemented in the circuit and therefore also the errors resulting from 1Q gates. In our circuits, there are roughly five times more 1Q gates with DD than without. One might expect that since 1Q gate errors are much smaller than two-qubit (2Q) gate errors on H2 ($\sim3\times 10^{-5}$ compared to $\sim6\times 10^{-4}$ for the $U_{ZZ}(0.5)$ gate used throughout this manuscript), the effect of 1Q gate errors would be negligible. While this is likely true for purely incoherent errors, this is not true for coherent under-rotation or over-rotation errors, which can add coherently between the DD pulses in a single Trotter layer. \diff{By running test circuits in which the transport of a full Trotter circuit was mimicked but only DD gates were applied, we found experimentally that our 1Q gate errors are indeed partially coherent, with a typical over/under rotation error on the order of $\delta\theta\sim 5\times 10^{-3}$ radians. Such an error equates to only about a $\delta\theta^2/6\sim 4\times 10^6$ average infidelity contribution to a 1Q gate, considerably less than our typical 1Q gate infidelity. However, such errors accrue coherently withing a Trotter step, effectively leading to a shift of the transverse field strength $\delta h\sim 4\delta\theta/(2dt)\sim 4\times 10^{-2}$. Numerical estimates of the transverse field susceptibility of $\langle Z_{\rm tot}^2\rangle$ using finite-size extrapolation indicated that such a shift would induce an observable error $\langle Z_{\rm tot}^2\rangle\sim 0.04$, considerably larger than our statistical error bars (and observed experimentally in Trotter circuits without the mitigation technique described below).}

We reduce the impact of coherent 1Q gate errors by alternating the phase of the $X$ gates performed during DD. The 1Q gate native to our hardware is
\begin{align}
U_{1q}(\theta,\phi)=e^{-i(X\cos \phi + Y\sin \phi )\theta/2},
\end{align}
which performs a rotation by $0 \leq \theta \leq \pi$ about a Bloch vector in the $xy$ plane specified by the angle $0 \leq \phi < 2\pi$. (The $U_X(\varphi)$ gate in our Trotter circuit, shown in 
 Fig.~1 in the manuscript, is implemented as $U_X(\varphi)=U_{1q}(\varphi,0)$.) An $X$ gate used in DD can be implemented in two physically inequivalent ways: $U_{1q}(\pi,0)=-iX$ and $U_{1q}(\pi,\pi) = +iX$. While these two gates differ by a logically-irrelevant global phase, \textit{coherent errors} on the two gates due to systematic under- or over-rotations are not logically equivalent. During DD, we alternate the phase of each $X$ gate on each qubit between $U_{1q}(\pi,0)$ and $U_{1q}(\pi,\pi)$ to approximately cancel such errors. 

\subsection{Randomized compiling} \label{sec:rc_implementation}

Not all coherent errors can be reduced through cancellation using DD or similar methods. For example, coherent errors of the form $U_{ZZ}(\theta)\rightarrow U_{ZZ}(\theta+\delta\theta)$ cannot be eliminated in this way. In addition, DD is not perfect and coherent memory errors or coherent 1Q gate errors still exist in the circuit. Given the hard-to-predict nature of coherent errors and their empirically large effects on observables, we convert coherent errors into incoherent errors using randomized compiling, or Pauli twirling~\cite{Wallman2016}.

In standard randomized compiling, one replaces a quantum circuit with an ensemble of logically equivalent random circuits, with specified Clifford circuit elements $U$ conjugated by random Paulis $P$ ($U \rightarrow PUP'$ where $P'=U^{\dagger}PU$). The effect of this randomization is to convert the error channel of $U$ into a purely incoherent stochastic Pauli channel $\mathcal{E}(\rho)=\sum_j p_j P_j \rho P_j$. For our circuits, we choose the circuit element $U$ to randomize over to be two layers of $U_{ZZ}(\theta)$ gates, or half of a Trotter step. We made this choice so that randomized compiling can be performed in parallel with DD: Every two gate layers, the coherent memory error is first approximately canceled by the $X$ gates from DD, then any remaining coherent memory errors as well as the coherent 2Q gate errors are made incoherent by randomized compiling.
However, $U_{ZZ}(\theta)$ is non-Clifford,
so the standard randomized compiling procedure needs to be modified.

In our randomized compiling procedure, random $N$-qubit Paulis $P$ are chosen so that $I,X,Y,Z$ are chosen with probability $1/4$ on each qubit, these Paulis are inserted as $P^2=I$ before every other layer of 2Q gates, then one of the two Paulis $P$ is ``pushed through'' two layers of 2Q gates. Depending on the random Pauli chosen, the 2Q gates will be modified. In H2, the gates $U_{ZZ}(\theta)$ and $U_{ZZ}(-\theta)$ can both be implemented directly in hardware and are physically inequivalent gates with different error models. When the random Pauli $P$ anticommutes with $Z\otimes Z$, pushing through the Pauli changes the sign of the gate: $U_{ZZ}(\theta)P=PU_{ZZ}(-\theta)$; when $P$ commutes with $Z\otimes Z$, the $U_{ZZ}(\theta)$ gate remains unchanged. Since $P$ is equally likely to commute or anticommute, each 2Q gate in the randomized circuit is equally likely to be $U_{ZZ}(\theta)$ or $U_{ZZ}(-\theta)$. We interpret this randomization as partially converting the original 2Q gate error model for $U_{ZZ}(\theta)$ into a stochastic Pauli error model corresponding to the Pauli twirl of the average of $U_{ZZ}(\theta)$ and $U_{ZZ}(-\theta)$. We note that this procedure cannot remove all coherent errors. For example,
if a coherent over-rotation by angle $\theta_{\epsilon}$ in $U_{ZZ}(\theta)$ is correlated with an under-rotation by angle $\theta_{\epsilon}$ in $U_{ZZ}(-\theta)$, then this error will still be present. We measure this averaged error model experimentally using the procedure described in Sec.~\ref{sec: Pauli learning}, and use it for zero-noise extrapolation as discussed in Secs.\,\ref{sec:zne_implementation} and \ref{sec:zne_postprocess}.

During preparation of this manuscript,
we became aware of Ref.~\cite{P.Santos2024},
which gives a theoretical treatment of this randomized compiling procedure for non-Clifford multi-qubit gates.

\subsection{Leakage detection} \label{sec:ld_implementation}

\begin{figure}[!t]
\centering
\includegraphics[width=0.8\columnwidth]{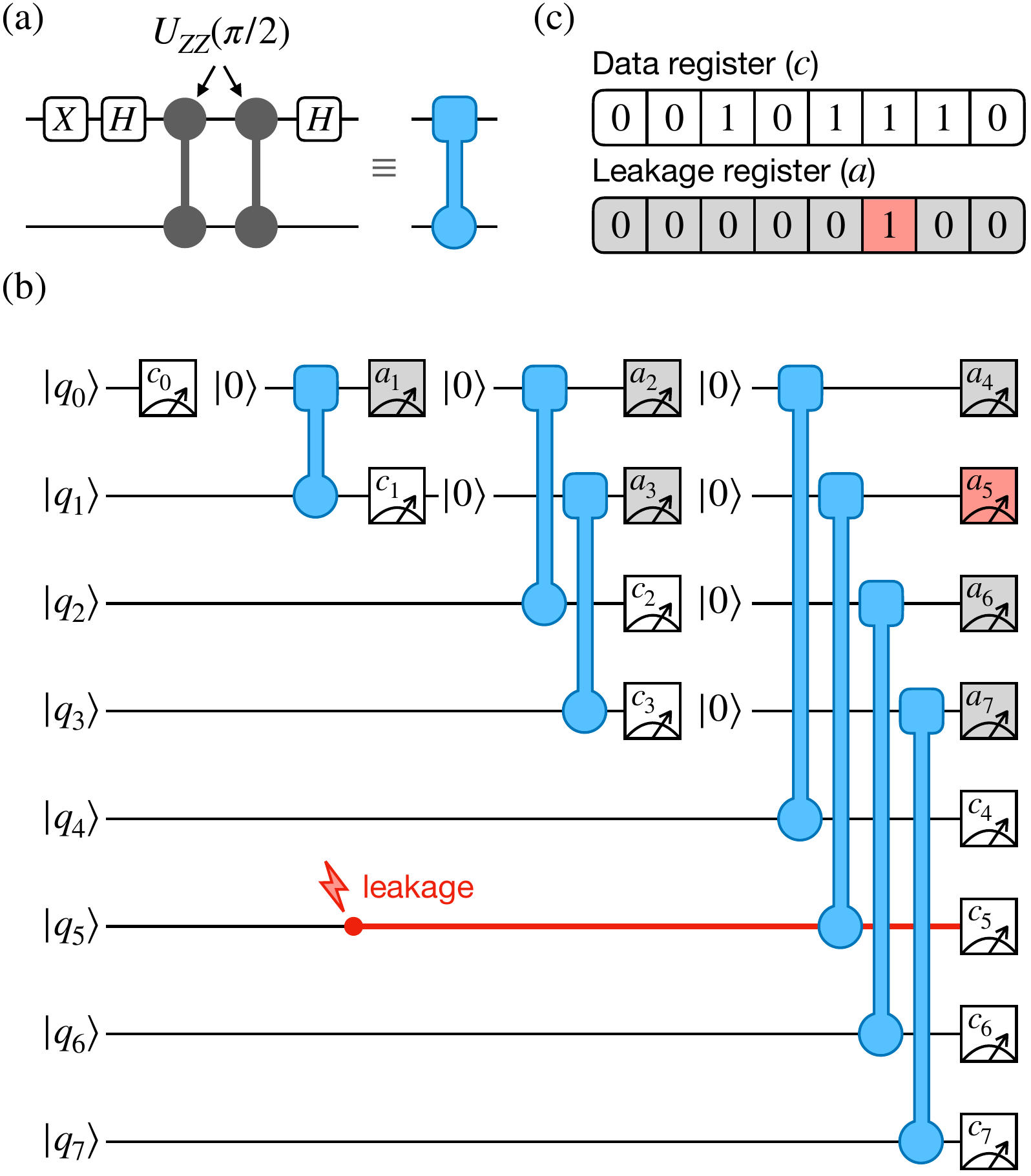}
\caption{(a) This leakage detection gadget either applies $X$ or does nothing on the top (ancilla) qubit depending on whether the bottom (data) qubit has or has not leaked, such that leakage events on the data are mapped to $1$ measurement outcomes for an ancilla initialized in $\ket{0}$. (b) If all available qubits are utilized in a circuit, one must be measured without leakage detection, but the remaining $N-1$ can then be leakage detected prior to measurement using a $\sim\log(N)$ depth circuit by reusing measured qubits. (c) The results of the $Z$-basis measurements and leakage detection results are stored in two separate classical registers, $c$ and $a$, respectively.
\label{fig:LD}}
\end{figure}

Leakage errors on H2, which occur primarily during 2Q gates and qubit idling, can have significant impact on a circuit's results. In this device, when a qubit is leaked gates applied to that qubit have no effect (act as identity) and measurements on that qubit register as $\ket{1}$. The effect of leakage can be mitigated using a circuit-level leakage detection gadget \cite{Stricker_2020,Moses2023}. As discussed in Ref.\,\cite{Moses2023} and shown in \fref{fig:LD}(a), the gadget involves two 2Q gates (so introduces additional 2Q gate noise) and an ancilla qubit and can detect whether a qubit is leaked. By reusing qubits with mid-circuit measurements and resets, one can use no ancilla and a $\sim\log_2 N$ depth circuit to detect leakage on $N-1$ qubits, as shown in \fref{fig:LD}(b,c) for the case of $N=8$.

The standard use of error detection is to perform post-selection, i.e., to discard data with any detected leakage errors. This technique is not scalable and produces accurate noiseless observables estimates at the cost of an exponential in system size post-selection overhead. We instead use zero-noise regression (ZNR) a more scalable technique that we introduce in \sref{sec:znr} -- similar in spirit to zero-noise extrapolation, that uses heralded errors along with a heuristic fitting procedure to estimate the zero-noise limit of an observable. 

In addition to ZNR, we also attempt to heuristically reduce ``false-positive'' leakage detections in our data. If a data qubit $q$ is actually leaked, then its measurement result should be $\ket{1}_q$ and its corresponding ancilla qubit $a$ should also be measured as $\ket{1}_a$. If the measurement result is $\ket{0}_q\ket{1}_a$ then there are three possible explanations. (1) This is a true leakage event, where $q$ leaked, but then the $q$ measurement outcome was flipped due to a measurement error. (2) This is a ``false-positive,'' where $q$ did not leak and was supposed to be measured as $0$, but the $a$ measurement result was flipped due to a two-qubit gate error or measurement error in the leakage-detection gadget. (3) This is a ``false-positive,'' where $q$ did not leak and was supposed to be measured as $1$, but both the $q$ and $a$ measurement results were flipped due to a two-qubit gate error in the leakage-detection gadget. Reason (2) is the most likely possibility, $\sim 2$ times more likely than (1) and $\sim 5$ times more likely than (3) for the specific errors in H2. Therefore, we do not count such events as actual leakage events in our error detection protocol. Note, however, that simply counting the $\ket{0}_q\ket{1}_a$ result as ``not-leaked'' will lead to a bias in the $q$ measurement results towards $0$. This is because a $\ket{1}_q\ket{1}_a$ measurement that would be marked as ``leaked'' under this protocol, can also result from the situation where $q$ did not leak and was supposed to measure $1$ and a single leakage-detection-gadget error flipped the measurement on $a$. This situation occurs about as often as the ``false-positive'' outcome in (2) (assuming $q$ is approximately equally likely to measure $0$ as $1$), though is not recorded in the same way, leading to the bias. In an attempt to heuristically remove this bias, whenever a $\ket{0}_q\ket{1}_a$ is recorded in a shot, then the $q$ measurement result is replaced by the maximally mixed state, i.e., $\ket{0}_q\bra{0}_q \rightarrow I_q/2$, so that observables like $Z_q Z_{q'}$ will have expectation value $0$ for that shot. Similarly, whenever $\ket{1}_q\ket{1}_a$ is recorded in a shot, $q$ is replaced by the maximally mixed state as well, which is important for ZNR (see Sec.~\ref{sec:znr}).

\subsection{Pauli insertion for zero-noise extrapolation} \label{sec:zne_implementation}

For the zero-noise extrapolation (ZNE) technique, one needs to amplify the hardware errors in a quantum circuit in order to perform an extrapolation to the zero-noise limit. In Ref.\,\cite{ibm_utility}, the error amplification was performed by learning a stochastic Pauli noise model and then randomly inserting Paulis according to their probability of occurring in the amplified noise model. This procedure has the benefit of not increasing the 2Q gate count in the circuit, thereby maintaining the circuit depth and memory error, though it requires learning the noise model accurately. 

In our implementation of ZNE, we learn a stochastic Pauli noise model for the average of the $U_{ZZ}(+2Jdt)$ and $U_{ZZ}(-2Jdt)$ gates using a variant of cycle benchmarking, as discussed in Sec.~\ref{sec: Pauli learning}, which assigns a probability $p_j$ for each two-qubit Pauli $P_j$. In our error-amplified circuits, before each $U_{ZZ}(\pm 2Jdt)$ gate we randomly insert the Pauli $P_j$ with probability $(\alpha-1) p_j$, resulting in $\alpha$ times more 2Q gate errors than the unamplified circuits. For a particular circuit, we execute $S$ shots total on the device split between two data points used in ZNE: $rS$ shots of the original unamplified circuit and $(1-r)S$ shots of the error-amplified circuit. These two points are used to extrapolate the zero-noise value of the $\langle Z_{\rm tot}^2 \rangle$ observable using an exponential fit. Evidence for minimal bias of the exponential fitting, as well as details of how to optimize the choice of $(r,\alpha)$ in ZNE in order to minimize the variance on the extrapolated observable, are discussed in Sec.~\ref{sec:zne_postprocess}.

For the random Pauli insertion to accurately represent an amplification of the 2Q gate error model, ZNE needs to be combined with randomized compiling around each 2Q gate that converts the error model to a stochastic Pauli channel. The randomized compiling that we implement, described in Sec.~\ref{sec:rc_implementation}, has a slightly different effect; it converts the error model of two layers of 2Q gates into a stochastic Pauli channel. Nonetheless, we expect that this randomization has the desired effect of improving the performance of ZNE, even if imperfectly.

\subsection{The error-mitigated Trotter circuit}

\begin{figure}[!t]
\centering
\includegraphics[width=1.0\columnwidth]{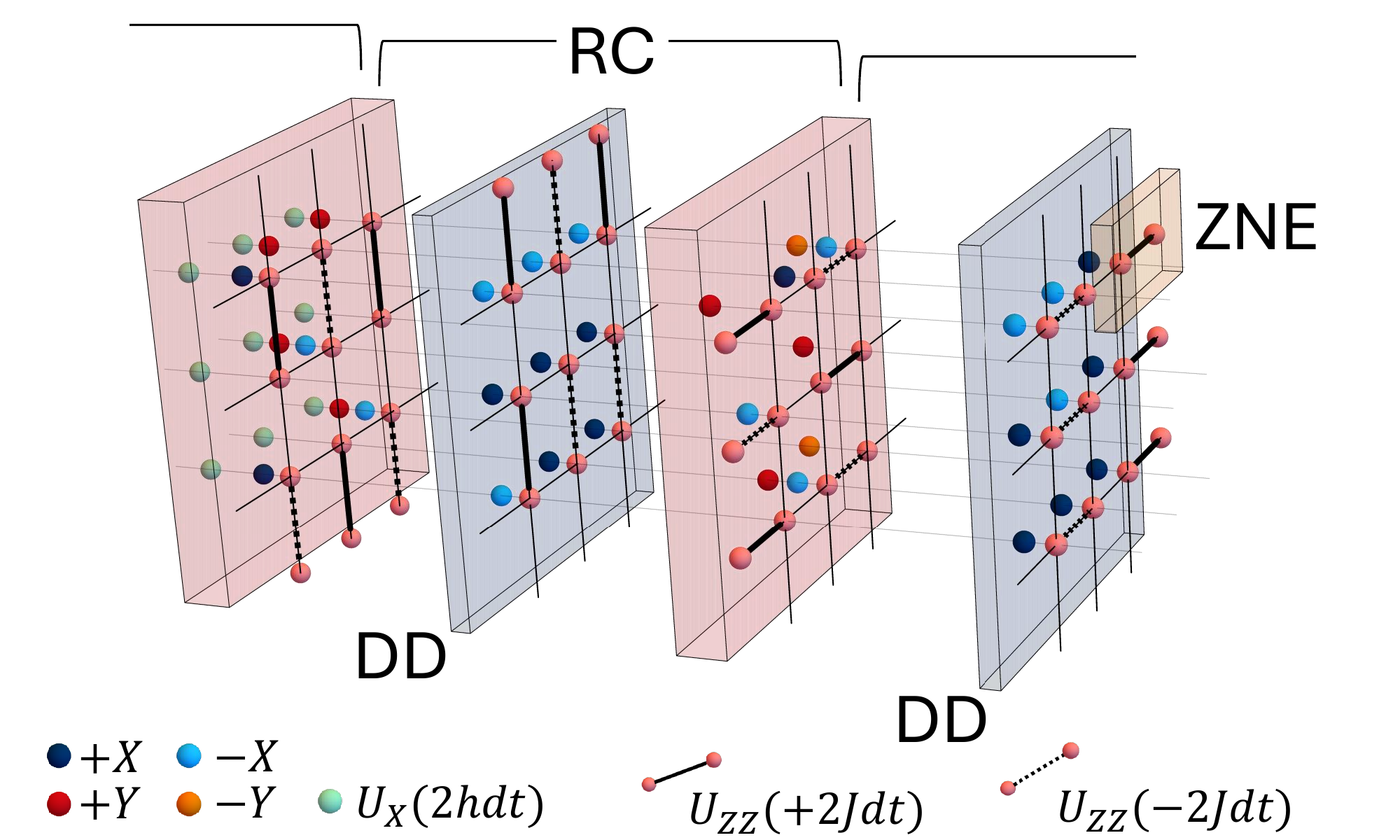}
\caption{A Trotter step of the error-mitigated Trotter circuit. Dynamical decoupling (DD), which involves periodic insertions of $\pm X$ gates, is performed to reduce coherent memory errors. Randomized compiling (RC), or Pauli twirling, is performed around every two layers of two-qubit gates (half of a Trotter step) to make coherent memory and gate errors incoherent. Paulis are inserted for zero-noise extrapolation (ZNE) before each two-qubit gate. Details are described in the text.
\label{fig:rc}}
\end{figure}

There are additional important details to consider when the error mitigation methods discussed above -- dynamical decoupling, randomized compiling, leakage detection, and Pauli insertion for ZNE -- are combined together into the same circuit.

Dynamical decoupling and randomized compiling both introduce many single-qubit gates into the error-mitigated circuit, which add gate errors and additional ion transport. To reduce the number of single-qubit gates, the $X$ Paulis from DD and the random Paulis from randomized compiling, including randomized compiling on adjacent layers of 2Q gates, are compiled together into new Paulis. Note that we always implement a Pauli $Z\propto XY$ using a physical $Y$ followed by a physical $X$ gate, rather than using a software $Z$ gate \cite{Moses2023}, to ensure that the Pauli is physically happening at the correct point in time in the circuit. In order to cancel the coherent errors from the compiled 1Q gates, we use a modified form of the phase-alternating trick discussed in Sec.~\ref{sec:dd_implementation}. With the native $U_{1q}(\theta,\phi)$ gate, we can implement both $X$ and $Y$ Paulis with different phases: $X\propto U_{1q}(\pi,0),-X\propto U_{1q}(\pi,\pi),Y\propto U_{1q}(\pi,\pi/2),-Y\propto U_{1q}(\pi,-\pi/2)$ As we build the circuit, for each qubit we keep track of which phase to use ($\phi=0,\pi$) for the next Pauli $X$ or $Y$ gate. To cancel the coherent errors on the $U_{1q}$ gates, we follow the rules that: after each $Y$ Pauli, the phases of subsequent $X$ and $Y$ gates flips and after each $X$ Pauli, the phases of subsequent $X$ gates flips. These rules account for the fact that $X$ and $Y$ anticommute so that they change the phase of the gates if they are passed through one another: $\pm XY= \mp YX$ \footnote{After the low-temperature and hydrodynamics quench experiments concluded, we found that the optimal way to arrange the phases is to instead follow the rule: after each $X$ and $Y$ Pauli, the phases of subsequent $X$ and $Y$ gates flips. This is based on the fact that $\pm Xe^{+i\delta\theta_x X}Ye^{+i\delta\theta_y Y}= \mp Ye^{-i\delta\theta_y Y}Xe^{-i\delta\theta_x X} + O(\delta\theta_x \delta\theta_y)$. The intermediate-temperature and triangular lattice quenches used the optimal approach. The original approach still heuristically cancels the coherent error, though slightly suboptimally.}.

The error-mitigated circuits are random, with random Paulis used in both randomized compiling and the Pauli insertion for ZNE. The software stack and control systems on H2 can produce in real-time random numbers using a classical pseudo-random number generator (PRNG) that can be used in the quantum circuit. Using this feature, we insert random Paulis for ZNE by using random bits generated from the PRNG along with conditional single-qubit gates. This allows us to have per-shot randomization of our circuits without large compiler overheads for each randomization. For technical reasons, we only used the PRNG for the Pauli insertion for ZNE. The randomization for randomized compiling was performed in a standard way, with each random circuit separately compiled and executed for a few shots on the device. The final error-mitigated circuit implementing all of the error mitigation protocols discussed above is depicted in Fig.~\ref{fig:rc}.

\section{Error model learning and error mitigation implemented in post-processing} \label{sec:postprocess}

\subsection{Pauli error learning}\label{sec: Pauli learning}

A Pauli error channel is given by
\begin{equation}
\mathcal{E}(\rho) = \sum_j p_j P_j\rho P_j,
\end{equation}
where $p_j$ is the probability of Pauli error $P_j$ occurring.
Each Pauli operator is an eigenvector of the channel:
\begin{equation}
\mathcal{E}(P_j)=f_j P_j,
\end{equation}
where the eigenvalues (also called Pauli fidelities)
are related to the Pauli error probabilities by
\begin{align}\label{eq: Hadamard transform}
f_i &= \sum_j(-1)^{\braket{i,j}}p_j,\notag\\
p_j &= \frac{1}{d^2}\sum_i (-1)^{\braket{i,j}}f_i,
\end{align}
where $d$ is the Hilbert space dimension and $\braket{i,j}$ equals 0 or 1 depending on whether $P_i$ commutes or anti-commutes with $P_j$.
To estimate a Pauli error channel associated with the $U_{ZZ}(\pm2Jdt)$ gates used in our circuits,
we use a variation of cycle benchmarking~\cite{Erhard2019} for non-Clifford gates,
which is similar to the procedure in Ref.~\cite{Layden2024}.
In the standard cycle benchmarking protocol,
a twirl over the Pauli group is performed to project the error channel per cycle into a Pauli error channel.
In our case the gate $U_{ZZ}(\pm2Jdt)$ is non-Clifford,
and we implement a pseudo-twirl by conjugating each $U_{ZZ}$
by a uniformly random Pauli and flipping the sign of $U_{ZZ}$ rotation angle if the sampled Pauli anti-commutes with $ZZ$.
Assuming that the positive and negative angle gates
have the same error rates in the stochastic Pauli sector,
the resulting average error channel per 2Q gate will approximately equal a Pauli error channel composed with a coherent $ZZ$ rotation.

\begin{figure}[!t]
\centering
\includegraphics[width=1.0\columnwidth]{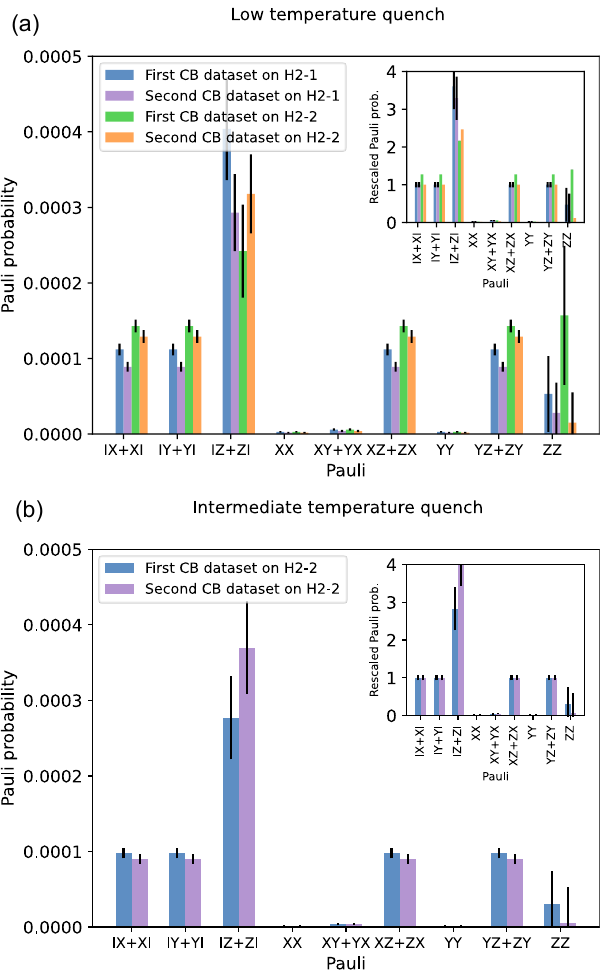}
\caption{The Pauli error probabilities $p_j$ for each of the two-qubit Paulis $P_j$ for the sign-averaged $U_{ZZ}(\pm 0.5)$ two-qubit gate, as measured by the cycle benchmarking procedure described in Sec.~\ref{sec: Pauli learning}, for the (a) low-temperature and (b) intermediate-temperature quenches. For each quench and each machine used (H2-1 and H2-2 for the low-temperature quench and H2-2 for the intermediate-temperature quench), two cycle benchmarking data sets were taken, one before and one after the set of quantum simulation experiments. Insets: the Pauli probabilities rescaled by the Pauli probability for $IX+XI$.
\label{fig:cb_data}}
\end{figure}

\begin{figure*}[!t]
\centering
\includegraphics[width=2.05\columnwidth]{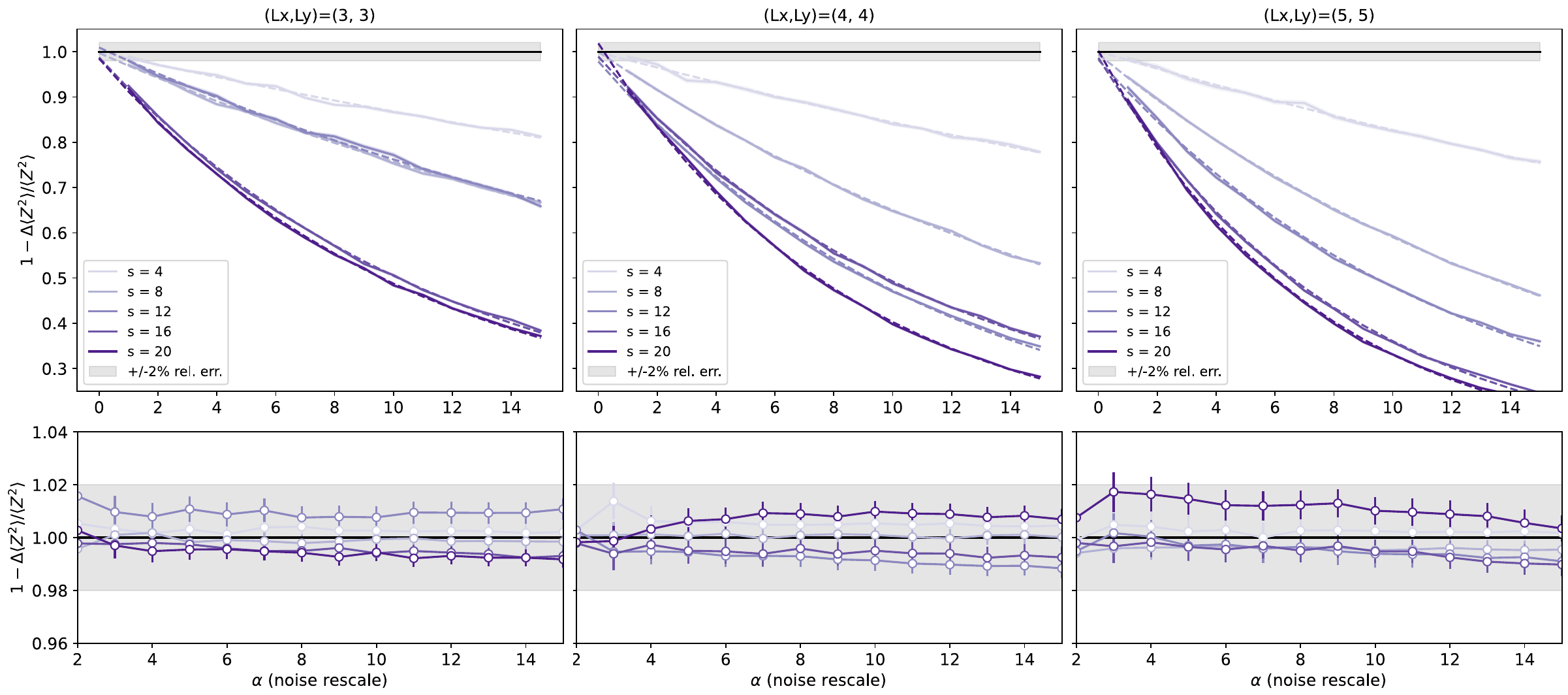}
\caption{Exponentiality of observable errors as a function of noise rate. (Top row) We see that over a range of simulable system sizes and a range of circuit depths relevant for the data presented, decay of the observable $\langle Z^2\rangle$ as a function of noise is accurately modeled as an exponential. Exponential fits to the collection of numerical data over the entire noise range explored generally produces no more than a $2\%$ bias.  (Bottom row) Extrapolations from just two noise-values, the raw value and $p$ indicated on the x-axis. As can be seen, exponential fits to two points yield remarkably unbiased estimates over the entire range of noise levels plotted.
\label{fig:exp_ZNE}}
\end{figure*}

By preparing an initial Pauli operator $P_j$,
applying the pseudo-twirl of the $U_{ZZ}$ gate $l$ times,
and measuring in the $P_j$ basis,
the expectation value of $P_j$ as a function of $l$
is given by
\begin{equation}\label{eq: CB decays}
\mathbb{E}(P_j)(l)= \begin{cases}
                    A[f_j]^l \quad\quad\quad\quad\,\,\, [P_j, ZZ] = 0\\
                    A[f_j]^l \cos(\theta_{\epsilon}l) \quad\{P_j, ZZ\}=0
                    \end{cases}.
\end{equation}
Here $\theta_{\epsilon}$ is a coherent $ZZ$ error angle,
and $A$ a state preparation and measurement (SPAM) parameter. The quantity $[f_j]=(f_jf_{u(j)})^{1/2}$ is the symmetrized Pauli fidelity, with $u(j)$ defined by $U_{ZZ}(\pi/2)P_jU_{ZZ}(\pi/2)^{\dagger}=P_{u(j)}$.
According to the theory of Pauli error learnability~\cite{Chen2023},
only the quantities $[f_j]$, rather than the individual $f_j$,
can be learned in a SPAM independent way.
In our cycle benchmarking experiments we assume $f_j=f_{u(j)}$ for $j\ne u(j)$.
The operator $P_j$ is not a physical state (that is, a density matrix),
but the preparation of $P_j$ can be simulated by preparing the 4 eigenstates of $P_j$ and combining the results linearly.
We choose sequence lengths $l\in\{4, 80, 160\}$ and run 5 circuits with 100 shots for each sequence length and each Pauli eigenstate preparation.
The circuits are performed in parallel across the 4 active gate zones in H2 (with different randomizations in each gate zone).
The empirical expectation value decay curves are best-fit to Eq.~\eqref{eq: CB decays} and the Pauli error probabilities $p_j$ are computed from Eq.~\eqref{eq: Hadamard transform}.
The error probabilities are then averaged across the 4 gate zones to obtain the Pauli error model used for the Pauli insertion used in ZNE.

Figure~\ref{fig:cb_data} shows the results of the cycle benchmarking protocol for the $U_{ZZ}(\pm2Jdt)$ gate on both the H2-1 and H2-2 quantum computers. For both machines, we perform two sets of cycle benchmarking experiments before and after quantum simulation experiments in order to test the stability of the error model. We find that while the total scale of the error model (i.e., the process infidelity or average gate infidelity) can drift over time, the relative Pauli probabilities stay constant up to error bars. This motivates a rescaling procedure described in Sec.~\ref{sec:zne_postprocess}. All benchmarking circuits used in this work used the leakage detection gadget in \fref{fig:LD}(a) at the end of the circuit and had their results post-selected on no detected leakage errors. This ensured that the learned error model used in ZNE (Sec.~\ref{sec:zne_postprocess}) excludes the leakage errors, which were accounted for separately with the leakage-detection based ZNR procedure described in Sec.~\ref{sec:znr}.

\subsection{Optimization of zero-noise extrapolation} \label{sec:zne_postprocess}

Extrapolation to the zero-noise limit (ZNE) from data with varying levels of noise was originally proposed in \cite{PhysRevLett.119.180509}, and has been explored in many contexts. The general idea is that the expectation value of a given operator is an implicit function of the noise level $p$, $\mathcal{O}(p) = \langle O\rangle_{{\rm noise~level}=p}$. Given a physical noise level $p_0$, amplification of the noise allows one to learn the function $\mathcal{O}(p>p_0)$, from which one can attempt to extrapolate the desired noiseless value $\mathcal{O}(p=0)$. In the limit of very little noise, a linear extrapolation should be unbiased under fairly mild assumptions.

\begin{figure}[!t]
\centering
\includegraphics[width=0.9\columnwidth]{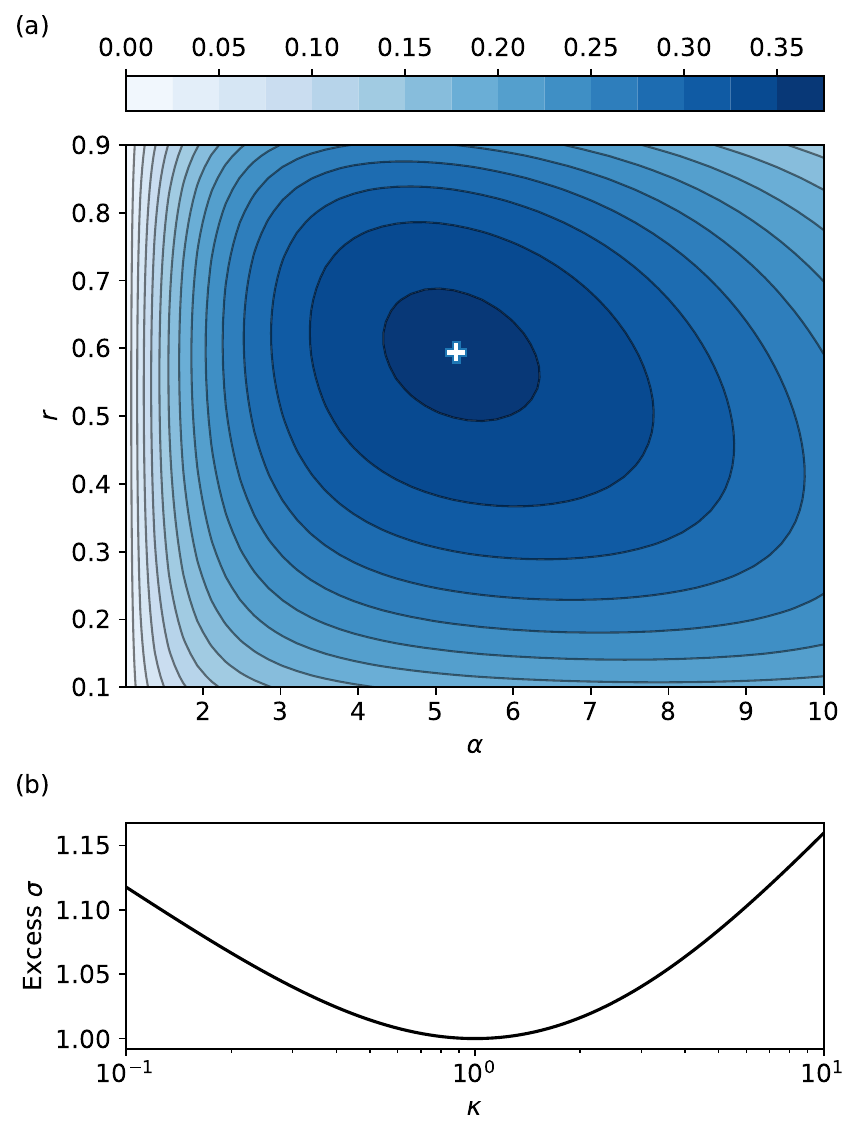}
\caption{(a) Dependence of ZNE estimate variance (inverse standard deviation is plotted in arbitrary units) on the choice of both the injected noise level $p_1$ relative to the bare noise level $p_0$ ($\alpha=p_1/p_0$) and the distribution of samples between the two noise levels $r$, shown here at $(\kappa,\zeta) = (1,0.3)$.  Since $\kappa$ is \emph{a priori} unknown (though generally expected to be reasonably close to unity), we always assume $\kappa=1$ when choosing optimal parameters $(\alpha,r)$. Panel (b) shows (again for $\zeta=0.3$) by how much the standard deviation on the ZNE extrapolation using this procedure is in excess of the true optimum for a range of $\kappa$, showing that very little penalty is paid unless $\kappa$ is very large or small compared to unity.
\label{fig:ZNE_opt}}
\end{figure}

Given a physical noise rate $p_0$, the choice of an amplified noise level (or several) from which to perform ZNE is subject to competing interests.  If one is fortunate to have a low-enough physical noise level $p_0$ to be safely in the linear-response regime, increasing the amplified noise level $p_{1}$ relative to $p_0$ decreases the variance on the estimate of $\mathcal{O}_{\rm noiseless}\equiv \mathcal{O}(0)$ [assuming one holds fixed the variance on the estimates of $\mathcal{O}_0\equiv \mathcal{O}(p_0)$ and $\mathcal{O}_1\equiv \mathcal{O}(p_1)$]. However, choosing $p_1$ too large relative to $p_0$ inevitably causes the breakdown of linearity.

In \fref{fig:exp_ZNE}, we show from numerics on a range of system sizes that the response of our computed observable $\langle Z_{\rm tot}^2\rangle$ to noise is to a very good approximation exponential over a large range of noise.  Assuming exponentiality of $\mathcal{Z}_{\rm tot}^2(p)\equiv\langle{Z_{\rm tot}^2}\rangle_{{\rm noise~level}=p}$ with respect to $p$ enables $\mathcal{Z}_{\rm tot}^2(0)$ to be extrapolated from just two values of $p$ (i.e. the raw data and a single noise-amplified data set), and also dictates a choice of noise level $p_1$ that minimizes the variance on the noiseless estimate. In particular, suppose that
\begin{align}
\mathcal{Z}^2(p) = \mathcal{Z}^2(0)e^{-b p}.
\end{align}
The noiseless value $\mathcal{Z}^2(0)$ extrapolated from two measurements $\mathcal{Z}^2_0$ and $\mathcal{Z}^2_1$ at noise levels $p_0$ and $p_1$ is an implicit function of those two values, and its variance is given by
\begin{align}
\Delta^2\mathcal{Z}(0)=\frac{\delta \mathcal{Z}^2(0)}{\delta \mathcal{Z}^2_0}\Delta\mathcal{Z}^2_0+\frac{\delta \mathcal{Z}^2(0)}{\delta \mathcal{Z}^2_1}\Delta\mathcal{Z}^2_1,
\end{align}
where $\Delta \mathcal{Z}^2_{0(1)}$ is the variance on the estimate of $\mathcal{Z}^2_{0(1)}$. Denoting the fraction of samples taken at noise level $p_0$ by $r$, the variance on the estimate will be proportional to
\begin{align}
\label{supeq:variance}
\frac{\delta \mathcal{Z}^2(0)}{\delta \mathcal{Z}^2_0}\frac{1}{r}+\frac{\delta \mathcal{Z}^2(0)}{\delta \mathcal{Z}^2_1}\frac{\kappa}{1-r},
\end{align}
where $\kappa$ is an implicit function of $\mathcal{Z}^2$ characterizing how the variance on the estimate of $\mathcal{Z}^2$ changes as a function of $\mathcal{Z}^2$ at fixed sample count. Determining this dependence \emph{a priori} would be a challenging numerical problem (essentially requiring noisy simulations of the dynamics). However, for now we just treat $\kappa$ as an unspecified constant, and we will see that our ultimate conclusions are only weakly dependent on its value.  Equation~(\ref{supeq:variance}) depends on both $r$ and the ratio $p_1/p_0\equiv \alpha$, and its minimization can be shown to require
\begin{align}
\label{supeq:opt_vals_a}
\alpha &= \frac{1+\gamma_{\kappa}+\zeta}{\zeta},\\
\label{supeq:opt_vals_r}
r&=1-\frac{\zeta}{(\gamma_{\kappa}+\zeta)(1+\zeta)}.
\end{align}
Here $\zeta= b\times p_0$ characterizes the response to the physical level of noise, and $\gamma_{\kappa}\equiv W(1/[e\sqrt{\kappa}])$, with $W$ the Lambert W-function.  This solution can be formulated independently of $b$ and purely in terms of an optimal ratio for the bare and noise amplified observable estimates, $R\equiv \mathcal{Z}_1^2/\mathcal{Z}_0^2$, as
\begin{align}
R=\frac{1}{\exp\big(1+\gamma_{\kappa}\big)}.
\end{align}
Figure \ref{fig:ZNE_opt}(a) shows the dependence of inverse standard-deviation of the extrapolated estimate on the parameters $\alpha$ and $r$, with the optimal point from Eqs.\,(\ref{supeq:opt_vals_a},\ref{supeq:opt_vals_r}) shown as a white ``+''. Figure \ref{fig:ZNE_opt}b shows how much the standard-deviation of the estimate based on parameters optimized for $\kappa=1$ is in excess of its true optimal value for $\kappa$ varying over a wide range ($+/-$ one order of magnitude relative to the nominal value of $1$), demonstrating that in practice the dependence of estimate variance on noise level can be ignored with minimal overhead; all of our noise level choices are made assuming $\kappa=1$.

\begin{figure}[!t]
\centering
\includegraphics[width=1.0\columnwidth]{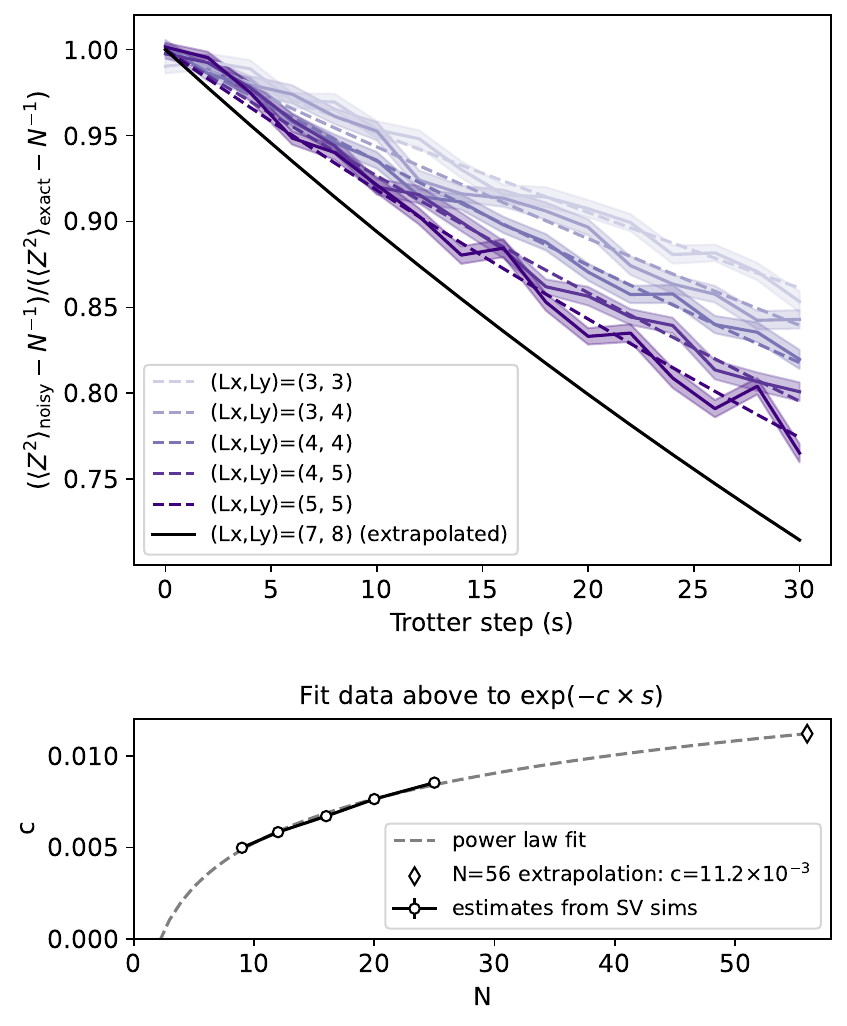}
\caption{Determining the response of $\langle Z_{\rm tot}^2\rangle$ to noise from finite-size extrapolation. (a) Noisy simulations using a stochastic Pauli error model (learned from cycle benchmarking) for our two-qubit gate, here for the benchmark quench shown in Fig.\,2c in the manuscript. Decay of the observable is roughly exponential in time, and fitting to an exponential $e^{-c s}$ yields the decay constant $c$ on system sizes from 9 up to 25 qubits.  We then extrapolate $c$ (using a power law fit) to $N=56$, and use the resulting constant to estimate the decay rate $\zeta$ used for the ZNE parameter optimizations. The same procedure is followed for the quench shown in Figs.\,2(d,e) of the manuscript, and yielded a fit constant $c=8.7\times 10^{-3}$
\label{fig:easy_quench_alpha}.}
\end{figure}

Determination of the optimal values of $\alpha$ and $r$ requires a-prior knowledge of $\zeta$, characterizing the response of the observable in question to noise.  We use the heuristic that the decay exponent is linearly proportional to circuit depth, $\zeta = c\times s$, and estimate the constant $c$ by fitting noisy simulations over a range of system sizes and then performing finite-size extrapolation to the $7\times8$ lattice for which data is reported in the manuscript (\fref{fig:easy_quench_alpha}). Note that these finite-size extrapolations \emph{do not} inform estimates of observables from the quantum data, but rather they inform details of the ZNE procedure that ultimately impact the \emph{variance} on that estimate. Therefore, errors in these extrapolations ultimately lead our estimate variances in ZNE to be weakly suboptimal, but do not impact the estimates themselves. \ddiff{Since $c$ characterizes the low-depth response of a $k-$local observable to local noise, we expect on very general (light-cone-related) grounds that $c$ should saturate to a constant as $N$ grows, implying that the difficulty of error mitigation should not grow indefinitely with system size for quenches at fixed time. However, we were not able to perform noisy simulations over a broad enough range of system sizes to reliably infer what that saturated value might be for the observables considered in this work, or precisely how quickly the error-mitigation sampling overheads would grow with $N>56$ (before eventually saturating).}

Instead of optimizing $r$ and $\alpha$ together, one can instead determine the optimal value of $r$ for a fixed $\alpha$:
\begin{align}
r = \frac{e^{2\zeta}\alpha^2 - e^{(\alpha+1)\zeta}\alpha \sqrt{\kappa}}{e^{2\zeta}\alpha^2 - e^{2\alpha\zeta}\kappa}. \label{eq:r_opt_fixedalpha}
\end{align}
Eq.~\eqref{eq:r_opt_fixedalpha} is useful when the full $(r,\alpha)$ optimization in Eq.~\eqref{supeq:opt_vals_a} results in a very large $\alpha$. While using a large value of $\alpha$ is not necessarily problematic, we chose to artificially restrict $\alpha$ when it lies outside the range of noise levels ($\alpha\leq 15$) that we studied numerically. We note that large $\alpha$ values are optimal at small circuit depths, where the noisy observable is very similar to its noiseless value, and reducing the size of $\alpha$ in these cases has minimal impact (simply because ZNE is not modifying the noisy estimate that much in the first place).

\begin{figure}[!t]
\centering
\includegraphics[width=1.0\columnwidth]{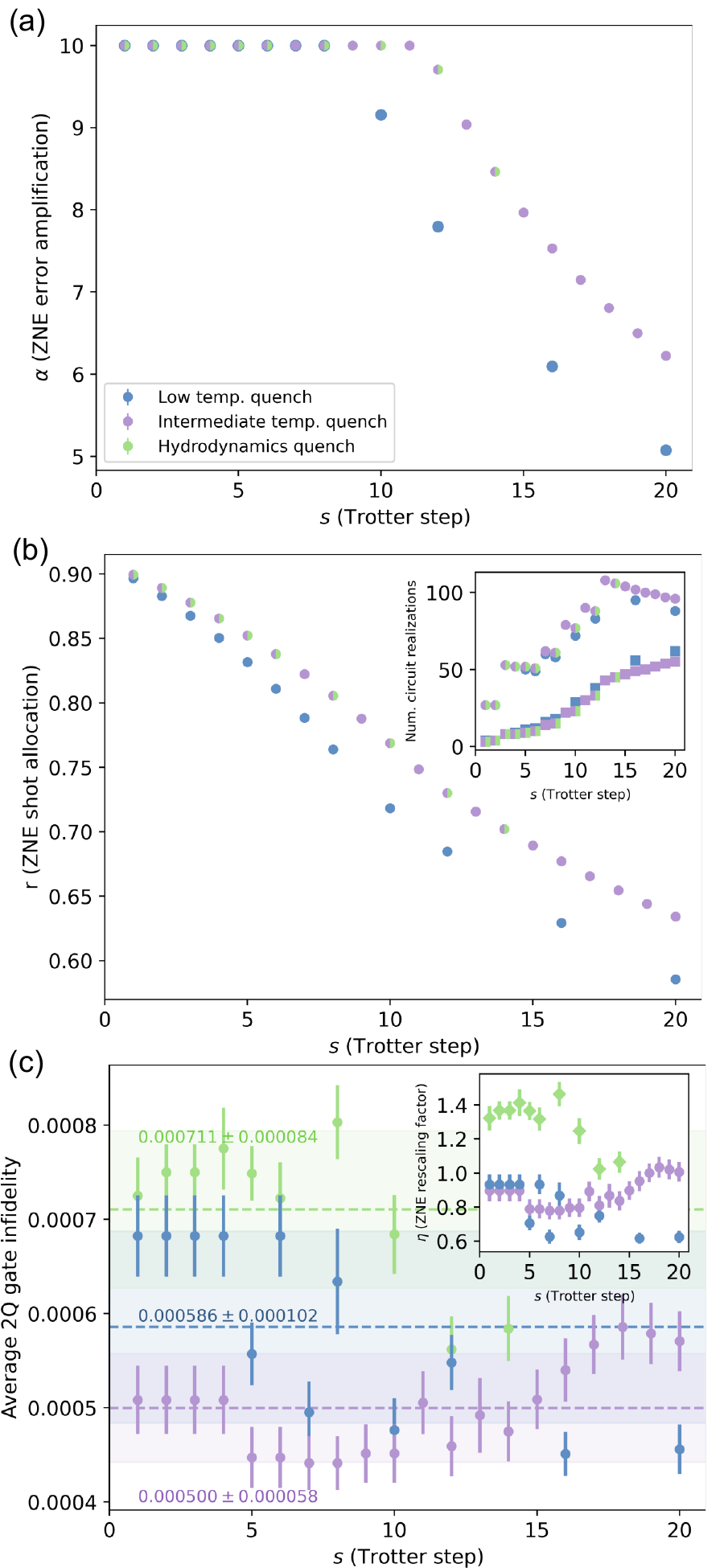}
\caption{(a) The ZNE error amplification factor $\alpha$ used in the experiments. (b) The fraction of shots $r$ to allocate to the unamplified error circuits in ZNE. Inset: The number of unique randomized compiling circuit realizations for the unamplified (circle) and amplified (square) circuits in ZNE. (c) The average two-qubit (2Q) gate infidelity for the $U_{ZZ}(\pm 2Jdt)=U_{ZZ}(\mp 0.5)$ gate at each Trotter step of each experimental quench considered. Inset: The ZNE rescaling factor $\eta$ for each quench. 
\label{fig:zne_details}}
\end{figure}

In our quantum simulation experiments, we performed a quench at low temperature, one at intermediate temperature, and one demonstrating hydrodynamic behavior. For these three quenches, we used the $\alpha$ and $r$ values shown in Fig.~\ref{fig:zne_details}(a)-(b). The $(r,\alpha)$ values were obtained using Eqs.~(\ref{supeq:opt_vals_a})~and~(\ref{supeq:opt_vals_r}) whenever the optimal $\alpha < \alpha_c$ and were otherwise replaced by $(r,\alpha_c)$ with $r$ determined by \eref{eq:r_opt_fixedalpha} with fixed $\alpha=\alpha_c$. The $\alpha_c=10$ cutoff was chosen based on the numerical evidence in Fig.~\ref{fig:exp_ZNE} that showed clear exponential decay with $\alpha$ for $\alpha \leq 15$.

The low-temperature quench covered the Trotter steps $s=1,2,3,4,5,6,7,8,10,12,16,20$; the intermediate-temperature quench covered steps $s=1,2,3,\ldots,19,20$; and the hydrodynamics quench covered steps $s=1,2,3,4,5,6,8,10,12,14$. Most of the low-temperature quench circuits were run on the H2-1 quantum computer, except for Trotter steps $s=5,7$ for the low- and intermediate-temperature quenches, which were run on H2-2 -- a quantum computer with the same architecture and nearly identical performance as the H2-1 device. All of the hydrodynamics quench circuits were run on H2-1 and all of the intermediate-temperature quench circuits were run on H2-2. Before each set of quench experiments run a particular device, we performed arbitrary-angle cycle benchmarking (see Fig.~\ref{fig:cb_data} and Sec.~\ref{sec: Pauli learning}) to learn the error model of our $U_{ZZ}(\pm 2Jdt)$ two-qubit gates (averaged over both signs) for the Pauli error insertion used in ZNE. Before the low-temperature (hydrodynamics) quench, the cycle benchmarking circuits on H2-1 measured a two-qubit average gate infidelity of $7.3(7)\times 10^{-4}$ ($5.5(5)\times 10^{-4}$). Before the low-temperature quench run on H2-2, the cycle benchmarking measured an infidelity of $7.9(9)\times 10^{-4}$ before the quench simulations on that machine. Before the intermediate-temperature quench run on H2-2 the cycle benchmarking measured an infidelity of $5.7(6)\times 10^{-4}$. Note that for a fixed Trotter step at each quench, there were many error amplified and unamplified circuits, each a random circuit with a different realization of randomized compiling and inserted Pauli errors (see Secs.~\ref{sec:rc_implementation}~and~\ref{sec:zne_implementation} for details). We chose to execute the specific number of random circuit instances shown in the inset of Fig.~\ref{fig:zne_details}(b), which were chosen to maximize circuit throughput on H2-1 and H2-2.

An added complication to performing ZNE by inserting Pauli errors is that 
the two-qubit gate error model can drift with time, causing the error amplification value $\alpha$ to drift during the experiment away from its intended value. To ensure that we are able to record and account for any potential drift, we interleaved arbitrary-angle randomized benchmarking experiments (an instance of direct RB \cite{proctor2019}) with our quantum simulation experiments to keep a running record of the average infidelity of the two-qubit gate. The average gate infidelity during each Trotter step of each quench is shown in Fig.~\ref{fig:zne_details}(c). Over the course of all of our experiments, we find that the $U_{ZZ}(\pm 0.5)$ two-qubit gate infidelity is $6(1)\times 10^{-4}$, $5.0(6) \times 10^{-4}$, and $7.1(8)\times 10^{-4}$ during the low-temperature, intermediate-temperature, and hydrodynamics quench experiments [marked as the dashed lines and shaded regions in Fig.~\ref{fig:zne_details}(c)]. The difference in gate performance for the experiments can be attributed to the experiments being performed at different times, with the intermediate-temperature quench in particular being taken most recently in January 2026 after many hardware improvements had been implemented. Using the law of total variance, the error bar for this gate infidelity is $\sqrt{\textrm{Var}(\{\mu_1,\ldots,\mu_M\}) + \frac{1}{M}\sum_j \sigma_j^2}$, where $\mu_j$ and $\sigma_j$ are the average gate infidelity and uncertainty on the average obtained in benchmarking experiment $j$ ($1 \leq j \leq M$), which accounts for both the drift in the means between experiments and the statistical variance inherent in each experiment.

In the inset of Fig.~\ref{fig:zne_details}(c), we show the gate infidelity divided by the original gate infidelities measured by cycle benchmarking, denoted as $\eta$. If the gate infidelity does not drift then $\eta=1$ should hold, however, we see clear drift in $\eta$ over the course of our experiments. We use $\eta$ in our ZNE calculations by rescaling our error amplification parameter
\begin{align}
\alpha' = \frac{\alpha - 1}{\eta} + 1, \label{eq:zne_alpha}
\end{align}
and using the rescaled $\alpha'$ in our fits. Note that there is uncertainty $\delta \alpha'$ in the quantity $\alpha'$ due to uncertainty in the measured gate infidelity obtained from benchmarking. 

\subsection{Zero-noise regression} \label{sec:znr}

Here we introduce zero-noise regression (ZNR), a technique for extracting improved statistical estimates for the zero-noise limit of observables. This method, inspired by zero-noise extrapolation, can be applied whenever error detection information is available for each shot of a quantum circuit. For our circuits, this error detection information is the number of leakage errors detected by the leakage detection gadget (see Sec.~\ref{sec:ld_implementation}). Other forms of error detection can also be used with ZNR, such as quantum error detection (possible in error detecting and error correcting codes), hardware-level error detection protocols, or symmetry-based error detection. In all forms, ZNR can significantly reduce the statistical uncertainty on the estimated zero-noise limit of an observable compared with the exponentially costly and non-scalable post-selection procedure typically performed with error detection.

\begin{figure}[!t]
\centering
\includegraphics[width=1.0\columnwidth]{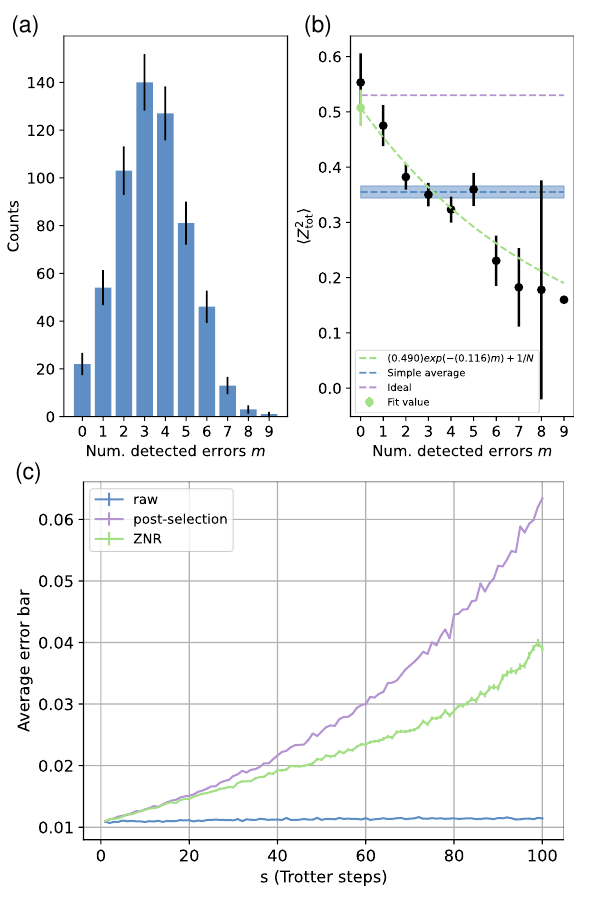}
\caption{An illustration of ZNR using a synthetic toy dataset. (a) A histogram of the number of detected errors $m$ in each shot. (b) The observable expectation value for fixed numbers of detected errors. The ZNR estimate is the green point at $m=0$ obtained from an exponential fit (green dashed line). (c) The error bars for the observables obtained by ZNR, post-selection, and raw data without error detection processing versus Trotter step for the toy dataset.
\label{fig:znr}}
\end{figure}

Figure~\ref{fig:znr} illustrates ZNR schematically with a synthetic toy dataset meant to roughly correspond to the low-temperature quench experiment performed in the main text. In this toy dataset, we generate 600 shots, with each shot having $m$ errors drawn from a binomial distribution with error probability per qubit of $p=(6 \times 10^{-4})s$ for Trotter step $s$. For simplicity, the observables are assumed to decay exponentially with number of detected errors $O_m=ae^{-bm} + 1/N$, with $a=0.53,b=0.152,N=56$. To simulate statistical noise in experiments, for each shot with $m$ errors we sample the observable from a Gaussian distribution $\mathcal{N}(O_m,\sigma)$ with $\sigma=2/\sqrt{N}$. The results in Fig.~\ref{fig:znr}(a,b) are for $s=100$ Trotter steps.

When performing error detection, one can bin each shot of a quantum circuit by how many errors were detected in that shot, as illustrated in Fig.~\ref{fig:znr}(a). When post-selecting on error-free data one only uses the $m=0$ bin, which for deep or wide circuits can contain a small fraction of the total data collected. When detecting errors, one can also compute observables binned by the number of detected errors, as in Fig.~\ref{fig:znr}(b). Similar to ZNE, one can perform a heuristic fit, such as an exponential fit, of the observable versus number of detected errors to obtain the zero-noise limit value of the observable.  Importantly, this fit uses \emph{all} of the data rather than only the $m=0$ bin data used in post-selection. Therefore, it can produce reduced statistical error bars on the zero-noise limit observable. This can be seen by comparing the ZNR estimate (red point) compared to the post-selection estimate (black point at $m=0$) in Fig.~\ref{fig:znr}(b). Note that similar to ZNE this procedure can produce a biased estimate of the zero-noise observable if the fit is poor. The benefit of ZNR is most pronounced in the large-error limit when the typical number of detected errors is significantly above zero and the post-selection signal vanishes, as shown in Fig.~\ref{fig:znr}(c). In the small-error limit, most of the data is in the $m=0$ bin and ZNR's output is approximately the output of post-selection. In our experiments, we are in the small-error limit so that ZNR produces essentially the same results as post-selection, though with slightly reduced error bars. Nonetheless, we still use ZNR since it is a more scalable procedure than post-selection.

In our implementation of ZNR, we use errors detected by the leakage detection gadget and fit the data to an exponential. This fit is performed using weighted least-squares regression, with weights $1/\sigma_m^2$ where $\sigma_m$ is the standard error on the observable with $m$ errors. Only $m$-bins containing more than 20 shots are used in the fit. If a fit fails, then the observable in the smallest $m$ bin is returned. We perform the leakage ZNR before the ZNE discussed in Sec.~\ref{sec:zne_postprocess} in order to mitigate the effects of leakage before the non-leakage-error extrapolation. We first perform leakage ZNR separately on the circuits without 2Q gate error amplification and those with the error amplification to measure leakage-mitigated observables. The unamplified and amplified leakage-mitigated observables are then used in the ZNE extrapolation to obtain the final processed observables that have both leakage and non-leakage gate errors mitigated.

Note that ZNR can be extended in a number of ways. For example, the spatial information of errors can be used to improve the fitting procedure. Similarly, error detection can be performed more often in time, which provides additional temporal information that can also improve the fitting. We expect this type of additional information can only improve the ZNR estimate and reduce its error bars relative to post-selection.

\section{Overview of experimental details} \label{sec:exp_detail}

In this work, we perform four sets of two-dimensional Ising model experiments:  (a) an $8 \times 7$ square lattice experiment demonstrating prethermalization at low-temperature and (b) intermediate-temperature (see Fig.~2 in the manuscript), (c) a $14 \times 4$ square lattice experiment demonstrating emergent hydrodynamics (see Fig.~3 in the manuscript), and (d) an antiferromagnetic $9 \times 6$ triangular lattice experiment demonstrating an emergent gauge theory (see Fig.~4 in the manuscript). Here we provide a high-level summary of the resources used in these experiments and how error bars were computed, using the error mitigation strategies outlined in the previous sections.

In experiments (a),(b), and (c), $S=600$ shots are executed at each Trotter step. These shots are split among two types of circuits used in zero-noise extrapolation (ZNE), unamplified circuits with no Paulis inserted and noise-amplified circuits with Paulis inserted. For the unamplified (noise-amplified) circuits, we execute $rS$ ($(1-r)S$) shots on the quantum computer, with different $r$ (and noise amplification rates $\alpha$) chosen at each Trotter step as shown in \fref{fig:zne_details}. For both unamplified and noise-amplified circuits, we execute many random circuits (see \fref{fig:zne_details}(b) inset) that have different patterns of random Paulis used for the randomized compiling (see \sref{sec:rc_implementation}). For noise-amplified circuits, each shot has a different random realization of inserted Paulis used for ZNE (see \sref{sec:znr}). Before, after, and during the Trotter steps, benchmarking circuits are run to keep track of 2Q gate infidelity in order to properly account for drifts in the effective ZNE error amplification rate $\alpha'$ (see \eref{eq:zne_alpha}). Observables presented in Fig.~2~and~3 in the manuscript also utilize the zero-noise regression (ZNR) technique, which is used to mitigate leakage errors by using a circuit-level leakage detection gadget (see \sref{sec:ld_implementation}). Experiment (a) is performed on both the H2-1 and H2-2 quantum computers, (b) only on H2-2, and (c) only on H2-1.

In experiment (d), performed entirely on H2-2, we execute our simulations slightly differently. We use $S=200$ shots at each Trotter step and use standard leakage post-selection instead of leakage ZNR (without attempting to remove false positive leakage detections; see \sref{sec:ld_implementation}). We again use the same ZNE procedure, though for these circuits we fix the shot allocation to $r=0.5$ and noise amplification to $\alpha=4$ for all Trotter steps for simplicity (even though this is suboptimal). We randomly interleave arbitrary-angle randomized benchmarking circuits with all of the shots at different Trotter steps, measuring an average gate infidelity of $3.3\times 10^{-4}$ and an effective error amplification rate of $\alpha'=8.2$ for all Trotter steps. We use groups of 5 shots for each randomized compiling circuit realization and again have each shot of the noise-amplified circuits use a different random circuit realization for ZNE.

All observable error bars $\delta O$ reported in this work are obtained by bootstrap resampling. In bootstrap resampling, the data set of interest is sampled with replacement many times to generate many resampled data sets with the same size as the original data set. An observable can then be computed for each resampled data set and the standard deviation of that observable among the resampled values is an estimate of the standard error on the mean for the observable in the original data set. Computing certain observable can require non-trivial processing, such as fitting in the case of observables computed using ZNE or ZNR or for the measurement of the diffusion constant shown in the inset of Fig.~3(b) in the manuscript. In these cases, the entire processing is performed on each resampled data set in the bootstrap resampling to obtain error bars that accurately reflect the propagation of errors through the multi-step workflow.

In addition, for an observable $O$ computed using ZNE in experiments (a) and (b), that observable's bootstrapped error bar is modified to 
\begin{align*}
\delta O' =  \textrm{max} &\left\{\sqrt{|O(\alpha') - O(\alpha'-\delta\alpha')|^2 + [\delta O(\alpha'-\delta\alpha')]^2},\right. \\
&\left.\sqrt{|O(\alpha') - O(\alpha'+\delta\alpha')|^2 + [\delta O(\alpha'+\delta\alpha')]^2}\right\} .
\end{align*}
This error bar accounts for the uncertainty in the ZNE rescaling parameter $\alpha'$, measured from benchmarking data (see discussion in \sref{sec:zne_postprocess} above \eref{eq:zne_alpha}).

\section{Finite-temperature results} \label{sec:thermal_observables}

To gain insights into the dynamics at intermediate times, we employ the Floquet theory for prethermalization~\cite{KUWAHARA201696, HO2023169297}. For a high-frequency Floquet drive $\Omega = 2\pi/dt \gg J$, with $J$ the typical energy scale of the Hamiltonian terms, there exists an effective Hamiltonian $H_\mathrm{eff}$ which is approximately conserved for an exponentially long time scale $\sim \exp(\Omega/J)$. Heating is strongly suppressed for these times. Generically, the time evolution with this effective Hamiltonian will lead to a thermal state that locally looks like $\rho_\mathrm{pre} \sim e^{-\beta_\mathrm{eff}H_\mathrm{eff}}$. The inverse effective temperature $\beta_\mathrm{eff}$ is determined by the energy of the initial state, such that $\mathrm{Tr}(\rho_\mathrm{pre} H_\mathrm{eff})=\bra{\psi(t)}H_\mathrm{eff}\ket{\psi(t)} \approx \bra{\psi(0)}H_\mathrm{eff}\ket{\psi(0)}$. Moreover, if the Hamiltonian for the Floquet drive only consists of short-ranged terms, then also local observables are determined by $H_\mathrm{eff}$, e.g. for a local $A$: $\bra{\psi(t)}A\ket{\psi(t)} \approx \bra{\psi(0)}e^{iH_\mathrm{eff}t}Ae^{-iH_\mathrm{eff}t}\ket{\psi(0)}$ which approaches the thermal value $\mathrm{Tr}(\rho_\mathrm{pre} A)$ for long enough times.

The effective Hamiltonian can be computed explicitly with the Floquet-Magnus expansion in the limit of large frequencies~\cite{Bukov_2015, KUWAHARA201696, HO2023169297}:
\begin{equation}
    H_\mathrm{F} = \sum_{n=0}^\infty H^{(n)} dt^n.
\end{equation}
In general, the Floquet-Magnus expansion yields an asymptotic series, that (in the thermodynamic limit) diverges due to the non-perturbative nature of heating. Nevertheless, computing the first few terms of the series gives a good approximation of the effective Hamiltonian for times when heating is suppressed.

The zeroth order Floquet-Magnus Hamiltonian is given by the time-averaged Hamiltonian $H^{(0)}=1/dt\int_0^{dt} dt_1 H(t_1)=H_{ZZ} + H_X$. Since we use a second-order Trotter scheme in the Floquet drive, $H^{(1)}$ is vanishing. The second-order Floquet-Magnus Hamiltonian can be computed from nested commutators~\cite{Blanes_2009}:
\begin{align}
    H^{(2)} =& \frac{1}{n!dt} \int_0^{dt} dt_1\int_0^{t_1}dt_2 \int_0^{t_2} dt_3\Big( [H(t_1), [H(t_2), H(t_3)]] \nonumber \\ &+ [H(t_3), [H(t_2), H(t_1)]] \Big) \nonumber \\
    = & \frac{dt^2h^2J}{3} \sum_{\braket{ij}}Z_i Z_j - \frac{dt^2h^2J}{3} \sum_{\braket{ij}} Y_i Y_j  \nonumber \\ &- \frac{4dt^2J^2h}{3}\sum_{i} X_i -\frac{2dt^2 hJ^2}{3} \sum_{\braket{imj}} Z_iX_mZ_j, \label{eq:Magnus_Hamiltonian}
\end{align}
where the last sum is taken over all three neighboring sites connected by bonds $(i,m)$ and $(m,j)$.
For $J<0$, the second-order correction strengthens the ferromagnetic term and weakens the transverse field, therefore pushing the quench more towards the ordered phase.

To evaluate the thermal expectation values $\braket{Z^2_\mathrm{tot}}=\mathrm{Tr}(\rho_\mathrm{pre} Z^2_\mathrm{tot})$, we employ a matrix product state (MPS) purification ansatz~\cite{Verstraete2004_mpdo, Barthel2009}. There, the physical Hilbert space $\mathcal{H}^p$ is enlarged by an ancilla Hilbert space $\mathcal{H}^a$ of the same dimension. Then a physical density matrix $\rho$ can be obtained by tracing out the ancilla degrees of freedom of a pure state $\ket{\Psi} \in \mathcal{H}^p \otimes \mathcal{H}^a$:
\begin{equation}
    \rho = \mathrm{Tr}_a\ket{\Psi} \bra{\Psi}.
\end{equation}

\begin{figure}[!t]
\centering
\includegraphics[width=1.0\columnwidth]{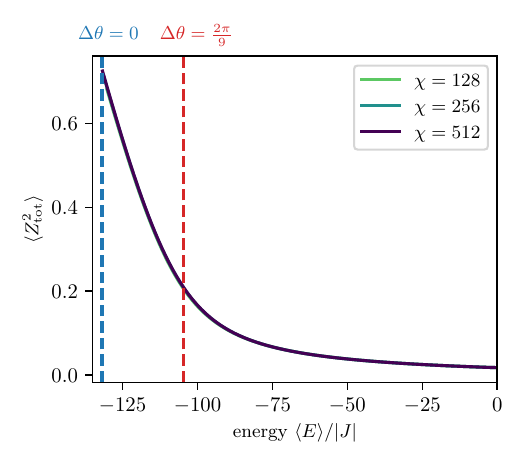}
\caption{Thermal expectation values for $\braket{Z^2_\mathrm{tot}}$ from MPS purification. Starting from an infinite temperature state ($\braket{E}=0$), we apply an imaginary time evolution with $H_\mathrm{eff}$ according to Eq.~\eqref{eq:Magnus_Hamiltonian} and parameters as in the main text, $J<0, h=2|J|, dt=0.25 |J|$ on a $8\times7$ torus. The dashed lines represent energies of the initial state for the low $\Delta \theta = 0$ and intermediate temperature $\Delta \theta = 2\pi/9$  quenches. Curves for different bond dimensions are on top of each other.
\label{fig:thermal_purification}}
\end{figure}

In practice, we start with a state at infinite temperature. In the purification picture, this is a product state of singlets, entangling the physical and ancilla degrees of freedom. We then cool down this state to finite temperatures with an imaginary time evolution $\ket{\Psi(\beta)} = e^{-\beta H_\mathrm{eff}/2}\ket{\Psi(\beta=0)}$ to obtain $\rho(\beta) \sim e^{-\beta H_\mathrm{eff}}$. Since we are simulating a 2D system with MPS, we need to include long-range terms in the matrix product operator (MPO) representation of the effective Hamiltonian. Therefore, we use the $W^\mathrm{II}$ method~\cite{Zaletel2014} to perform the imaginary time evolution, where $e^{-\delta H_\mathrm{eff}}$ is efficiently written as an MPO. Here we consider the same parameters as in the main text.
To directly read off the expectation values for the initial states considered in the main text, we plot $\braket{Z^2_\mathrm{tot}}$ as a function of the energy $\braket{E}$; see Fig.~\ref{fig:thermal_purification}. The results are converged already for bond dimensions of up to $\chi=512$.

\begin{figure}[!t]
\centering
\includegraphics[width=1.0\columnwidth]{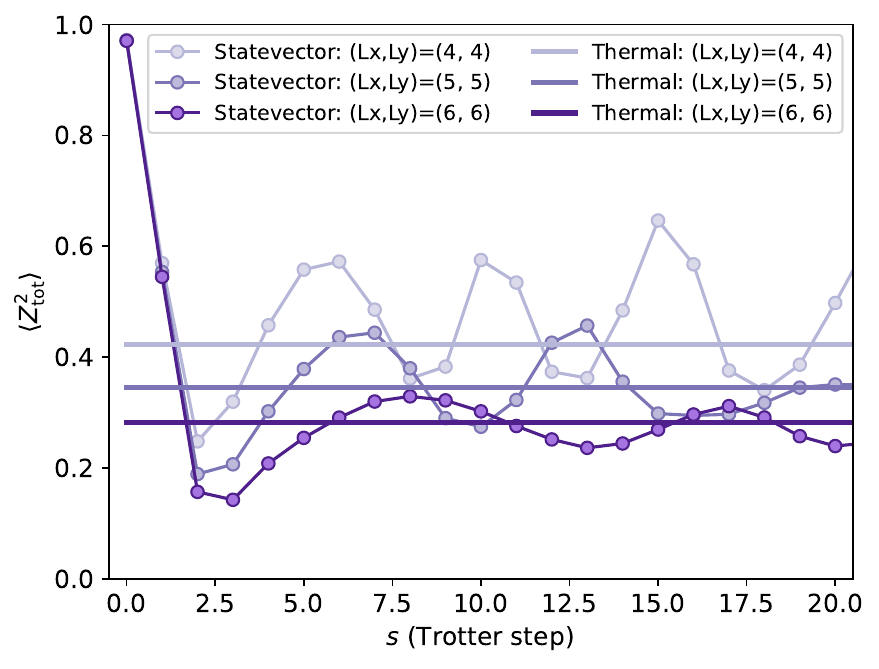}
\caption{Exact time evolution (circles) compared to thermal results using an MPS purification ansatz (horizontal solid lines) for various system sizes. Each curve uses the same quench parameters $(h/|J|,|J|dt,\Delta\theta)=(2,0.25,2\pi/9)$ [the same quench parameters as Figs.\,2(d,e) of the manuscript]. For these quench parameters (and for the system sizes accessible via full state vector simulation), we see that the late time dynamics tends to equilibrate close to the thermal value associated with the energy of the initial state with respect to the perturbative Floquet-Magnus Hamiltonian.
\label{fig:QMC_dynamics_comp}}
\end{figure}

For the Floquet time evolution, we expect that $\braket{Z^2_\mathrm{tot}}$ approaches the thermal values in the prethermal regime. To confirm this, we compare the dynamics for small system sizes, accessible with exact statevector simulations, to the thermal values from the MPS purification; Fig.~\ref{fig:QMC_dynamics_comp}. We find oscillations around the thermal expectation values. These finite-size oscillations decrease with increasing system size. Similarly, for the $7\times8$ torus considered in the main text, the thermal value provides a good estimate for the values approached by the Floquet time evolution; see Fig.\,2 in the main text.

\section{MPS simulations} \label{supp:MPS simulations}

In this section we present the classical simulation results of the digitized TFIM obtained with 1D matrix-product-state (MPS)~\cite{fannes1992finitely, ostlund1995thermodynamic, vidal2003efficient} methods.
The results are accompanied by a quantitative analysis of the extrapolation method used, from which we conclude that, for the observable of interest [$Z_{\rm tot}^2$ from Eq.\,(4) of the manuscript], extrapolating in simulation fidelity is more reliable than extrapolating in bond dimension -- an observation that likely holds more generally beyond the scope of this work.  Furthermore, based on comparisons to exact simulations for smaller systems, we include a heuristic confidence interval on the extrapolations. The zero-noise extrapolated expectation values obtained from H2 are found to lie within this confidence interval at the early times for which it can be reliably established (i.e., before the MPS fidelity becomes so low that the extrapolations are completely unpredictable, up to $s\approx 8$ for the intermediate quench).

The simulations of the digitized TFIM are performed at the quantum-circuit level, taking the quantum circuit implementing Trotterized time evolution as a starting point, and using a DMRG~\cite{white1992,schollwock2011,ayral2023density} routine to optimize the MPS tensors. We use Fermioniq's tensor network emulator \href{https://www.fermioniq.com/ava}{Ava} running on a single NVIDIA Grace-Hopper GH200 to perform the MPS simulations.

\subsection{Simulation method}

\subsubsection{Setup}

The input for each simulation is the initial state $\ket{\Psi_0} = \ket{\Psi}$ from Eq.\,(3) of the manuscript, and a circuit $C$ implementing (second-order) Trotterized time-evolution of the TFIM with Ising coupling $J=-1$ and transverse field $h = 2$ using a Trotter step size of $dt = 0.25$. 
In order to perform the MPS simulation, we first decompose the circuit $C$ into a sequence of subcircuits: $U_1, U_2, \ldots, U_K$ such that $C = U_K U_{K-1} \cdots U_1$, where for each $k \in \{1, \ldots, K\}$, $U_k$ is an $N$-qubit unitary that comprises several gates. 
The $k$-th intermediate state obtained after applying the first $k$ subcircuits exactly is denoted by
\begin{equation*}
    \ket{\Psi_k} = U_k U_{k-1} \cdots U_1 \ket{\Psi_0} \, .
\end{equation*}
The output of the simulation is a sequence of MPSs $\{\ket{\Phi_k}\}_{k=0}^{K}$ that are (ideally good) approximations to the exact intermediate states $\{\ket{\Psi_k}\}_{k=0}^{K}$. 

The MPSs $\{\ket{\Phi_k}\}_{k=0}^{K}$ are obtained as follows. 
For $k=0$, we have $\ket{\Phi_0} = \ket{\Psi_0}$ because $\ket{\Psi_0}$ is itself a product state. 
Next, at each step $k>0$, given $\ket{\Phi_{k-1}}$ we use the  DMRG routine from~\cite{ayral2023density} to obtain $\ket{\Phi_k}$ by maximizing the $k$-th \emph{partial fidelity} $f_k$ given by
\begin{equation*}
f_k = \left| \bra{\Phi_k} U_k \ket{\Phi_{k-1}}  \right|^2 \, ,
\end{equation*}
under the constraint that $\bra{\Phi_k}\Phi_k\rangle=1$. 
We proceed subcircuit by subcircuit until the final MPS $\ket{\Phi_K}$ is obtained, which is our approximation to the state $\ket{\Psi_K}$ prepared by the circuit $C$ (applied to the initial state $\ket{\Psi_0}$). 
When decomposing the circuit $C$ into subcircuits $U_1, U_2, \ldots, U_K$, we make sure that the subcircuit splitting respects the Trotter decomposition of the circuit~\footnote{In the sense that for every Trotter time $s$, there is a $k\in \{1, \ldots, K\}$ such that $U_k U_{k-1} \cdots U_1$ implements the first $s$ Trotter steps.}, so that we can collect all the required expectation values at intermediate Trotter steps within one single simulation (by computing expectation values of the relevant intermediate states).

\begin{figure*}
    \centering
    \includegraphics[width=1.\linewidth]{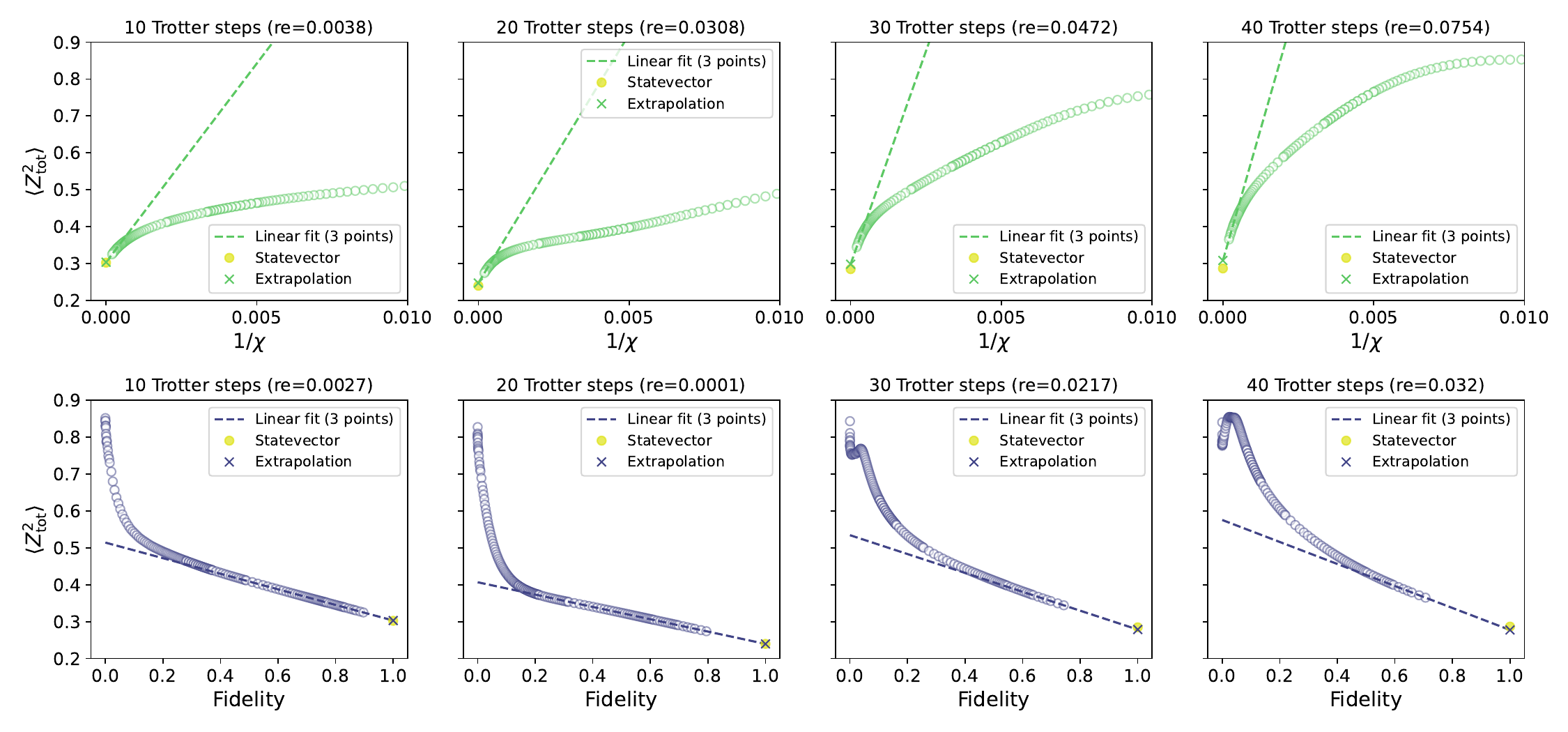}
    \caption{Expectation value of $Z^2_{\rm tot}$ as a function of one-over-bond-dimension $1/ \chi$ (top) or fidelity $F_{\text{sim}}$ (bottom) plotted for several values of the bond dimension for a $6 \times 6$ lattice at $\Delta \theta = 2\pi / 9$ (with the relative error (re) on the extrapolation displayed in the plot titles). To obtain an estimate of the true expectation value, we extrapolate  one-over-bond-dimension to zero or fidelity to one using a straight line fit through the three highest bond-dimension data points, resulting in the extrapolated expectation value (cross) that can be compared to the exact expectation value (yellow dot).}
    \label{fig:mps-fid-vs-one-over-bdim}
\end{figure*}

\subsubsection{Fidelity}
If every subcircuit is simulated with partial fidelity $f_k = 1$, then $\ket{\Phi_k} = \ket{\Psi_k}$ for every $k$, in which case the simulation is exact. 
When any of the partial fidelities is strictly less than one (as is often the case), the simulation becomes approximate.
The accuracy of the simulation at step $k$ (which signifies how accurately the first $k$ subcircuits have been simulated) is given by the $k$-th \emph{true fidelity}
\begin{equation*}
F_{\text{true}}(k) = \left| \bra{\Phi_k} \Psi_{k}\rangle  \right|^2 \, .
\end{equation*}
It is in general not possible to compute this quantity exactly. 
However, the $k$-th \emph{simulation fidelity}, defined by
\begin{equation*}
F_{\text{sim}}(k) = \prod_{k'=1}^k f_{k'}
\end{equation*}
can be computed exactly. 

The simulation fidelity is not guaranteed to be a good proxy for the true fidelity, nor a lower bound thereof.
% \footnote{We do have the following recursive lower bound on $F_k$ given by $\sqrt{F_k} \geq \sqrt{F_{k-1} f_k} - \sqrt{1 - F_{k-1}} \sqrt{1 - f_k}$. In practice, this bound is too loose to be informative.}.
However, it turns out that the simulation fidelity approximates the true fidelity very well in practice~\cite{zhou2020,ayral2023density,thompson2025}.
Regardless, the simulation fidelity will only be used for estimating expectation values of observables (by means of extrapolation), which will then be compared to the expectation values obtained from the actual quantum hardware.
In the next section, we argue by means of comparison to exact results for small systems that the simulation fidelity is a suitable quantity to perform reliable extrapolations in. 

In the remainder of the MPS section of the Appendix, \emph{fidelity} will refer to simulation fidelity.

\subsubsection{Extrapolations}

The fidelity can be controlled by varying the \emph{bond dimension} $\chi$, which determines both the amount of classical resources used for the MPS simulation as well as the amount of entanglement the MPS can capture. 
For highly entangling circuits (e.g.~those that generate volume-law states) an exponentially large (in qubit number) bond dimension is needed to capture the state prepared by the circuit exactly. 
Given limited classical computational resources, we use a maximum bond dimension that allows the simulations to be carried out in no more than a few days, and then attempt to extrapolate the obtained results to the $\chi \rightarrow \infty$ limit. 

% To change the figure to a pdf, just rename the extension (to pdf)
\begin{figure*}
    \includegraphics[width=2.08\columnwidth]{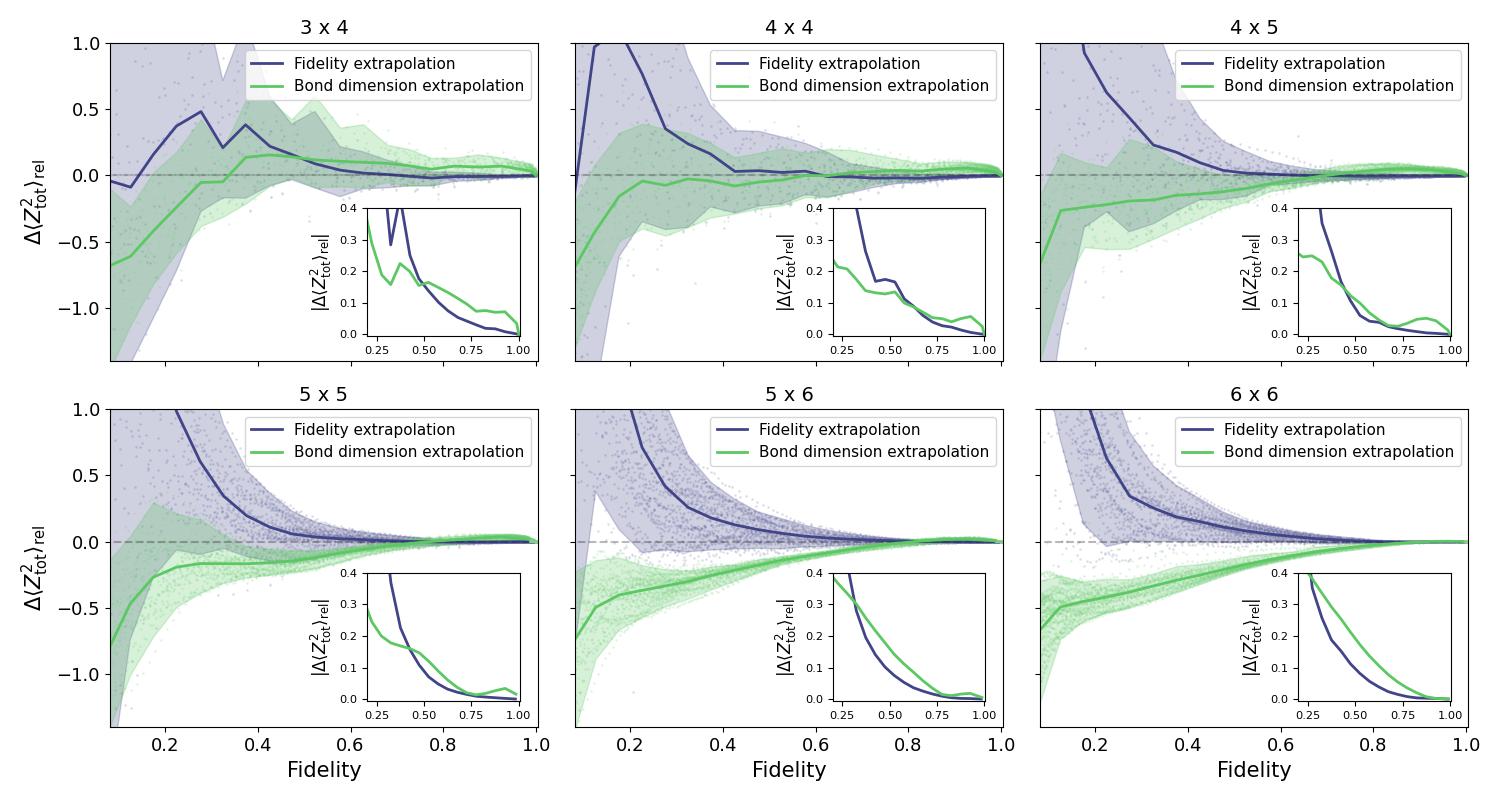}
    \caption{Relative error $\left(\langle Z_{\rm tot}^2 \rangle - \overline{\langle Z_{\rm tot}^2 \rangle}\right)/ \left| \langle Z_{\rm tot}^2 \rangle \right|$ between the exact value $\langle Z_{\rm tot}^2 \rangle$ and an estimate $\overline{\langle Z_{\rm tot}^2 \rangle}$ from either fidelity extrapolation (blue) or bond dimension extrapolation (green) plotted against fidelity for $\Delta \theta=2\pi/9$ with varying system size. \emph{Fidelity} here means the fidelity of the highest bond-dimension point used for the extrapolation. The main plots show all simulation results (individual dots) for 1 up to 40 Trotter steps together with the average relative errors (straight lines) obtained by binning the data points by fidelity using bins of width 0.05. The shaded areas signify a 5th - 95th percentile window within each fidelity bin. The embedded plots show the mean values of the absolute relative errors against fidelity.}
    \label{fig:mps-fid-vs-bdim-extrap}
\end{figure*}

In practice, we simulate the circuit $C$ acting on the initial state $\ket{\Psi_0} = \ket{\Psi(\theta)}$ for a range of (increasing) bond dimensions $\chi_1, \chi_2, \ldots, \chi_{m}$.
For every $k \in \{1, \ldots, K\}$, this yields a sequence $\{\ket{\Phi_k^{\chi_j}}\}_{j=1}^m$ of MPSs that represent increasingly more accurate approximations to the true intermediate state $\ket{\Psi_k}$.
To estimate the expectation value $\bra{\Psi_k} O \ket{\Psi_k}$ of an observable $O$ at some intermediate step $k$ corresponding to a Trotter time step of interest, we then perform an extrapolation on the data points $\{\bra{\Phi_k^{\chi_j}} O \ket{\Phi_k^{\chi_{j}}}\}_{j=1}^m$.

Figure~\ref{fig:mps-fid-vs-one-over-bdim} (top) shows the expectation value of $Z_{\rm tot}^2$ as a function of one over the bond dimension ($1/\chi$) with respect to the state obtained after $s = 10$, $20$, $30$ and $40$ Trotter steps at $\Delta \theta = 2\pi / 9$ for a $6 \times 6$ lattice.
As the bond dimension increases to infinity, the fidelity increases monotonically to one.
\fref{fig:mps-fid-vs-one-over-bdim} (bottom) shows the same quantities as above as functions of fidelity.
In both top and bottom, the extrapolated expectation value of $Z_{\rm tot}^2$, marked by the cross, is obtained via a straight-line fit through the three highest bond-dimension / fidelity data points (we always use a straight line fit through the three largest bond dimension / fidelity data points for extrapolations).
For reference, the exact expectation value (obtained with a statevector simulation) is given by the yellow dot.
Both top and bottom figures have been generated using the same list of bond dimensions.
Comparing both extrapolation methods in \fref{fig:mps-fid-vs-one-over-bdim}, we observe that fidelity extrapolation yields a more accurate result.

By looking at the dataset of all simulations run for $\Delta \theta = 2\pi / 9$ on system sizes up to $6 \times 6$, we find that the observation from \fref{fig:mps-fid-vs-one-over-bdim} holds more generally: extrapolating fidelity $F_{\text{sim}} \rightarrow 1$ provides more reliable extrapolations than $1/\chi \rightarrow 0$ does~\footnote{The fact that fidelity allows for more accurate extrapolations than bond dimension does make intuitive sense: the fidelity contains information on the accuracy of the simulation, whereas the bond dimension merely specifies the amount of classical resources used.}.
This can be seen in \fref{fig:mps-fid-vs-bdim-extrap}, which shows the relative error obtained by either fidelity or bond dimension extrapolation compared to the exact value of $\langle Z_{\rm tot}^2 \rangle$ obtained via a statevector simulation at $\Delta \theta = 2\pi / 9$ for varying system sizes.
Every single point in the scatter plots of \fref{fig:mps-fid-vs-bdim-extrap} represents a set of simulations for a list of bond dimensions $\chi_1, \chi_2, \ldots, \chi_m$, where the x-value is the fidelity obtained by the (highest-bond-dimension) $\chi_m$ simulation, and the y-value is the relative error $\left(\langle Z_{\rm tot}^2 \rangle - \overline{\langle Z_{\rm tot}^2 \rangle}\right)/ \left|\langle Z_{\rm tot}^2 \rangle \right|$ between the exact value $\langle Z_{\rm tot}^2 \rangle$ and an estimate $\overline{\langle Z_{\rm tot}^2 \rangle}$ obtained using either $1/\chi$- (green) or fidelity- (blue) extrapolation via a straight line fit through the $\chi_{m-2}$, $\chi_{m-1}$ and $\chi_m$ points.
The solid lines in \fref{fig:mps-fid-vs-bdim-extrap} represent the mean of the relative errors obtained by binning the data points according to their fidelities, using a bin width of 0.05.
All points in the shaded areas lie within the 5th - 95th percentile window of the errors obtained within each bin. 
The embedded plots in \fref{fig:mps-fid-vs-bdim-extrap} show the means of the \textit{absolute} relative errors as a function of fidelity. 

From the main plots in \fref{fig:mps-fid-vs-bdim-extrap} we can see that extrapolations in one over bond dimension tend to overestimate the value of $\langle Z_{\rm tot}^2 \rangle$, and extrapolations in fidelity tend to underestimate its value. From the embedded plots in \fref{fig:mps-fid-vs-bdim-extrap} we can conclude that for simulations for which the highest fidelity data point used for the straight line fit has a fidelity of more than $\sim 0.65$ (i.e.~to the right of where the blue and green lines cross), fidelity extrapolation is on average more accurate than bond-dimension extrapolation. 
Moreover, we find that as the system size grows, this cross-over point moves leftward, suggesting that fidelity extrapolation becomes more accurate with increasing system size compared to bond dimension extrapolation.
Left of where the blue and green lines cross, bond-dimension extrapolation actually performs better than fidelity extrapolation. However, in this regime, the extrapolations themselves are inaccurate (as can be seen from the relative errors). Given the fidelities obtained for the quench simulations shown in Figs.\,(\ref{fig:mps-theta-pi-over-12-trotter},\ref{fig:mps-theta-pi-over-36-trotter}), we believe fidelity extrapolation likely performs best for the quenches of interest, and so have obtained all extrapolated estimates in this manner.

\subsection{Results}
\begin{table}[]
    \centering
    \begin{tabular}{l|l|l|l|l|l|l|l|l|l}
    $\chi$ & 256 & 512 & 1024 & 1500& 2048 & 2500 & 3000 & 3500& 4000 \\
    \hline
    $\Delta \theta = 0$ & 0.36 & 0.43 & 1.03 & 2.34 & 5.12 & 8.84 & 14.18 & 22.26 & 32.35 \\
    \hline
    $\Delta \theta = 2\pi / 9$ & 0.38 & 0.49 & 1.36 & 2.82 & 5.81 & 10.12 & 16.13 & 24.98 & 36.43 \\
    \end{tabular}
    \caption{Runtimes (in hours) for the entire 40-Trotter step simulation (including the computation of the expectation values of all single-site $Z$ and all two-site $ZZ$ observables at every Trotter step) for different values of the bond dimension $\chi$, obtained using Fermioniq's tensor network emulator \href{https://www.fermioniq.com/ava}{Ava} running on a single NVIDIA Grace-Hopper GH200.}
    \label{tab:mps-runtimes}
\end{table}

Having established the extrapolation method used for the estimation of expectation values, in this section we present our results for $N=56$ qubits for the low-temperature ($\Delta \theta=0$) and intermediate-temperature ($\Delta \theta=2\pi/9$) quenches. For both quenches we use the following list of bond dimensions: $\chi = 256$, $512$, $1024$, $1500$, $2048$, $2500$, $3000$, $3500$, $4000$. For every value of $\chi$, we perform a single simulation encompassing 40 Trotter steps, where we compute expectation values after every Trotter step.
The runtimes of the simulations can be found in Table \ref{tab:mps-runtimes}.

\subsubsection{Low-temperature quench: $\Delta \theta = 0$}

\begin{figure}
    \centering
    \includegraphics[width=1.0\linewidth]{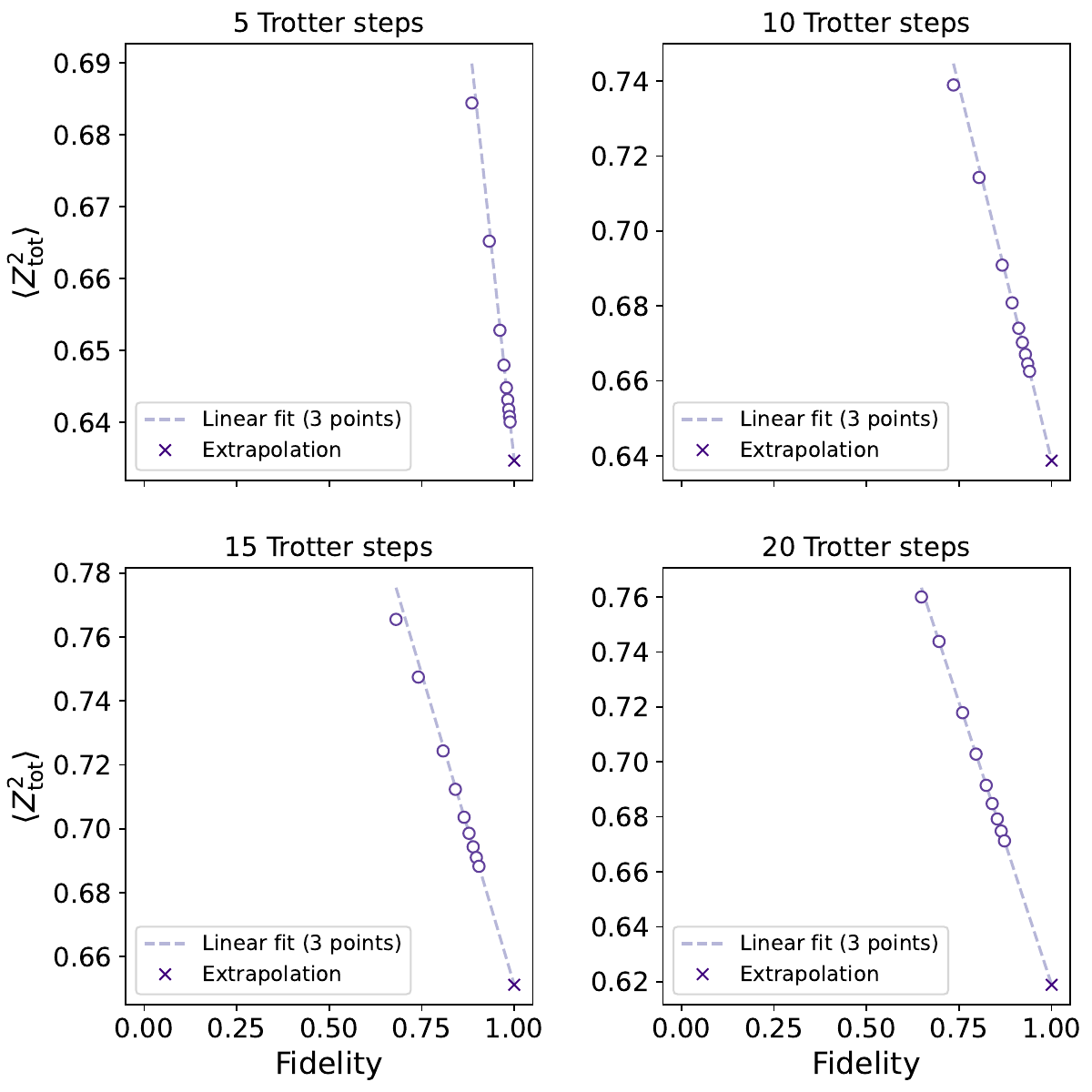}
    \caption{Expectation value of $Z_{\rm tot}^2$ as a function of fidelity for $s = 5$, $10$, $15$, $20$ Trotter steps for $\Delta \theta=0$ for fixed values of the bond dimension $\chi = 256$, $512$, $1024$, $1500$, $2048$, $2500$, $3000$, $3500$, $4000$ (circles) together with the extrapolated value (cross) obtained by a straight line fit through the $\chi = 3000$, $3500$, $4000$ data points.}
    \label{fig:mps-theta-pi-over-12-extrap}
\end{figure}

\begin{figure}
    \includegraphics[width=1.0\linewidth]{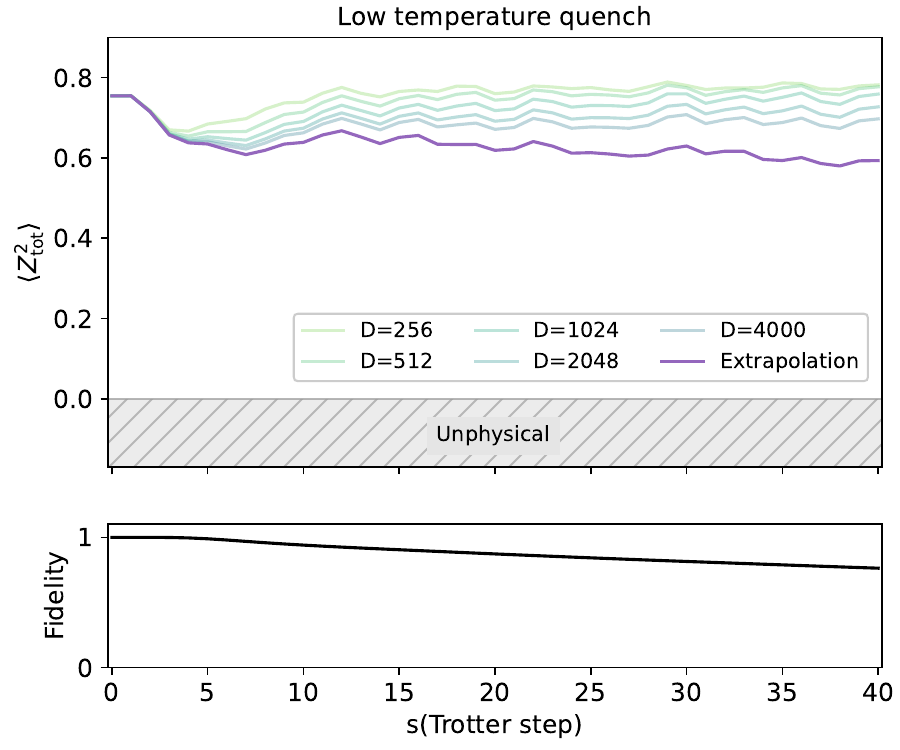}
    \caption{$\langle Z^2_{\rm tot} \rangle$ for $\Delta \theta = 0$ for a number of values of $\chi$ together with its (fidelity-to-one) extrapolated value (top) and fidelity of the corresponding highest-bond-dimension $\chi=4000$ MPS simulation (bottom) as a function of the number of Trotter steps.}
    \label{fig:mps-theta-pi-over-12-trotter}
\end{figure}

Figure~\ref{fig:mps-theta-pi-over-12-extrap} shows the expectation value of $Z_{\rm tot}^2$ as a function of fidelity for several Trotter step times ($s = 5$, $10$, $15$, $20$). As the number of Trotter steps increases, the fidelity of the simulations decreases (due to entanglement being built up by the circuit).
Figure~\ref{fig:mps-theta-pi-over-12-trotter} shows the expectation value of $Z_{\rm tot}^2$ for fixed values of $\chi$ together with the fidelity-to-one extrapolated value as a function of number of Trotter steps.
Because the highest bond dimension $\chi=4000$ simulations retain a relatively high fidelity for all Trotter times simulated, reliable extrapolations can be made up to $s=40$ Trotter steps. 

\subsubsection{Intermediate-temperature quench: $\Delta \theta = 2\pi/9$}

\begin{figure}
    \centering
    \includegraphics[width=1.0\linewidth]{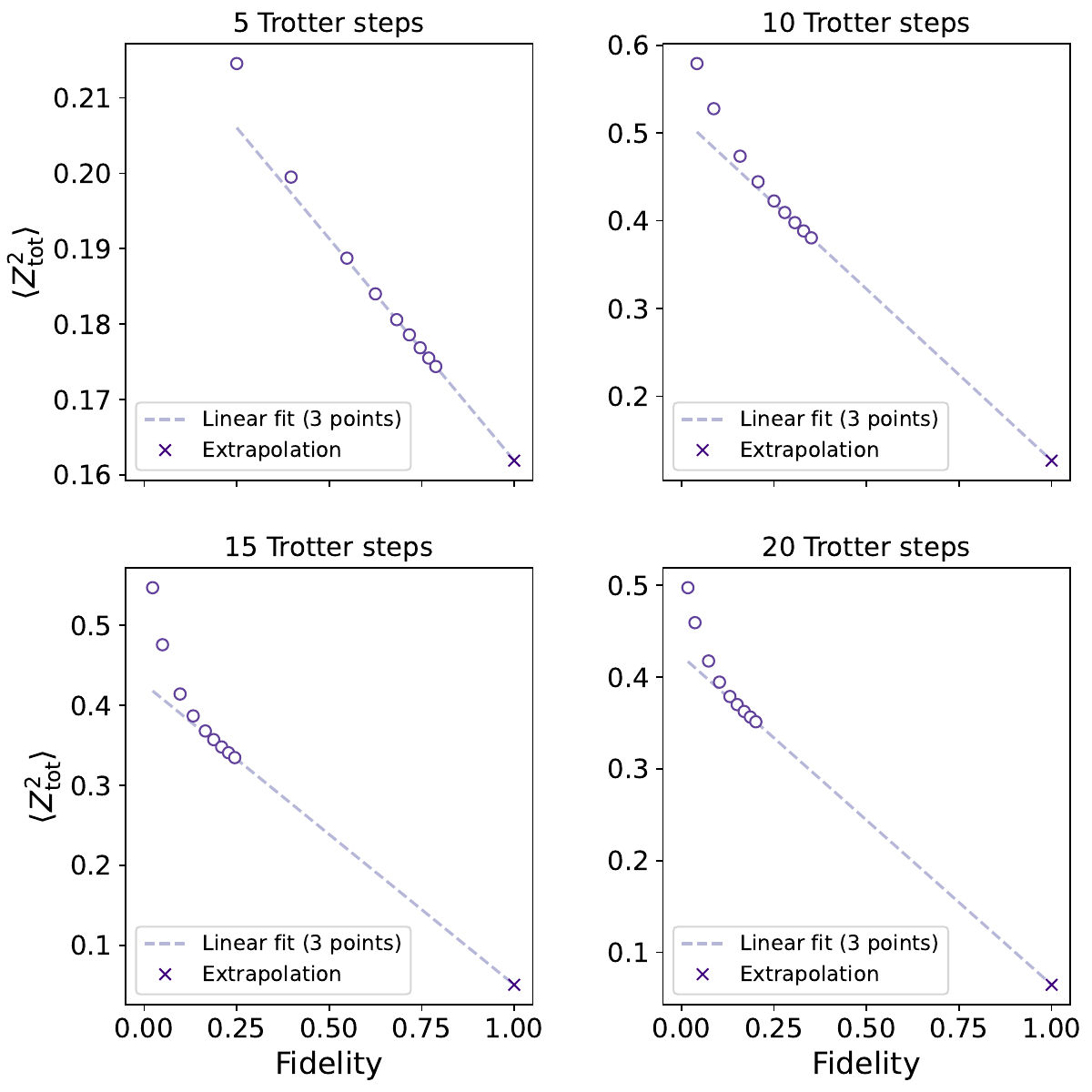}
    \caption{Expectation value of $Z_{\rm tot}^2$ as a function of fidelity for $s = 5$, $10$, $15$, $20$ Trotter steps for $\Delta \theta= 2\pi/9$ for fixed values of the bond dimension $\chi = 256$, $512$, $1024$, $1500$, $2048$, $2500$, $3000$, $3500$, $4000$ (circles) together with the extrapolated value (cross) obtained by a straight line fit through the $\chi = 3000$, $3500$, $4000$ data points.}
    \label{fig:mps-theta-pi-over-36-extrap}
\end{figure}

Figure~\ref{fig:mps-theta-pi-over-36-extrap} shows the expectation value of $Z_{\rm tot}^2$ as a function of fidelity for several Trotter step times ($s = 5$, $10$, $15$, $20$) for $\Delta \theta = 2\pi/9$. 
Compared to the low-temperature quench, the fidelity of the simulations decreases much faster with increasing number of Trotter steps. Still, for a small number of Trotter steps, the extrapolations can be deemed reliable. 
However, from $s \geq 10$, we see that the extrapolation has to do quite some work. 
Moreover, in \fref{fig:mps-theta-pi-over-36-trotter}, which shows the expectation value of $Z_{\rm tot}^2$ for fixed values of $\chi$ (solid lines) as well as the fidelity-to-one extrapolated value (solid-dashed purple line) as a function of number of Trotter steps, we observe that at $s = 16$ the extrapolated value becomes negative, which is unphysical for the observable $Z_{\rm tot}^2$ and therefore a clear sign that the extrapolation is no longer reliable beyond this time step.

Using data obtained from simulations on smaller systems, we can add a (heuristic) notion of accuracy to the extrapolated values. 
In \fref{fig:mps-fid-vs-bdim-extrap} we observe a clear relationship between the fidelity (of the highest bond dimension simulation performed) and the relative error on $\langle Z_{\rm tot}^2 \rangle$.  
Focusing on the largest $6 \times 6$ simulations, which are closest to and therefore hopefully representative of the $N=56$ qubit simulations, we obtain an average relative error on the extrapolation as a function of fidelity (given by the blue line in the $6 \times 6$ plot of \fref{fig:mps-fid-vs-bdim-extrap}), together with a 5th - 95th percentile window around this same error (visualized by the shaded blue region in the $6 \times 6$ plot of \fref{fig:mps-fid-vs-bdim-extrap}).

The shaded purple region in \fref{fig:mps-theta-pi-over-36-trotter} signifies a confidence interval around the extrapolated value of $\langle Z_{\rm tot}^2 \rangle$ based on the 5th - 95th percentile window of the relative errors of the $6 \times 6$ simulations.
The confidence region is not shown beyond $s=10$, because once the estimated errors become too large and unpredictable (at the low fidelities obtained) the confidence interval itself becomes less meaningful. The same extrapolated value and corresponding confidence region are also shown in Fig.\,2d of the manuscript.
Comparing to the zero-noise extrapolated data from H2 displayed in Fig.\,2d of the manuscript, we see that the zero-noise extrapolated data points consistently lie within the shaded purple confidence region.

\begin{figure}
    \includegraphics[width=1.0\linewidth]{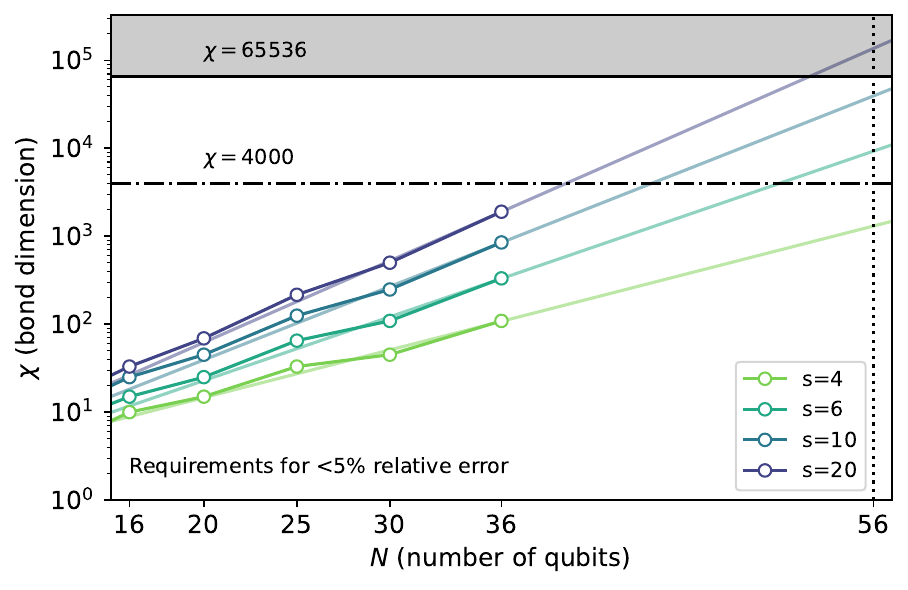}
    \caption{By comparing exact and MPS-based numerical simulations of the intermediate-temperature quench at $\Delta\theta = 2\pi/9$ for system sizes up to $6\times 6$, we can estimate the bond dimension of an MPS simulation necessary to determine (using extrapolation from the MPS data up to the reported bond dimension) observables for a $7\times 8$ system with a given accuracy.}
    \label{fig:err_thresholds}
\end{figure}

\begin{figure}
    \includegraphics[width=1.0\linewidth]{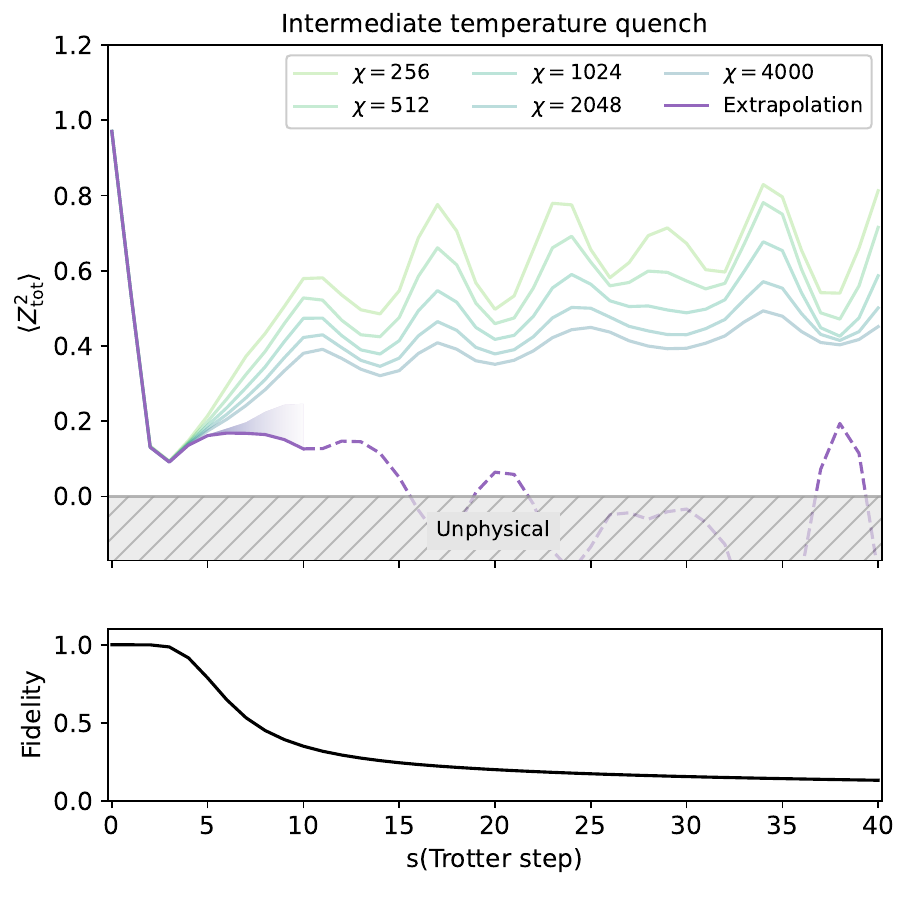}
    \caption{Simulation data for $\Delta \theta = 2\pi/9$ as a function of the number of Trotter steps. The top plot displays $\langle Z_{\rm tot}^2 \rangle$ for a number of values for $\chi$ (solid lines) together with its fidelity-to-one extrapolated value (solid and dashed purple line) and a corresponding confidence region (shaded purple area) -- for a detailed description see the main text. The bottom plot shows the fidelity of the highest-bond-dimension $\chi=4000$ MPS simulation. The confidence region is not shown once the fidelity drops below 0.4 (signified by the purple line changing from solid to dashed) because we are not confident in the confidence intervals at lower fidelities.}
    \label{fig:mps-theta-pi-over-36-trotter}
\end{figure}

The above analysis lets us determine what bond dimension is required for ZTE to work within some threshold accuracy. By obtaining such results for various simulable system sizes (where the accuracy can be assessed), we can extrapolate bond dimension requirements for achieving a given accuracy on the $7\times 8$ intermediate quench reported in the manuscript.  Figure \ref{fig:err_thresholds} shows the result of such an extrapolation. Bond dimensions of more than $2^{16}=65,536$ are likely necessary to obtain less than $5\%$ relative error on the observable $\langle Z_{\rm tot}^2\rangle$ for all Trotter steps $s\leq 20$ (with $\chi =2^{16}$ being the largest bond dimension used in any simulation that we are aware of \cite{PRXQuantum.4.010317}).

\begin{figure}
    \includegraphics[width=1.0\linewidth]{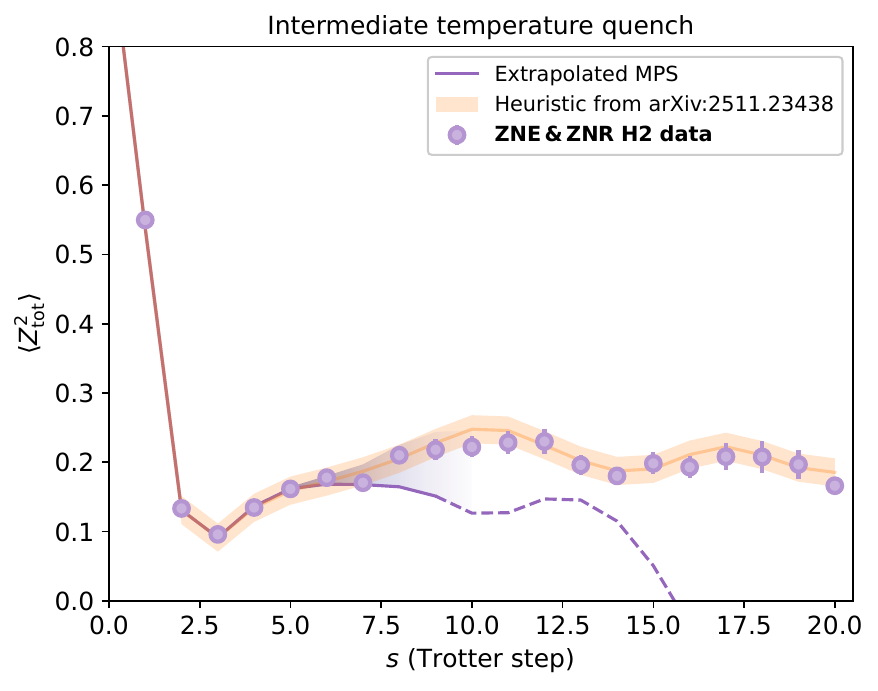}
    \caption{Comparison of our data for the intermediate-temperature quench (purple data points, see main text) to ZTE results (purple line) and the rescaled-MPS heuristics of Ref. \cite{mandra2025heuristic}. The orange region indicates an estimated worst-case confidence region (see Ref. \cite{mandra2025heuristic} for details).}
    \label{fig:google_mps}
\end{figure}

The above extrapolation procedure can be applied directly to MPS data at a large system size, and does not make use of any knowledge of the exact dynamics (with the exception of our confidence-region estimation, which is based on small-scale statevector simulations). Recently, we became aware of a heuristic method to rescale MPS observables in a system of a given size in a manner that is \emph{learned} from exact results at nearby (but small enough to be simulated exactly) system sizes \cite{mandra2025heuristic}.  Figure \ref{fig:google_mps} shows remarkably good agreement between this newly introduced heuristic---trained on exact diagonalization results for system sizes up to $(L_x,L_y)=(6, 8)$---and our data from H2 (rescaled using ZNE and ZNR).

\section{2D tensor network simulations}

\begin{figure*}[pth]
    \centering
    \includegraphics[width=0.95\linewidth]{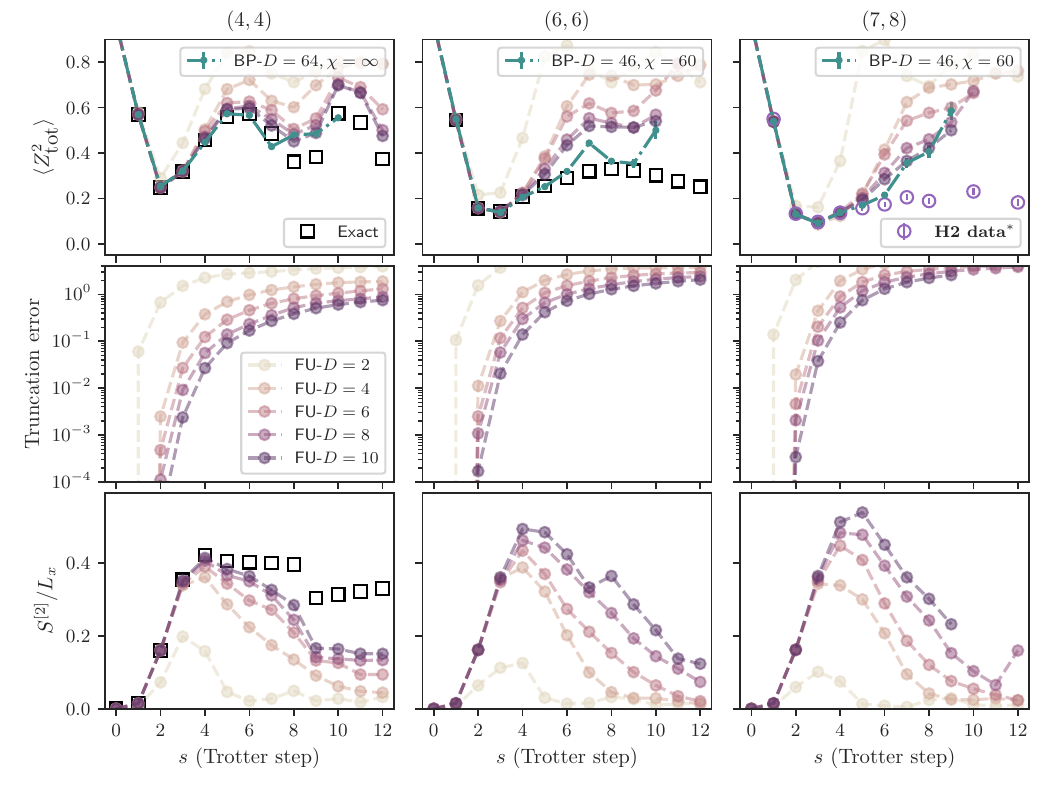}
    \caption{
    The expectation value of $\langle Z_\textrm{tot}^2 \rangle$ from the OBC-PEPS to simulate PBC states using the FU algorithm and the BP algorithm.
    In addition, we compute the total truncation error as the accumulated truncation error after each gate applications in the second row.
    In the third row, we compute the rescaled R\'enyi-$2$ entanglement entropy $S^{[2]} / L_x$ at half cut using an estimation constructed from the boundary MPO.
    The H2 data shown here is with ZNE + ZNR.
    }
    \label{fig:obc_peps_fu_bp}
\end{figure*}

In this section, we present classical simulation results based on 2D tensor network methods to benchmark their performance against H2 quantum hardware. Specifically, we consider the time evolution algorithms for quantum states in the Schr\"odinger picture using projected entangled pair states (PEPS)~\cite{niggemann1997quantum,nishino1998density,sierra1998density,verstraete2004renormalization} and the time evolution algorithms for observables in the Heisenberg picture using time-dependent projected entangled pair operators (PEPO)~\cite{liao2023simulation,Tomislav:2024}.

Consider the full tensor network diagram representing the expectation value of an observable for a state evolved by a given number of Trotter steps $s$,
\begin{equation}
    \langle O \rangle = \langle \Psi | \left(\prod_{t=1}^{s} \mathscr{U}_t^\dagger\right) O \left(\prod_{t=1}^{s} \mathscr{U}_t\right) | \Psi \rangle .
\end{equation}
PEPS and PEPO simulations correspond to two distinct contraction strategies of this equation, carried out in opposite directions: inwardly and outwardly, respectively. To achieve this, we leverage two canonical compression algorithms -- Belief Propagation (BP)~\cite{AlAr21} and the Full Update (FU) method~\cite{JoOrViVeCi08} -- to gradually evolve the PEPS/PEPO, and ultimately estimate the observable.

\subsection{PEPS simulation with the Full Update algorithm \label{sec: PEPS_FU}}

Two common time evolution algorithms based on the Suzuki-Trotter decomposition of the time evolution operator into local gates and local tensor updates are (i) the Full Update (FU)~\cite{JoOrViVeCi08} and (ii) Simple Update (SU)~\cite{JiWeXi08}. The two algorithms mainly differ in how the environmental tensor $N$ is approximated. The SU algorithm can be viewed as performing the approximate contraction of the environment with a product state, and the FU algorithms perform controlled approximate environment contractions either using the boundary matrix product operators (MPO)~\cite{verstraete2004renormalization} or the corner transfer matrix methods~\cite{baxter1968dimers,nishino1996corner,fishman2018faster}.

Here, we consider the combination of the FU algorithm with the boundary MPO contraction and a finite PEPS ansatz with \emph{open boundary conditions} (OBC), where each tensor is independent.
The FU algorithm mainly follows the description in the literature~\cite{LuCiBa14a,LuCiBa14b}. We perform the FU algorithm to evolve the PEPS under the application of two-qubit $U=e^{-idtH_{ZZ}}$ gates, while the single-qubit $e^{-idt H_X/2}$ gates can be applied on each tensor directly without increasing the bond dimension.
The boundary contraction is performed from the left and from the right. Then, a column of tensors is evolved with the gate set $\{U_{1,2},U_{2,3},\cdots,U_{L_y-1,L_y}\}$ one two-qubit gate after another using the reduced tensor update over two sites~\cite{LuCiBa14a,LuCiBa14b}.
To address the periodic boundary condition, we then apply the gate set $\{\{\text{SWAPS-upward}\},U_{L_y-1,L_y},\{\text{SWAPS-downward}\} \}$
where 
\begin{equation*}
    \{\text{SWAPS-upward}\} = \{\text{SWAP}_{1,2},\cdots,\text{SWAP}_{L_y-2,L_y-1}\}
\end{equation*}
is a sequence of SWAP gates to move the first site upward and 
\begin{equation*}
    \{\text{SWAPS-downward}\} = \{\text{SWAP}_{L_y-2,L_y-1}\cdots \text{SWAP}_{1,2} \}
\end{equation*}
a sequence of SWAP gates to move the first site back downward.
After we have updated all tensors in all columns, we then perform the same updates on all rows.

We can obtain an estimate of the truncation error by keeping track of the target tensor $\tilde{\Theta} = U\Theta^{[0]}$, where $U$ is a two-qubit gate and $\Theta^{[0]} = a^{[0]}_L a^{[0]}_R$ with $a^{[0]}_L$ and $a^{[0]}_R$ being the original reduced tensors.
We denote the updated reduced tensors resulting from the $i$-th iteration of the alternating least squares procedure as $a_L^{[i]}$ and $a_R^{[i]}$, which form the tensor $\Theta^{[i]} = a^{[i]}_L a^{[i]}_R$. The truncation error is approximated by $d =  \left(\tilde{\Theta} -\Theta^{[i]} \right)^\dagger N_\Theta \left( \tilde{\Theta} -\Theta^{[i]} \right)$ with the approximation arising from the approximate nature of the boundary contraction. By keeping track of the truncation error of each gate application, we can compute an estimate of the total truncation error as the accumulated sum of the errors.

The result of the FU simulation is given in Fig.~\ref{fig:obc_peps_fu_bp}. The computational complexity of the FU algorithm is $\mathcal{O}(D^6 \chi^2 + D^4 \chi^3)$ where $D$ is the PEPS bond dimension and $\chi$ is the bond dimension of the boundary MPO.  We manage to perform simulations up to $D=10$ and $\chi=128$ where the convergence is checked with $\chi=192$. We see that increasing the bond dimension $D$ consistently improves the result, reducing the deviation from the expected correct values.

We can compare the $\langle Z_\textrm{tot}^2 \rangle$ values to the exact state vector simulation results for small system sizes for all steps and to the H2 data with ZNE + ZNR for $7 \times 8$ system.
The results with the largest bond dimension $D=10$ start to deviate notably from Trotter step 5 onwards for all system sizes.
The deviation has a similar trend as MPS with respect to overestimating the $\langle Z_\textrm{tot}^2 \rangle$ value. 
Separating the system into two by a horizontal cut at $L_y/2$, the half-cut R\'enyi-2 entanglement entropy $S^{[2]}$ can be estimated with the reduced density matrix formed by the boundary MPO.
For states with 2D area-law entanglement, the rescaled entanglement entropies $S^{[2]}/L_x$ should remain constant with increasing system sizes.
In Fig.~\ref{fig:obc_peps_fu_bp}, we observe a similar convergence trend of entanglement entropies compared to $\langle Z_\textrm{tot}^2 \rangle$, where data converged nicely until Trotter step 3.
The entanglement entropies with the largest bond dimension $D=10$ also deviate notably from Trotter step 5 onwards for the $4 \times 4$ system.
The maximal values of the rescaled entanglement entropies for system sizes $6 \times 6$ and $7 \times 8$ exceed the maximal value for system size $4 \times 4$, indicating the entanglements are beyond area law.

The main reason for why the FU PEPS algorithm struggles is that it is restricted in practice to relatively low bond dimensions. While finite bond-dimension PEPS are often sufficient to describe ground states of gapped local Hamiltonians, and PEPS capture various physically relevant states with area-law quantum entanglement, this restriction makes it difficult to accurately represent the finite-energy density quenches studied here since they lead to volume-law entanglement as observed in Fig.\,2(b) of the manuscript.

One potential way forward is to utilize the ansatz in an unconventional way that goes to larger bond dimensions by performing approximate updates, e.g., SU. In the next section, we discuss the Belief Propagation update approach, which is a class of methods generalizing the SU approach.
Another effect we have to consider is that using an OBC-PEPS to simulate periodic boundary conditions (PBC) dynamics itself has a significant overhead in the bond dimension, as the correlations have to be carried from one of the open ends to the other.
A heuristic argument based on collapsing the PBC-PEPS to an OBC-PEPS suggests that it requires a bond dimension $D^2$ OBC-PEPS to represent the same state corresponding to a bond dimension $D$ PBC-PEPS.
Again, we discuss this issue and consider this effect of the ansatz in the coming section.

\begin{figure}[!t]
\centering
\includegraphics[width=1.0\columnwidth]{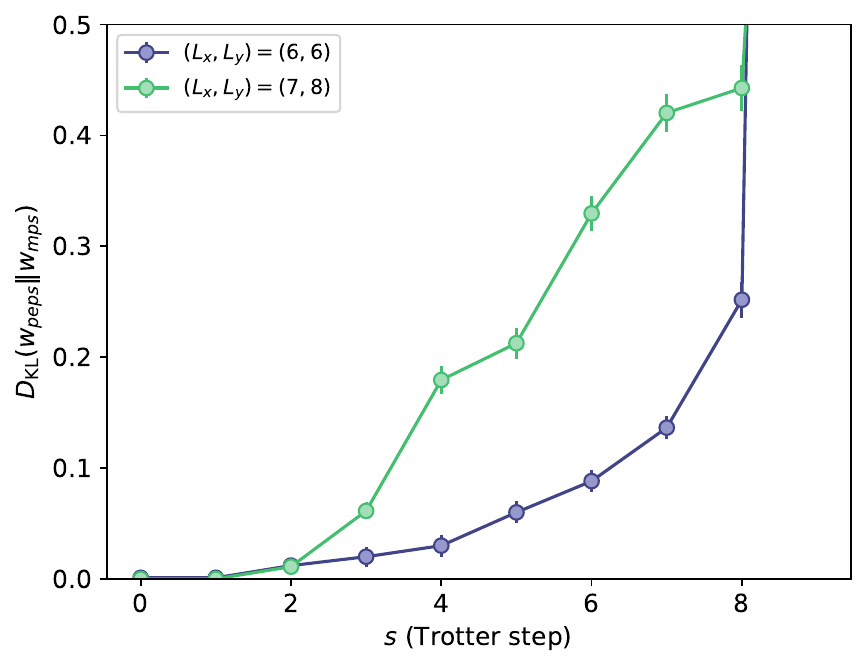}
\caption{ Kullback-Leibler divergence between PBC-PEPS with $D=20$ and MPS with $D=520$ for different system sizes. 
\label{fig:dkl}}
\end{figure}

\subsection{PEPS simulation with the Belief Propagation algorithm} \label{supp:PEPS_BP}

Belief Propagation (BP) algorithms~\cite{AlAr21, tindall2023gauging}, when combined with tensor network techniques, have emerged as a leading classical approach for simulating quantum dynamics, as demonstrated in various contexts~\cite{tindall2024efficient,Tomislav:2024, tindall2025dynamics}. These methods exploit BP’s efficient message-passing framework to approximate tree-like quantum correlations within tensor networks, facilitating efficient contraction and compression.

The BP compression is fundamentally similar to the SU method, as it performs compression using a locally approximated environment for each bond. The key difference lies in its improved adaptability to the original tensors, eliminating the need to form intermediate tensors. We employ the $2$-norm BP compression~\cite{Tomislav:2024} to efficiently evolve the PEPS. As demonstrated below, the unitary $\mathscr{U}$ (red, represented as a PBC-PEPO with $D=2$) acting on the PBC-PEPS (blue) is compressed back to its original form, maintaining the same bond dimension $D$ at each Trotter step,
\begin{align} 
\label{EQ:tn3d}
\diagramb{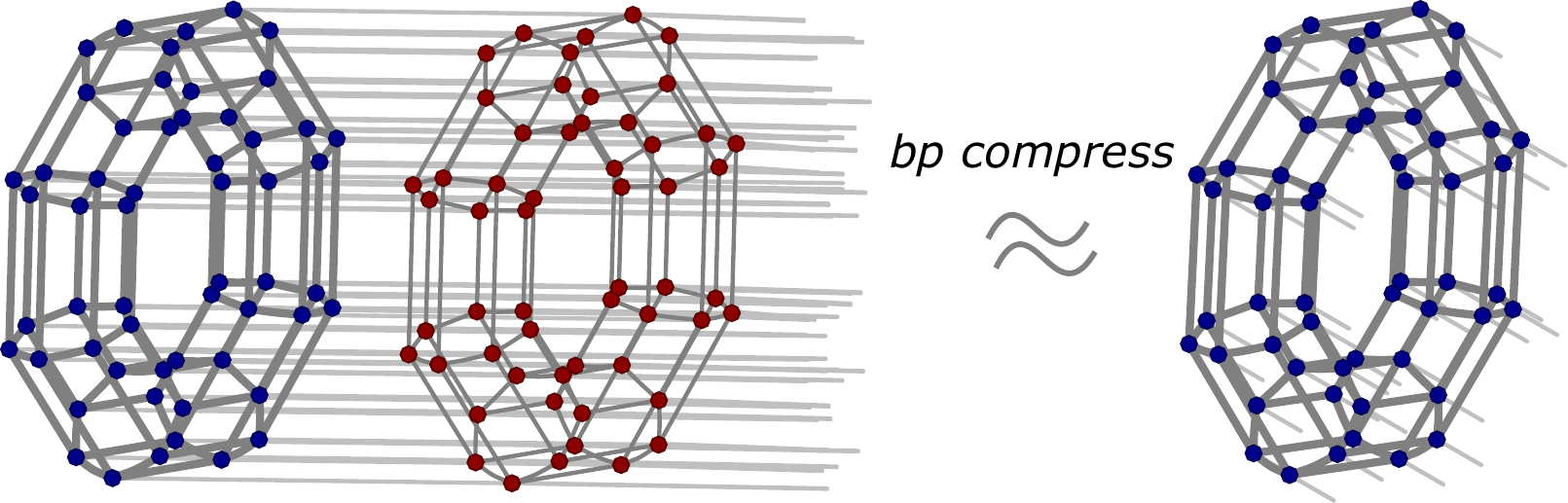}.
\end{align}
The compression is achieved by running the BP algorithm once to compute the BP messages (environment tensors), while simultaneously calculating and integrating the projections to compress the tensor network object. As noted, it differs from the SU compression, a gate-by-gate technique, by being more computationally efficient. However, we have not observed a significant difference in accuracy. The computational cost of the $2$-norm BP compression scales as $\sim \mathcal{O}(D^5)$ for the square lattice studied here. 

\begin{figure*}[!t]
\centering
\includegraphics[width=2.1\columnwidth]{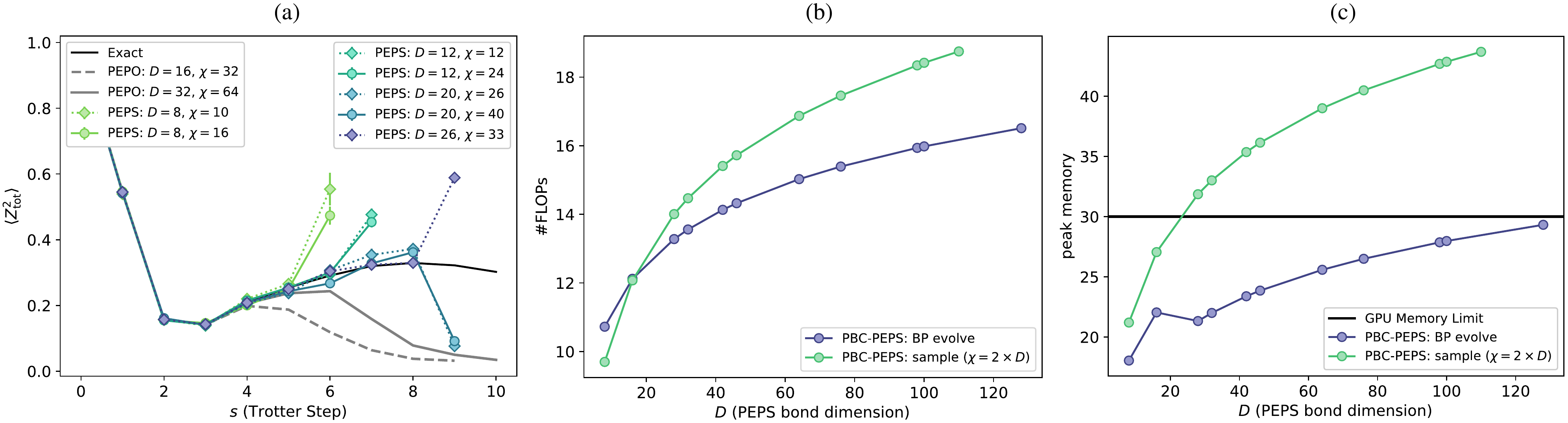}
\caption{
(a) $\langle Z_\textrm{tot}^2 \rangle$ estimation using PBC-PEPS and PBC-PEPO for a $6 \times 6$ system using a variety of bond dimensions $D$ and boundary bond dimensions $\chi$. The PEPS data points have been obtained using a sample size ranging from $2000$ to $3000$. (b), (c) Cost estimation of the PEPS algorithm utilized alongside the BP algorithm and sampling methods. The FLOPs and peak memory (the number of elements in the largest intermediate tensor) are reported on logarithmic scales with base 10 and base 2, respectively.  } 
\label{fig:peps_z2_cost}
\end{figure*}

Using BP, we can efficiently obtain the time-evolved PEPS with a considerably larger bond dimension $D$ than is possible with the FU method. However, the main challenge lies in measuring the observable $\langle Z_\textrm{tot}^2 \rangle$, which remains the most difficult task in the process. While several strategies exist, we explored the loop series expansion~\cite{EvEtAl24} and clustering approximation~\cite{Lubasch:2014}. However, we ultimately achieved the best results using Monte Carlo sampling. The latter procedure works as follows: for a given Trotter step, we obtain an approximate MPS wavefunction with bond dimension $D=520$ (as described in Sec.~\ref{supp:MPS simulations}), and then efficiently produce samples $|x \rangle$ (in the computational basis) with probability $w_{\text{MPS}}(x)$ using perfect sampling~\cite{ferris2012perfect}. We then estimate the true probability $w_{\rm PEPS}(x)=|\bra{x}\Psi_{\rm PEPS}\rangle|^2$ by approximately contracting the associated tensor network, and use such probabilities to form a Monte Carlo estimate of observables. This method provides an unbiased estimate of an observable $\bra{\Psi_{\rm PEPS}}O\ket{\Psi_{\rm PEPS}}$ (assuming no approximation error in computing $|\bra{x}\Psi_{\rm PEPS}\rangle|^2$) even if the MPS forms a poor approximation to $\ket{\Psi_{\rm PEPS}}$, but the quality of the MPS facilitates a degree of importance sampling (and therefore improves estimate variances). Note that we expect $w_{\text{MPS}}(x)$ to serve as a reasonably good approximation of the true PEPS probability distribution $w_{\text{PEPS}}(x)$ at short times, thereby reducing sample overhead due to effective importance sampling. Figure \ref{fig:dkl} quantifies the quality of the distribution $w_{\rm MPS}(x)$ with respect to $w_{\rm PEPS}(x)$ via the Kullback-Leibler (KL) divergence $D_{\rm KL}(w_{\rm MPS}||w_{\rm PEPS})$. Small values of $D_{\rm KL}$ at small $s$ indicate  good overlap between $w_{\rm MPS}$ and $w_{\rm PEPS}$, implying effective importance sampling. Larger values of $D_{\rm KL}$ at larger $s$ indicate the breakdown of importance sampling, leading to larger estimate variances.

Measuring the observable using importance sampling is significantly simpler than direct  estimation (requiring double-layer PBC-PEPS approximate contraction), as it only requires the contraction of a single-layer PBC-PEPS,
\begin{align} 
\label{EQ:tn3da}
\diagram{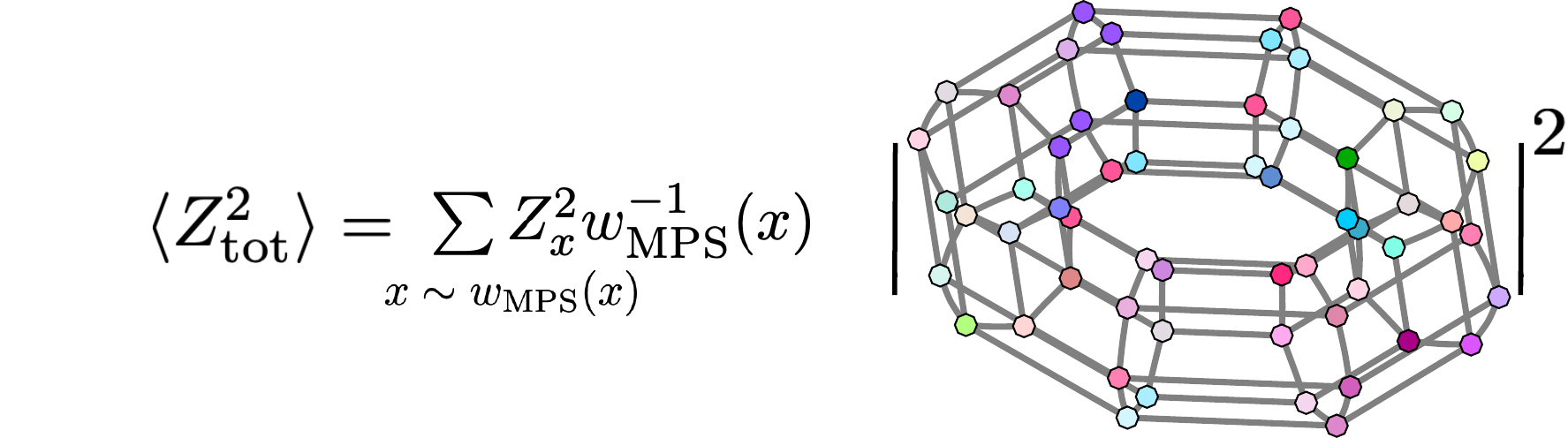}
\end{align}
Here, $Z_x^2 = \langle x | Z_\textrm{tot}^2|x\rangle$ is simple to compute as the samples $\ket{x}$ are in the computational basis (i.e., each $\ket{x}$ is an eigenstate of all $Z_j$). The notation $x\sim w_{\rm MPS}(x)$ in the sum subscript implies that the sum samples $x$ from the distribution $w_{\rm MPS}(x)$, and the single-layer PBC-PEPS diagram in \eref{EQ:tn3da} represents the amplitude $\langle x \ket{\Psi_{\rm PEPS}}$ [its square represents $w_{\rm PEPS}(x)$]. We estimate this diagram using a \emph{hyper-optimized approximate contraction} approach, controlled by a boundary bond dimension $\chi$~\cite{gray:2024}. Empirically, it is found that a boundary bond dimension $\chi = 2D$ is sufficient to produce converged results, comparable to the sampling overhead (see \fref{fig:peps_z2_cost}a).

The extent to which the bond dimension $D$ can be increased (assuming $\chi = 2D$) is examined in \fref{fig:peps_z2_cost}(b-c), presenting the required floating-point operations (FLOPSs) and peak memory usage---defined as the number of elements in the largest intermediate tensor--- needed for the BP algorithm and single-layer contraction \cite{Gray:2021, gray2018quimb}. While the time evolution of the PBC-PEPS can be performed with a high bond dimension ($D \sim 100$), the primary computational bottleneck arises from the memory requirements for contracting the single-layer PEPS---scales like $D^6$ assuming $\chi \sim D$---which are ultimately constrained by available GPU resources. Empirically, on the single NVIDIA A100 GPU with 40 GB of memory used for our PEPS calculations, we find that the maximum practical peak memory usage is approximately $2^{30}$. This effectively limits our computations to a bond dimension of $D \sim 22-26$. \fref{fig:peps_z2_cost}(a) presents benchmarking results for the intermediate temperature quench in a $6 \times 6$ system, where we have exact results from state vector simulations.
We find that for this system size we can obtain accurate results up to about $s=8$ Trotter steps by pushing the PBC-PEPS bond dimension to its limit on the hardware we had access to.

\subsection{PEPO simulation with the Belief Propagation algorithm} \label{supp:PEPO}

Similarly to PBC-PEPS, we can instead evolve a PBC-PEPO under Heisenberg evolution. The starting point is to express the observable $Z_\mathrm{tot}^2$ as a PEPO operator (with bond dimension $D=3$) and then apply the unitary operator to evolve it to a desired Trotter step $s$: $\mathscr{U}^{s\dagger} Z_\mathrm{tot}^2 \mathscr{U}^s$. The PBC-PEPO is compressed back to its original form after each application of one Trotter step unitary $\mathscr{U}$, by using a $2$-norm BP compression scheme, shown here,
\begin{align} 
\label{EQ:tn3d}
\diagrams{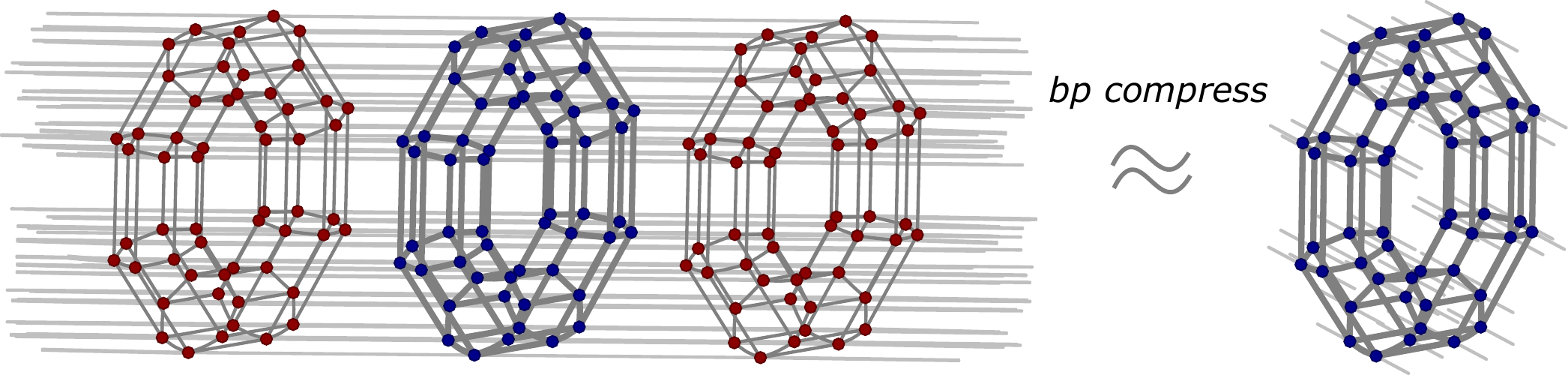}.
\end{align}
Note that the red PBC-PEPO represents the unitary $\mathscr{U}$ while the blue PBC-PEPO corresponds to the time-evolved observable. We run the BP algorithm and once it converges to a fixed point, the resulting projectors are used to compress it back into a PBC-PEPO. Despite being computationally more expensive than PEPS evolution due to the double-layer picture, measuring the observable is much easier, as it only requires contracting a single-layer tensor network, as shown here:
\begin{align} 
\label{EQ:tn3d}
\diagramf{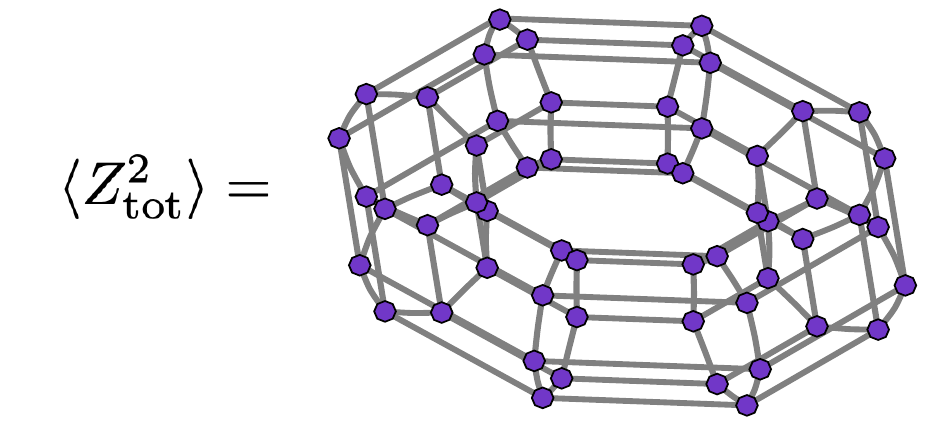}.
\end{align}
We similarly employ a \emph{hyper-optimized approximate contraction} method to estimate its value and, consequently, the observable. The computational cost and accuracy of PBC-PEPO are presented in \fref{fig:peps_z2_cost}.

\begin{figure}[!t]
\centering
\includegraphics[width=1.0\columnwidth]{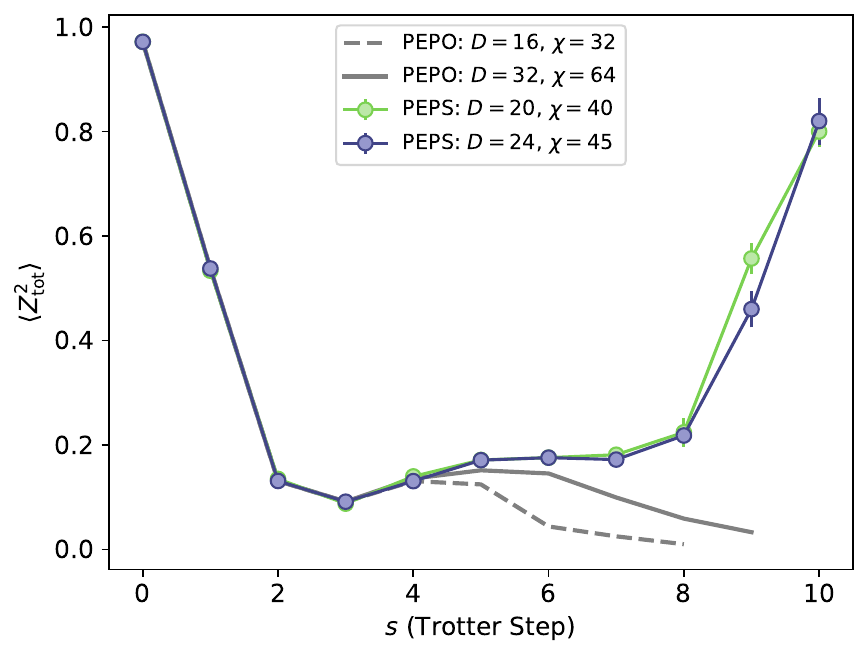}
\caption{PBC-PEPS and PBC-PEPO simulation results for $\Delta \theta = 2\pi/9$ as a function of the number of Trotter steps, shown for system size $(L_x,L_y) = (7,8)$.
\label{fig:peps_tn_z2}}
\end{figure}

We present the PBC-PEPS and PBC-PEPO results achieved for a $7 \times 8$ system in \fref{fig:peps_tn_z2}. This suggests that both methods agree up to Trotter step $6$, after which the PBC-PEPO method declines, while the PBC-PEPS method remains accurate up to Trotter steps $7-8$. This divergence follows the expected pattern, similar to the behavior observed for the $6 \times 6$ system size. 

It is possible to combine the Schr\"odinger picture evolution of the state of the previous section, with the Heisenberg picture evolution of the operator here. We could then obtain a mixed PEPS-PEPO representation for the sum of the maximum evolution times accessible to the PEPS and PEPO separately. However, there remains the challenge of whether we can estimate the observable accurately in this representation at the maximum time, which we leave as a topic for further study.

\section{Sparse Pauli dynamics} \label{supp:SPD}

The classical simulation methods considered here are based on the Heisenberg-picture evolution of the observable operator in the Pauli basis. We employ a sparse representation
\begin{equation}
    O = \sum_{P} a_P P
    \label{eq:spd_obs_operator}
\end{equation}
of the observable operator by storing only non-zero coefficients $a_P$ and the corresponding Pauli operators $P$. The evolution under a Pauli rotation gate $U_{\sigma}(\theta) = \exp(-i \theta \sigma / 2)$ is given by
\begin{equation}
    U_{\sigma}(\theta)^{\dagger} P U_{\sigma}(\theta) = \begin{cases}
        P, \quad [P, \sigma] = 0, \\
        \cos(\theta) P + i \sin(\theta) \sigma P, \quad \{P, \sigma\} = 0.
    \end{cases}
    \label{eq:spd_evolution}
\end{equation}
In other words, for every Pauli operator that does not commute with the rotation gate, we generate a new Pauli operator $\sigma P$. This can lead to a fast growth of the sum in Eq.~\eqref{eq:spd_obs_operator}. Here, we consider two strategies for truncating the representation of the observable operator. Firstly, sparse Pauli dynamics uses a threshold-based ($\delta$) truncation, where after each rotation gate we discard Pauli operators with $|a_P|<\delta$. Secondly, we use a Pauli weight truncation where we also discard Pauli operators whose weight (the number of non-identity Pauli matrices) is greater than a predefined value $w$. Truncation based on small coefficients is a simple heuristic that attempts to minimally damage the backpropagated observable and is at the base of, e.g., Refs.~\cite{Tomislav:2024,beguvsic2024real}, and generalizes the so-called Clifford perturbation or near-Clifford truncations from Refs.~\cite{beguvsic2023simulating,lerch2024efficient}. Truncation based on Pauli weight on the other hand is more invasive to the propagating observable, potentially truncating Pauli operators with relatively large coefficients, but it aims to keep expectation values intact. It has been proven to yield efficient average-case simulation of expectation values in random quantum circuits~\cite{angrisani2024classically} and noisy circuits \cite{cirstoiu2024fourier, fontana2023classical}, and has been used as a heuristic for quantum dynamics~\cite{rudolph2023classical} and quantum machine learning simulations~\cite{bermejo2024quantum}.

Given that our methods scale unfavorably with the number of applied rotation gates, we minimize this number by rewriting the circuit as
\begin{eqnarray}
    U(t) &=& \left[ U_X\left(\frac{dt}{2}\right) U_{ZZ}(dt) U_X\left(\frac{dt}{2}\right) \right]^t \nonumber\\ 
      &=& U_X\left(\frac{dt}{2}\right) \left[ U_{ZZ}(dt) U_X(dt) \right]^t U_X\left(-\frac{dt}{2}\right) \nonumber\\
      &=& U_X\left(\frac{dt}{2}\right) U'(t) U_X\left(-\frac{dt}{2}\right),
\end{eqnarray}
and evaluate the observable expectation value as
\begin{equation}
    \langle O \rangle_t = \langle \psi | U'(t)^{\dagger} O_X U'(t) | \psi \rangle,
\end{equation}
where $O_X = U_X(\frac{dt}{2})^{\dagger} O U_X(\frac{dt}{2})$ is the rotated observable and $|\psi\rangle = U_X(-\frac{dt}{2}) |0^{\otimes n} \rangle$ is a product state.

\begin{figure*}[pth]
    \centering
    \includegraphics[width=\linewidth]{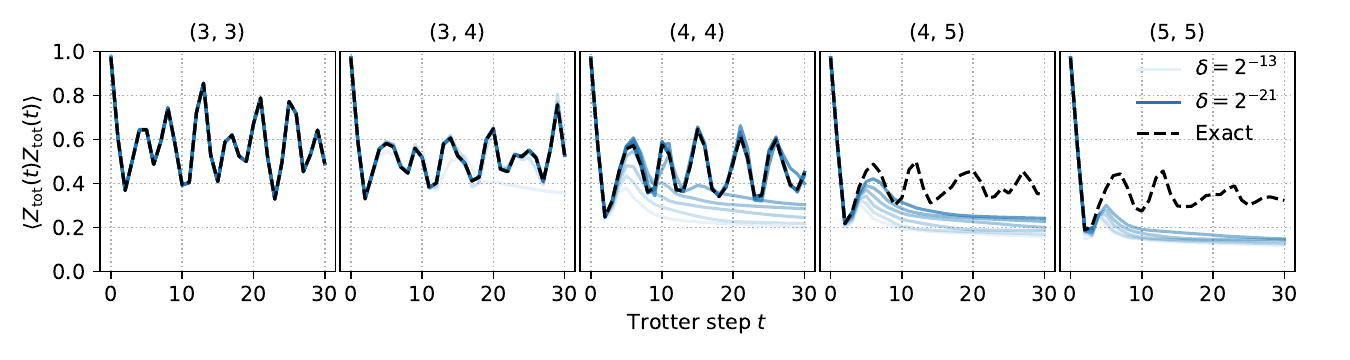}
    \includegraphics[width=\linewidth]{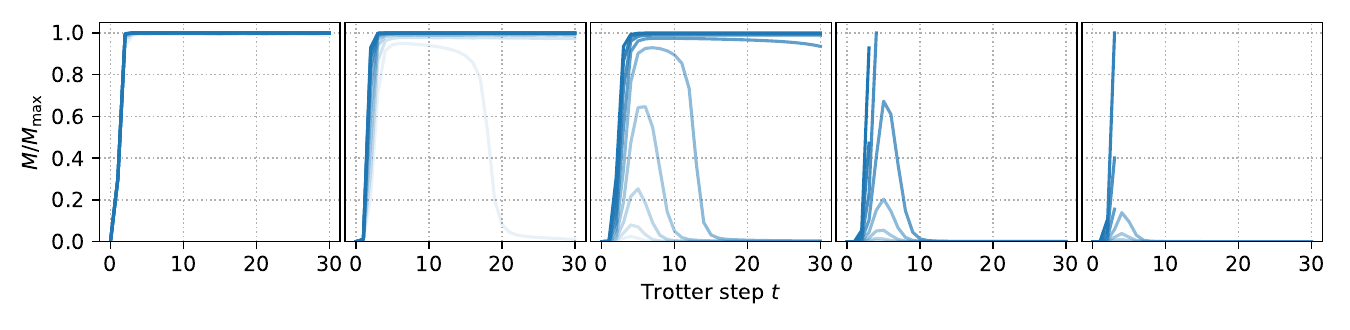}
    \caption{Sparse Pauli dynamics for small benchmark problems with lattice sizes ranging from $3\times 3$ to $5 \times 5$, calculated with different values of threshold $\delta = 2^{-13}, 2^{-14}, \dots, 2^{-21}$. Top: Comparison between sparse Pauli dynamics and exact results. Bottom: Number of Pauli operators ($M$) after each Trotter step divided by the maximum number of Paulis ($M_{\text{max}}$) over all thresholds for a given system size. These maximal values are $4119$ $(3\times 3)$, $184318$ $(3\times 4)$, $34028934$ $(4\times 4)$, $764375216$ $(4\times 5)$, $871811155$ $(5\times 5)$.}
    \label{fig:spd_benchmarks}
\end{figure*}

We further augment our sparse Pauli dynamics by a symmetrization technique that applies to the circuits at hand. Specifically, we note that the $X$- and $ZZ$-rotation layers, as well as the initial state $|\psi\rangle$, are invariant to symmetries of the rectangular lattice with periodic boundary conditions: translations along $x$ and $y$ axes, inversion, and reflections with respect to $x$ and $y$ axes ($N_{\text{symm}} = 4N$). Therefore, for each Pauli operator $P$, $\langle P \rangle_t = \langle P' \rangle_t$, where $P$ and $P'$ are related by these symmetries. Given that the expectation value is linear in the observable operator, we have
\begin{equation}
    \langle O \rangle_t = \sum_P a_P \langle P \rangle = \sum_{P'} a_{P'} \langle P' \rangle,
\end{equation}
where the coefficients are unchanged. Importantly, for different Pauli operators we can apply different symmetries, which motivates us to transform all symmetry-equivalent Pauli operators $\mathcal{P}_{\text{symm}}$ into one:
\begin{equation}
    \sum_{P \in \mathcal{P}_{\text{symm}}} a_P P \rightarrow \left(\sum_{P \in \mathcal{P}_{\text{symm}}} a_P \right) P'.
    \label{eq:spd_symmetrization}
\end{equation}
In this way, multiple symmetry-equivalent Paulis are not stored separately but only as one Pauli operator, leading to memory and time savings by up to a factor of $N_{\text{symm}}$. In our implementation, each Pauli operator is represented by a sequence of bits in unsigned integer types. As an example, one can encode a Pauli string $P$ via bitstrings $z$ and $x$,
\begin{equation}
    P = \prod_{i=0}^{N-1} Z_i^{z_i} X_i^{x_i},
\end{equation}
where $z_i$ and $x_i$ are each encoded as bits of unsigned 64-bit integers. For each Pauli, we can find the index permutation corresponding to a symmetry transformation that minimizes these integers. This will transform all symmetry-equivalent Pauli operators into a unique Pauli. Therefore, after transforming each Pauli in the representation of the observable, we only need to merge the coefficients of Paulis that appear multiple times.

For completeness, we list below the index permutations associated with different symmetries. These are written out for 2D lattice site indices $(i_x, i_y)$, which can be converted to the unfolded index $i = i_x + i_y N_x$, where $i_{x,y} \in \{0, 1, \dots, N_{x,y}-1\}$ and $i \in \{0, 1, \dots, N-1\}$.
\begin{eqnarray}
    &&\text{Translation:}\quad (i_x, i_y) \rightarrow (i_x + t_x, i_y + t_y) \nonumber \\
    &&\text{Inversion:}\quad (i_x, i_y) \rightarrow (N_x-i_x, N_y-i_y) \nonumber \\
    &&\text{$x$-reflection:}\quad (i_x, i_y) \rightarrow (i_x, N_y-i_y) \nonumber \\
    &&\text{$y$-reflection:}\quad (i_x, i_y) \rightarrow (N_x-i_x, i_y) \nonumber
\end{eqnarray}

\begin{figure*}[pth]
    \centering
    \includegraphics[width=0.8\linewidth]{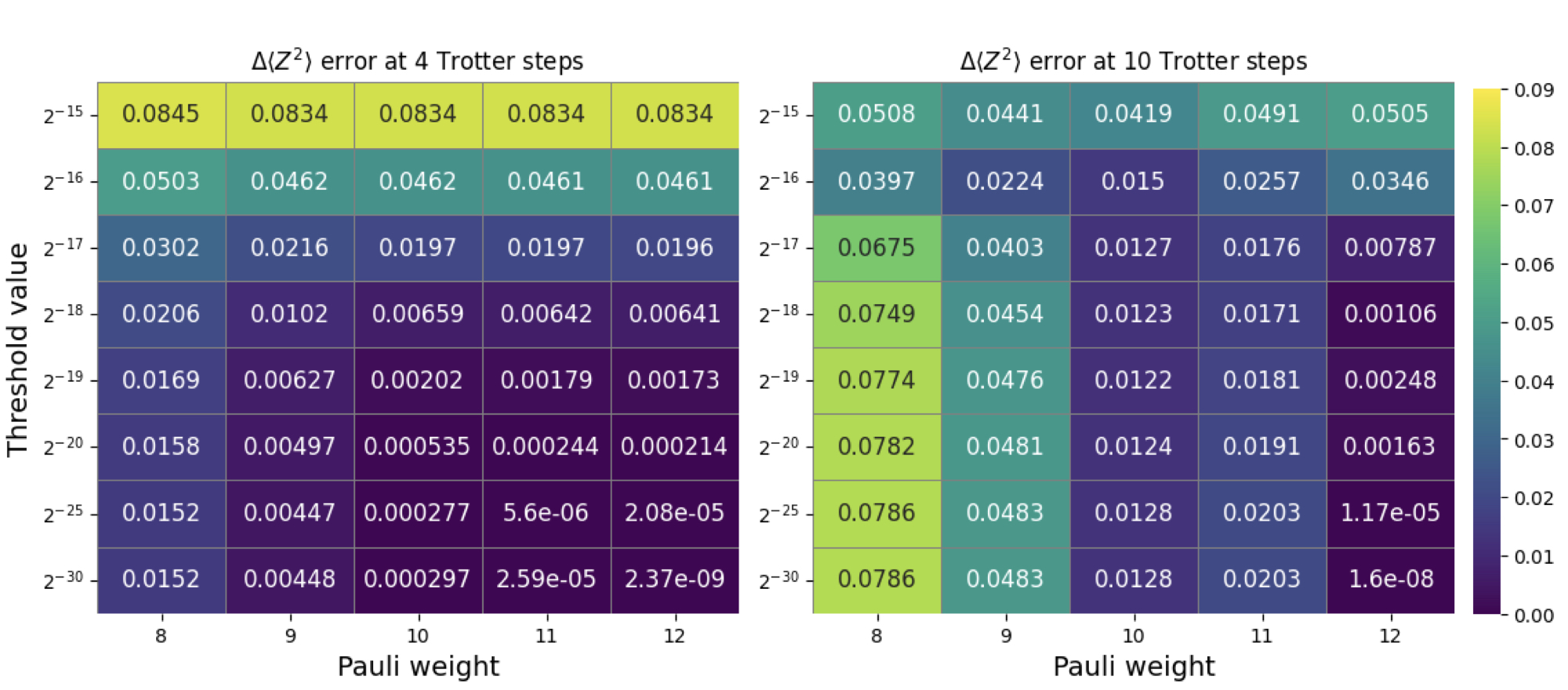}
    \caption{Absolute approximation error between  the exact value of $\langle Z_{tot}(t)^2\rangle$ on a  lattice of size $3\times4$ and Pauli-based classical simulations with truncation strategies involving Pauli weight and coefficient threshold.}
    \label{fig:weight_truncation_trade-off}
\end{figure*}

Before analyzing the full $7 \times 8$ model, we first benchmarked the method on smaller lattices ranging from $3\times 3$ to $5 \times 5$ (Fig.~\ref{fig:spd_benchmarks}), where exact results can be easily computed. Although sparse Pauli dynamics captures exact dynamics in the few smallest examples, it struggles at lattice sizes $4\times 5$ and $5 \times 5$. In fact, we find that the number of Pauli operators in these systems grows quickly with $t$ and saturates to the full size of the operator Hilbert space, which is smaller than $4^N$ due to symmetries but still scales exponentially with the system size. In other words, simulations where the Hilbert space is too large simply cannot be converged except for the first few time steps. We attribute the fast growth to the fact that the rotation gates in the studied circuits very far from Clifford points. Namely, the $X$-rotations with $\theta = 1.0$ are not far from $\pi/4\approx 0.785$, a point at which the branching of Eq.~\eqref{eq:spd_evolution} leads to equally weighted Pauli operators.

We also investigate truncations based on the Pauli weight of the propagated Pauli observables using the \href{https://github.com/MSRudolph/PauliPropagation.jl}{PauliPropagation.jl} package. A weight cut-off can allow us to use a similar or lower coefficient threshold and reduce memory constraints. This type of truncation is motivated by previous work showing that low-weight Pauli operators give the dominant contribution to noisy observables. In the noiseless case, the effectiveness of a weight cut-off is problem-specific. We illustrate in Fig. ~\ref{fig:weight_truncation_trade-off} the absolute approximation error in the observable $\<Z^2\>$ that can be achieved using different coefficient thresholds and weight cut-offs. Particularly at late times, it appears that any weight lower than the total number of sites results in errors significantly higher than without it. This implies that for these studied circuits, weight truncation is not particularly effective at reducing the runtime and number of operators while maintaining low error.
\begin{figure*}[t]
    \centering
    \includegraphics[width=0.49\linewidth]{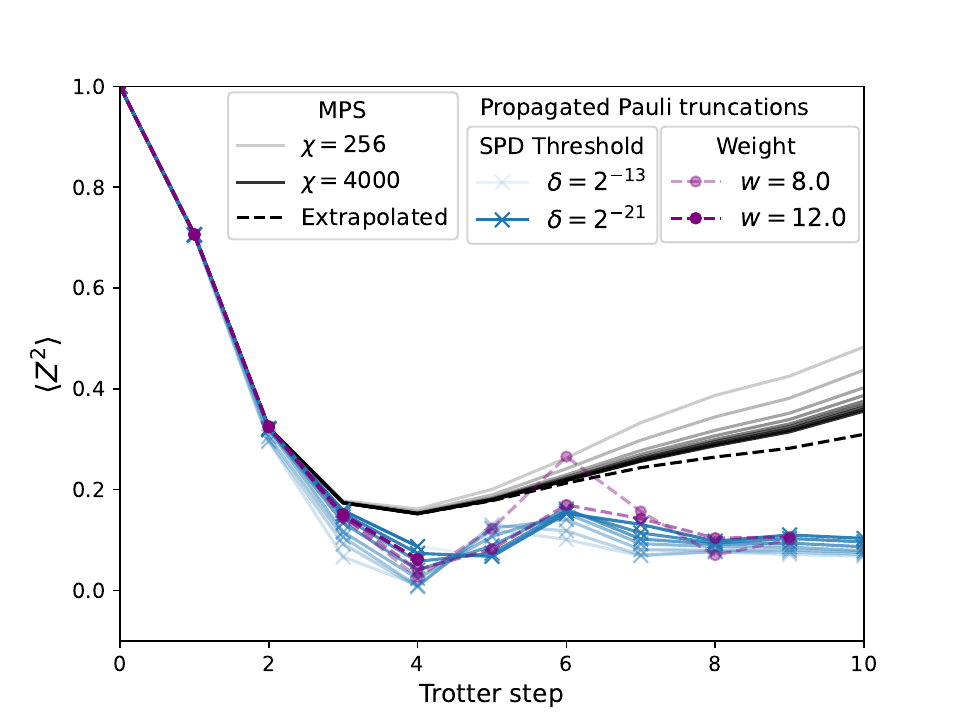}
   \hspace{-0.5cm}
    \includegraphics[width=0.49\linewidth]{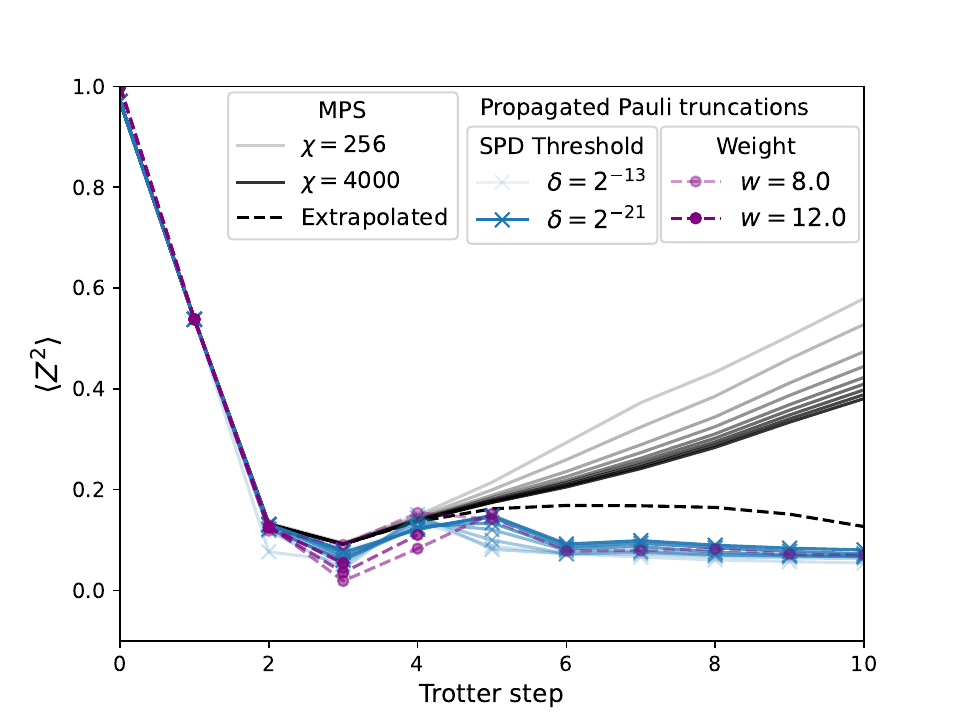}
    
    \caption{Pauli-based simulations for different truncations with coefficient threshold $\delta$ and optional Pauli weight cut-off $w$ compared against extrapolated MPS results for the 
    $7\times 8$ square lattice. (Left) A quench at parameters $(h/|J|,dt\times |J|,\Delta\theta)=(2,0.2,-\pi/6)$.  (Right) The intermediate temperature quench $(h/|J|,dt\times |J|,\Delta\theta)=(2,0.25, 2\pi/9)$. The coefficient truncation thresholds range across 
    $ \log_2\delta \in \{-13,-14,\dots,-21\}$. We test additional truncation by Pauli weight with cut-offs $w\in\{8,9,...,12\}$ at a lower coefficient threshold of $4\cdot 10^{-7}$. 
    } 
    \label{fig:spd_n56}
\end{figure*}

We now look at the $N=56$ system, where we compare sparse Pauli dynamics against the extrapolated MPS expectation values, see Fig.~\ref{fig:spd_n56} for the results.
Sparse Pauli dynamics deviate already at $t=3$ and is inaccurate at later times beyond $t=5$ (for the intermediate quench). For reference, the simulation at $\delta=2^{-20}$ takes about 51h using 16 CPUs and up to 1.4~TB of memory, generating around 4 billion Pauli operators. Using the same resources, the simulation at $\delta = 2^{-21}$ runs out of memory after $t=4$. We tested if the simulation time can be extended using a lower coefficient truncation at or below $\delta = 2^{-21}$ and a weight cut-off up to $12$.  This has not led to a significant reduction in memory resources needed for a good approximation. \diff{Since the Pauli weight-based truncation has the same effect on the simulation cost as the introduction of depolarizing noise, our results imply that noisy circuit sparse Pauli simulations cannot be used to perform zero-noise extrapolation as done in the experiments.} The simulations with weight truncation were executed with the \href{https://github.com/MSRudolph/PauliPropagation.jl}{PauliPropagation.jl} package on a single CPU thread, resulting in shorter runtimes but larger errors at any fixed coefficient truncation threshold.

% Revised part
\diff{
Finally, we consider extrapolating SPD results, to allow for a fair comparison against the extrapolated MPS results. First, we extrapolate our results linearly using the three most accurate points (three smallest thresholds). This can be done either with respect to the threshold or with respect to the Frobenius norm of the operator, $F=\sqrt{\text{Tr}[O_t^\dagger O_t]/2^N}$, a quantity that is conserved during unitary evolution and is similar to the simulation fidelity introduced for the MPS computations. The Frobenius norm is scaled by its value at time zero. Due to the symmetrization technique presented above, which effectively merges symmetry-equivalent Pauli operators (Eq. (\ref{eq:spd_symmetrization})), this norm is not conserved even in the zero-threshold limit. However, we can estimate the level of truncation during unitary dynamics by
\begin{equation}
    \tilde{f} = \prod_k \tilde{f}_k = \prod_k \frac{F(\text{after } U_k)}{F(\text{before } U_k)},
    \label{eq:spd_f}
\end{equation}
where $k$ runs over all layers between which we employ the symmetrization step and $U_k \in \{U_X, U_{ZZ}\}$ are the corresponding unitaries.
}

\begin{figure*}[t]
    \centering
    \includegraphics[width=0.48\linewidth]{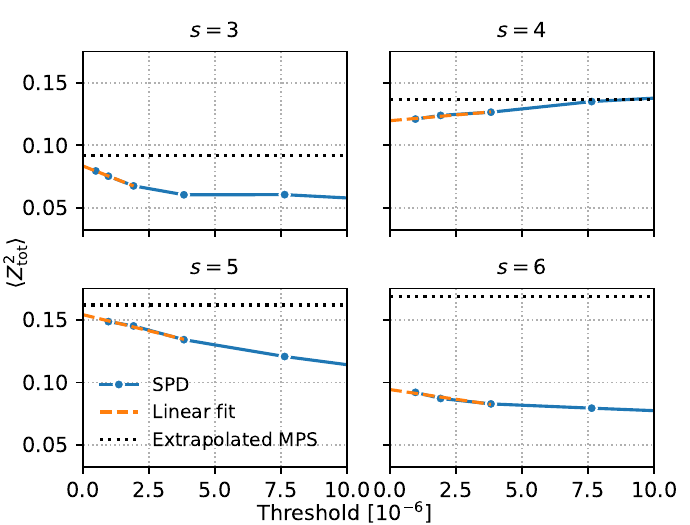}
    \includegraphics[width=0.48\linewidth]{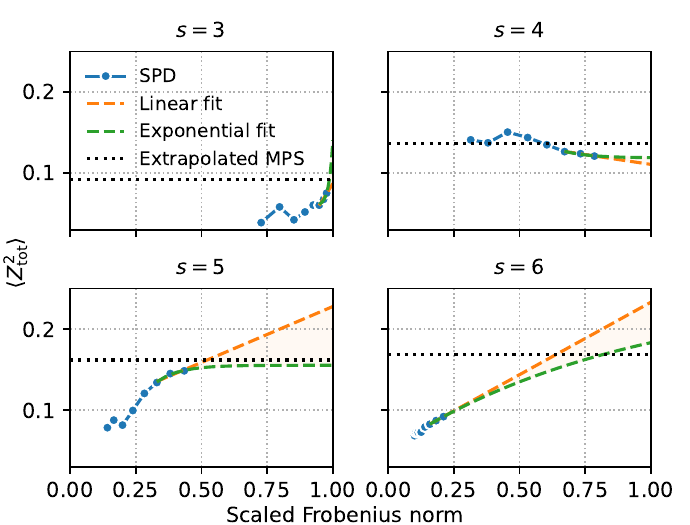}
    \caption{\diff{Zero-truncation extrapolation of SPD results for the intermediate temperature quench at Trotter steps $s=3, 4, 5, 6$. (Left) Linear extrapolation with respect to threshold parameter ($\delta \rightarrow 0$ limit). (Right) Linear and exponential extrapolation with respect to the scaled Frobenius norm (Eq.~(\ref{eq:spd_f}), $\tilde{f} \rightarrow 1$ limit).}}
    \label{fig:spd_extrapolation}
\end{figure*}

\diff{
The results are shown in Fig. \ref{fig:spd_extrapolation} for Trotter steps $s=3, 4, 5, 6$. We find that already at step $s=4$, the observable is insufficiently converged and the extrapolation result is less accurate than the non-extrapolated estimate. At steps $s=5$ and $s=6$, the observable expectation value is converging in the right direction, so the extrapolations generally improve the results, with threshold-based extrapolation underestimating and fidelity-based extrapolation overestimating the final value. Alternatively, we can estimate the error of our extrapolation by comparing two different fits. For this, we performed fidelity-based extrapolation using an exponential fit $f(x) = a\exp(bx) + c$ applied to the last four points. The error can then be estimated as a difference between the linear and exponential extrapolation values (shaded area in Fig. \ref{fig:spd_extrapolation}, right). This error estimate agrees well with the true extrapolation error at later time steps ($s=5,6$). Still, the overall errors are greater than those found in the extrapolation of MPS simulation data.
}

\section{Neural network quantum states}
\label{sec:nqs}

Recent works~\cite{Schmitt2020, schmitt2022quantum} have demonstrated the viability of neural-network quantum state (NQS)~\cite{carleo2017solving}, when combined with time-dependent variational Monte Carlo (tVMC)~\cite{carleo2012localization}, as a competitive tool for simulating quantum many-body dynamics compared to the state of the art infinite PEPS simulation~\cite{czarnik2019time,dziarmaga2022simulation}.
These works have served as concrete benchmarks and stimulated the development of new algorithms in simulating dynamics with NQS, for example, the projected time-dependent variational Monte Carlo (ptVMC) method~\cite{kochkov2018variational,medvidovic2021classical,gutierrez2022real,sinibaldi2023unbiasing,barison2021efficient,donatella2023dynamics} and the time-dependent neural quantum state (t-NQS)~\cite{Sinibaldi2024,VandeWalle2024} method.

To classically simulate the quantum circuit considered in this work, we choose the CNN architecture and evolve it using tVMC algorithm. Precisely, we use the Group-CNN (GCNN)~\cite{roth2021group} implementation of the NetKet library \cite{netket3:2022}, which symmetrizes the networks according to all the symmetries of the lattice. Hence, on a $(L, L)$ square lattice, a GCNN is a CNN with an additional $D_4$ symmetry and on a $(L, L')$ rectangular lattice it is a CNN with an additional $D_2$ symmetry.
This is exactly the setup considered in~\cite{Schmitt2020} where the translation and point group symmetries are taken into account.
The tVMC algorithm is realized by the default implementation of the time-dependent variational principle (TDVP) function in NetKet.
For the integration of the TDVP equation, we use the adaptive second- and third-order Runge-Kutta (RK23) scheme implemented in NetKet.

The architecture choice is motivated by the previous success in simulating quantum quenches~\cite{Schmitt2020,schmitt2022quantum}.
The other important deciding factors are the symmetries of the system.
The setup we considered in this work has the translation symmetry but explicitly breaks the $\mathbb{Z}_2$ symmetry flipping all spins. The translation symmetry supports the usage of CNN.
The broken $\mathbb{Z}_2$ symmetry suggests the usage of an activation function that is not even~\footnote{A single hidden-layer CNN with an even activation function, e.g., {poly6} or $\ln(\cosh( z))$, and zero bias leads to an $\mathbb{Z}_2$ symmetrized ansatz with respect to flipping all the spins.
Since the $\mathbb{Z}_2$ symmetry of the Ising model is broken by our initial state, we use the {poly5} activation function, which is an odd function, in all our calculations.}.
As a result, we consider the {poly5} activation function, which is the Taylor expansion of $\tanh{z}$ to the fifth order.
It is the derivative of the {poly6} activation function, which is the Taylor expansion of $\ln(\cosh( z))$ up to the sixth order~\footnote{
A single hidden-layer CNN with the $\ln(\cosh( z))$ activation function is exactly equivalent to the symmetrized RBM.
A single hidden-layer CNN with the {poly6} activation function serves as an holomorphic ansatz closed to the symmetrized RBM and has been shown to give competitive results to the state of the art tensor network simulation~\cite{Schmitt2020}.}.
We would like to note that the results obtained with CNN on the Ising model are now routinely used to benchmark the new advancements of NQS methods~\cite{VandeWalle2024,Gravina2024}.

\begin{figure*}[!t]
    \centering\includegraphics[width=2.05\columnwidth]{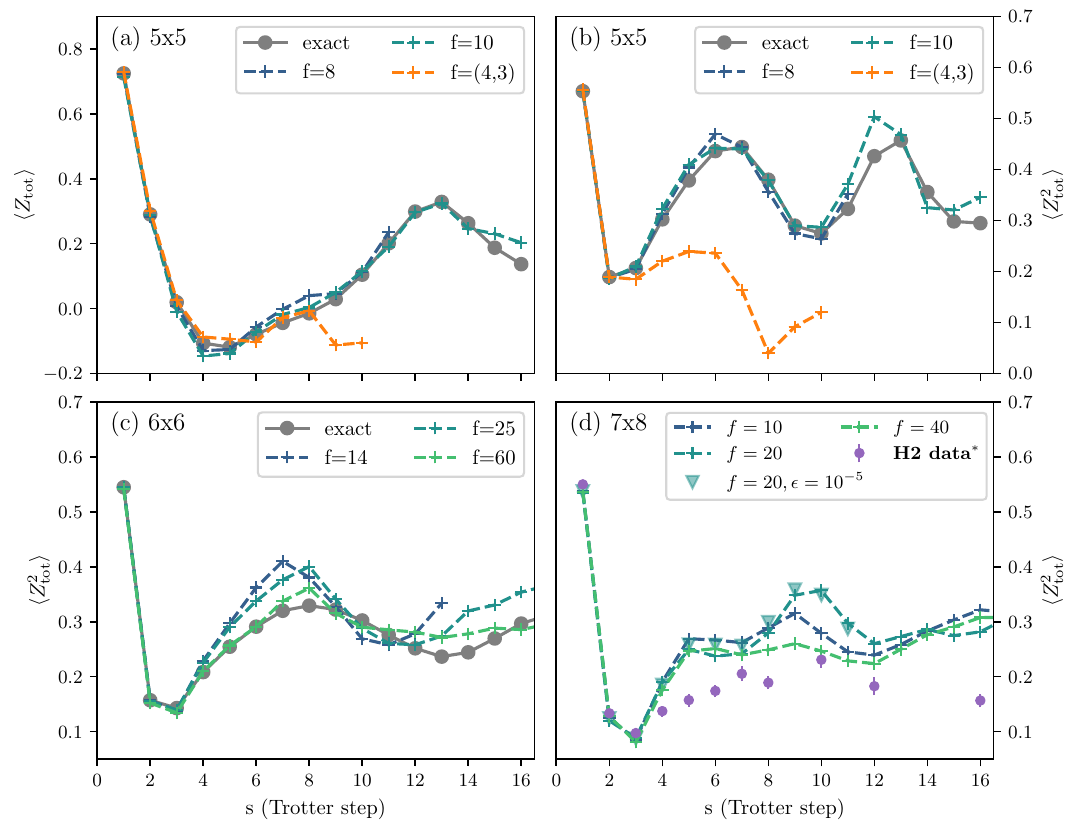}
    \caption{Expectation values of (a) the total magnetization for the $5\times 5$ system, (b) the total magnetization squared for the $5\times 5$ system, (c) the total magnetization squared for the $6\times 6$ system, and (d) the total magnetization squared for the $7\times 8$ system. These values were obtained using neural quantum state simulations for the intermediate-temperature quench with a single-layer and double-layer CNN. Different numbers of features were used for the single-layer CNN, while (4,3) features were used for the double-layer CNN in panels (a) and (b). A tolerance $\epsilon = 5 \times 10^{-5}$ was chosen for the adaptive integration scheme, with a minimum time step of $5 \times 10^{-3}$ and a maximum time step of $5 \times 10^{-2}$, except for the turquoise triangles in panel (d), where the tolerance was $\epsilon = 1 \times 10^{-5}$ and the minimum time step was $1 \times 10^{-3}$. All simulations used $10^4$ Monte Carlo samples.
    The H2 data shown here is with ZNE + ZNR.
    }
    \label{fig:neural_networks}
\end{figure*}

We first present the results for the $5\times 5$ system with periodic boundary conditions, alongside the exact results, in Fig.~\ref{fig:neural_networks}, panels (a) and (b). For the single-layer CNN, we observe that increasing the number of features reduces the error in the observables.
%In contrast,
The two-layer CNN requires significantly longer computational times and yields less accurate results. Consequently, we focus on the single-layer architecture for larger system sizes. We also observe that single-site observables are easier to capture accurately than two-point correlation functions. This is particularly evident in panels (a) and (b) of Fig.~\ref{fig:neural_networks}. The simulation with (4,3) features for the $5 \times 5$ system yields reasonably accurate results for the magnetization up to time step 8, whereas the values obtained for the magnetization squared show significant deviations as early as time step 4.

As the system size increases, we find that more features are needed to accurately capture the dynamics of the observables to the same accuracy.
The results for the $6\times 6$ system are shown in Fig.~\ref{fig:neural_networks}, panel (c). We observe that a single-layer CNN with 60 features achieves reasonable accuracy up to time step 6.

Finally, the results for the $7\times8$ system are shown in Fig.~\ref{fig:neural_networks}, panel (d). Here, we varied the number of features used from $f=10$ up to $f=40$ only due to the increasing computation cost. The runtime of $f=40$ simulation up to $s=19$ Trotter steps is roughly one week with a single NVIDIA A100 GPU.   
We found that with the maximal number of features considered, $f=40$, NQS only achieves qualitatively converged results to 3 time steps.
It is worth noting that though the NQS results deviate from the correct value earlier compared to the tensor-network methods reported in previous sections, the NQS results appear to remain qualitatively reasonable even once they are no longer quantitatively correct.
We attribute this to the fact that one part of the energy,~i.e., $H_{ZZ}$ or $H_X$, is exactly conserved under t-VMC method when the $U_{ZZ}$ or $U_X$ gate are applied.

For all system sizes considered, a tolerance $\epsilon = 5 \times 10^{-5}$ was chosen for the adaptive integration scheme, with a minimum time step of $5 \times 10^{-3}$ and a maximum time step of $5 \times 10^{-2}$. All simulations used $10^4$ Monte Carlo samples. To assess the influence of the simulation hyper-parameters, we ran the $7\times8$ case with 20 features, decreasing the integration tolerance and the minimum time step by a factor of 5. We found that the magnetization squared values remained almost identical up to time step 4, with only minor deviations at later times. This result gives us confidence in our choice of time step and number of Monte Carlo samples.

Lastly, we would like to comment on the potential pitfalls that NQS simulations are subject to. Specifically, Monte Carlo sampling can introduce bias when the wave function has amplitudes close to zero~\cite{sinibaldi2023unbiasing}, which can show up when measurements are involved.
It is also often hard to distinguish the restrictions imposed by the representation power of the network from those arising due to the optimization scheme.
Overall, performing time evolution with NQS is a rapidly evolving field.
While we have benchmarked against a specific combination of ansatz and algorithm choice, it would be interesting to compare our results with those obtained using different architectures and time evolution methods recently proposed or specifically designed for the purpose of simulating quantum circuits~\cite{Gravina2024, sinibaldi2023unbiasing, Sinibaldi2024,VandeWalle2024}. We envision that the quantum data and simulation data provided here can serve as a useful benchmark for testing new algorithms and variational wavefunctions.

\begin{thebibliography}{117}%
\makeatletter
\providecommand \@ifxundefined [1]{%
 \@ifx{#1\undefined}
}%
\providecommand \@ifnum [1]{%
 \ifnum #1\expandafter \@firstoftwo
 \else \expandafter \@secondoftwo
 \fi
}%
\providecommand \@ifx [1]{%
 \ifx #1\expandafter \@firstoftwo
 \else \expandafter \@secondoftwo
 \fi
}%
\providecommand \natexlab [1]{#1}%
\providecommand \enquote  [1]{``#1''}%
\providecommand \bibnamefont  [1]{#1}%
\providecommand \bibfnamefont [1]{#1}%
\providecommand \citenamefont [1]{#1}%
\providecommand \href@noop [0]{\@secondoftwo}%
\providecommand \href [0]{\begingroup \@sanitize@url \@href}%
\providecommand \@href[1]{\@@startlink{#1}\@@href}%
\providecommand \@@href[1]{\endgroup#1\@@endlink}%
\providecommand \@sanitize@url [0]{\catcode `\\12\catcode `\$12\catcode
  `\&12\catcode `\#12\catcode `\^12\catcode `\_12\catcode `\%12\relax}%
\providecommand \@@startlink[1]{}%
\providecommand \@@endlink[0]{}%
\providecommand \url  [0]{\begingroup\@sanitize@url \@url }%
\providecommand \@url [1]{\endgroup\@href {#1}{\urlprefix }}%
\providecommand \urlprefix  [0]{URL }%
\providecommand \Eprint [0]{\href }%
\providecommand \doibase [0]{https://doi.org/}%
\providecommand \selectlanguage [0]{\@gobble}%
\providecommand \bibinfo  [0]{\@secondoftwo}%
\providecommand \bibfield  [0]{\@secondoftwo}%
\providecommand \translation [1]{[#1]}%
\providecommand \BibitemOpen [0]{}%
\providecommand \bibitemStop [0]{}%
\providecommand \bibitemNoStop [0]{.\EOS\space}%
\providecommand \EOS [0]{\spacefactor3000\relax}%
\providecommand \BibitemShut  [1]{\csname bibitem#1\endcsname}%
\let\auto@bib@innerbib\@empty
%</preamble>
\bibitem [{\citenamefont {D'Alessio}\ and\ \citenamefont
  {Rigol}(2014)}]{PhysRevX.4.041048}%
  \BibitemOpen
  \bibfield  {author} {\bibinfo {author} {\bibfnamefont {L.}~\bibnamefont
  {D'Alessio}}\ and\ \bibinfo {author} {\bibfnamefont {M.}~\bibnamefont
  {Rigol}},\ }\bibfield  {title} {\bibinfo {title} {Long-time behavior of
  isolated periodically driven interacting lattice systems},\ }\href
  {https://doi.org/10.1103/PhysRevX.4.041048} {\bibfield  {journal} {\bibinfo
  {journal} {Phys. Rev. X}\ }\textbf {\bibinfo {volume} {4}},\ \bibinfo {pages}
  {041048} (\bibinfo {year} {2014})}\BibitemShut {NoStop}%
\bibitem [{\citenamefont {Bukov}\ \emph
  {et~al.}(2015{\natexlab{a}})\citenamefont {Bukov}, \citenamefont
  {Gopalakrishnan}, \citenamefont {Knap},\ and\ \citenamefont
  {Demler}}]{Bukov2015}%
  \BibitemOpen
  \bibfield  {author} {\bibinfo {author} {\bibfnamefont {M.}~\bibnamefont
  {Bukov}}, \bibinfo {author} {\bibfnamefont {S.}~\bibnamefont
  {Gopalakrishnan}}, \bibinfo {author} {\bibfnamefont {M.}~\bibnamefont
  {Knap}},\ and\ \bibinfo {author} {\bibfnamefont {E.}~\bibnamefont {Demler}},\
  }\bibfield  {title} {\bibinfo {title} {Prethermal floquet steady states and
  instabilities in the periodically driven, weakly interacting bose-hubbard
  model},\ }\href {https://doi.org/10.1103/physrevlett.115.205301} {\bibfield
  {journal} {\bibinfo  {journal} {Phys. Rev. Lett.}\ }\textbf {\bibinfo
  {volume} {115}},\ \bibinfo {pages} {205301} (\bibinfo {year}
  {2015}{\natexlab{a}})}\BibitemShut {NoStop}%
\bibitem [{\citenamefont {Kuwahara}\ \emph {et~al.}(2016)\citenamefont
  {Kuwahara}, \citenamefont {Mori},\ and\ \citenamefont
  {Saito}}]{KUWAHARA201696}%
  \BibitemOpen
  \bibfield  {author} {\bibinfo {author} {\bibfnamefont {T.}~\bibnamefont
  {Kuwahara}}, \bibinfo {author} {\bibfnamefont {T.}~\bibnamefont {Mori}},\
  and\ \bibinfo {author} {\bibfnamefont {K.}~\bibnamefont {Saito}},\ }\bibfield
   {title} {\bibinfo {title} {Floquet--magnus theory and generic transient
  dynamics in periodically driven many-body quantum systems},\ }\href
  {https://doi.org/https://doi.org/10.1016/j.aop.2016.01.012} {\bibfield
  {journal} {\bibinfo  {journal} {Annals of Physics}\ }\textbf {\bibinfo
  {volume} {367}},\ \bibinfo {pages} {96} (\bibinfo {year} {2016})}\BibitemShut
  {NoStop}%
\bibitem [{\citenamefont {Abanin}\ \emph {et~al.}(2017)\citenamefont {Abanin},
  \citenamefont {De~Roeck}, \citenamefont {Ho},\ and\ \citenamefont
  {Huveneers}}]{PhysRevB.95.014112}%
  \BibitemOpen
  \bibfield  {author} {\bibinfo {author} {\bibfnamefont {D.~A.}\ \bibnamefont
  {Abanin}}, \bibinfo {author} {\bibfnamefont {W.}~\bibnamefont {De~Roeck}},
  \bibinfo {author} {\bibfnamefont {W.~W.}\ \bibnamefont {Ho}},\ and\ \bibinfo
  {author} {\bibfnamefont {F.~m.~c.}\ \bibnamefont {Huveneers}},\ }\bibfield
  {title} {\bibinfo {title} {{Effective Hamiltonians, prethermalization, and
  slow energy absorption in periodically driven many-body systems}},\ }\href
  {https://doi.org/10.1103/PhysRevB.95.014112} {\bibfield  {journal} {\bibinfo
  {journal} {Phys. Rev. B}\ }\textbf {\bibinfo {volume} {95}},\ \bibinfo
  {pages} {014112} (\bibinfo {year} {2017})}\BibitemShut {NoStop}%
\bibitem [{\citenamefont {Heyl}\ \emph {et~al.}(2019)\citenamefont {Heyl},
  \citenamefont {Hauke},\ and\ \citenamefont
  {Zoller}}]{doi:10.1126/sciadv.aau8342}%
  \BibitemOpen
  \bibfield  {author} {\bibinfo {author} {\bibfnamefont {M.}~\bibnamefont
  {Heyl}}, \bibinfo {author} {\bibfnamefont {P.}~\bibnamefont {Hauke}},\ and\
  \bibinfo {author} {\bibfnamefont {P.}~\bibnamefont {Zoller}},\ }\bibfield
  {title} {\bibinfo {title} {Quantum localization bounds trotter errors in
  digital quantum simulation},\ }\href {https://doi.org/10.1126/sciadv.aau8342}
  {\bibfield  {journal} {\bibinfo  {journal} {Science Advances}\ }\textbf
  {\bibinfo {volume} {5}},\ \bibinfo {pages} {eaau8342} (\bibinfo {year}
  {2019})}\BibitemShut {NoStop}%
\bibitem [{\citenamefont {Weidinger}\ and\ \citenamefont
  {Knap}(2017)}]{Weidinger2017}%
  \BibitemOpen
  \bibfield  {author} {\bibinfo {author} {\bibfnamefont {S.~A.}\ \bibnamefont
  {Weidinger}}\ and\ \bibinfo {author} {\bibfnamefont {M.}~\bibnamefont
  {Knap}},\ }\bibfield  {title} {\bibinfo {title} {Floquet prethermalization
  and regimes of heating in a periodically driven, interacting quantum
  system},\ }\href {https://doi.org/10.1038/srep45382} {\bibfield  {journal}
  {\bibinfo  {journal} {Scientific Reports}\ }\textbf {\bibinfo {volume} {7}},\
  \bibinfo {pages} {45382} (\bibinfo {year} {2017})}\BibitemShut {NoStop}%
\bibitem [{\citenamefont {Ho}\ \emph {et~al.}(2023)\citenamefont {Ho},
  \citenamefont {Mori}, \citenamefont {Abanin},\ and\ \citenamefont {{Dalla
  Torre}}}]{HO2023169297}%
  \BibitemOpen
  \bibfield  {author} {\bibinfo {author} {\bibfnamefont {W.~W.}\ \bibnamefont
  {Ho}}, \bibinfo {author} {\bibfnamefont {T.}~\bibnamefont {Mori}}, \bibinfo
  {author} {\bibfnamefont {D.~A.}\ \bibnamefont {Abanin}},\ and\ \bibinfo
  {author} {\bibfnamefont {E.~G.}\ \bibnamefont {{Dalla Torre}}},\ }\bibfield
  {title} {\bibinfo {title} {Quantum and classical floquet prethermalization},\
  }\href {https://doi.org/https://doi.org/10.1016/j.aop.2023.169297} {\bibfield
   {journal} {\bibinfo  {journal} {Annals of Physics}\ }\textbf {\bibinfo
  {volume} {454}},\ \bibinfo {pages} {169297} (\bibinfo {year}
  {2023})}\BibitemShut {NoStop}%
\bibitem [{\citenamefont {D'Alessio}\ \emph {et~al.}(2016)\citenamefont
  {D'Alessio}, \citenamefont {Kafri}, \citenamefont {Polkovnikov},\ and\
  \citenamefont {Rigol}}]{dallesio_ETH}%
  \BibitemOpen
  \bibfield  {author} {\bibinfo {author} {\bibfnamefont {L.}~\bibnamefont
  {D'Alessio}}, \bibinfo {author} {\bibfnamefont {Y.}~\bibnamefont {Kafri}},
  \bibinfo {author} {\bibfnamefont {A.}~\bibnamefont {Polkovnikov}},\ and\
  \bibinfo {author} {\bibfnamefont {M.}~\bibnamefont {Rigol}},\ }\bibfield
  {title} {\bibinfo {title} {From quantum chaos and eigenstate thermalization
  to statistical mechanics and thermodynamics},\ }\href
  {https://doi.org/10.1080/00018732.2016.1198134} {\bibfield  {journal}
  {\bibinfo  {journal} {Advances in Physics}\ }\textbf {\bibinfo {volume}
  {65}},\ \bibinfo {pages} {239} (\bibinfo {year} {2016})}\BibitemShut
  {NoStop}%
\bibitem [{\citenamefont {Moses}\ \emph {et~al.}(2023)\citenamefont {Moses},
  \citenamefont {Baldwin}, \citenamefont {Allman}, \citenamefont {Ancona},
  \citenamefont {Ascarrunz}, \citenamefont {Barnes}, \citenamefont
  {Bartolotta}, \citenamefont {Bjork}, \citenamefont {Blanchard}, \citenamefont
  {Bohn}, \citenamefont {Bohnet}, \citenamefont {Brown}, \citenamefont
  {Burdick}, \citenamefont {Burton}, \citenamefont {Campbell}, \citenamefont
  {Campora}, \citenamefont {Carron}, \citenamefont {Chambers}, \citenamefont
  {Chan}, \citenamefont {Chen}, \citenamefont {Chernoguzov}, \citenamefont
  {Chertkov}, \citenamefont {Colina}, \citenamefont {Curtis}, \citenamefont
  {Daniel}, \citenamefont {DeCross}, \citenamefont {Deen}, \citenamefont
  {Delaney}, \citenamefont {Dreiling}, \citenamefont {Ertsgaard}, \citenamefont
  {Esposito}, \citenamefont {Estey}, \citenamefont {Fabrikant}, \citenamefont
  {Figgatt}, \citenamefont {Foltz}, \citenamefont {Foss-Feig}, \citenamefont
  {Francois}, \citenamefont {Gaebler}, \citenamefont {Gatterman}, \citenamefont
  {Gilbreth}, \citenamefont {Giles}, \citenamefont {Glynn}, \citenamefont
  {Hall}, \citenamefont {Hankin}, \citenamefont {Hansen}, \citenamefont
  {Hayes}, \citenamefont {Higashi}, \citenamefont {Hoffman}, \citenamefont
  {Horning}, \citenamefont {Hout}, \citenamefont {Jacobs}, \citenamefont
  {Johansen}, \citenamefont {Jones}, \citenamefont {Karcz}, \citenamefont
  {Klein}, \citenamefont {Lauria}, \citenamefont {Lee}, \citenamefont {Liefer},
  \citenamefont {Lu}, \citenamefont {Lucchetti}, \citenamefont {Lytle},
  \citenamefont {Malm}, \citenamefont {Matheny}, \citenamefont {Mathewson},
  \citenamefont {Mayer}, \citenamefont {Miller}, \citenamefont {Mills},
  \citenamefont {Neyenhuis}, \citenamefont {Nugent}, \citenamefont {Olson},
  \citenamefont {Parks}, \citenamefont {Price}, \citenamefont {Price},
  \citenamefont {Pugh}, \citenamefont {Ransford}, \citenamefont {Reed},
  \citenamefont {Roman}, \citenamefont {Rowe}, \citenamefont {Ryan-Anderson},
  \citenamefont {Sanders}, \citenamefont {Sedlacek}, \citenamefont {Shevchuk},
  \citenamefont {Siegfried}, \citenamefont {Skripka}, \citenamefont {Spaun},
  \citenamefont {Sprenkle}, \citenamefont {Stutz}, \citenamefont {Swallows},
  \citenamefont {Tobey}, \citenamefont {Tran}, \citenamefont {Tran},
  \citenamefont {Vogt}, \citenamefont {Volin}, \citenamefont {Walker},
  \citenamefont {Zolot},\ and\ \citenamefont {Pino}}]{Moses2023}%
  \BibitemOpen
  \bibfield  {author} {\bibinfo {author} {\bibfnamefont {S.~A.}\ \bibnamefont
  {Moses}}, \bibinfo {author} {\bibfnamefont {C.~H.}\ \bibnamefont {Baldwin}},
  \bibinfo {author} {\bibfnamefont {M.~S.}\ \bibnamefont {Allman}}, \bibinfo
  {author} {\bibfnamefont {R.}~\bibnamefont {Ancona}}, \bibinfo {author}
  {\bibfnamefont {L.}~\bibnamefont {Ascarrunz}}, \bibinfo {author}
  {\bibfnamefont {C.}~\bibnamefont {Barnes}}, \bibinfo {author} {\bibfnamefont
  {J.}~\bibnamefont {Bartolotta}}, \bibinfo {author} {\bibfnamefont
  {B.}~\bibnamefont {Bjork}}, \bibinfo {author} {\bibfnamefont
  {P.}~\bibnamefont {Blanchard}}, \bibinfo {author} {\bibfnamefont
  {M.}~\bibnamefont {Bohn}}, \bibinfo {author} {\bibfnamefont {J.~G.}\
  \bibnamefont {Bohnet}}, \bibinfo {author} {\bibfnamefont {N.~C.}\
  \bibnamefont {Brown}}, \bibinfo {author} {\bibfnamefont {N.~Q.}\ \bibnamefont
  {Burdick}}, \bibinfo {author} {\bibfnamefont {W.~C.}\ \bibnamefont {Burton}},
  \bibinfo {author} {\bibfnamefont {S.~L.}\ \bibnamefont {Campbell}}, \bibinfo
  {author} {\bibfnamefont {J.~P.}\ \bibnamefont {Campora}}, \bibinfo {author}
  {\bibfnamefont {C.}~\bibnamefont {Carron}}, \bibinfo {author} {\bibfnamefont
  {J.}~\bibnamefont {Chambers}}, \bibinfo {author} {\bibfnamefont {J.~W.}\
  \bibnamefont {Chan}}, \bibinfo {author} {\bibfnamefont {Y.~H.}\ \bibnamefont
  {Chen}}, \bibinfo {author} {\bibfnamefont {A.}~\bibnamefont {Chernoguzov}},
  \bibinfo {author} {\bibfnamefont {E.}~\bibnamefont {Chertkov}}, \bibinfo
  {author} {\bibfnamefont {J.}~\bibnamefont {Colina}}, \bibinfo {author}
  {\bibfnamefont {J.~P.}\ \bibnamefont {Curtis}}, \bibinfo {author}
  {\bibfnamefont {R.}~\bibnamefont {Daniel}}, \bibinfo {author} {\bibfnamefont
  {M.}~\bibnamefont {DeCross}}, \bibinfo {author} {\bibfnamefont
  {D.}~\bibnamefont {Deen}}, \bibinfo {author} {\bibfnamefont {C.}~\bibnamefont
  {Delaney}}, \bibinfo {author} {\bibfnamefont {J.~M.}\ \bibnamefont
  {Dreiling}}, \bibinfo {author} {\bibfnamefont {C.~T.}\ \bibnamefont
  {Ertsgaard}}, \bibinfo {author} {\bibfnamefont {J.}~\bibnamefont {Esposito}},
  \bibinfo {author} {\bibfnamefont {B.}~\bibnamefont {Estey}}, \bibinfo
  {author} {\bibfnamefont {M.}~\bibnamefont {Fabrikant}}, \bibinfo {author}
  {\bibfnamefont {C.}~\bibnamefont {Figgatt}}, \bibinfo {author} {\bibfnamefont
  {C.}~\bibnamefont {Foltz}}, \bibinfo {author} {\bibfnamefont
  {M.}~\bibnamefont {Foss-Feig}}, \bibinfo {author} {\bibfnamefont
  {D.}~\bibnamefont {Francois}}, \bibinfo {author} {\bibfnamefont {J.~P.}\
  \bibnamefont {Gaebler}}, \bibinfo {author} {\bibfnamefont {T.~M.}\
  \bibnamefont {Gatterman}}, \bibinfo {author} {\bibfnamefont {C.~N.}\
  \bibnamefont {Gilbreth}}, \bibinfo {author} {\bibfnamefont {J.}~\bibnamefont
  {Giles}}, \bibinfo {author} {\bibfnamefont {E.}~\bibnamefont {Glynn}},
  \bibinfo {author} {\bibfnamefont {A.}~\bibnamefont {Hall}}, \bibinfo {author}
  {\bibfnamefont {A.~M.}\ \bibnamefont {Hankin}}, \bibinfo {author}
  {\bibfnamefont {A.}~\bibnamefont {Hansen}}, \bibinfo {author} {\bibfnamefont
  {D.}~\bibnamefont {Hayes}}, \bibinfo {author} {\bibfnamefont
  {B.}~\bibnamefont {Higashi}}, \bibinfo {author} {\bibfnamefont {I.~M.}\
  \bibnamefont {Hoffman}}, \bibinfo {author} {\bibfnamefont {B.}~\bibnamefont
  {Horning}}, \bibinfo {author} {\bibfnamefont {J.~J.}\ \bibnamefont {Hout}},
  \bibinfo {author} {\bibfnamefont {R.}~\bibnamefont {Jacobs}}, \bibinfo
  {author} {\bibfnamefont {J.}~\bibnamefont {Johansen}}, \bibinfo {author}
  {\bibfnamefont {L.}~\bibnamefont {Jones}}, \bibinfo {author} {\bibfnamefont
  {J.}~\bibnamefont {Karcz}}, \bibinfo {author} {\bibfnamefont
  {T.}~\bibnamefont {Klein}}, \bibinfo {author} {\bibfnamefont
  {P.}~\bibnamefont {Lauria}}, \bibinfo {author} {\bibfnamefont
  {P.}~\bibnamefont {Lee}}, \bibinfo {author} {\bibfnamefont {D.}~\bibnamefont
  {Liefer}}, \bibinfo {author} {\bibfnamefont {S.~T.}\ \bibnamefont {Lu}},
  \bibinfo {author} {\bibfnamefont {D.}~\bibnamefont {Lucchetti}}, \bibinfo
  {author} {\bibfnamefont {C.}~\bibnamefont {Lytle}}, \bibinfo {author}
  {\bibfnamefont {A.}~\bibnamefont {Malm}}, \bibinfo {author} {\bibfnamefont
  {M.}~\bibnamefont {Matheny}}, \bibinfo {author} {\bibfnamefont
  {B.}~\bibnamefont {Mathewson}}, \bibinfo {author} {\bibfnamefont
  {K.}~\bibnamefont {Mayer}}, \bibinfo {author} {\bibfnamefont {D.~B.}\
  \bibnamefont {Miller}}, \bibinfo {author} {\bibfnamefont {M.}~\bibnamefont
  {Mills}}, \bibinfo {author} {\bibfnamefont {B.}~\bibnamefont {Neyenhuis}},
  \bibinfo {author} {\bibfnamefont {L.}~\bibnamefont {Nugent}}, \bibinfo
  {author} {\bibfnamefont {S.}~\bibnamefont {Olson}}, \bibinfo {author}
  {\bibfnamefont {J.}~\bibnamefont {Parks}}, \bibinfo {author} {\bibfnamefont
  {G.~N.}\ \bibnamefont {Price}}, \bibinfo {author} {\bibfnamefont
  {Z.}~\bibnamefont {Price}}, \bibinfo {author} {\bibfnamefont
  {M.}~\bibnamefont {Pugh}}, \bibinfo {author} {\bibfnamefont {A.}~\bibnamefont
  {Ransford}}, \bibinfo {author} {\bibfnamefont {A.~P.}\ \bibnamefont {Reed}},
  \bibinfo {author} {\bibfnamefont {C.}~\bibnamefont {Roman}}, \bibinfo
  {author} {\bibfnamefont {M.}~\bibnamefont {Rowe}}, \bibinfo {author}
  {\bibfnamefont {C.}~\bibnamefont {Ryan-Anderson}}, \bibinfo {author}
  {\bibfnamefont {S.}~\bibnamefont {Sanders}}, \bibinfo {author} {\bibfnamefont
  {J.}~\bibnamefont {Sedlacek}}, \bibinfo {author} {\bibfnamefont
  {P.}~\bibnamefont {Shevchuk}}, \bibinfo {author} {\bibfnamefont
  {P.}~\bibnamefont {Siegfried}}, \bibinfo {author} {\bibfnamefont
  {T.}~\bibnamefont {Skripka}}, \bibinfo {author} {\bibfnamefont
  {B.}~\bibnamefont {Spaun}}, \bibinfo {author} {\bibfnamefont {R.~T.}\
  \bibnamefont {Sprenkle}}, \bibinfo {author} {\bibfnamefont {R.~P.}\
  \bibnamefont {Stutz}}, \bibinfo {author} {\bibfnamefont {M.}~\bibnamefont
  {Swallows}}, \bibinfo {author} {\bibfnamefont {R.~I.}\ \bibnamefont {Tobey}},
  \bibinfo {author} {\bibfnamefont {A.}~\bibnamefont {Tran}}, \bibinfo {author}
  {\bibfnamefont {T.}~\bibnamefont {Tran}}, \bibinfo {author} {\bibfnamefont
  {E.}~\bibnamefont {Vogt}}, \bibinfo {author} {\bibfnamefont {C.}~\bibnamefont
  {Volin}}, \bibinfo {author} {\bibfnamefont {J.}~\bibnamefont {Walker}},
  \bibinfo {author} {\bibfnamefont {A.~M.}\ \bibnamefont {Zolot}},\ and\
  \bibinfo {author} {\bibfnamefont {J.~M.}\ \bibnamefont {Pino}},\ }\bibfield
  {title} {\bibinfo {title} {A race-track trapped-ion quantum processor},\
  }\href {https://doi.org/10.1103/PhysRevX.13.041052} {\bibfield  {journal}
  {\bibinfo  {journal} {Phys. Rev. X}\ }\textbf {\bibinfo {volume} {13}},\
  \bibinfo {pages} {041052} (\bibinfo {year} {2023})}\BibitemShut {NoStop}%
\bibitem [{\citenamefont {DeCross}\ \emph {et~al.}(2025)\citenamefont
  {DeCross}, \citenamefont {Haghshenas}, \citenamefont {Liu}, \citenamefont
  {Rinaldi}, \citenamefont {Gray}, \citenamefont {Alexeev}, \citenamefont
  {Baldwin}, \citenamefont {Bartolotta}, \citenamefont {Bohn}, \citenamefont
  {Chertkov}, \citenamefont {Cline}, \citenamefont {Colina}, \citenamefont
  {DelVento}, \citenamefont {Dreiling}, \citenamefont {Foltz}, \citenamefont
  {Gaebler}, \citenamefont {Gatterman}, \citenamefont {Gilbreth}, \citenamefont
  {Giles}, \citenamefont {Gresh}, \citenamefont {Hall}, \citenamefont {Hankin},
  \citenamefont {Hansen}, \citenamefont {Hewitt}, \citenamefont {Hoffman},
  \citenamefont {Holliman}, \citenamefont {Hutson}, \citenamefont {Jacobs},
  \citenamefont {Johansen}, \citenamefont {Lee}, \citenamefont {Lehman},
  \citenamefont {Lucchetti}, \citenamefont {Lykov}, \citenamefont {Madjarov},
  \citenamefont {Mathewson}, \citenamefont {Mayer}, \citenamefont {Mills},
  \citenamefont {Niroula}, \citenamefont {Pino}, \citenamefont {Roman},
  \citenamefont {Schecter}, \citenamefont {Siegfried}, \citenamefont {Tiemann},
  \citenamefont {Volin}, \citenamefont {Walker}, \citenamefont {Shaydulin},
  \citenamefont {Pistoia}, \citenamefont {Moses}, \citenamefont {Hayes},
  \citenamefont {Neyenhuis}, \citenamefont {Stutz},\ and\ \citenamefont
  {Foss-Feig}}]{decross2024computational}%
  \BibitemOpen
  \bibfield  {author} {\bibinfo {author} {\bibfnamefont {M.}~\bibnamefont
  {DeCross}}, \bibinfo {author} {\bibfnamefont {R.}~\bibnamefont {Haghshenas}},
  \bibinfo {author} {\bibfnamefont {M.}~\bibnamefont {Liu}}, \bibinfo {author}
  {\bibfnamefont {E.}~\bibnamefont {Rinaldi}}, \bibinfo {author} {\bibfnamefont
  {J.}~\bibnamefont {Gray}}, \bibinfo {author} {\bibfnamefont {Y.}~\bibnamefont
  {Alexeev}}, \bibinfo {author} {\bibfnamefont {C.~H.}\ \bibnamefont
  {Baldwin}}, \bibinfo {author} {\bibfnamefont {J.~P.}\ \bibnamefont
  {Bartolotta}}, \bibinfo {author} {\bibfnamefont {M.}~\bibnamefont {Bohn}},
  \bibinfo {author} {\bibfnamefont {E.}~\bibnamefont {Chertkov}}, \bibinfo
  {author} {\bibfnamefont {J.}~\bibnamefont {Cline}}, \bibinfo {author}
  {\bibfnamefont {J.}~\bibnamefont {Colina}}, \bibinfo {author} {\bibfnamefont
  {D.}~\bibnamefont {DelVento}}, \bibinfo {author} {\bibfnamefont {J.~M.}\
  \bibnamefont {Dreiling}}, \bibinfo {author} {\bibfnamefont {C.}~\bibnamefont
  {Foltz}}, \bibinfo {author} {\bibfnamefont {J.~P.}\ \bibnamefont {Gaebler}},
  \bibinfo {author} {\bibfnamefont {T.~M.}\ \bibnamefont {Gatterman}}, \bibinfo
  {author} {\bibfnamefont {C.~N.}\ \bibnamefont {Gilbreth}}, \bibinfo {author}
  {\bibfnamefont {J.}~\bibnamefont {Giles}}, \bibinfo {author} {\bibfnamefont
  {D.}~\bibnamefont {Gresh}}, \bibinfo {author} {\bibfnamefont
  {A.}~\bibnamefont {Hall}}, \bibinfo {author} {\bibfnamefont {A.}~\bibnamefont
  {Hankin}}, \bibinfo {author} {\bibfnamefont {A.}~\bibnamefont {Hansen}},
  \bibinfo {author} {\bibfnamefont {N.}~\bibnamefont {Hewitt}}, \bibinfo
  {author} {\bibfnamefont {I.}~\bibnamefont {Hoffman}}, \bibinfo {author}
  {\bibfnamefont {C.}~\bibnamefont {Holliman}}, \bibinfo {author}
  {\bibfnamefont {R.~B.}\ \bibnamefont {Hutson}}, \bibinfo {author}
  {\bibfnamefont {T.}~\bibnamefont {Jacobs}}, \bibinfo {author} {\bibfnamefont
  {J.}~\bibnamefont {Johansen}}, \bibinfo {author} {\bibfnamefont {P.~J.}\
  \bibnamefont {Lee}}, \bibinfo {author} {\bibfnamefont {E.}~\bibnamefont
  {Lehman}}, \bibinfo {author} {\bibfnamefont {D.}~\bibnamefont {Lucchetti}},
  \bibinfo {author} {\bibfnamefont {D.}~\bibnamefont {Lykov}}, \bibinfo
  {author} {\bibfnamefont {I.~S.}\ \bibnamefont {Madjarov}}, \bibinfo {author}
  {\bibfnamefont {B.}~\bibnamefont {Mathewson}}, \bibinfo {author}
  {\bibfnamefont {K.}~\bibnamefont {Mayer}}, \bibinfo {author} {\bibfnamefont
  {M.}~\bibnamefont {Mills}}, \bibinfo {author} {\bibfnamefont
  {P.}~\bibnamefont {Niroula}}, \bibinfo {author} {\bibfnamefont {J.~M.}\
  \bibnamefont {Pino}}, \bibinfo {author} {\bibfnamefont {C.}~\bibnamefont
  {Roman}}, \bibinfo {author} {\bibfnamefont {M.}~\bibnamefont {Schecter}},
  \bibinfo {author} {\bibfnamefont {P.~E.}\ \bibnamefont {Siegfried}}, \bibinfo
  {author} {\bibfnamefont {B.~G.}\ \bibnamefont {Tiemann}}, \bibinfo {author}
  {\bibfnamefont {C.}~\bibnamefont {Volin}}, \bibinfo {author} {\bibfnamefont
  {J.}~\bibnamefont {Walker}}, \bibinfo {author} {\bibfnamefont
  {R.}~\bibnamefont {Shaydulin}}, \bibinfo {author} {\bibfnamefont
  {M.}~\bibnamefont {Pistoia}}, \bibinfo {author} {\bibfnamefont {S.~A.}\
  \bibnamefont {Moses}}, \bibinfo {author} {\bibfnamefont {D.}~\bibnamefont
  {Hayes}}, \bibinfo {author} {\bibfnamefont {B.}~\bibnamefont {Neyenhuis}},
  \bibinfo {author} {\bibfnamefont {R.~P.}\ \bibnamefont {Stutz}},\ and\
  \bibinfo {author} {\bibfnamefont {M.}~\bibnamefont {Foss-Feig}},\ }\bibfield
  {title} {\bibinfo {title} {Computational power of random quantum circuits in
  arbitrary geometries},\ }\href {https://doi.org/10.1103/PhysRevX.15.021052}
  {\bibfield  {journal} {\bibinfo  {journal} {Phys. Rev. X}\ }\textbf {\bibinfo
  {volume} {15}},\ \bibinfo {pages} {021052} (\bibinfo {year}
  {2025})}\BibitemShut {NoStop}%
\bibitem [{\citenamefont {Moessner}\ \emph {et~al.}(2000)\citenamefont
  {Moessner}, \citenamefont {Sondhi},\ and\ \citenamefont
  {Chandra}}]{PhysRevLett.84.4457}%
  \BibitemOpen
  \bibfield  {author} {\bibinfo {author} {\bibfnamefont {R.}~\bibnamefont
  {Moessner}}, \bibinfo {author} {\bibfnamefont {S.~L.}\ \bibnamefont
  {Sondhi}},\ and\ \bibinfo {author} {\bibfnamefont {P.}~\bibnamefont
  {Chandra}},\ }\bibfield  {title} {\bibinfo {title} {Two-dimensional periodic
  frustrated ising models in a transverse field},\ }\href
  {https://doi.org/10.1103/PhysRevLett.84.4457} {\bibfield  {journal} {\bibinfo
   {journal} {Phys. Rev. Lett.}\ }\textbf {\bibinfo {volume} {84}},\ \bibinfo
  {pages} {4457} (\bibinfo {year} {2000})}\BibitemShut {NoStop}%
\bibitem [{\citenamefont {Moessner}\ and\ \citenamefont
  {Sondhi}(2001)}]{PhysRevB.63.224401}%
  \BibitemOpen
  \bibfield  {author} {\bibinfo {author} {\bibfnamefont {R.}~\bibnamefont
  {Moessner}}\ and\ \bibinfo {author} {\bibfnamefont {S.~L.}\ \bibnamefont
  {Sondhi}},\ }\bibfield  {title} {\bibinfo {title} {Ising models of quantum
  frustration},\ }\href {https://doi.org/10.1103/PhysRevB.63.224401} {\bibfield
   {journal} {\bibinfo  {journal} {Phys. Rev. B}\ }\textbf {\bibinfo {volume}
  {63}},\ \bibinfo {pages} {224401} (\bibinfo {year} {2001})}\BibitemShut
  {NoStop}%
\bibitem [{\citenamefont {Savary}\ and\ \citenamefont
  {Balents}(2016)}]{savary2016quantum}%
  \BibitemOpen
  \bibfield  {author} {\bibinfo {author} {\bibfnamefont {L.}~\bibnamefont
  {Savary}}\ and\ \bibinfo {author} {\bibfnamefont {L.}~\bibnamefont
  {Balents}},\ }\bibfield  {title} {\bibinfo {title} {Quantum spin liquids: a
  review},\ }\href@noop {} {\bibfield  {journal} {\bibinfo  {journal} {Reports
  on Progress in Physics}\ }\textbf {\bibinfo {volume} {80}},\ \bibinfo {pages}
  {016502} (\bibinfo {year} {2016})}\BibitemShut {NoStop}%
\bibitem [{\citenamefont {Lloyd}(1996)}]{doi:10.1126/science.273.5278.1073}%
  \BibitemOpen
  \bibfield  {author} {\bibinfo {author} {\bibfnamefont {S.}~\bibnamefont
  {Lloyd}},\ }\bibfield  {title} {\bibinfo {title} {Universal quantum
  simulators},\ }\href {https://doi.org/10.1126/science.273.5278.1073}
  {\bibfield  {journal} {\bibinfo  {journal} {Science}\ }\textbf {\bibinfo
  {volume} {273}},\ \bibinfo {pages} {1073} (\bibinfo {year}
  {1996})}\BibitemShut {NoStop}%
\bibitem [{\citenamefont {Daley}\ \emph {et~al.}(2022)\citenamefont {Daley},
  \citenamefont {Bloch}, \citenamefont {Kokail}, \citenamefont {Flannigan},
  \citenamefont {Pearson}, \citenamefont {Troyer},\ and\ \citenamefont
  {Zoller}}]{daley_qa_review}%
  \BibitemOpen
  \bibfield  {author} {\bibinfo {author} {\bibfnamefont {A.~J.}\ \bibnamefont
  {Daley}}, \bibinfo {author} {\bibfnamefont {I.}~\bibnamefont {Bloch}},
  \bibinfo {author} {\bibfnamefont {C.}~\bibnamefont {Kokail}}, \bibinfo
  {author} {\bibfnamefont {S.}~\bibnamefont {Flannigan}}, \bibinfo {author}
  {\bibfnamefont {N.}~\bibnamefont {Pearson}}, \bibinfo {author} {\bibfnamefont
  {M.}~\bibnamefont {Troyer}},\ and\ \bibinfo {author} {\bibfnamefont
  {P.}~\bibnamefont {Zoller}},\ }\bibfield  {title} {\bibinfo {title}
  {Practical quantum advantage in quantum simulation},\ }\href
  {https://doi.org/10.1038/s41586-022-04940-6} {\bibfield  {journal} {\bibinfo
  {journal} {Nature}\ }\textbf {\bibinfo {volume} {607}},\ \bibinfo {pages}
  {667} (\bibinfo {year} {2022})}\BibitemShut {NoStop}%
\bibitem [{\citenamefont {Hild}\ \emph {et~al.}(2014)\citenamefont {Hild},
  \citenamefont {Fukuhara}, \citenamefont {Schau\ss{}}, \citenamefont {Zeiher},
  \citenamefont {Knap}, \citenamefont {Demler}, \citenamefont {Bloch},\ and\
  \citenamefont {Gross}}]{PhysRevLett.113.147205}%
  \BibitemOpen
  \bibfield  {author} {\bibinfo {author} {\bibfnamefont {S.}~\bibnamefont
  {Hild}}, \bibinfo {author} {\bibfnamefont {T.}~\bibnamefont {Fukuhara}},
  \bibinfo {author} {\bibfnamefont {P.}~\bibnamefont {Schau\ss{}}}, \bibinfo
  {author} {\bibfnamefont {J.}~\bibnamefont {Zeiher}}, \bibinfo {author}
  {\bibfnamefont {M.}~\bibnamefont {Knap}}, \bibinfo {author} {\bibfnamefont
  {E.}~\bibnamefont {Demler}}, \bibinfo {author} {\bibfnamefont
  {I.}~\bibnamefont {Bloch}},\ and\ \bibinfo {author} {\bibfnamefont
  {C.}~\bibnamefont {Gross}},\ }\bibfield  {title} {\bibinfo {title}
  {Far-from-equilibrium spin transport in heisenberg quantum magnets},\ }\href
  {https://doi.org/10.1103/PhysRevLett.113.147205} {\bibfield  {journal}
  {\bibinfo  {journal} {Phys. Rev. Lett.}\ }\textbf {\bibinfo {volume} {113}},\
  \bibinfo {pages} {147205} (\bibinfo {year} {2014})}\BibitemShut {NoStop}%
\bibitem [{\citenamefont {Bernien}\ \emph {et~al.}(2017)\citenamefont
  {Bernien}, \citenamefont {Schwartz}, \citenamefont {Keesling}, \citenamefont
  {Levine}, \citenamefont {Omran}, \citenamefont {Pichler}, \citenamefont
  {Choi}, \citenamefont {Zibrov}, \citenamefont {Endres}, \citenamefont
  {Greiner}, \citenamefont {Vuleti{\'c}},\ and\ \citenamefont
  {Lukin}}]{bernien_2017}%
  \BibitemOpen
  \bibfield  {author} {\bibinfo {author} {\bibfnamefont {H.}~\bibnamefont
  {Bernien}}, \bibinfo {author} {\bibfnamefont {S.}~\bibnamefont {Schwartz}},
  \bibinfo {author} {\bibfnamefont {A.}~\bibnamefont {Keesling}}, \bibinfo
  {author} {\bibfnamefont {H.}~\bibnamefont {Levine}}, \bibinfo {author}
  {\bibfnamefont {A.}~\bibnamefont {Omran}}, \bibinfo {author} {\bibfnamefont
  {H.}~\bibnamefont {Pichler}}, \bibinfo {author} {\bibfnamefont
  {S.}~\bibnamefont {Choi}}, \bibinfo {author} {\bibfnamefont {A.~S.}\
  \bibnamefont {Zibrov}}, \bibinfo {author} {\bibfnamefont {M.}~\bibnamefont
  {Endres}}, \bibinfo {author} {\bibfnamefont {M.}~\bibnamefont {Greiner}},
  \bibinfo {author} {\bibfnamefont {V.}~\bibnamefont {Vuleti{\'c}}},\ and\
  \bibinfo {author} {\bibfnamefont {M.~D.}\ \bibnamefont {Lukin}},\ }\bibfield
  {title} {\bibinfo {title} {Probing many-body dynamics on a 51-atom quantum
  simulator},\ }\href {https://doi.org/10.1038/nature24622} {\bibfield
  {journal} {\bibinfo  {journal} {Nature}\ }\textbf {\bibinfo {volume} {551}},\
  \bibinfo {pages} {579} (\bibinfo {year} {2017})}\BibitemShut {NoStop}%
\bibitem [{\citenamefont {Zhang}\ \emph {et~al.}(2017)\citenamefont {Zhang},
  \citenamefont {Pagano}, \citenamefont {Hess}, \citenamefont {Kyprianidis},
  \citenamefont {Becker}, \citenamefont {Kaplan}, \citenamefont {Gorshkov},
  \citenamefont {Gong},\ and\ \citenamefont {Monroe}}]{zhang_2017}%
  \BibitemOpen
  \bibfield  {author} {\bibinfo {author} {\bibfnamefont {J.}~\bibnamefont
  {Zhang}}, \bibinfo {author} {\bibfnamefont {G.}~\bibnamefont {Pagano}},
  \bibinfo {author} {\bibfnamefont {P.~W.}\ \bibnamefont {Hess}}, \bibinfo
  {author} {\bibfnamefont {A.}~\bibnamefont {Kyprianidis}}, \bibinfo {author}
  {\bibfnamefont {P.}~\bibnamefont {Becker}}, \bibinfo {author} {\bibfnamefont
  {H.}~\bibnamefont {Kaplan}}, \bibinfo {author} {\bibfnamefont {A.~V.}\
  \bibnamefont {Gorshkov}}, \bibinfo {author} {\bibfnamefont {Z.~X.}\
  \bibnamefont {Gong}},\ and\ \bibinfo {author} {\bibfnamefont
  {C.}~\bibnamefont {Monroe}},\ }\bibfield  {title} {\bibinfo {title}
  {Observation of a many-body dynamical phase transition with a 53-qubit
  quantum simulator},\ }\href {https://doi.org/10.1038/nature24654} {\bibfield
  {journal} {\bibinfo  {journal} {Nature}\ }\textbf {\bibinfo {volume} {551}},\
  \bibinfo {pages} {601} (\bibinfo {year} {2017})}\BibitemShut {NoStop}%
\bibitem [{\citenamefont {Joshi}\ \emph {et~al.}(2022)\citenamefont {Joshi},
  \citenamefont {Kranzl}, \citenamefont {Schuckert}, \citenamefont {Lovas},
  \citenamefont {Maier}, \citenamefont {Blatt}, \citenamefont {Knap},\ and\
  \citenamefont {Roos}}]{Joshi_2022_hydro}%
  \BibitemOpen
  \bibfield  {author} {\bibinfo {author} {\bibfnamefont {M.~K.}\ \bibnamefont
  {Joshi}}, \bibinfo {author} {\bibfnamefont {F.}~\bibnamefont {Kranzl}},
  \bibinfo {author} {\bibfnamefont {A.}~\bibnamefont {Schuckert}}, \bibinfo
  {author} {\bibfnamefont {I.}~\bibnamefont {Lovas}}, \bibinfo {author}
  {\bibfnamefont {C.}~\bibnamefont {Maier}}, \bibinfo {author} {\bibfnamefont
  {R.}~\bibnamefont {Blatt}}, \bibinfo {author} {\bibfnamefont
  {M.}~\bibnamefont {Knap}},\ and\ \bibinfo {author} {\bibfnamefont {C.~F.}\
  \bibnamefont {Roos}},\ }\bibfield  {title} {\bibinfo {title} {Observing
  emergent hydrodynamics in a long-range quantum magnet},\ }\href
  {https://doi.org/10.1126/science.abk2400} {\bibfield  {journal} {\bibinfo
  {journal} {Science}\ }\textbf {\bibinfo {volume} {376}},\ \bibinfo {pages}
  {720–724} (\bibinfo {year} {2022})}\BibitemShut {NoStop}%
\bibitem [{\citenamefont {Andersen}\ \emph {et~al.}(2025)\citenamefont
  {Andersen}, \citenamefont {Astrakhantsev}, \citenamefont {Karamlou},
  \citenamefont {Berndtsson}, \citenamefont {Motruk}, \citenamefont {Szasz},
  \citenamefont {Gross}, \citenamefont {Schuckert}, \citenamefont {Westerhout},
  \citenamefont {Zhang}, \citenamefont {Forati}, \citenamefont {Rossi},
  \citenamefont {Kobrin}, \citenamefont {Paolo}, \citenamefont {Klots},
  \citenamefont {Drozdov}, \citenamefont {Kurilovich}, \citenamefont
  {Petukhov}, \citenamefont {Ioffe}, \citenamefont {Elben}, \citenamefont
  {Rath}, \citenamefont {Vitale}, \citenamefont {Vermersch}, \citenamefont
  {Acharya}, \citenamefont {Beni}, \citenamefont {Anderson}, \citenamefont
  {Ansmann}, \citenamefont {Arute}, \citenamefont {Arya}, \citenamefont
  {Asfaw}, \citenamefont {Atalaya}, \citenamefont {Ballard}, \citenamefont
  {Bardin}, \citenamefont {Bengtsson}, \citenamefont {Bilmes}, \citenamefont
  {Bortoli}, \citenamefont {Bourassa}, \citenamefont {Bovaird}, \citenamefont
  {Brill}, \citenamefont {Broughton}, \citenamefont {Browne}, \citenamefont
  {Buchea}, \citenamefont {Buckley}, \citenamefont {Buell}, \citenamefont
  {Burger}, \citenamefont {Burkett}, \citenamefont {Bushnell}, \citenamefont
  {Cabrera}, \citenamefont {Campero}, \citenamefont {Chang}, \citenamefont
  {Chen}, \citenamefont {Chiaro}, \citenamefont {Claes}, \citenamefont
  {Cleland}, \citenamefont {Cogan}, \citenamefont {Collins}, \citenamefont
  {Conner}, \citenamefont {Courtney}, \citenamefont {Crook}, \citenamefont
  {Das}, \citenamefont {Debroy}, \citenamefont {Lorenzo}, \citenamefont
  {Barba}, \citenamefont {Demura}, \citenamefont {Donohoe}, \citenamefont
  {Dunsworth}, \citenamefont {Earle}, \citenamefont {Eickbusch}, \citenamefont
  {Elbag}, \citenamefont {Elzouka}, \citenamefont {Erickson}, \citenamefont
  {Faoro}, \citenamefont {Fatemi}, \citenamefont {Ferreira}, \citenamefont
  {Burgos}, \citenamefont {Fowler}, \citenamefont {Foxen}, \citenamefont
  {Ganjam}, \citenamefont {Gasca}, \citenamefont {Giang}, \citenamefont
  {Gidney}, \citenamefont {Gilboa}, \citenamefont {Giustina}, \citenamefont
  {Gosula}, \citenamefont {Dau}, \citenamefont {Graumann}, \citenamefont
  {Greene}, \citenamefont {Habegger}, \citenamefont {Hamilton}, \citenamefont
  {Hansen}, \citenamefont {Harrigan}, \citenamefont {Harrington}, \citenamefont
  {Heslin}, \citenamefont {Heu}, \citenamefont {Hill}, \citenamefont
  {Hoffmann}, \citenamefont {Huang}, \citenamefont {Huang}, \citenamefont
  {Huff}, \citenamefont {Huggins}, \citenamefont {Isakov}, \citenamefont
  {Jeffrey}, \citenamefont {Jiang}, \citenamefont {Jones}, \citenamefont
  {Jordan}, \citenamefont {Joshi}, \citenamefont {Juhas}, \citenamefont
  {Kafri}, \citenamefont {Kang}, \citenamefont {Kechedzhi}, \citenamefont
  {Khaire}, \citenamefont {Khattar}, \citenamefont {Khezri}, \citenamefont
  {Kieferov{\'a}}, \citenamefont {Kim}, \citenamefont {Kitaev}, \citenamefont
  {Klimov}, \citenamefont {Korotkov}, \citenamefont {Kostritsa}, \citenamefont
  {Kreikebaum}, \citenamefont {Landhuis}, \citenamefont {Langley},
  \citenamefont {Laptev}, \citenamefont {Lau}, \citenamefont {Guevel},
  \citenamefont {Ledford}, \citenamefont {Lee}, \citenamefont {Lee},
  \citenamefont {Lensky}, \citenamefont {Lester}, \citenamefont {Li},
  \citenamefont {Lill}, \citenamefont {Liu}, \citenamefont {Livingston},
  \citenamefont {Locharla}, \citenamefont {Lundahl}, \citenamefont {Lunt},
  \citenamefont {Madhuk}, \citenamefont {Maloney}, \citenamefont {Mandr{\`a}},
  \citenamefont {Martin}, \citenamefont {Martin}, \citenamefont {Martin},
  \citenamefont {Maxfield}, \citenamefont {McClean}, \citenamefont {McEwen},
  \citenamefont {Meeks}, \citenamefont {Miao}, \citenamefont {Mieszala},
  \citenamefont {Molina}, \citenamefont {Montazeri}, \citenamefont {Morvan},
  \citenamefont {Movassagh}, \citenamefont {Neill}, \citenamefont {Nersisyan},
  \citenamefont {Newman}, \citenamefont {Nguyen}, \citenamefont {Nguyen},
  \citenamefont {Ni}, \citenamefont {Niu}, \citenamefont {Oliver},
  \citenamefont {Ottosson}, \citenamefont {Pizzuto}, \citenamefont {Potter},
  \citenamefont {Pritchard}, \citenamefont {Pryadko}, \citenamefont {Quintana},
  \citenamefont {Reagor}, \citenamefont {Rhodes}, \citenamefont {Roberts},
  \citenamefont {Rocque}, \citenamefont {Rosenberg}, \citenamefont {Rubin},
  \citenamefont {Saei}, \citenamefont {Sankaragomathi}, \citenamefont
  {Satzinger}, \citenamefont {Schurkus}, \citenamefont {Schuster},
  \citenamefont {Shearn}, \citenamefont {Shorter}, \citenamefont {Shutty},
  \citenamefont {Shvarts}, \citenamefont {Sivak}, \citenamefont {Skruzny},
  \citenamefont {Small}, \citenamefont {Smith}, \citenamefont {Springer},
  \citenamefont {Sterling}, \citenamefont {Suchard}, \citenamefont {Szalay},
  \citenamefont {Sztein}, \citenamefont {Thor}, \citenamefont {Torres},
  \citenamefont {Torunbalci}, \citenamefont {Vaishnav}, \citenamefont
  {Vdovichev}, \citenamefont {Villalonga}, \citenamefont {Heidweiller},
  \citenamefont {Waltman}, \citenamefont {Wang}, \citenamefont {White},
  \citenamefont {Wong}, \citenamefont {Woo}, \citenamefont {Xing},
  \citenamefont {Yao}, \citenamefont {Yeh}, \citenamefont {Ying}, \citenamefont
  {Yoo}, \citenamefont {Yosri}, \citenamefont {Young}, \citenamefont {Zalcman},
  \citenamefont {Zhu}, \citenamefont {Zobrist}, \citenamefont {Neven},
  \citenamefont {Babbush}, \citenamefont {Boixo}, \citenamefont {Hilton},
  \citenamefont {Lucero}, \citenamefont {Megrant}, \citenamefont {Kelly},
  \citenamefont {Chen}, \citenamefont {Smelyanskiy}, \citenamefont {Vidal},
  \citenamefont {Roushan}, \citenamefont {L{\"a}uchli}, \citenamefont
  {Abanin},\ and\ \citenamefont {Mi}}]{andersen2024thermalization}%
  \BibitemOpen
  \bibfield  {author} {\bibinfo {author} {\bibfnamefont {T.~I.}\ \bibnamefont
  {Andersen}}, \bibinfo {author} {\bibfnamefont {N.}~\bibnamefont
  {Astrakhantsev}}, \bibinfo {author} {\bibfnamefont {A.~H.}\ \bibnamefont
  {Karamlou}}, \bibinfo {author} {\bibfnamefont {J.}~\bibnamefont
  {Berndtsson}}, \bibinfo {author} {\bibfnamefont {J.}~\bibnamefont {Motruk}},
  \bibinfo {author} {\bibfnamefont {A.}~\bibnamefont {Szasz}}, \bibinfo
  {author} {\bibfnamefont {J.~A.}\ \bibnamefont {Gross}}, \bibinfo {author}
  {\bibfnamefont {A.}~\bibnamefont {Schuckert}}, \bibinfo {author}
  {\bibfnamefont {T.}~\bibnamefont {Westerhout}}, \bibinfo {author}
  {\bibfnamefont {Y.}~\bibnamefont {Zhang}}, \bibinfo {author} {\bibfnamefont
  {E.}~\bibnamefont {Forati}}, \bibinfo {author} {\bibfnamefont
  {D.}~\bibnamefont {Rossi}}, \bibinfo {author} {\bibfnamefont
  {B.}~\bibnamefont {Kobrin}}, \bibinfo {author} {\bibfnamefont {A.~D.}\
  \bibnamefont {Paolo}}, \bibinfo {author} {\bibfnamefont {A.~R.}\ \bibnamefont
  {Klots}}, \bibinfo {author} {\bibfnamefont {I.}~\bibnamefont {Drozdov}},
  \bibinfo {author} {\bibfnamefont {V.}~\bibnamefont {Kurilovich}}, \bibinfo
  {author} {\bibfnamefont {A.}~\bibnamefont {Petukhov}}, \bibinfo {author}
  {\bibfnamefont {L.~B.}\ \bibnamefont {Ioffe}}, \bibinfo {author}
  {\bibfnamefont {A.}~\bibnamefont {Elben}}, \bibinfo {author} {\bibfnamefont
  {A.}~\bibnamefont {Rath}}, \bibinfo {author} {\bibfnamefont {V.}~\bibnamefont
  {Vitale}}, \bibinfo {author} {\bibfnamefont {B.}~\bibnamefont {Vermersch}},
  \bibinfo {author} {\bibfnamefont {R.}~\bibnamefont {Acharya}}, \bibinfo
  {author} {\bibfnamefont {L.~A.}\ \bibnamefont {Beni}}, \bibinfo {author}
  {\bibfnamefont {K.}~\bibnamefont {Anderson}}, \bibinfo {author}
  {\bibfnamefont {M.}~\bibnamefont {Ansmann}}, \bibinfo {author} {\bibfnamefont
  {F.}~\bibnamefont {Arute}}, \bibinfo {author} {\bibfnamefont
  {K.}~\bibnamefont {Arya}}, \bibinfo {author} {\bibfnamefont {A.}~\bibnamefont
  {Asfaw}}, \bibinfo {author} {\bibfnamefont {J.}~\bibnamefont {Atalaya}},
  \bibinfo {author} {\bibfnamefont {B.}~\bibnamefont {Ballard}}, \bibinfo
  {author} {\bibfnamefont {J.~C.}\ \bibnamefont {Bardin}}, \bibinfo {author}
  {\bibfnamefont {A.}~\bibnamefont {Bengtsson}}, \bibinfo {author}
  {\bibfnamefont {A.}~\bibnamefont {Bilmes}}, \bibinfo {author} {\bibfnamefont
  {G.}~\bibnamefont {Bortoli}}, \bibinfo {author} {\bibfnamefont
  {A.}~\bibnamefont {Bourassa}}, \bibinfo {author} {\bibfnamefont
  {J.}~\bibnamefont {Bovaird}}, \bibinfo {author} {\bibfnamefont
  {L.}~\bibnamefont {Brill}}, \bibinfo {author} {\bibfnamefont
  {M.}~\bibnamefont {Broughton}}, \bibinfo {author} {\bibfnamefont {D.~A.}\
  \bibnamefont {Browne}}, \bibinfo {author} {\bibfnamefont {B.}~\bibnamefont
  {Buchea}}, \bibinfo {author} {\bibfnamefont {B.~B.}\ \bibnamefont {Buckley}},
  \bibinfo {author} {\bibfnamefont {D.~A.}\ \bibnamefont {Buell}}, \bibinfo
  {author} {\bibfnamefont {T.}~\bibnamefont {Burger}}, \bibinfo {author}
  {\bibfnamefont {B.}~\bibnamefont {Burkett}}, \bibinfo {author} {\bibfnamefont
  {N.}~\bibnamefont {Bushnell}}, \bibinfo {author} {\bibfnamefont
  {A.}~\bibnamefont {Cabrera}}, \bibinfo {author} {\bibfnamefont
  {J.}~\bibnamefont {Campero}}, \bibinfo {author} {\bibfnamefont {H.~S.}\
  \bibnamefont {Chang}}, \bibinfo {author} {\bibfnamefont {Z.}~\bibnamefont
  {Chen}}, \bibinfo {author} {\bibfnamefont {B.}~\bibnamefont {Chiaro}},
  \bibinfo {author} {\bibfnamefont {J.}~\bibnamefont {Claes}}, \bibinfo
  {author} {\bibfnamefont {A.~Y.}\ \bibnamefont {Cleland}}, \bibinfo {author}
  {\bibfnamefont {J.}~\bibnamefont {Cogan}}, \bibinfo {author} {\bibfnamefont
  {R.}~\bibnamefont {Collins}}, \bibinfo {author} {\bibfnamefont
  {P.}~\bibnamefont {Conner}}, \bibinfo {author} {\bibfnamefont
  {W.}~\bibnamefont {Courtney}}, \bibinfo {author} {\bibfnamefont {A.~L.}\
  \bibnamefont {Crook}}, \bibinfo {author} {\bibfnamefont {S.}~\bibnamefont
  {Das}}, \bibinfo {author} {\bibfnamefont {D.~M.}\ \bibnamefont {Debroy}},
  \bibinfo {author} {\bibfnamefont {L.~D.}\ \bibnamefont {Lorenzo}}, \bibinfo
  {author} {\bibfnamefont {A.~D.~T.}\ \bibnamefont {Barba}}, \bibinfo {author}
  {\bibfnamefont {S.}~\bibnamefont {Demura}}, \bibinfo {author} {\bibfnamefont
  {P.}~\bibnamefont {Donohoe}}, \bibinfo {author} {\bibfnamefont
  {A.}~\bibnamefont {Dunsworth}}, \bibinfo {author} {\bibfnamefont
  {C.}~\bibnamefont {Earle}}, \bibinfo {author} {\bibfnamefont
  {A.}~\bibnamefont {Eickbusch}}, \bibinfo {author} {\bibfnamefont {A.~M.}\
  \bibnamefont {Elbag}}, \bibinfo {author} {\bibfnamefont {M.}~\bibnamefont
  {Elzouka}}, \bibinfo {author} {\bibfnamefont {C.}~\bibnamefont {Erickson}},
  \bibinfo {author} {\bibfnamefont {L.}~\bibnamefont {Faoro}}, \bibinfo
  {author} {\bibfnamefont {R.}~\bibnamefont {Fatemi}}, \bibinfo {author}
  {\bibfnamefont {V.~S.}\ \bibnamefont {Ferreira}}, \bibinfo {author}
  {\bibfnamefont {L.~F.}\ \bibnamefont {Burgos}}, \bibinfo {author}
  {\bibfnamefont {A.~G.}\ \bibnamefont {Fowler}}, \bibinfo {author}
  {\bibfnamefont {B.}~\bibnamefont {Foxen}}, \bibinfo {author} {\bibfnamefont
  {S.}~\bibnamefont {Ganjam}}, \bibinfo {author} {\bibfnamefont
  {R.}~\bibnamefont {Gasca}}, \bibinfo {author} {\bibfnamefont
  {W.}~\bibnamefont {Giang}}, \bibinfo {author} {\bibfnamefont
  {C.}~\bibnamefont {Gidney}}, \bibinfo {author} {\bibfnamefont
  {D.}~\bibnamefont {Gilboa}}, \bibinfo {author} {\bibfnamefont
  {M.}~\bibnamefont {Giustina}}, \bibinfo {author} {\bibfnamefont
  {R.}~\bibnamefont {Gosula}}, \bibinfo {author} {\bibfnamefont {A.~G.}\
  \bibnamefont {Dau}}, \bibinfo {author} {\bibfnamefont {D.}~\bibnamefont
  {Graumann}}, \bibinfo {author} {\bibfnamefont {A.}~\bibnamefont {Greene}},
  \bibinfo {author} {\bibfnamefont {S.}~\bibnamefont {Habegger}}, \bibinfo
  {author} {\bibfnamefont {M.~C.}\ \bibnamefont {Hamilton}}, \bibinfo {author}
  {\bibfnamefont {M.}~\bibnamefont {Hansen}}, \bibinfo {author} {\bibfnamefont
  {M.~P.}\ \bibnamefont {Harrigan}}, \bibinfo {author} {\bibfnamefont {S.~D.}\
  \bibnamefont {Harrington}}, \bibinfo {author} {\bibfnamefont
  {S.}~\bibnamefont {Heslin}}, \bibinfo {author} {\bibfnamefont
  {P.}~\bibnamefont {Heu}}, \bibinfo {author} {\bibfnamefont {G.}~\bibnamefont
  {Hill}}, \bibinfo {author} {\bibfnamefont {M.~R.}\ \bibnamefont {Hoffmann}},
  \bibinfo {author} {\bibfnamefont {H.~Y.}\ \bibnamefont {Huang}}, \bibinfo
  {author} {\bibfnamefont {T.}~\bibnamefont {Huang}}, \bibinfo {author}
  {\bibfnamefont {A.}~\bibnamefont {Huff}}, \bibinfo {author} {\bibfnamefont
  {W.~J.}\ \bibnamefont {Huggins}}, \bibinfo {author} {\bibfnamefont {S.~V.}\
  \bibnamefont {Isakov}}, \bibinfo {author} {\bibfnamefont {E.}~\bibnamefont
  {Jeffrey}}, \bibinfo {author} {\bibfnamefont {Z.}~\bibnamefont {Jiang}},
  \bibinfo {author} {\bibfnamefont {C.}~\bibnamefont {Jones}}, \bibinfo
  {author} {\bibfnamefont {S.}~\bibnamefont {Jordan}}, \bibinfo {author}
  {\bibfnamefont {C.}~\bibnamefont {Joshi}}, \bibinfo {author} {\bibfnamefont
  {P.}~\bibnamefont {Juhas}}, \bibinfo {author} {\bibfnamefont
  {D.}~\bibnamefont {Kafri}}, \bibinfo {author} {\bibfnamefont
  {H.}~\bibnamefont {Kang}}, \bibinfo {author} {\bibfnamefont {K.}~\bibnamefont
  {Kechedzhi}}, \bibinfo {author} {\bibfnamefont {T.}~\bibnamefont {Khaire}},
  \bibinfo {author} {\bibfnamefont {T.}~\bibnamefont {Khattar}}, \bibinfo
  {author} {\bibfnamefont {M.}~\bibnamefont {Khezri}}, \bibinfo {author}
  {\bibfnamefont {M.}~\bibnamefont {Kieferov{\'a}}}, \bibinfo {author}
  {\bibfnamefont {S.}~\bibnamefont {Kim}}, \bibinfo {author} {\bibfnamefont
  {A.}~\bibnamefont {Kitaev}}, \bibinfo {author} {\bibfnamefont
  {P.}~\bibnamefont {Klimov}}, \bibinfo {author} {\bibfnamefont {A.~N.}\
  \bibnamefont {Korotkov}}, \bibinfo {author} {\bibfnamefont {F.}~\bibnamefont
  {Kostritsa}}, \bibinfo {author} {\bibfnamefont {J.~M.}\ \bibnamefont
  {Kreikebaum}}, \bibinfo {author} {\bibfnamefont {D.}~\bibnamefont
  {Landhuis}}, \bibinfo {author} {\bibfnamefont {B.~W.}\ \bibnamefont
  {Langley}}, \bibinfo {author} {\bibfnamefont {P.}~\bibnamefont {Laptev}},
  \bibinfo {author} {\bibfnamefont {K.~M.}\ \bibnamefont {Lau}}, \bibinfo
  {author} {\bibfnamefont {L.~L.}\ \bibnamefont {Guevel}}, \bibinfo {author}
  {\bibfnamefont {J.}~\bibnamefont {Ledford}}, \bibinfo {author} {\bibfnamefont
  {J.}~\bibnamefont {Lee}}, \bibinfo {author} {\bibfnamefont {K.~W.}\
  \bibnamefont {Lee}}, \bibinfo {author} {\bibfnamefont {Y.~D.}\ \bibnamefont
  {Lensky}}, \bibinfo {author} {\bibfnamefont {B.~J.}\ \bibnamefont {Lester}},
  \bibinfo {author} {\bibfnamefont {W.~Y.}\ \bibnamefont {Li}}, \bibinfo
  {author} {\bibfnamefont {A.~T.}\ \bibnamefont {Lill}}, \bibinfo {author}
  {\bibfnamefont {W.}~\bibnamefont {Liu}}, \bibinfo {author} {\bibfnamefont
  {W.~P.}\ \bibnamefont {Livingston}}, \bibinfo {author} {\bibfnamefont
  {A.}~\bibnamefont {Locharla}}, \bibinfo {author} {\bibfnamefont
  {D.}~\bibnamefont {Lundahl}}, \bibinfo {author} {\bibfnamefont
  {A.}~\bibnamefont {Lunt}}, \bibinfo {author} {\bibfnamefont {S.}~\bibnamefont
  {Madhuk}}, \bibinfo {author} {\bibfnamefont {A.}~\bibnamefont {Maloney}},
  \bibinfo {author} {\bibfnamefont {S.}~\bibnamefont {Mandr{\`a}}}, \bibinfo
  {author} {\bibfnamefont {L.~S.}\ \bibnamefont {Martin}}, \bibinfo {author}
  {\bibfnamefont {O.}~\bibnamefont {Martin}}, \bibinfo {author} {\bibfnamefont
  {S.}~\bibnamefont {Martin}}, \bibinfo {author} {\bibfnamefont
  {C.}~\bibnamefont {Maxfield}}, \bibinfo {author} {\bibfnamefont {J.~R.}\
  \bibnamefont {McClean}}, \bibinfo {author} {\bibfnamefont {M.}~\bibnamefont
  {McEwen}}, \bibinfo {author} {\bibfnamefont {S.}~\bibnamefont {Meeks}},
  \bibinfo {author} {\bibfnamefont {K.~C.}\ \bibnamefont {Miao}}, \bibinfo
  {author} {\bibfnamefont {A.}~\bibnamefont {Mieszala}}, \bibinfo {author}
  {\bibfnamefont {S.}~\bibnamefont {Molina}}, \bibinfo {author} {\bibfnamefont
  {S.}~\bibnamefont {Montazeri}}, \bibinfo {author} {\bibfnamefont
  {A.}~\bibnamefont {Morvan}}, \bibinfo {author} {\bibfnamefont
  {R.}~\bibnamefont {Movassagh}}, \bibinfo {author} {\bibfnamefont
  {C.}~\bibnamefont {Neill}}, \bibinfo {author} {\bibfnamefont
  {A.}~\bibnamefont {Nersisyan}}, \bibinfo {author} {\bibfnamefont
  {M.}~\bibnamefont {Newman}}, \bibinfo {author} {\bibfnamefont
  {A.}~\bibnamefont {Nguyen}}, \bibinfo {author} {\bibfnamefont
  {M.}~\bibnamefont {Nguyen}}, \bibinfo {author} {\bibfnamefont {C.~H.}\
  \bibnamefont {Ni}}, \bibinfo {author} {\bibfnamefont {M.~Y.}\ \bibnamefont
  {Niu}}, \bibinfo {author} {\bibfnamefont {W.~D.}\ \bibnamefont {Oliver}},
  \bibinfo {author} {\bibfnamefont {K.}~\bibnamefont {Ottosson}}, \bibinfo
  {author} {\bibfnamefont {A.}~\bibnamefont {Pizzuto}}, \bibinfo {author}
  {\bibfnamefont {R.}~\bibnamefont {Potter}}, \bibinfo {author} {\bibfnamefont
  {O.}~\bibnamefont {Pritchard}}, \bibinfo {author} {\bibfnamefont {L.~P.}\
  \bibnamefont {Pryadko}}, \bibinfo {author} {\bibfnamefont {C.}~\bibnamefont
  {Quintana}}, \bibinfo {author} {\bibfnamefont {M.~J.}\ \bibnamefont
  {Reagor}}, \bibinfo {author} {\bibfnamefont {D.~M.}\ \bibnamefont {Rhodes}},
  \bibinfo {author} {\bibfnamefont {G.}~\bibnamefont {Roberts}}, \bibinfo
  {author} {\bibfnamefont {C.}~\bibnamefont {Rocque}}, \bibinfo {author}
  {\bibfnamefont {E.}~\bibnamefont {Rosenberg}}, \bibinfo {author}
  {\bibfnamefont {N.~C.}\ \bibnamefont {Rubin}}, \bibinfo {author}
  {\bibfnamefont {N.}~\bibnamefont {Saei}}, \bibinfo {author} {\bibfnamefont
  {K.}~\bibnamefont {Sankaragomathi}}, \bibinfo {author} {\bibfnamefont
  {K.~J.}\ \bibnamefont {Satzinger}}, \bibinfo {author} {\bibfnamefont {H.~F.}\
  \bibnamefont {Schurkus}}, \bibinfo {author} {\bibfnamefont {C.}~\bibnamefont
  {Schuster}}, \bibinfo {author} {\bibfnamefont {M.~J.}\ \bibnamefont
  {Shearn}}, \bibinfo {author} {\bibfnamefont {A.}~\bibnamefont {Shorter}},
  \bibinfo {author} {\bibfnamefont {N.}~\bibnamefont {Shutty}}, \bibinfo
  {author} {\bibfnamefont {V.}~\bibnamefont {Shvarts}}, \bibinfo {author}
  {\bibfnamefont {V.}~\bibnamefont {Sivak}}, \bibinfo {author} {\bibfnamefont
  {J.}~\bibnamefont {Skruzny}}, \bibinfo {author} {\bibfnamefont
  {S.}~\bibnamefont {Small}}, \bibinfo {author} {\bibfnamefont {W.~C.}\
  \bibnamefont {Smith}}, \bibinfo {author} {\bibfnamefont {S.}~\bibnamefont
  {Springer}}, \bibinfo {author} {\bibfnamefont {G.}~\bibnamefont {Sterling}},
  \bibinfo {author} {\bibfnamefont {J.}~\bibnamefont {Suchard}}, \bibinfo
  {author} {\bibfnamefont {M.}~\bibnamefont {Szalay}}, \bibinfo {author}
  {\bibfnamefont {A.}~\bibnamefont {Sztein}}, \bibinfo {author} {\bibfnamefont
  {D.}~\bibnamefont {Thor}}, \bibinfo {author} {\bibfnamefont {A.}~\bibnamefont
  {Torres}}, \bibinfo {author} {\bibfnamefont {M.~M.}\ \bibnamefont
  {Torunbalci}}, \bibinfo {author} {\bibfnamefont {A.}~\bibnamefont
  {Vaishnav}}, \bibinfo {author} {\bibfnamefont {S.}~\bibnamefont {Vdovichev}},
  \bibinfo {author} {\bibfnamefont {B.}~\bibnamefont {Villalonga}}, \bibinfo
  {author} {\bibfnamefont {C.~V.}\ \bibnamefont {Heidweiller}}, \bibinfo
  {author} {\bibfnamefont {S.}~\bibnamefont {Waltman}}, \bibinfo {author}
  {\bibfnamefont {S.~X.}\ \bibnamefont {Wang}}, \bibinfo {author}
  {\bibfnamefont {T.}~\bibnamefont {White}}, \bibinfo {author} {\bibfnamefont
  {K.}~\bibnamefont {Wong}}, \bibinfo {author} {\bibfnamefont {B.~W.~K.}\
  \bibnamefont {Woo}}, \bibinfo {author} {\bibfnamefont {C.}~\bibnamefont
  {Xing}}, \bibinfo {author} {\bibfnamefont {Z.~J.}\ \bibnamefont {Yao}},
  \bibinfo {author} {\bibfnamefont {P.}~\bibnamefont {Yeh}}, \bibinfo {author}
  {\bibfnamefont {B.}~\bibnamefont {Ying}}, \bibinfo {author} {\bibfnamefont
  {J.}~\bibnamefont {Yoo}}, \bibinfo {author} {\bibfnamefont {N.}~\bibnamefont
  {Yosri}}, \bibinfo {author} {\bibfnamefont {G.}~\bibnamefont {Young}},
  \bibinfo {author} {\bibfnamefont {A.}~\bibnamefont {Zalcman}}, \bibinfo
  {author} {\bibfnamefont {N.}~\bibnamefont {Zhu}}, \bibinfo {author}
  {\bibfnamefont {N.}~\bibnamefont {Zobrist}}, \bibinfo {author} {\bibfnamefont
  {H.}~\bibnamefont {Neven}}, \bibinfo {author} {\bibfnamefont
  {R.}~\bibnamefont {Babbush}}, \bibinfo {author} {\bibfnamefont
  {S.}~\bibnamefont {Boixo}}, \bibinfo {author} {\bibfnamefont
  {J.}~\bibnamefont {Hilton}}, \bibinfo {author} {\bibfnamefont
  {E.}~\bibnamefont {Lucero}}, \bibinfo {author} {\bibfnamefont
  {A.}~\bibnamefont {Megrant}}, \bibinfo {author} {\bibfnamefont
  {J.}~\bibnamefont {Kelly}}, \bibinfo {author} {\bibfnamefont
  {Y.}~\bibnamefont {Chen}}, \bibinfo {author} {\bibfnamefont {V.}~\bibnamefont
  {Smelyanskiy}}, \bibinfo {author} {\bibfnamefont {G.}~\bibnamefont {Vidal}},
  \bibinfo {author} {\bibfnamefont {P.}~\bibnamefont {Roushan}}, \bibinfo
  {author} {\bibfnamefont {A.~M.}\ \bibnamefont {L{\"a}uchli}}, \bibinfo
  {author} {\bibfnamefont {D.~A.}\ \bibnamefont {Abanin}},\ and\ \bibinfo
  {author} {\bibfnamefont {X.}~\bibnamefont {Mi}},\ }\bibfield  {title}
  {\bibinfo {title} {Thermalization and criticality on an analogue--digital
  quantum simulator},\ }\href {https://doi.org/10.1038/s41586-024-08460-3}
  {\bibfield  {journal} {\bibinfo  {journal} {Nature}\ }\textbf {\bibinfo
  {volume} {638}},\ \bibinfo {pages} {79} (\bibinfo {year} {2025})}\BibitemShut
  {NoStop}%
\bibitem [{\citenamefont {Wienand}\ \emph {et~al.}(2024)\citenamefont
  {Wienand}, \citenamefont {Karch}, \citenamefont {Impertro}, \citenamefont
  {Schweizer}, \citenamefont {McCulloch}, \citenamefont {Vasseur},
  \citenamefont {Gopalakrishnan}, \citenamefont {Aidelsburger},\ and\
  \citenamefont {Bloch}}]{Wienand_2024_hydro}%
  \BibitemOpen
  \bibfield  {author} {\bibinfo {author} {\bibfnamefont {J.~F.}\ \bibnamefont
  {Wienand}}, \bibinfo {author} {\bibfnamefont {S.}~\bibnamefont {Karch}},
  \bibinfo {author} {\bibfnamefont {A.}~\bibnamefont {Impertro}}, \bibinfo
  {author} {\bibfnamefont {C.}~\bibnamefont {Schweizer}}, \bibinfo {author}
  {\bibfnamefont {E.}~\bibnamefont {McCulloch}}, \bibinfo {author}
  {\bibfnamefont {R.}~\bibnamefont {Vasseur}}, \bibinfo {author} {\bibfnamefont
  {S.}~\bibnamefont {Gopalakrishnan}}, \bibinfo {author} {\bibfnamefont
  {M.}~\bibnamefont {Aidelsburger}},\ and\ \bibinfo {author} {\bibfnamefont
  {I.}~\bibnamefont {Bloch}},\ }\bibfield  {title} {\bibinfo {title} {Emergence
  of fluctuating hydrodynamics in chaotic quantum systems},\ }\href
  {https://doi.org/10.1038/s41567-024-02611-z} {\bibfield  {journal} {\bibinfo
  {journal} {Nature Physics}\ }\textbf {\bibinfo {volume} {20}},\ \bibinfo
  {pages} {1732–1737} (\bibinfo {year} {2024})}\BibitemShut {NoStop}%
\bibitem [{\citenamefont {King}\ \emph {et~al.}(2025)\citenamefont {King},
  \citenamefont {Nocera}, \citenamefont {Rams}, \citenamefont {Dziarmaga},
  \citenamefont {Wiersema}, \citenamefont {Bernoudy}, \citenamefont {Raymond},
  \citenamefont {Kaushal}, \citenamefont {Heinsdorf}, \citenamefont {Harris},
  \citenamefont {Boothby}, \citenamefont {Altomare}, \citenamefont {Asad},
  \citenamefont {Berkley}, \citenamefont {Boschnak}, \citenamefont {Chern},
  \citenamefont {Christiani}, \citenamefont {Cibere}, \citenamefont {Connor},
  \citenamefont {Dehn}, \citenamefont {Deshpande}, \citenamefont {Ejtemaee},
  \citenamefont {Farre}, \citenamefont {Hamer}, \citenamefont {Hoskinson},
  \citenamefont {Huang}, \citenamefont {Johnson}, \citenamefont {Kortas},
  \citenamefont {Ladizinsky}, \citenamefont {Lanting}, \citenamefont {Lai},
  \citenamefont {Li}, \citenamefont {MacDonald}, \citenamefont {Marsden},
  \citenamefont {McGeoch}, \citenamefont {Molavi}, \citenamefont {Oh},
  \citenamefont {Neufeld}, \citenamefont {Norouzpour}, \citenamefont
  {Pasvolsky}, \citenamefont {Poitras}, \citenamefont {Poulin-Lamarre},
  \citenamefont {Prescott}, \citenamefont {Reis}, \citenamefont {Rich},
  \citenamefont {Samani}, \citenamefont {Sheldan}, \citenamefont {Smirnov},
  \citenamefont {Sterpka}, \citenamefont {Clavera}, \citenamefont {Tsai},
  \citenamefont {Volkmann}, \citenamefont {Whiticar}, \citenamefont
  {Whittaker}, \citenamefont {Wilkinson}, \citenamefont {Yao}, \citenamefont
  {Yi}, \citenamefont {Sandvik}, \citenamefont {Alvarez}, \citenamefont
  {Melko}, \citenamefont {Carrasquilla}, \citenamefont {Franz},\ and\
  \citenamefont {Amin}}]{king2024computational}%
  \BibitemOpen
  \bibfield  {author} {\bibinfo {author} {\bibfnamefont {A.~D.}\ \bibnamefont
  {King}}, \bibinfo {author} {\bibfnamefont {A.}~\bibnamefont {Nocera}},
  \bibinfo {author} {\bibfnamefont {M.~M.}\ \bibnamefont {Rams}}, \bibinfo
  {author} {\bibfnamefont {J.}~\bibnamefont {Dziarmaga}}, \bibinfo {author}
  {\bibfnamefont {R.}~\bibnamefont {Wiersema}}, \bibinfo {author}
  {\bibfnamefont {W.}~\bibnamefont {Bernoudy}}, \bibinfo {author}
  {\bibfnamefont {J.}~\bibnamefont {Raymond}}, \bibinfo {author} {\bibfnamefont
  {N.}~\bibnamefont {Kaushal}}, \bibinfo {author} {\bibfnamefont
  {N.}~\bibnamefont {Heinsdorf}}, \bibinfo {author} {\bibfnamefont
  {R.}~\bibnamefont {Harris}}, \bibinfo {author} {\bibfnamefont
  {K.}~\bibnamefont {Boothby}}, \bibinfo {author} {\bibfnamefont
  {F.}~\bibnamefont {Altomare}}, \bibinfo {author} {\bibfnamefont
  {M.}~\bibnamefont {Asad}}, \bibinfo {author} {\bibfnamefont {A.~J.}\
  \bibnamefont {Berkley}}, \bibinfo {author} {\bibfnamefont {M.}~\bibnamefont
  {Boschnak}}, \bibinfo {author} {\bibfnamefont {K.}~\bibnamefont {Chern}},
  \bibinfo {author} {\bibfnamefont {H.}~\bibnamefont {Christiani}}, \bibinfo
  {author} {\bibfnamefont {S.}~\bibnamefont {Cibere}}, \bibinfo {author}
  {\bibfnamefont {J.}~\bibnamefont {Connor}}, \bibinfo {author} {\bibfnamefont
  {M.~H.}\ \bibnamefont {Dehn}}, \bibinfo {author} {\bibfnamefont
  {R.}~\bibnamefont {Deshpande}}, \bibinfo {author} {\bibfnamefont
  {S.}~\bibnamefont {Ejtemaee}}, \bibinfo {author} {\bibfnamefont
  {P.}~\bibnamefont {Farre}}, \bibinfo {author} {\bibfnamefont
  {K.}~\bibnamefont {Hamer}}, \bibinfo {author} {\bibfnamefont
  {E.}~\bibnamefont {Hoskinson}}, \bibinfo {author} {\bibfnamefont
  {S.}~\bibnamefont {Huang}}, \bibinfo {author} {\bibfnamefont {M.~W.}\
  \bibnamefont {Johnson}}, \bibinfo {author} {\bibfnamefont {S.}~\bibnamefont
  {Kortas}}, \bibinfo {author} {\bibfnamefont {E.}~\bibnamefont {Ladizinsky}},
  \bibinfo {author} {\bibfnamefont {T.}~\bibnamefont {Lanting}}, \bibinfo
  {author} {\bibfnamefont {T.}~\bibnamefont {Lai}}, \bibinfo {author}
  {\bibfnamefont {R.}~\bibnamefont {Li}}, \bibinfo {author} {\bibfnamefont
  {A.~J.~R.}\ \bibnamefont {MacDonald}}, \bibinfo {author} {\bibfnamefont
  {G.}~\bibnamefont {Marsden}}, \bibinfo {author} {\bibfnamefont {C.~C.}\
  \bibnamefont {McGeoch}}, \bibinfo {author} {\bibfnamefont {R.}~\bibnamefont
  {Molavi}}, \bibinfo {author} {\bibfnamefont {T.}~\bibnamefont {Oh}}, \bibinfo
  {author} {\bibfnamefont {R.}~\bibnamefont {Neufeld}}, \bibinfo {author}
  {\bibfnamefont {M.}~\bibnamefont {Norouzpour}}, \bibinfo {author}
  {\bibfnamefont {J.}~\bibnamefont {Pasvolsky}}, \bibinfo {author}
  {\bibfnamefont {P.}~\bibnamefont {Poitras}}, \bibinfo {author} {\bibfnamefont
  {G.}~\bibnamefont {Poulin-Lamarre}}, \bibinfo {author} {\bibfnamefont
  {T.}~\bibnamefont {Prescott}}, \bibinfo {author} {\bibfnamefont
  {M.}~\bibnamefont {Reis}}, \bibinfo {author} {\bibfnamefont {C.}~\bibnamefont
  {Rich}}, \bibinfo {author} {\bibfnamefont {M.}~\bibnamefont {Samani}},
  \bibinfo {author} {\bibfnamefont {B.}~\bibnamefont {Sheldan}}, \bibinfo
  {author} {\bibfnamefont {A.}~\bibnamefont {Smirnov}}, \bibinfo {author}
  {\bibfnamefont {E.}~\bibnamefont {Sterpka}}, \bibinfo {author} {\bibfnamefont
  {B.~T.}\ \bibnamefont {Clavera}}, \bibinfo {author} {\bibfnamefont
  {N.}~\bibnamefont {Tsai}}, \bibinfo {author} {\bibfnamefont {M.}~\bibnamefont
  {Volkmann}}, \bibinfo {author} {\bibfnamefont {A.~M.}\ \bibnamefont
  {Whiticar}}, \bibinfo {author} {\bibfnamefont {J.~D.}\ \bibnamefont
  {Whittaker}}, \bibinfo {author} {\bibfnamefont {W.}~\bibnamefont
  {Wilkinson}}, \bibinfo {author} {\bibfnamefont {J.}~\bibnamefont {Yao}},
  \bibinfo {author} {\bibfnamefont {T.~J.}\ \bibnamefont {Yi}}, \bibinfo
  {author} {\bibfnamefont {A.~W.}\ \bibnamefont {Sandvik}}, \bibinfo {author}
  {\bibfnamefont {G.}~\bibnamefont {Alvarez}}, \bibinfo {author} {\bibfnamefont
  {R.~G.}\ \bibnamefont {Melko}}, \bibinfo {author} {\bibfnamefont
  {J.}~\bibnamefont {Carrasquilla}}, \bibinfo {author} {\bibfnamefont
  {M.}~\bibnamefont {Franz}},\ and\ \bibinfo {author} {\bibfnamefont {M.~H.}\
  \bibnamefont {Amin}},\ }\bibfield  {title} {\bibinfo {title}
  {Beyond-classical computation in quantum simulation},\ }\href
  {https://doi.org/10.1126/science.ado6285} {\bibfield  {journal} {\bibinfo
  {journal} {Science}\ }\textbf {\bibinfo {volume} {388}},\ \bibinfo {pages}
  {199} (\bibinfo {year} {2025})}\BibitemShut {NoStop}%
\bibitem [{\citenamefont {Kim}\ \emph {et~al.}(2023)\citenamefont {Kim},
  \citenamefont {Eddins}, \citenamefont {Anand}, \citenamefont {Wei},
  \citenamefont {van~den Berg}, \citenamefont {Rosenblatt}, \citenamefont
  {Nayfeh}, \citenamefont {Wu}, \citenamefont {Zaletel}, \citenamefont
  {Temme},\ and\ \citenamefont {Kandala}}]{ibm_utility}%
  \BibitemOpen
  \bibfield  {author} {\bibinfo {author} {\bibfnamefont {Y.}~\bibnamefont
  {Kim}}, \bibinfo {author} {\bibfnamefont {A.}~\bibnamefont {Eddins}},
  \bibinfo {author} {\bibfnamefont {S.}~\bibnamefont {Anand}}, \bibinfo
  {author} {\bibfnamefont {K.~X.}\ \bibnamefont {Wei}}, \bibinfo {author}
  {\bibfnamefont {E.}~\bibnamefont {van~den Berg}}, \bibinfo {author}
  {\bibfnamefont {S.}~\bibnamefont {Rosenblatt}}, \bibinfo {author}
  {\bibfnamefont {H.}~\bibnamefont {Nayfeh}}, \bibinfo {author} {\bibfnamefont
  {Y.}~\bibnamefont {Wu}}, \bibinfo {author} {\bibfnamefont {M.}~\bibnamefont
  {Zaletel}}, \bibinfo {author} {\bibfnamefont {K.}~\bibnamefont {Temme}},\
  and\ \bibinfo {author} {\bibfnamefont {A.}~\bibnamefont {Kandala}},\
  }\bibfield  {title} {\bibinfo {title} {Evidence for the utility of quantum
  computing before fault tolerance},\ }\href
  {https://doi.org/10.1038/s41586-023-06096-3} {\bibfield  {journal} {\bibinfo
  {journal} {Nature}\ }\textbf {\bibinfo {volume} {618}},\ \bibinfo {pages}
  {500} (\bibinfo {year} {2023})}\BibitemShut {NoStop}%
\bibitem [{\citenamefont {Cochran}\ \emph {et~al.}(2025)\citenamefont
  {Cochran}, \citenamefont {Jobst}, \citenamefont {Rosenberg}, \citenamefont
  {Lensky}, \citenamefont {Gyawali}, \citenamefont {Eassa}, \citenamefont
  {Will}, \citenamefont {Szasz}, \citenamefont {Abanin}, \citenamefont
  {Acharya}, \citenamefont {Aghababaie~Beni}, \citenamefont {Andersen},
  \citenamefont {Ansmann}, \citenamefont {Arute}, \citenamefont {Arya},
  \citenamefont {Asfaw}, \citenamefont {Atalaya}, \citenamefont {Babbush},
  \citenamefont {Ballard}, \citenamefont {Bardin}, \citenamefont {Bengtsson},
  \citenamefont {Bilmes}, \citenamefont {Bourassa}, \citenamefont {Bovaird},
  \citenamefont {Broughton}, \citenamefont {Browne}, \citenamefont {Buchea},
  \citenamefont {Buckley}, \citenamefont {Burger}, \citenamefont {Burkett},
  \citenamefont {Bushnell}, \citenamefont {Cabrera}, \citenamefont {Campero},
  \citenamefont {Chang}, \citenamefont {Chen}, \citenamefont {Chiaro},
  \citenamefont {Claes}, \citenamefont {Cleland}, \citenamefont {Cogan},
  \citenamefont {Collins}, \citenamefont {Conner}, \citenamefont {Courtney},
  \citenamefont {Crook}, \citenamefont {Curtin}, \citenamefont {Das},
  \citenamefont {Demura}, \citenamefont {De~Lorenzo}, \citenamefont {Di~Paolo},
  \citenamefont {Donohoe}, \citenamefont {Drozdov}, \citenamefont {Dunsworth},
  \citenamefont {Eickbusch}, \citenamefont {Elbag}, \citenamefont {Elzouka},
  \citenamefont {Erickson}, \citenamefont {Ferreira}, \citenamefont {Burgos},
  \citenamefont {Forati}, \citenamefont {Fowler}, \citenamefont {Foxen},
  \citenamefont {Ganjam}, \citenamefont {Gasca}, \citenamefont {Genois},
  \citenamefont {Giang}, \citenamefont {Gilboa}, \citenamefont {Gosula},
  \citenamefont {Grajales~Dau}, \citenamefont {Graumann}, \citenamefont
  {Greene}, \citenamefont {Gross}, \citenamefont {Habegger}, \citenamefont
  {Hansen}, \citenamefont {Harrigan}, \citenamefont {Harrington}, \citenamefont
  {Heu}, \citenamefont {Higgott}, \citenamefont {Hilton}, \citenamefont
  {Huang}, \citenamefont {Huff}, \citenamefont {Huggins}, \citenamefont
  {Jeffrey}, \citenamefont {Jiang}, \citenamefont {Jones}, \citenamefont
  {Joshi}, \citenamefont {Juhas}, \citenamefont {Kafri}, \citenamefont {Kang},
  \citenamefont {Karamlou}, \citenamefont {Kechedzhi}, \citenamefont {Khaire},
  \citenamefont {Khattar}, \citenamefont {Khezri}, \citenamefont {Kim},
  \citenamefont {Klimov}, \citenamefont {Kobrin}, \citenamefont {Korotkov},
  \citenamefont {Kostritsa}, \citenamefont {Kreikebaum}, \citenamefont
  {Kurilovich}, \citenamefont {Landhuis}, \citenamefont {Lange-Dei},
  \citenamefont {Langley}, \citenamefont {Lau}, \citenamefont {Ledford},
  \citenamefont {Lee}, \citenamefont {Lester}, \citenamefont {Le~Guevel},
  \citenamefont {Li}, \citenamefont {Lill}, \citenamefont {Livingston},
  \citenamefont {Locharla}, \citenamefont {Lundahl}, \citenamefont {Lunt},
  \citenamefont {Madhuk}, \citenamefont {Maloney}, \citenamefont {Mandr{\`a}},
  \citenamefont {Martin}, \citenamefont {Martin}, \citenamefont {Maxfield},
  \citenamefont {McClean}, \citenamefont {McEwen}, \citenamefont {Meeks},
  \citenamefont {Megrant}, \citenamefont {Miao}, \citenamefont {Molavi},
  \citenamefont {Molina}, \citenamefont {Montazeri}, \citenamefont {Movassagh},
  \citenamefont {Neill}, \citenamefont {Newman}, \citenamefont {Nguyen},
  \citenamefont {Nguyen}, \citenamefont {Ni}, \citenamefont {Ottosson},
  \citenamefont {Pizzuto}, \citenamefont {Potter}, \citenamefont {Pritchard},
  \citenamefont {Quintana}, \citenamefont {Ramachandran}, \citenamefont
  {Reagor}, \citenamefont {Rhodes}, \citenamefont {Roberts}, \citenamefont
  {Sankaragomathi}, \citenamefont {Satzinger}, \citenamefont {Schurkus},
  \citenamefont {Shearn}, \citenamefont {Shorter}, \citenamefont {Shutty},
  \citenamefont {Shvarts}, \citenamefont {Sivak}, \citenamefont {Small},
  \citenamefont {Smith}, \citenamefont {Springer}, \citenamefont {Sterling},
  \citenamefont {Suchard}, \citenamefont {Sztein}, \citenamefont {Thor},
  \citenamefont {Torunbalci}, \citenamefont {Vaishnav}, \citenamefont {Vargas},
  \citenamefont {Vdovichev}, \citenamefont {Vidal}, \citenamefont
  {Vollgraff~Heidweiller}, \citenamefont {Waltman}, \citenamefont {Wang},
  \citenamefont {Ware}, \citenamefont {White}, \citenamefont {Wong},
  \citenamefont {Woo}, \citenamefont {Xing}, \citenamefont {Yao}, \citenamefont
  {Yeh}, \citenamefont {Ying}, \citenamefont {Yoo}, \citenamefont {Yosri},
  \citenamefont {Young}, \citenamefont {Zalcman}, \citenamefont {Zhang},
  \citenamefont {Zhu}, \citenamefont {Zobrist}, \citenamefont {Boixo},
  \citenamefont {Kelly}, \citenamefont {Lucero}, \citenamefont {Chen},
  \citenamefont {Smelyanskiy}, \citenamefont {Neven}, \citenamefont
  {Gammon-Smith}, \citenamefont {Pollmann}, \citenamefont {Knap},\ and\
  \citenamefont {Roushan}}]{cochran2024visualizing}%
  \BibitemOpen
  \bibfield  {author} {\bibinfo {author} {\bibfnamefont {T.~A.}\ \bibnamefont
  {Cochran}}, \bibinfo {author} {\bibfnamefont {B.}~\bibnamefont {Jobst}},
  \bibinfo {author} {\bibfnamefont {E.}~\bibnamefont {Rosenberg}}, \bibinfo
  {author} {\bibfnamefont {Y.~D.}\ \bibnamefont {Lensky}}, \bibinfo {author}
  {\bibfnamefont {G.}~\bibnamefont {Gyawali}}, \bibinfo {author} {\bibfnamefont
  {N.}~\bibnamefont {Eassa}}, \bibinfo {author} {\bibfnamefont
  {M.}~\bibnamefont {Will}}, \bibinfo {author} {\bibfnamefont {A.}~\bibnamefont
  {Szasz}}, \bibinfo {author} {\bibfnamefont {D.}~\bibnamefont {Abanin}},
  \bibinfo {author} {\bibfnamefont {R.}~\bibnamefont {Acharya}}, \bibinfo
  {author} {\bibfnamefont {L.}~\bibnamefont {Aghababaie~Beni}}, \bibinfo
  {author} {\bibfnamefont {T.~I.}\ \bibnamefont {Andersen}}, \bibinfo {author}
  {\bibfnamefont {M.}~\bibnamefont {Ansmann}}, \bibinfo {author} {\bibfnamefont
  {F.}~\bibnamefont {Arute}}, \bibinfo {author} {\bibfnamefont
  {K.}~\bibnamefont {Arya}}, \bibinfo {author} {\bibfnamefont {A.}~\bibnamefont
  {Asfaw}}, \bibinfo {author} {\bibfnamefont {J.}~\bibnamefont {Atalaya}},
  \bibinfo {author} {\bibfnamefont {R.}~\bibnamefont {Babbush}}, \bibinfo
  {author} {\bibfnamefont {B.}~\bibnamefont {Ballard}}, \bibinfo {author}
  {\bibfnamefont {J.~C.}\ \bibnamefont {Bardin}}, \bibinfo {author}
  {\bibfnamefont {A.}~\bibnamefont {Bengtsson}}, \bibinfo {author}
  {\bibfnamefont {A.}~\bibnamefont {Bilmes}}, \bibinfo {author} {\bibfnamefont
  {A.}~\bibnamefont {Bourassa}}, \bibinfo {author} {\bibfnamefont
  {J.}~\bibnamefont {Bovaird}}, \bibinfo {author} {\bibfnamefont
  {M.}~\bibnamefont {Broughton}}, \bibinfo {author} {\bibfnamefont {D.~A.}\
  \bibnamefont {Browne}}, \bibinfo {author} {\bibfnamefont {B.}~\bibnamefont
  {Buchea}}, \bibinfo {author} {\bibfnamefont {B.~B.}\ \bibnamefont {Buckley}},
  \bibinfo {author} {\bibfnamefont {T.}~\bibnamefont {Burger}}, \bibinfo
  {author} {\bibfnamefont {B.}~\bibnamefont {Burkett}}, \bibinfo {author}
  {\bibfnamefont {N.}~\bibnamefont {Bushnell}}, \bibinfo {author}
  {\bibfnamefont {A.}~\bibnamefont {Cabrera}}, \bibinfo {author} {\bibfnamefont
  {J.}~\bibnamefont {Campero}}, \bibinfo {author} {\bibfnamefont {H.~S.}\
  \bibnamefont {Chang}}, \bibinfo {author} {\bibfnamefont {Z.}~\bibnamefont
  {Chen}}, \bibinfo {author} {\bibfnamefont {B.}~\bibnamefont {Chiaro}},
  \bibinfo {author} {\bibfnamefont {J.}~\bibnamefont {Claes}}, \bibinfo
  {author} {\bibfnamefont {A.~Y.}\ \bibnamefont {Cleland}}, \bibinfo {author}
  {\bibfnamefont {J.}~\bibnamefont {Cogan}}, \bibinfo {author} {\bibfnamefont
  {R.}~\bibnamefont {Collins}}, \bibinfo {author} {\bibfnamefont
  {P.}~\bibnamefont {Conner}}, \bibinfo {author} {\bibfnamefont
  {W.}~\bibnamefont {Courtney}}, \bibinfo {author} {\bibfnamefont {A.~L.}\
  \bibnamefont {Crook}}, \bibinfo {author} {\bibfnamefont {B.}~\bibnamefont
  {Curtin}}, \bibinfo {author} {\bibfnamefont {S.}~\bibnamefont {Das}},
  \bibinfo {author} {\bibfnamefont {S.}~\bibnamefont {Demura}}, \bibinfo
  {author} {\bibfnamefont {L.}~\bibnamefont {De~Lorenzo}}, \bibinfo {author}
  {\bibfnamefont {A.}~\bibnamefont {Di~Paolo}}, \bibinfo {author}
  {\bibfnamefont {P.}~\bibnamefont {Donohoe}}, \bibinfo {author} {\bibfnamefont
  {I.}~\bibnamefont {Drozdov}}, \bibinfo {author} {\bibfnamefont
  {A.}~\bibnamefont {Dunsworth}}, \bibinfo {author} {\bibfnamefont
  {A.}~\bibnamefont {Eickbusch}}, \bibinfo {author} {\bibfnamefont {A.~M.}\
  \bibnamefont {Elbag}}, \bibinfo {author} {\bibfnamefont {M.}~\bibnamefont
  {Elzouka}}, \bibinfo {author} {\bibfnamefont {C.}~\bibnamefont {Erickson}},
  \bibinfo {author} {\bibfnamefont {V.~S.}\ \bibnamefont {Ferreira}}, \bibinfo
  {author} {\bibfnamefont {L.~F.}\ \bibnamefont {Burgos}}, \bibinfo {author}
  {\bibfnamefont {E.}~\bibnamefont {Forati}}, \bibinfo {author} {\bibfnamefont
  {A.~G.}\ \bibnamefont {Fowler}}, \bibinfo {author} {\bibfnamefont
  {B.}~\bibnamefont {Foxen}}, \bibinfo {author} {\bibfnamefont
  {S.}~\bibnamefont {Ganjam}}, \bibinfo {author} {\bibfnamefont
  {R.}~\bibnamefont {Gasca}}, \bibinfo {author} {\bibfnamefont
  {{\'E}.}~\bibnamefont {Genois}}, \bibinfo {author} {\bibfnamefont
  {W.}~\bibnamefont {Giang}}, \bibinfo {author} {\bibfnamefont
  {D.}~\bibnamefont {Gilboa}}, \bibinfo {author} {\bibfnamefont
  {R.}~\bibnamefont {Gosula}}, \bibinfo {author} {\bibfnamefont
  {A.}~\bibnamefont {Grajales~Dau}}, \bibinfo {author} {\bibfnamefont
  {D.}~\bibnamefont {Graumann}}, \bibinfo {author} {\bibfnamefont
  {A.}~\bibnamefont {Greene}}, \bibinfo {author} {\bibfnamefont {J.~A.}\
  \bibnamefont {Gross}}, \bibinfo {author} {\bibfnamefont {S.}~\bibnamefont
  {Habegger}}, \bibinfo {author} {\bibfnamefont {M.}~\bibnamefont {Hansen}},
  \bibinfo {author} {\bibfnamefont {M.~P.}\ \bibnamefont {Harrigan}}, \bibinfo
  {author} {\bibfnamefont {S.~D.}\ \bibnamefont {Harrington}}, \bibinfo
  {author} {\bibfnamefont {P.}~\bibnamefont {Heu}}, \bibinfo {author}
  {\bibfnamefont {O.}~\bibnamefont {Higgott}}, \bibinfo {author} {\bibfnamefont
  {J.}~\bibnamefont {Hilton}}, \bibinfo {author} {\bibfnamefont {H.~Y.}\
  \bibnamefont {Huang}}, \bibinfo {author} {\bibfnamefont {A.}~\bibnamefont
  {Huff}}, \bibinfo {author} {\bibfnamefont {W.}~\bibnamefont {Huggins}},
  \bibinfo {author} {\bibfnamefont {E.}~\bibnamefont {Jeffrey}}, \bibinfo
  {author} {\bibfnamefont {Z.}~\bibnamefont {Jiang}}, \bibinfo {author}
  {\bibfnamefont {C.}~\bibnamefont {Jones}}, \bibinfo {author} {\bibfnamefont
  {C.}~\bibnamefont {Joshi}}, \bibinfo {author} {\bibfnamefont
  {P.}~\bibnamefont {Juhas}}, \bibinfo {author} {\bibfnamefont
  {D.}~\bibnamefont {Kafri}}, \bibinfo {author} {\bibfnamefont
  {H.}~\bibnamefont {Kang}}, \bibinfo {author} {\bibfnamefont {A.~H.}\
  \bibnamefont {Karamlou}}, \bibinfo {author} {\bibfnamefont {K.}~\bibnamefont
  {Kechedzhi}}, \bibinfo {author} {\bibfnamefont {T.}~\bibnamefont {Khaire}},
  \bibinfo {author} {\bibfnamefont {T.}~\bibnamefont {Khattar}}, \bibinfo
  {author} {\bibfnamefont {M.}~\bibnamefont {Khezri}}, \bibinfo {author}
  {\bibfnamefont {S.}~\bibnamefont {Kim}}, \bibinfo {author} {\bibfnamefont
  {P.}~\bibnamefont {Klimov}}, \bibinfo {author} {\bibfnamefont
  {B.}~\bibnamefont {Kobrin}}, \bibinfo {author} {\bibfnamefont
  {A.}~\bibnamefont {Korotkov}}, \bibinfo {author} {\bibfnamefont
  {F.}~\bibnamefont {Kostritsa}}, \bibinfo {author} {\bibfnamefont
  {J.}~\bibnamefont {Kreikebaum}}, \bibinfo {author} {\bibfnamefont
  {V.}~\bibnamefont {Kurilovich}}, \bibinfo {author} {\bibfnamefont
  {D.}~\bibnamefont {Landhuis}}, \bibinfo {author} {\bibfnamefont
  {T.}~\bibnamefont {Lange-Dei}}, \bibinfo {author} {\bibfnamefont
  {B.}~\bibnamefont {Langley}}, \bibinfo {author} {\bibfnamefont {K.~M.}\
  \bibnamefont {Lau}}, \bibinfo {author} {\bibfnamefont {J.}~\bibnamefont
  {Ledford}}, \bibinfo {author} {\bibfnamefont {K.}~\bibnamefont {Lee}},
  \bibinfo {author} {\bibfnamefont {B.}~\bibnamefont {Lester}}, \bibinfo
  {author} {\bibfnamefont {L.}~\bibnamefont {Le~Guevel}}, \bibinfo {author}
  {\bibfnamefont {W.}~\bibnamefont {Li}}, \bibinfo {author} {\bibfnamefont
  {A.~T.}\ \bibnamefont {Lill}}, \bibinfo {author} {\bibfnamefont
  {W.}~\bibnamefont {Livingston}}, \bibinfo {author} {\bibfnamefont
  {A.}~\bibnamefont {Locharla}}, \bibinfo {author} {\bibfnamefont
  {D.}~\bibnamefont {Lundahl}}, \bibinfo {author} {\bibfnamefont
  {A.}~\bibnamefont {Lunt}}, \bibinfo {author} {\bibfnamefont {S.}~\bibnamefont
  {Madhuk}}, \bibinfo {author} {\bibfnamefont {A.}~\bibnamefont {Maloney}},
  \bibinfo {author} {\bibfnamefont {S.}~\bibnamefont {Mandr{\`a}}}, \bibinfo
  {author} {\bibfnamefont {L.}~\bibnamefont {Martin}}, \bibinfo {author}
  {\bibfnamefont {O.}~\bibnamefont {Martin}}, \bibinfo {author} {\bibfnamefont
  {C.}~\bibnamefont {Maxfield}}, \bibinfo {author} {\bibfnamefont
  {J.}~\bibnamefont {McClean}}, \bibinfo {author} {\bibfnamefont
  {M.}~\bibnamefont {McEwen}}, \bibinfo {author} {\bibfnamefont
  {S.}~\bibnamefont {Meeks}}, \bibinfo {author} {\bibfnamefont
  {A.}~\bibnamefont {Megrant}}, \bibinfo {author} {\bibfnamefont
  {K.}~\bibnamefont {Miao}}, \bibinfo {author} {\bibfnamefont {R.}~\bibnamefont
  {Molavi}}, \bibinfo {author} {\bibfnamefont {S.}~\bibnamefont {Molina}},
  \bibinfo {author} {\bibfnamefont {S.}~\bibnamefont {Montazeri}}, \bibinfo
  {author} {\bibfnamefont {R.}~\bibnamefont {Movassagh}}, \bibinfo {author}
  {\bibfnamefont {C.}~\bibnamefont {Neill}}, \bibinfo {author} {\bibfnamefont
  {M.}~\bibnamefont {Newman}}, \bibinfo {author} {\bibfnamefont
  {A.}~\bibnamefont {Nguyen}}, \bibinfo {author} {\bibfnamefont
  {M.}~\bibnamefont {Nguyen}}, \bibinfo {author} {\bibfnamefont {C.~H.}\
  \bibnamefont {Ni}}, \bibinfo {author} {\bibfnamefont {K.}~\bibnamefont
  {Ottosson}}, \bibinfo {author} {\bibfnamefont {A.}~\bibnamefont {Pizzuto}},
  \bibinfo {author} {\bibfnamefont {R.}~\bibnamefont {Potter}}, \bibinfo
  {author} {\bibfnamefont {O.}~\bibnamefont {Pritchard}}, \bibinfo {author}
  {\bibfnamefont {C.}~\bibnamefont {Quintana}}, \bibinfo {author}
  {\bibfnamefont {G.}~\bibnamefont {Ramachandran}}, \bibinfo {author}
  {\bibfnamefont {M.}~\bibnamefont {Reagor}}, \bibinfo {author} {\bibfnamefont
  {D.}~\bibnamefont {Rhodes}}, \bibinfo {author} {\bibfnamefont
  {G.}~\bibnamefont {Roberts}}, \bibinfo {author} {\bibfnamefont
  {K.}~\bibnamefont {Sankaragomathi}}, \bibinfo {author} {\bibfnamefont
  {K.}~\bibnamefont {Satzinger}}, \bibinfo {author} {\bibfnamefont
  {H.}~\bibnamefont {Schurkus}}, \bibinfo {author} {\bibfnamefont
  {M.}~\bibnamefont {Shearn}}, \bibinfo {author} {\bibfnamefont
  {A.}~\bibnamefont {Shorter}}, \bibinfo {author} {\bibfnamefont
  {N.}~\bibnamefont {Shutty}}, \bibinfo {author} {\bibfnamefont
  {V.}~\bibnamefont {Shvarts}}, \bibinfo {author} {\bibfnamefont
  {V.}~\bibnamefont {Sivak}}, \bibinfo {author} {\bibfnamefont
  {S.}~\bibnamefont {Small}}, \bibinfo {author} {\bibfnamefont {W.~C.}\
  \bibnamefont {Smith}}, \bibinfo {author} {\bibfnamefont {S.}~\bibnamefont
  {Springer}}, \bibinfo {author} {\bibfnamefont {G.}~\bibnamefont {Sterling}},
  \bibinfo {author} {\bibfnamefont {J.}~\bibnamefont {Suchard}}, \bibinfo
  {author} {\bibfnamefont {A.}~\bibnamefont {Sztein}}, \bibinfo {author}
  {\bibfnamefont {D.}~\bibnamefont {Thor}}, \bibinfo {author} {\bibfnamefont
  {M.}~\bibnamefont {Torunbalci}}, \bibinfo {author} {\bibfnamefont
  {A.}~\bibnamefont {Vaishnav}}, \bibinfo {author} {\bibfnamefont
  {J.}~\bibnamefont {Vargas}}, \bibinfo {author} {\bibfnamefont
  {S.}~\bibnamefont {Vdovichev}}, \bibinfo {author} {\bibfnamefont
  {G.}~\bibnamefont {Vidal}}, \bibinfo {author} {\bibfnamefont
  {C.}~\bibnamefont {Vollgraff~Heidweiller}}, \bibinfo {author} {\bibfnamefont
  {S.}~\bibnamefont {Waltman}}, \bibinfo {author} {\bibfnamefont {S.~X.}\
  \bibnamefont {Wang}}, \bibinfo {author} {\bibfnamefont {B.}~\bibnamefont
  {Ware}}, \bibinfo {author} {\bibfnamefont {T.}~\bibnamefont {White}},
  \bibinfo {author} {\bibfnamefont {K.}~\bibnamefont {Wong}}, \bibinfo {author}
  {\bibfnamefont {B.~W.~K.}\ \bibnamefont {Woo}}, \bibinfo {author}
  {\bibfnamefont {C.}~\bibnamefont {Xing}}, \bibinfo {author} {\bibfnamefont
  {Z.~J.}\ \bibnamefont {Yao}}, \bibinfo {author} {\bibfnamefont
  {P.}~\bibnamefont {Yeh}}, \bibinfo {author} {\bibfnamefont {B.}~\bibnamefont
  {Ying}}, \bibinfo {author} {\bibfnamefont {J.}~\bibnamefont {Yoo}}, \bibinfo
  {author} {\bibfnamefont {N.}~\bibnamefont {Yosri}}, \bibinfo {author}
  {\bibfnamefont {G.}~\bibnamefont {Young}}, \bibinfo {author} {\bibfnamefont
  {A.}~\bibnamefont {Zalcman}}, \bibinfo {author} {\bibfnamefont
  {Y.}~\bibnamefont {Zhang}}, \bibinfo {author} {\bibfnamefont
  {N.}~\bibnamefont {Zhu}}, \bibinfo {author} {\bibfnamefont {N.}~\bibnamefont
  {Zobrist}}, \bibinfo {author} {\bibfnamefont {S.}~\bibnamefont {Boixo}},
  \bibinfo {author} {\bibfnamefont {J.}~\bibnamefont {Kelly}}, \bibinfo
  {author} {\bibfnamefont {E.}~\bibnamefont {Lucero}}, \bibinfo {author}
  {\bibfnamefont {Y.}~\bibnamefont {Chen}}, \bibinfo {author} {\bibfnamefont
  {V.}~\bibnamefont {Smelyanskiy}}, \bibinfo {author} {\bibfnamefont
  {H.}~\bibnamefont {Neven}}, \bibinfo {author} {\bibfnamefont
  {A.}~\bibnamefont {Gammon-Smith}}, \bibinfo {author} {\bibfnamefont
  {F.}~\bibnamefont {Pollmann}}, \bibinfo {author} {\bibfnamefont
  {M.}~\bibnamefont {Knap}},\ and\ \bibinfo {author} {\bibfnamefont
  {P.}~\bibnamefont {Roushan}},\ }\bibfield  {title} {\bibinfo {title}
  {Visualizing dynamics of charges and strings in (2 + 1)d lattice gauge
  theories},\ }\href {https://doi.org/10.1038/s41586-025-08999-9} {\bibfield
  {journal} {\bibinfo  {journal} {Nature}\ }\textbf {\bibinfo {volume} {642}},\
  \bibinfo {pages} {315} (\bibinfo {year} {2025})}\BibitemShut {NoStop}%
\bibitem [{\citenamefont {Will}\ \emph {et~al.}(2025)\citenamefont {Will},
  \citenamefont {Cochran}, \citenamefont {Rosenberg}, \citenamefont {Jobst},
  \citenamefont {Eassa}, \citenamefont {Roushan}, \citenamefont {Knap},
  \citenamefont {Gammon-Smith},\ and\ \citenamefont
  {Pollmann}}]{will2025probing}%
  \BibitemOpen
  \bibfield  {author} {\bibinfo {author} {\bibfnamefont {M.}~\bibnamefont
  {Will}}, \bibinfo {author} {\bibfnamefont {T.~A.}\ \bibnamefont {Cochran}},
  \bibinfo {author} {\bibfnamefont {E.}~\bibnamefont {Rosenberg}}, \bibinfo
  {author} {\bibfnamefont {B.}~\bibnamefont {Jobst}}, \bibinfo {author}
  {\bibfnamefont {N.~M.}\ \bibnamefont {Eassa}}, \bibinfo {author}
  {\bibfnamefont {P.}~\bibnamefont {Roushan}}, \bibinfo {author} {\bibfnamefont
  {M.}~\bibnamefont {Knap}}, \bibinfo {author} {\bibfnamefont {A.}~\bibnamefont
  {Gammon-Smith}},\ and\ \bibinfo {author} {\bibfnamefont {F.}~\bibnamefont
  {Pollmann}},\ }\bibfield  {title} {\bibinfo {title} {Probing non-equilibrium
  topological order on a quantum processor},\ }\href
  {https://doi.org/10.1038/s41586-025-09456-3} {\bibfield  {journal} {\bibinfo
  {journal} {Nature}\ }\textbf {\bibinfo {volume} {645}},\ \bibinfo {pages}
  {348} (\bibinfo {year} {2025})}\BibitemShut {NoStop}%
\bibitem [{\citenamefont {Childs}\ \emph {et~al.}(2021)\citenamefont {Childs},
  \citenamefont {Su}, \citenamefont {Tran}, \citenamefont {Wiebe},\ and\
  \citenamefont {Zhu}}]{PhysRevX.11.011020}%
  \BibitemOpen
  \bibfield  {author} {\bibinfo {author} {\bibfnamefont {A.~M.}\ \bibnamefont
  {Childs}}, \bibinfo {author} {\bibfnamefont {Y.}~\bibnamefont {Su}}, \bibinfo
  {author} {\bibfnamefont {M.~C.}\ \bibnamefont {Tran}}, \bibinfo {author}
  {\bibfnamefont {N.}~\bibnamefont {Wiebe}},\ and\ \bibinfo {author}
  {\bibfnamefont {S.}~\bibnamefont {Zhu}},\ }\bibfield  {title} {\bibinfo
  {title} {Theory of trotter error with commutator scaling},\ }\href
  {https://doi.org/10.1103/PhysRevX.11.011020} {\bibfield  {journal} {\bibinfo
  {journal} {Phys. Rev. X}\ }\textbf {\bibinfo {volume} {11}},\ \bibinfo
  {pages} {011020} (\bibinfo {year} {2021})}\BibitemShut {NoStop}%
\bibitem [{\citenamefont {Arute}\ \emph {et~al.}(2019)\citenamefont {Arute},
  \citenamefont {Arya}, \citenamefont {Babbush}, \citenamefont {Bacon},
  \citenamefont {Bardin}, \citenamefont {Barends}, \citenamefont {Biswas},
  \citenamefont {Boixo}, \citenamefont {Brandao}, \citenamefont {Buell},
  \citenamefont {Burkett}, \citenamefont {Chen}, \citenamefont {Chen},
  \citenamefont {Chiaro}, \citenamefont {Collins}, \citenamefont {Courtney},
  \citenamefont {Dunsworth}, \citenamefont {Farhi}, \citenamefont {Foxen},
  \citenamefont {Fowler}, \citenamefont {Gidney}, \citenamefont {Giustina},
  \citenamefont {Graff}, \citenamefont {Guerin}, \citenamefont {Habegger},
  \citenamefont {Harrigan}, \citenamefont {Hartmann}, \citenamefont {Ho},
  \citenamefont {Hoffmann}, \citenamefont {Huang}, \citenamefont {Humble},
  \citenamefont {Isakov}, \citenamefont {Jeffrey}, \citenamefont {Jiang},
  \citenamefont {Kafri}, \citenamefont {Kechedzhi}, \citenamefont {Kelly},
  \citenamefont {Klimov}, \citenamefont {Knysh}, \citenamefont {Korotkov},
  \citenamefont {Kostritsa}, \citenamefont {Landhuis}, \citenamefont
  {Lindmark}, \citenamefont {Lucero}, \citenamefont {Lyakh}, \citenamefont
  {Mandr{\`a}}, \citenamefont {McClean}, \citenamefont {McEwen}, \citenamefont
  {Megrant}, \citenamefont {Mi}, \citenamefont {Michielsen}, \citenamefont
  {Mohseni}, \citenamefont {Mutus}, \citenamefont {Naaman}, \citenamefont
  {Neeley}, \citenamefont {Neill}, \citenamefont {Niu}, \citenamefont {Ostby},
  \citenamefont {Petukhov}, \citenamefont {Platt}, \citenamefont {Quintana},
  \citenamefont {Rieffel}, \citenamefont {Roushan}, \citenamefont {Rubin},
  \citenamefont {Sank}, \citenamefont {Satzinger}, \citenamefont {Smelyanskiy},
  \citenamefont {Sung}, \citenamefont {Trevithick}, \citenamefont
  {Vainsencher}, \citenamefont {Villalonga}, \citenamefont {White},
  \citenamefont {Yao}, \citenamefont {Yeh}, \citenamefont {Zalcman},
  \citenamefont {Neven},\ and\ \citenamefont {Martinis}}]{arute_2019}%
  \BibitemOpen
  \bibfield  {author} {\bibinfo {author} {\bibfnamefont {F.}~\bibnamefont
  {Arute}}, \bibinfo {author} {\bibfnamefont {K.}~\bibnamefont {Arya}},
  \bibinfo {author} {\bibfnamefont {R.}~\bibnamefont {Babbush}}, \bibinfo
  {author} {\bibfnamefont {D.}~\bibnamefont {Bacon}}, \bibinfo {author}
  {\bibfnamefont {J.~C.}\ \bibnamefont {Bardin}}, \bibinfo {author}
  {\bibfnamefont {R.}~\bibnamefont {Barends}}, \bibinfo {author} {\bibfnamefont
  {R.}~\bibnamefont {Biswas}}, \bibinfo {author} {\bibfnamefont
  {S.}~\bibnamefont {Boixo}}, \bibinfo {author} {\bibfnamefont {F.~G. S.~L.}\
  \bibnamefont {Brandao}}, \bibinfo {author} {\bibfnamefont {D.~A.}\
  \bibnamefont {Buell}}, \bibinfo {author} {\bibfnamefont {B.}~\bibnamefont
  {Burkett}}, \bibinfo {author} {\bibfnamefont {Y.}~\bibnamefont {Chen}},
  \bibinfo {author} {\bibfnamefont {Z.}~\bibnamefont {Chen}}, \bibinfo {author}
  {\bibfnamefont {B.}~\bibnamefont {Chiaro}}, \bibinfo {author} {\bibfnamefont
  {R.}~\bibnamefont {Collins}}, \bibinfo {author} {\bibfnamefont
  {W.}~\bibnamefont {Courtney}}, \bibinfo {author} {\bibfnamefont
  {A.}~\bibnamefont {Dunsworth}}, \bibinfo {author} {\bibfnamefont
  {E.}~\bibnamefont {Farhi}}, \bibinfo {author} {\bibfnamefont
  {B.}~\bibnamefont {Foxen}}, \bibinfo {author} {\bibfnamefont
  {A.}~\bibnamefont {Fowler}}, \bibinfo {author} {\bibfnamefont
  {C.}~\bibnamefont {Gidney}}, \bibinfo {author} {\bibfnamefont
  {M.}~\bibnamefont {Giustina}}, \bibinfo {author} {\bibfnamefont
  {R.}~\bibnamefont {Graff}}, \bibinfo {author} {\bibfnamefont
  {K.}~\bibnamefont {Guerin}}, \bibinfo {author} {\bibfnamefont
  {S.}~\bibnamefont {Habegger}}, \bibinfo {author} {\bibfnamefont {M.~P.}\
  \bibnamefont {Harrigan}}, \bibinfo {author} {\bibfnamefont {M.~J.}\
  \bibnamefont {Hartmann}}, \bibinfo {author} {\bibfnamefont {A.}~\bibnamefont
  {Ho}}, \bibinfo {author} {\bibfnamefont {M.}~\bibnamefont {Hoffmann}},
  \bibinfo {author} {\bibfnamefont {T.}~\bibnamefont {Huang}}, \bibinfo
  {author} {\bibfnamefont {T.~S.}\ \bibnamefont {Humble}}, \bibinfo {author}
  {\bibfnamefont {S.~V.}\ \bibnamefont {Isakov}}, \bibinfo {author}
  {\bibfnamefont {E.}~\bibnamefont {Jeffrey}}, \bibinfo {author} {\bibfnamefont
  {Z.}~\bibnamefont {Jiang}}, \bibinfo {author} {\bibfnamefont
  {D.}~\bibnamefont {Kafri}}, \bibinfo {author} {\bibfnamefont
  {K.}~\bibnamefont {Kechedzhi}}, \bibinfo {author} {\bibfnamefont
  {J.}~\bibnamefont {Kelly}}, \bibinfo {author} {\bibfnamefont {P.~V.}\
  \bibnamefont {Klimov}}, \bibinfo {author} {\bibfnamefont {S.}~\bibnamefont
  {Knysh}}, \bibinfo {author} {\bibfnamefont {A.}~\bibnamefont {Korotkov}},
  \bibinfo {author} {\bibfnamefont {F.}~\bibnamefont {Kostritsa}}, \bibinfo
  {author} {\bibfnamefont {D.}~\bibnamefont {Landhuis}}, \bibinfo {author}
  {\bibfnamefont {M.}~\bibnamefont {Lindmark}}, \bibinfo {author}
  {\bibfnamefont {E.}~\bibnamefont {Lucero}}, \bibinfo {author} {\bibfnamefont
  {D.}~\bibnamefont {Lyakh}}, \bibinfo {author} {\bibfnamefont
  {S.}~\bibnamefont {Mandr{\`a}}}, \bibinfo {author} {\bibfnamefont {J.~R.}\
  \bibnamefont {McClean}}, \bibinfo {author} {\bibfnamefont {M.}~\bibnamefont
  {McEwen}}, \bibinfo {author} {\bibfnamefont {A.}~\bibnamefont {Megrant}},
  \bibinfo {author} {\bibfnamefont {X.}~\bibnamefont {Mi}}, \bibinfo {author}
  {\bibfnamefont {K.}~\bibnamefont {Michielsen}}, \bibinfo {author}
  {\bibfnamefont {M.}~\bibnamefont {Mohseni}}, \bibinfo {author} {\bibfnamefont
  {J.}~\bibnamefont {Mutus}}, \bibinfo {author} {\bibfnamefont
  {O.}~\bibnamefont {Naaman}}, \bibinfo {author} {\bibfnamefont
  {M.}~\bibnamefont {Neeley}}, \bibinfo {author} {\bibfnamefont
  {C.}~\bibnamefont {Neill}}, \bibinfo {author} {\bibfnamefont {M.~Y.}\
  \bibnamefont {Niu}}, \bibinfo {author} {\bibfnamefont {E.}~\bibnamefont
  {Ostby}}, \bibinfo {author} {\bibfnamefont {A.}~\bibnamefont {Petukhov}},
  \bibinfo {author} {\bibfnamefont {J.~C.}\ \bibnamefont {Platt}}, \bibinfo
  {author} {\bibfnamefont {C.}~\bibnamefont {Quintana}}, \bibinfo {author}
  {\bibfnamefont {E.~G.}\ \bibnamefont {Rieffel}}, \bibinfo {author}
  {\bibfnamefont {P.}~\bibnamefont {Roushan}}, \bibinfo {author} {\bibfnamefont
  {N.~C.}\ \bibnamefont {Rubin}}, \bibinfo {author} {\bibfnamefont
  {D.}~\bibnamefont {Sank}}, \bibinfo {author} {\bibfnamefont {K.~J.}\
  \bibnamefont {Satzinger}}, \bibinfo {author} {\bibfnamefont {V.}~\bibnamefont
  {Smelyanskiy}}, \bibinfo {author} {\bibfnamefont {K.~J.}\ \bibnamefont
  {Sung}}, \bibinfo {author} {\bibfnamefont {M.~D.}\ \bibnamefont
  {Trevithick}}, \bibinfo {author} {\bibfnamefont {A.}~\bibnamefont
  {Vainsencher}}, \bibinfo {author} {\bibfnamefont {B.}~\bibnamefont
  {Villalonga}}, \bibinfo {author} {\bibfnamefont {T.}~\bibnamefont {White}},
  \bibinfo {author} {\bibfnamefont {Z.~J.}\ \bibnamefont {Yao}}, \bibinfo
  {author} {\bibfnamefont {P.}~\bibnamefont {Yeh}}, \bibinfo {author}
  {\bibfnamefont {A.}~\bibnamefont {Zalcman}}, \bibinfo {author} {\bibfnamefont
  {H.}~\bibnamefont {Neven}},\ and\ \bibinfo {author} {\bibfnamefont {J.~M.}\
  \bibnamefont {Martinis}},\ }\bibfield  {title} {\bibinfo {title} {Quantum
  supremacy using a programmable superconducting processor},\ }\href
  {https://doi.org/10.1038/s41586-019-1666-5} {\bibfield  {journal} {\bibinfo
  {journal} {Nature}\ }\textbf {\bibinfo {volume} {574}},\ \bibinfo {pages}
  {505} (\bibinfo {year} {2019})}\BibitemShut {NoStop}%
\bibitem [{\citenamefont {Wu}\ \emph {et~al.}(2021)\citenamefont {Wu},
  \citenamefont {Bao}, \citenamefont {Cao}, \citenamefont {Chen}, \citenamefont
  {Chen}, \citenamefont {Chen}, \citenamefont {Chung}, \citenamefont {Deng},
  \citenamefont {Du}, \citenamefont {Fan}, \citenamefont {Gong}, \citenamefont
  {Guo}, \citenamefont {Guo}, \citenamefont {Guo}, \citenamefont {Han},
  \citenamefont {Hong}, \citenamefont {Huang}, \citenamefont {Huo},
  \citenamefont {Li}, \citenamefont {Li}, \citenamefont {Li}, \citenamefont
  {Li}, \citenamefont {Liang}, \citenamefont {Lin}, \citenamefont {Lin},
  \citenamefont {Qian}, \citenamefont {Qiao}, \citenamefont {Rong},
  \citenamefont {Su}, \citenamefont {Sun}, \citenamefont {Wang}, \citenamefont
  {Wang}, \citenamefont {Wu}, \citenamefont {Xu}, \citenamefont {Yan},
  \citenamefont {Yang}, \citenamefont {Yang}, \citenamefont {Ye}, \citenamefont
  {Yin}, \citenamefont {Ying}, \citenamefont {Yu}, \citenamefont {Zha},
  \citenamefont {Zhang}, \citenamefont {Zhang}, \citenamefont {Zhang},
  \citenamefont {Zhang}, \citenamefont {Zhao}, \citenamefont {Zhao},
  \citenamefont {Zhou}, \citenamefont {Zhu}, \citenamefont {Lu}, \citenamefont
  {Peng}, \citenamefont {Zhu},\ and\ \citenamefont
  {Pan}}]{PhysRevLett.127.180501}%
  \BibitemOpen
  \bibfield  {author} {\bibinfo {author} {\bibfnamefont {Y.}~\bibnamefont
  {Wu}}, \bibinfo {author} {\bibfnamefont {W.-S.}\ \bibnamefont {Bao}},
  \bibinfo {author} {\bibfnamefont {S.}~\bibnamefont {Cao}}, \bibinfo {author}
  {\bibfnamefont {F.}~\bibnamefont {Chen}}, \bibinfo {author} {\bibfnamefont
  {M.-C.}\ \bibnamefont {Chen}}, \bibinfo {author} {\bibfnamefont
  {X.}~\bibnamefont {Chen}}, \bibinfo {author} {\bibfnamefont {T.-H.}\
  \bibnamefont {Chung}}, \bibinfo {author} {\bibfnamefont {H.}~\bibnamefont
  {Deng}}, \bibinfo {author} {\bibfnamefont {Y.}~\bibnamefont {Du}}, \bibinfo
  {author} {\bibfnamefont {D.}~\bibnamefont {Fan}}, \bibinfo {author}
  {\bibfnamefont {M.}~\bibnamefont {Gong}}, \bibinfo {author} {\bibfnamefont
  {C.}~\bibnamefont {Guo}}, \bibinfo {author} {\bibfnamefont {C.}~\bibnamefont
  {Guo}}, \bibinfo {author} {\bibfnamefont {S.}~\bibnamefont {Guo}}, \bibinfo
  {author} {\bibfnamefont {L.}~\bibnamefont {Han}}, \bibinfo {author}
  {\bibfnamefont {L.}~\bibnamefont {Hong}}, \bibinfo {author} {\bibfnamefont
  {H.-L.}\ \bibnamefont {Huang}}, \bibinfo {author} {\bibfnamefont {Y.-H.}\
  \bibnamefont {Huo}}, \bibinfo {author} {\bibfnamefont {L.}~\bibnamefont
  {Li}}, \bibinfo {author} {\bibfnamefont {N.}~\bibnamefont {Li}}, \bibinfo
  {author} {\bibfnamefont {S.}~\bibnamefont {Li}}, \bibinfo {author}
  {\bibfnamefont {Y.}~\bibnamefont {Li}}, \bibinfo {author} {\bibfnamefont
  {F.}~\bibnamefont {Liang}}, \bibinfo {author} {\bibfnamefont
  {C.}~\bibnamefont {Lin}}, \bibinfo {author} {\bibfnamefont {J.}~\bibnamefont
  {Lin}}, \bibinfo {author} {\bibfnamefont {H.}~\bibnamefont {Qian}}, \bibinfo
  {author} {\bibfnamefont {D.}~\bibnamefont {Qiao}}, \bibinfo {author}
  {\bibfnamefont {H.}~\bibnamefont {Rong}}, \bibinfo {author} {\bibfnamefont
  {H.}~\bibnamefont {Su}}, \bibinfo {author} {\bibfnamefont {L.}~\bibnamefont
  {Sun}}, \bibinfo {author} {\bibfnamefont {L.}~\bibnamefont {Wang}}, \bibinfo
  {author} {\bibfnamefont {S.}~\bibnamefont {Wang}}, \bibinfo {author}
  {\bibfnamefont {D.}~\bibnamefont {Wu}}, \bibinfo {author} {\bibfnamefont
  {Y.}~\bibnamefont {Xu}}, \bibinfo {author} {\bibfnamefont {K.}~\bibnamefont
  {Yan}}, \bibinfo {author} {\bibfnamefont {W.}~\bibnamefont {Yang}}, \bibinfo
  {author} {\bibfnamefont {Y.}~\bibnamefont {Yang}}, \bibinfo {author}
  {\bibfnamefont {Y.}~\bibnamefont {Ye}}, \bibinfo {author} {\bibfnamefont
  {J.}~\bibnamefont {Yin}}, \bibinfo {author} {\bibfnamefont {C.}~\bibnamefont
  {Ying}}, \bibinfo {author} {\bibfnamefont {J.}~\bibnamefont {Yu}}, \bibinfo
  {author} {\bibfnamefont {C.}~\bibnamefont {Zha}}, \bibinfo {author}
  {\bibfnamefont {C.}~\bibnamefont {Zhang}}, \bibinfo {author} {\bibfnamefont
  {H.}~\bibnamefont {Zhang}}, \bibinfo {author} {\bibfnamefont
  {K.}~\bibnamefont {Zhang}}, \bibinfo {author} {\bibfnamefont
  {Y.}~\bibnamefont {Zhang}}, \bibinfo {author} {\bibfnamefont
  {H.}~\bibnamefont {Zhao}}, \bibinfo {author} {\bibfnamefont {Y.}~\bibnamefont
  {Zhao}}, \bibinfo {author} {\bibfnamefont {L.}~\bibnamefont {Zhou}}, \bibinfo
  {author} {\bibfnamefont {Q.}~\bibnamefont {Zhu}}, \bibinfo {author}
  {\bibfnamefont {C.-Y.}\ \bibnamefont {Lu}}, \bibinfo {author} {\bibfnamefont
  {C.-Z.}\ \bibnamefont {Peng}}, \bibinfo {author} {\bibfnamefont
  {X.}~\bibnamefont {Zhu}},\ and\ \bibinfo {author} {\bibfnamefont {J.-W.}\
  \bibnamefont {Pan}},\ }\bibfield  {title} {\bibinfo {title} {Strong quantum
  computational advantage using a superconducting quantum processor},\ }\href
  {https://doi.org/10.1103/PhysRevLett.127.180501} {\bibfield  {journal}
  {\bibinfo  {journal} {Phys. Rev. Lett.}\ }\textbf {\bibinfo {volume} {127}},\
  \bibinfo {pages} {180501} (\bibinfo {year} {2021})}\BibitemShut {NoStop}%
\bibitem [{\citenamefont {Rosenberg}\ \emph {et~al.}(2024)\citenamefont
  {Rosenberg}, \citenamefont {Andersen}, \citenamefont {Samajdar},
  \citenamefont {Petukhov}, \citenamefont {Hoke}, \citenamefont {Abanin},
  \citenamefont {Bengtsson}, \citenamefont {Drozdov}, \citenamefont {Erickson},
  \citenamefont {Klimov}, \citenamefont {Mi}, \citenamefont {Morvan},
  \citenamefont {Neeley}, \citenamefont {Neill}, \citenamefont {Acharya},
  \citenamefont {Allen}, \citenamefont {Anderson}, \citenamefont {Ansmann},
  \citenamefont {Arute}, \citenamefont {Arya}, \citenamefont {Asfaw},
  \citenamefont {Atalaya}, \citenamefont {Bardin}, \citenamefont {Bilmes},
  \citenamefont {Bortoli}, \citenamefont {Bourassa}, \citenamefont {Bovaird},
  \citenamefont {Brill}, \citenamefont {Broughton}, \citenamefont {Buckley},
  \citenamefont {Buell}, \citenamefont {Burger}, \citenamefont {Burkett},
  \citenamefont {Bushnell}, \citenamefont {Campero}, \citenamefont {Chang},
  \citenamefont {Chen}, \citenamefont {Chiaro}, \citenamefont {Chik},
  \citenamefont {Cogan}, \citenamefont {Collins}, \citenamefont {Conner},
  \citenamefont {Courtney}, \citenamefont {Crook}, \citenamefont {Curtin},
  \citenamefont {Debroy}, \citenamefont {Barba}, \citenamefont {Demura},
  \citenamefont {Di~Paolo}, \citenamefont {Dunsworth}, \citenamefont {Earle},
  \citenamefont {Faoro}, \citenamefont {Farhi}, \citenamefont {Fatemi},
  \citenamefont {Ferreira}, \citenamefont {Burgos}, \citenamefont {Forati},
  \citenamefont {Fowler}, \citenamefont {Foxen}, \citenamefont {Garcia},
  \citenamefont {Genois}, \citenamefont {Giang}, \citenamefont {Gidney},
  \citenamefont {Gilboa}, \citenamefont {Giustina}, \citenamefont {Gosula},
  \citenamefont {Dau}, \citenamefont {Gross}, \citenamefont {Habegger},
  \citenamefont {Hamilton}, \citenamefont {Hansen}, \citenamefont {Harrigan},
  \citenamefont {Harrington}, \citenamefont {Heu}, \citenamefont {Hill},
  \citenamefont {Hoffmann}, \citenamefont {Hong}, \citenamefont {Huang},
  \citenamefont {Huff}, \citenamefont {Huggins}, \citenamefont {Ioffe},
  \citenamefont {Isakov}, \citenamefont {Iveland}, \citenamefont {Jeffrey},
  \citenamefont {Jiang}, \citenamefont {Jones}, \citenamefont {Juhas},
  \citenamefont {Kafri}, \citenamefont {Khattar}, \citenamefont {Khezri},
  \citenamefont {Kieferov{\'a}}, \citenamefont {Kim}, \citenamefont {Kitaev},
  \citenamefont {Klots}, \citenamefont {Korotkov}, \citenamefont {Kostritsa},
  \citenamefont {Kreikebaum}, \citenamefont {Landhuis}, \citenamefont {Laptev},
  \citenamefont {Lau}, \citenamefont {Laws}, \citenamefont {Lee}, \citenamefont
  {Lee}, \citenamefont {Lensky}, \citenamefont {Lester}, \citenamefont {Lill},
  \citenamefont {Liu}, \citenamefont {Locharla}, \citenamefont {Mandr{\`a}},
  \citenamefont {Martin}, \citenamefont {Martin}, \citenamefont {McClean},
  \citenamefont {McEwen}, \citenamefont {Meeks}, \citenamefont {Miao},
  \citenamefont {Mieszala}, \citenamefont {Montazeri}, \citenamefont
  {Movassagh}, \citenamefont {Mruczkiewicz}, \citenamefont {Nersisyan},
  \citenamefont {Newman}, \citenamefont {Ng}, \citenamefont {Nguyen},
  \citenamefont {Nguyen}, \citenamefont {Niu}, \citenamefont {O'Brien},
  \citenamefont {Omonije}, \citenamefont {Opremcak}, \citenamefont {Potter},
  \citenamefont {Pryadko}, \citenamefont {Quintana}, \citenamefont {Rhodes},
  \citenamefont {Rocque}, \citenamefont {Rubin}, \citenamefont {Saei},
  \citenamefont {Sank}, \citenamefont {Sankaragomathi}, \citenamefont
  {Satzinger}, \citenamefont {Schurkus}, \citenamefont {Schuster},
  \citenamefont {Shearn}, \citenamefont {Shorter}, \citenamefont {Shutty},
  \citenamefont {Shvarts}, \citenamefont {Sivak}, \citenamefont {Skruzny},
  \citenamefont {Smith}, \citenamefont {Somma}, \citenamefont {Sterling},
  \citenamefont {Strain}, \citenamefont {Szalay}, \citenamefont {Thor},
  \citenamefont {Torres}, \citenamefont {Vidal}, \citenamefont {Villalonga},
  \citenamefont {Heidweiller}, \citenamefont {White}, \citenamefont {Woo},
  \citenamefont {Xing}, \citenamefont {Yao}, \citenamefont {Yeh}, \citenamefont
  {Yoo}, \citenamefont {Young}, \citenamefont {Zalcman}, \citenamefont {Zhang},
  \citenamefont {Zhu}, \citenamefont {Zobrist}, \citenamefont {Neven},
  \citenamefont {Babbush}, \citenamefont {Bacon}, \citenamefont {Boixo},
  \citenamefont {Hilton}, \citenamefont {Lucero}, \citenamefont {Megrant},
  \citenamefont {Kelly}, \citenamefont {Chen}, \citenamefont {Smelyanskiy},
  \citenamefont {Khemani}, \citenamefont {Gopalakrishnan}, \citenamefont
  {Prosen},\ and\ \citenamefont {Roushan}}]{rosenberg2024}%
  \BibitemOpen
  \bibfield  {author} {\bibinfo {author} {\bibfnamefont {E.}~\bibnamefont
  {Rosenberg}}, \bibinfo {author} {\bibfnamefont {T.~I.}\ \bibnamefont
  {Andersen}}, \bibinfo {author} {\bibfnamefont {R.}~\bibnamefont {Samajdar}},
  \bibinfo {author} {\bibfnamefont {A.}~\bibnamefont {Petukhov}}, \bibinfo
  {author} {\bibfnamefont {J.~C.}\ \bibnamefont {Hoke}}, \bibinfo {author}
  {\bibfnamefont {D.}~\bibnamefont {Abanin}}, \bibinfo {author} {\bibfnamefont
  {A.}~\bibnamefont {Bengtsson}}, \bibinfo {author} {\bibfnamefont {I.~K.}\
  \bibnamefont {Drozdov}}, \bibinfo {author} {\bibfnamefont {C.}~\bibnamefont
  {Erickson}}, \bibinfo {author} {\bibfnamefont {P.~V.}\ \bibnamefont
  {Klimov}}, \bibinfo {author} {\bibfnamefont {X.}~\bibnamefont {Mi}}, \bibinfo
  {author} {\bibfnamefont {A.}~\bibnamefont {Morvan}}, \bibinfo {author}
  {\bibfnamefont {M.}~\bibnamefont {Neeley}}, \bibinfo {author} {\bibfnamefont
  {C.}~\bibnamefont {Neill}}, \bibinfo {author} {\bibfnamefont
  {R.}~\bibnamefont {Acharya}}, \bibinfo {author} {\bibfnamefont
  {R.}~\bibnamefont {Allen}}, \bibinfo {author} {\bibfnamefont
  {K.}~\bibnamefont {Anderson}}, \bibinfo {author} {\bibfnamefont
  {M.}~\bibnamefont {Ansmann}}, \bibinfo {author} {\bibfnamefont
  {F.}~\bibnamefont {Arute}}, \bibinfo {author} {\bibfnamefont
  {K.}~\bibnamefont {Arya}}, \bibinfo {author} {\bibfnamefont {A.}~\bibnamefont
  {Asfaw}}, \bibinfo {author} {\bibfnamefont {J.}~\bibnamefont {Atalaya}},
  \bibinfo {author} {\bibfnamefont {J.~C.}\ \bibnamefont {Bardin}}, \bibinfo
  {author} {\bibfnamefont {A.}~\bibnamefont {Bilmes}}, \bibinfo {author}
  {\bibfnamefont {G.}~\bibnamefont {Bortoli}}, \bibinfo {author} {\bibfnamefont
  {A.}~\bibnamefont {Bourassa}}, \bibinfo {author} {\bibfnamefont
  {J.}~\bibnamefont {Bovaird}}, \bibinfo {author} {\bibfnamefont
  {L.}~\bibnamefont {Brill}}, \bibinfo {author} {\bibfnamefont
  {M.}~\bibnamefont {Broughton}}, \bibinfo {author} {\bibfnamefont {B.~B.}\
  \bibnamefont {Buckley}}, \bibinfo {author} {\bibfnamefont {D.~A.}\
  \bibnamefont {Buell}}, \bibinfo {author} {\bibfnamefont {T.}~\bibnamefont
  {Burger}}, \bibinfo {author} {\bibfnamefont {B.}~\bibnamefont {Burkett}},
  \bibinfo {author} {\bibfnamefont {N.}~\bibnamefont {Bushnell}}, \bibinfo
  {author} {\bibfnamefont {J.}~\bibnamefont {Campero}}, \bibinfo {author}
  {\bibfnamefont {H.~S.}\ \bibnamefont {Chang}}, \bibinfo {author}
  {\bibfnamefont {Z.}~\bibnamefont {Chen}}, \bibinfo {author} {\bibfnamefont
  {B.}~\bibnamefont {Chiaro}}, \bibinfo {author} {\bibfnamefont
  {D.}~\bibnamefont {Chik}}, \bibinfo {author} {\bibfnamefont {J.}~\bibnamefont
  {Cogan}}, \bibinfo {author} {\bibfnamefont {R.}~\bibnamefont {Collins}},
  \bibinfo {author} {\bibfnamefont {P.}~\bibnamefont {Conner}}, \bibinfo
  {author} {\bibfnamefont {W.}~\bibnamefont {Courtney}}, \bibinfo {author}
  {\bibfnamefont {A.~L.}\ \bibnamefont {Crook}}, \bibinfo {author}
  {\bibfnamefont {B.}~\bibnamefont {Curtin}}, \bibinfo {author} {\bibfnamefont
  {D.~M.}\ \bibnamefont {Debroy}}, \bibinfo {author} {\bibfnamefont {A.~D.~T.}\
  \bibnamefont {Barba}}, \bibinfo {author} {\bibfnamefont {S.}~\bibnamefont
  {Demura}}, \bibinfo {author} {\bibfnamefont {A.}~\bibnamefont {Di~Paolo}},
  \bibinfo {author} {\bibfnamefont {A.}~\bibnamefont {Dunsworth}}, \bibinfo
  {author} {\bibfnamefont {C.}~\bibnamefont {Earle}}, \bibinfo {author}
  {\bibfnamefont {L.}~\bibnamefont {Faoro}}, \bibinfo {author} {\bibfnamefont
  {E.}~\bibnamefont {Farhi}}, \bibinfo {author} {\bibfnamefont
  {R.}~\bibnamefont {Fatemi}}, \bibinfo {author} {\bibfnamefont {V.~S.}\
  \bibnamefont {Ferreira}}, \bibinfo {author} {\bibfnamefont {L.~F.}\
  \bibnamefont {Burgos}}, \bibinfo {author} {\bibfnamefont {E.}~\bibnamefont
  {Forati}}, \bibinfo {author} {\bibfnamefont {A.~G.}\ \bibnamefont {Fowler}},
  \bibinfo {author} {\bibfnamefont {B.}~\bibnamefont {Foxen}}, \bibinfo
  {author} {\bibfnamefont {G.}~\bibnamefont {Garcia}}, \bibinfo {author}
  {\bibfnamefont {{\'E}.}~\bibnamefont {Genois}}, \bibinfo {author}
  {\bibfnamefont {W.}~\bibnamefont {Giang}}, \bibinfo {author} {\bibfnamefont
  {C.}~\bibnamefont {Gidney}}, \bibinfo {author} {\bibfnamefont
  {D.}~\bibnamefont {Gilboa}}, \bibinfo {author} {\bibfnamefont
  {M.}~\bibnamefont {Giustina}}, \bibinfo {author} {\bibfnamefont
  {R.}~\bibnamefont {Gosula}}, \bibinfo {author} {\bibfnamefont {A.~G.}\
  \bibnamefont {Dau}}, \bibinfo {author} {\bibfnamefont {J.~A.}\ \bibnamefont
  {Gross}}, \bibinfo {author} {\bibfnamefont {S.}~\bibnamefont {Habegger}},
  \bibinfo {author} {\bibfnamefont {M.~C.}\ \bibnamefont {Hamilton}}, \bibinfo
  {author} {\bibfnamefont {M.}~\bibnamefont {Hansen}}, \bibinfo {author}
  {\bibfnamefont {M.~P.}\ \bibnamefont {Harrigan}}, \bibinfo {author}
  {\bibfnamefont {S.~D.}\ \bibnamefont {Harrington}}, \bibinfo {author}
  {\bibfnamefont {P.}~\bibnamefont {Heu}}, \bibinfo {author} {\bibfnamefont
  {G.}~\bibnamefont {Hill}}, \bibinfo {author} {\bibfnamefont {M.~R.}\
  \bibnamefont {Hoffmann}}, \bibinfo {author} {\bibfnamefont {S.}~\bibnamefont
  {Hong}}, \bibinfo {author} {\bibfnamefont {T.}~\bibnamefont {Huang}},
  \bibinfo {author} {\bibfnamefont {A.}~\bibnamefont {Huff}}, \bibinfo {author}
  {\bibfnamefont {W.~J.}\ \bibnamefont {Huggins}}, \bibinfo {author}
  {\bibfnamefont {L.~B.}\ \bibnamefont {Ioffe}}, \bibinfo {author}
  {\bibfnamefont {S.~V.}\ \bibnamefont {Isakov}}, \bibinfo {author}
  {\bibfnamefont {J.}~\bibnamefont {Iveland}}, \bibinfo {author} {\bibfnamefont
  {E.}~\bibnamefont {Jeffrey}}, \bibinfo {author} {\bibfnamefont
  {Z.}~\bibnamefont {Jiang}}, \bibinfo {author} {\bibfnamefont
  {C.}~\bibnamefont {Jones}}, \bibinfo {author} {\bibfnamefont
  {P.}~\bibnamefont {Juhas}}, \bibinfo {author} {\bibfnamefont
  {D.}~\bibnamefont {Kafri}}, \bibinfo {author} {\bibfnamefont
  {T.}~\bibnamefont {Khattar}}, \bibinfo {author} {\bibfnamefont
  {M.}~\bibnamefont {Khezri}}, \bibinfo {author} {\bibfnamefont
  {M.}~\bibnamefont {Kieferov{\'a}}}, \bibinfo {author} {\bibfnamefont
  {S.}~\bibnamefont {Kim}}, \bibinfo {author} {\bibfnamefont {A.}~\bibnamefont
  {Kitaev}}, \bibinfo {author} {\bibfnamefont {A.~R.}\ \bibnamefont {Klots}},
  \bibinfo {author} {\bibfnamefont {A.~N.}\ \bibnamefont {Korotkov}}, \bibinfo
  {author} {\bibfnamefont {F.}~\bibnamefont {Kostritsa}}, \bibinfo {author}
  {\bibfnamefont {J.~M.}\ \bibnamefont {Kreikebaum}}, \bibinfo {author}
  {\bibfnamefont {D.}~\bibnamefont {Landhuis}}, \bibinfo {author}
  {\bibfnamefont {P.}~\bibnamefont {Laptev}}, \bibinfo {author} {\bibfnamefont
  {K.~M.}\ \bibnamefont {Lau}}, \bibinfo {author} {\bibfnamefont
  {L.}~\bibnamefont {Laws}}, \bibinfo {author} {\bibfnamefont {J.}~\bibnamefont
  {Lee}}, \bibinfo {author} {\bibfnamefont {K.~W.}\ \bibnamefont {Lee}},
  \bibinfo {author} {\bibfnamefont {Y.~D.}\ \bibnamefont {Lensky}}, \bibinfo
  {author} {\bibfnamefont {B.~J.}\ \bibnamefont {Lester}}, \bibinfo {author}
  {\bibfnamefont {A.~T.}\ \bibnamefont {Lill}}, \bibinfo {author}
  {\bibfnamefont {W.}~\bibnamefont {Liu}}, \bibinfo {author} {\bibfnamefont
  {A.}~\bibnamefont {Locharla}}, \bibinfo {author} {\bibfnamefont
  {S.}~\bibnamefont {Mandr{\`a}}}, \bibinfo {author} {\bibfnamefont
  {O.}~\bibnamefont {Martin}}, \bibinfo {author} {\bibfnamefont
  {S.}~\bibnamefont {Martin}}, \bibinfo {author} {\bibfnamefont {J.~R.}\
  \bibnamefont {McClean}}, \bibinfo {author} {\bibfnamefont {M.}~\bibnamefont
  {McEwen}}, \bibinfo {author} {\bibfnamefont {S.}~\bibnamefont {Meeks}},
  \bibinfo {author} {\bibfnamefont {K.~C.}\ \bibnamefont {Miao}}, \bibinfo
  {author} {\bibfnamefont {A.}~\bibnamefont {Mieszala}}, \bibinfo {author}
  {\bibfnamefont {S.}~\bibnamefont {Montazeri}}, \bibinfo {author}
  {\bibfnamefont {R.}~\bibnamefont {Movassagh}}, \bibinfo {author}
  {\bibfnamefont {W.}~\bibnamefont {Mruczkiewicz}}, \bibinfo {author}
  {\bibfnamefont {A.}~\bibnamefont {Nersisyan}}, \bibinfo {author}
  {\bibfnamefont {M.}~\bibnamefont {Newman}}, \bibinfo {author} {\bibfnamefont
  {J.~H.}\ \bibnamefont {Ng}}, \bibinfo {author} {\bibfnamefont
  {A.}~\bibnamefont {Nguyen}}, \bibinfo {author} {\bibfnamefont
  {M.}~\bibnamefont {Nguyen}}, \bibinfo {author} {\bibfnamefont {M.~Y.}\
  \bibnamefont {Niu}}, \bibinfo {author} {\bibfnamefont {T.~E.}\ \bibnamefont
  {O'Brien}}, \bibinfo {author} {\bibfnamefont {S.}~\bibnamefont {Omonije}},
  \bibinfo {author} {\bibfnamefont {A.}~\bibnamefont {Opremcak}}, \bibinfo
  {author} {\bibfnamefont {R.}~\bibnamefont {Potter}}, \bibinfo {author}
  {\bibfnamefont {L.~P.}\ \bibnamefont {Pryadko}}, \bibinfo {author}
  {\bibfnamefont {C.}~\bibnamefont {Quintana}}, \bibinfo {author}
  {\bibfnamefont {D.~M.}\ \bibnamefont {Rhodes}}, \bibinfo {author}
  {\bibfnamefont {C.}~\bibnamefont {Rocque}}, \bibinfo {author} {\bibfnamefont
  {N.~C.}\ \bibnamefont {Rubin}}, \bibinfo {author} {\bibfnamefont
  {N.}~\bibnamefont {Saei}}, \bibinfo {author} {\bibfnamefont {D.}~\bibnamefont
  {Sank}}, \bibinfo {author} {\bibfnamefont {K.}~\bibnamefont
  {Sankaragomathi}}, \bibinfo {author} {\bibfnamefont {K.~J.}\ \bibnamefont
  {Satzinger}}, \bibinfo {author} {\bibfnamefont {H.~F.}\ \bibnamefont
  {Schurkus}}, \bibinfo {author} {\bibfnamefont {C.}~\bibnamefont {Schuster}},
  \bibinfo {author} {\bibfnamefont {M.~J.}\ \bibnamefont {Shearn}}, \bibinfo
  {author} {\bibfnamefont {A.}~\bibnamefont {Shorter}}, \bibinfo {author}
  {\bibfnamefont {N.}~\bibnamefont {Shutty}}, \bibinfo {author} {\bibfnamefont
  {V.}~\bibnamefont {Shvarts}}, \bibinfo {author} {\bibfnamefont
  {V.}~\bibnamefont {Sivak}}, \bibinfo {author} {\bibfnamefont
  {J.}~\bibnamefont {Skruzny}}, \bibinfo {author} {\bibfnamefont {W.~C.}\
  \bibnamefont {Smith}}, \bibinfo {author} {\bibfnamefont {R.~D.}\ \bibnamefont
  {Somma}}, \bibinfo {author} {\bibfnamefont {G.}~\bibnamefont {Sterling}},
  \bibinfo {author} {\bibfnamefont {D.}~\bibnamefont {Strain}}, \bibinfo
  {author} {\bibfnamefont {M.}~\bibnamefont {Szalay}}, \bibinfo {author}
  {\bibfnamefont {D.}~\bibnamefont {Thor}}, \bibinfo {author} {\bibfnamefont
  {A.}~\bibnamefont {Torres}}, \bibinfo {author} {\bibfnamefont
  {G.}~\bibnamefont {Vidal}}, \bibinfo {author} {\bibfnamefont
  {B.}~\bibnamefont {Villalonga}}, \bibinfo {author} {\bibfnamefont {C.~V.}\
  \bibnamefont {Heidweiller}}, \bibinfo {author} {\bibfnamefont
  {T.}~\bibnamefont {White}}, \bibinfo {author} {\bibfnamefont {B.~W.~K.}\
  \bibnamefont {Woo}}, \bibinfo {author} {\bibfnamefont {C.}~\bibnamefont
  {Xing}}, \bibinfo {author} {\bibfnamefont {Z.~J.}\ \bibnamefont {Yao}},
  \bibinfo {author} {\bibfnamefont {P.}~\bibnamefont {Yeh}}, \bibinfo {author}
  {\bibfnamefont {J.}~\bibnamefont {Yoo}}, \bibinfo {author} {\bibfnamefont
  {G.}~\bibnamefont {Young}}, \bibinfo {author} {\bibfnamefont
  {A.}~\bibnamefont {Zalcman}}, \bibinfo {author} {\bibfnamefont
  {Y.}~\bibnamefont {Zhang}}, \bibinfo {author} {\bibfnamefont
  {N.}~\bibnamefont {Zhu}}, \bibinfo {author} {\bibfnamefont {N.}~\bibnamefont
  {Zobrist}}, \bibinfo {author} {\bibfnamefont {H.}~\bibnamefont {Neven}},
  \bibinfo {author} {\bibfnamefont {R.}~\bibnamefont {Babbush}}, \bibinfo
  {author} {\bibfnamefont {D.}~\bibnamefont {Bacon}}, \bibinfo {author}
  {\bibfnamefont {S.}~\bibnamefont {Boixo}}, \bibinfo {author} {\bibfnamefont
  {J.}~\bibnamefont {Hilton}}, \bibinfo {author} {\bibfnamefont
  {E.}~\bibnamefont {Lucero}}, \bibinfo {author} {\bibfnamefont
  {A.}~\bibnamefont {Megrant}}, \bibinfo {author} {\bibfnamefont
  {J.}~\bibnamefont {Kelly}}, \bibinfo {author} {\bibfnamefont
  {Y.}~\bibnamefont {Chen}}, \bibinfo {author} {\bibfnamefont {V.}~\bibnamefont
  {Smelyanskiy}}, \bibinfo {author} {\bibfnamefont {V.}~\bibnamefont
  {Khemani}}, \bibinfo {author} {\bibfnamefont {S.}~\bibnamefont
  {Gopalakrishnan}}, \bibinfo {author} {\bibfnamefont {T.}~\bibnamefont
  {Prosen}},\ and\ \bibinfo {author} {\bibfnamefont {P.}~\bibnamefont
  {Roushan}},\ }\bibfield  {title} {\bibinfo {title} {Dynamics of magnetization
  at infinite temperature in a heisenberg spin chain},\ }\href
  {https://doi.org/10.1126/science.adi7877} {\bibfield  {journal} {\bibinfo
  {journal} {Science}\ }\textbf {\bibinfo {volume} {384}},\ \bibinfo {pages}
  {48} (\bibinfo {year} {2024})}\BibitemShut {NoStop}%
\bibitem [{\citenamefont {Ayral}\ \emph {et~al.}(2023)\citenamefont {Ayral},
  \citenamefont {Louvet}, \citenamefont {Zhou}, \citenamefont {Lambert},
  \citenamefont {Stoudenmire},\ and\ \citenamefont
  {Waintal}}]{ayral2023density}%
  \BibitemOpen
  \bibfield  {author} {\bibinfo {author} {\bibfnamefont {T.}~\bibnamefont
  {Ayral}}, \bibinfo {author} {\bibfnamefont {T.}~\bibnamefont {Louvet}},
  \bibinfo {author} {\bibfnamefont {Y.}~\bibnamefont {Zhou}}, \bibinfo {author}
  {\bibfnamefont {C.}~\bibnamefont {Lambert}}, \bibinfo {author} {\bibfnamefont
  {E.~M.}\ \bibnamefont {Stoudenmire}},\ and\ \bibinfo {author} {\bibfnamefont
  {X.}~\bibnamefont {Waintal}},\ }\bibfield  {title} {\bibinfo {title}
  {Density-matrix renormalization group algorithm for simulating quantum
  circuits with a finite fidelity},\ }\href
  {https://doi.org/10.1103/PRXQuantum.4.020304} {\bibfield  {journal} {\bibinfo
   {journal} {PRX Quantum}\ }\textbf {\bibinfo {volume} {4}},\ \bibinfo {pages}
  {020304} (\bibinfo {year} {2023})}\BibitemShut {NoStop}%
\bibitem [{\citenamefont {Begu{\v s}i{\'c}}\ \emph {et~al.}(2024)\citenamefont
  {Begu{\v s}i{\'c}}, \citenamefont {Gray},\ and\ \citenamefont
  {Chan}}]{Tomislav:2024}%
  \BibitemOpen
  \bibfield  {author} {\bibinfo {author} {\bibfnamefont {T.}~\bibnamefont
  {Begu{\v s}i{\'c}}}, \bibinfo {author} {\bibfnamefont {J.}~\bibnamefont
  {Gray}},\ and\ \bibinfo {author} {\bibfnamefont {G.~K.-L.}\ \bibnamefont
  {Chan}},\ }\bibfield  {title} {\bibinfo {title} {Fast and converged classical
  simulations of evidence for the utility of quantum computing before fault
  tolerance},\ }\href {https://doi.org/10.1126/sciadv.adk4321} {\bibfield
  {journal} {\bibinfo  {journal} {Science Advances}\ }\textbf {\bibinfo
  {volume} {10}},\ \bibinfo {pages} {eadk4321} (\bibinfo {year}
  {2024})}\BibitemShut {NoStop}%
\bibitem [{\citenamefont {Lubasch}\ \emph
  {et~al.}(2014{\natexlab{a}})\citenamefont {Lubasch}, \citenamefont {Cirac},\
  and\ \citenamefont {Ba\~nuls}}]{LuCiBa14a}%
  \BibitemOpen
  \bibfield  {author} {\bibinfo {author} {\bibfnamefont {M.}~\bibnamefont
  {Lubasch}}, \bibinfo {author} {\bibfnamefont {J.~I.}\ \bibnamefont {Cirac}},\
  and\ \bibinfo {author} {\bibfnamefont {M.-C.}\ \bibnamefont {Ba\~nuls}},\
  }\bibfield  {title} {\bibinfo {title} {Unifying projected entangled pair
  state contractions},\ }\href {https://doi.org/10.1088/1367-2630/16/3/033014}
  {\bibfield  {journal} {\bibinfo  {journal} {New J. Phys.}\ }\textbf {\bibinfo
  {volume} {16}},\ \bibinfo {pages} {033014} (\bibinfo {year}
  {2014}{\natexlab{a}})}\BibitemShut {NoStop}%
\bibitem [{\citenamefont {Schmitt}\ and\ \citenamefont
  {Heyl}(2020)}]{Schmitt2020}%
  \BibitemOpen
  \bibfield  {author} {\bibinfo {author} {\bibfnamefont {M.}~\bibnamefont
  {Schmitt}}\ and\ \bibinfo {author} {\bibfnamefont {M.}~\bibnamefont {Heyl}},\
  }\bibfield  {title} {\bibinfo {title} {Quantum many-body dynamics in two
  dimensions with artificial neural networks},\ }\href
  {https://doi.org/10.1103/PhysRevLett.125.100503} {\bibfield  {journal}
  {\bibinfo  {journal} {Phys. Rev. Lett.}\ }\textbf {\bibinfo {volume} {125}},\
  \bibinfo {pages} {100503} (\bibinfo {year} {2020})}\BibitemShut {NoStop}%
\bibitem [{\citenamefont {Begu\ifmmode \check{s}\else
  \v{s}\fi{}i\ifmmode~\acute{c}\else \'{c}\fi{}}\ and\ \citenamefont
  {Chan}(2025)}]{beguvsic2024real}%
  \BibitemOpen
  \bibfield  {author} {\bibinfo {author} {\bibfnamefont {T.}~\bibnamefont
  {Begu\ifmmode \check{s}\else \v{s}\fi{}i\ifmmode~\acute{c}\else \'{c}\fi{}}}\
  and\ \bibinfo {author} {\bibfnamefont {G.~K.-L.}\ \bibnamefont {Chan}},\
  }\bibfield  {title} {\bibinfo {title} {Real-time operator evolution in two
  and three dimensions via sparse pauli dynamics},\ }\href
  {https://doi.org/10.1103/PRXQuantum.6.020302} {\bibfield  {journal} {\bibinfo
   {journal} {PRX Quantum}\ }\textbf {\bibinfo {volume} {6}},\ \bibinfo {pages}
  {020302} (\bibinfo {year} {2025})}\BibitemShut {NoStop}%
\bibitem [{\citenamefont {Rudolph}\ \emph {et~al.}(2023)\citenamefont
  {Rudolph}, \citenamefont {Fontana}, \citenamefont {Holmes},\ and\
  \citenamefont {Cincio}}]{rudolph2023classical}%
  \BibitemOpen
  \bibfield  {author} {\bibinfo {author} {\bibfnamefont {M.~S.}\ \bibnamefont
  {Rudolph}}, \bibinfo {author} {\bibfnamefont {E.}~\bibnamefont {Fontana}},
  \bibinfo {author} {\bibfnamefont {Z.}~\bibnamefont {Holmes}},\ and\ \bibinfo
  {author} {\bibfnamefont {L.}~\bibnamefont {Cincio}},\ }\bibfield  {title}
  {\bibinfo {title} {Classical surrogate simulation of quantum systems with
  {LOWESA}},\ }\href {https://arxiv.org/abs/2308.09109} {\bibfield  {journal}
  {\bibinfo  {journal} {arXiv preprint arXiv:2308.09109}\ } (\bibinfo {year}
  {2023})}\BibitemShut {NoStop}%
\bibitem [{\citenamefont {Bermejo}\ \emph {et~al.}(2026)\citenamefont
  {Bermejo}, \citenamefont {Braccia}, \citenamefont {Rudolph}, \citenamefont
  {Holmes}, \citenamefont {Cincio},\ and\ \citenamefont
  {Cerezo}}]{bermejo2024quantum}%
  \BibitemOpen
  \bibfield  {author} {\bibinfo {author} {\bibfnamefont {P.}~\bibnamefont
  {Bermejo}}, \bibinfo {author} {\bibfnamefont {P.}~\bibnamefont {Braccia}},
  \bibinfo {author} {\bibfnamefont {M.~S.}\ \bibnamefont {Rudolph}}, \bibinfo
  {author} {\bibfnamefont {Z.}~\bibnamefont {Holmes}}, \bibinfo {author}
  {\bibfnamefont {L.}~\bibnamefont {Cincio}},\ and\ \bibinfo {author}
  {\bibfnamefont {M.}~\bibnamefont {Cerezo}},\ }\bibfield  {title} {\bibinfo
  {title} {Quantum convolutional neural networks are effectively classically
  simulable},\ }\href {https://doi.org/10.1103/8qt9-72ts} {\bibfield  {journal}
  {\bibinfo  {journal} {PRX Quantum}\ }\textbf {\bibinfo {volume} {7}},\
  \bibinfo {pages} {020304} (\bibinfo {year} {2026})}\BibitemShut {NoStop}%
\bibitem [{\citenamefont {Angrisani}\ \emph {et~al.}(2025)\citenamefont
  {Angrisani}, \citenamefont {Schmidhuber}, \citenamefont {Rudolph},
  \citenamefont {Cerezo}, \citenamefont {Holmes},\ and\ \citenamefont
  {Huang}}]{angrisani2024classically}%
  \BibitemOpen
  \bibfield  {author} {\bibinfo {author} {\bibfnamefont {A.}~\bibnamefont
  {Angrisani}}, \bibinfo {author} {\bibfnamefont {A.}~\bibnamefont
  {Schmidhuber}}, \bibinfo {author} {\bibfnamefont {M.~S.}\ \bibnamefont
  {Rudolph}}, \bibinfo {author} {\bibfnamefont {M.}~\bibnamefont {Cerezo}},
  \bibinfo {author} {\bibfnamefont {Z.}~\bibnamefont {Holmes}},\ and\ \bibinfo
  {author} {\bibfnamefont {H.-Y.}\ \bibnamefont {Huang}},\ }\bibfield  {title}
  {\bibinfo {title} {Classically estimating observables of noiseless quantum
  circuits},\ }\href {https://doi.org/10.1103/lh6x-7rc3} {\bibfield  {journal}
  {\bibinfo  {journal} {Phys. Rev. Lett.}\ }\textbf {\bibinfo {volume} {135}},\
  \bibinfo {pages} {170602} (\bibinfo {year} {2025})}\BibitemShut {NoStop}%
\bibitem [{\citenamefont {Cirstoiu}(2024)}]{cirstoiu2024fourier}%
  \BibitemOpen
  \bibfield  {author} {\bibinfo {author} {\bibfnamefont {C.}~\bibnamefont
  {Cirstoiu}},\ }\bibfield  {title} {\bibinfo {title} {A fourier analysis
  framework for approximate classical simulations of quantum circuits},\ }\href
  {https://doi.org/10.48550/arXiv.2410.13856} {\bibfield  {journal} {\bibinfo
  {journal} {arXiv preprint arXiv:2410.13856}\ } (\bibinfo {year}
  {2024})}\BibitemShut {NoStop}%
\bibitem [{Note1()}]{Note1}%
  \BibitemOpen
  \bibinfo {note} {For the intermediate-temperature quench discussed in this
  work, the Ising order parameter itself is small ($\langle Z_{\protect \rm
  tot}\rangle ^2\sim 0$) except at very early times, and $\langle Z_{\protect
  \rm tot}^2\rangle $ serves as a direct measure of order parameter
  fluctuations.}\BibitemShut {Stop}%
\bibitem [{fer()}]{fermioniq}%
  \BibitemOpen
  \href@noop {} {}\bibinfo {howpublished}
  {\url{https://www.fermioniq.com/ava}}\BibitemShut {NoStop}%
\bibitem [{\citenamefont {Temme}\ \emph {et~al.}(2017)\citenamefont {Temme},
  \citenamefont {Bravyi},\ and\ \citenamefont
  {Gambetta}}]{PhysRevLett.119.180509}%
  \BibitemOpen
  \bibfield  {author} {\bibinfo {author} {\bibfnamefont {K.}~\bibnamefont
  {Temme}}, \bibinfo {author} {\bibfnamefont {S.}~\bibnamefont {Bravyi}},\ and\
  \bibinfo {author} {\bibfnamefont {J.~M.}\ \bibnamefont {Gambetta}},\
  }\bibfield  {title} {\bibinfo {title} {Error mitigation for short-depth
  quantum circuits},\ }\href {https://doi.org/10.1103/PhysRevLett.119.180509}
  {\bibfield  {journal} {\bibinfo  {journal} {Phys. Rev. Lett.}\ }\textbf
  {\bibinfo {volume} {119}},\ \bibinfo {pages} {180509} (\bibinfo {year}
  {2017})}\BibitemShut {NoStop}%
\bibitem [{\citenamefont {Ganahl}\ \emph {et~al.}(2023)\citenamefont {Ganahl},
  \citenamefont {Beall}, \citenamefont {Hauru}, \citenamefont {Lewis},
  \citenamefont {Wojno}, \citenamefont {Yoo}, \citenamefont {Zou},\ and\
  \citenamefont {Vidal}}]{PRXQuantum.4.010317}%
  \BibitemOpen
  \bibfield  {author} {\bibinfo {author} {\bibfnamefont {M.}~\bibnamefont
  {Ganahl}}, \bibinfo {author} {\bibfnamefont {J.}~\bibnamefont {Beall}},
  \bibinfo {author} {\bibfnamefont {M.}~\bibnamefont {Hauru}}, \bibinfo
  {author} {\bibfnamefont {A.~G.}\ \bibnamefont {Lewis}}, \bibinfo {author}
  {\bibfnamefont {T.}~\bibnamefont {Wojno}}, \bibinfo {author} {\bibfnamefont
  {J.~H.}\ \bibnamefont {Yoo}}, \bibinfo {author} {\bibfnamefont
  {Y.}~\bibnamefont {Zou}},\ and\ \bibinfo {author} {\bibfnamefont
  {G.}~\bibnamefont {Vidal}},\ }\bibfield  {title} {\bibinfo {title} {Density
  matrix renormalization group with tensor processing units},\ }\href
  {https://doi.org/10.1103/PRXQuantum.4.010317} {\bibfield  {journal} {\bibinfo
   {journal} {PRX Quantum}\ }\textbf {\bibinfo {volume} {4}},\ \bibinfo {pages}
  {010317} (\bibinfo {year} {2023})}\BibitemShut {NoStop}%
\bibitem [{\citenamefont {Yang}\ \emph {et~al.}(2023)\citenamefont {Yang},
  \citenamefont {Christianen}, \citenamefont {Coll-Vinent}, \citenamefont
  {Smelyanskiy}, \citenamefont {Ba\~nuls}, \citenamefont {O'Brien},
  \citenamefont {Wild},\ and\ \citenamefont {Cirac}}]{PRXQuantum.4.030320}%
  \BibitemOpen
  \bibfield  {author} {\bibinfo {author} {\bibfnamefont {Y.}~\bibnamefont
  {Yang}}, \bibinfo {author} {\bibfnamefont {A.}~\bibnamefont {Christianen}},
  \bibinfo {author} {\bibfnamefont {S.}~\bibnamefont {Coll-Vinent}}, \bibinfo
  {author} {\bibfnamefont {V.}~\bibnamefont {Smelyanskiy}}, \bibinfo {author}
  {\bibfnamefont {M.~C.}\ \bibnamefont {Ba\~nuls}}, \bibinfo {author}
  {\bibfnamefont {T.~E.}\ \bibnamefont {O'Brien}}, \bibinfo {author}
  {\bibfnamefont {D.~S.}\ \bibnamefont {Wild}},\ and\ \bibinfo {author}
  {\bibfnamefont {J.~I.}\ \bibnamefont {Cirac}},\ }\bibfield  {title} {\bibinfo
  {title} {Simulating prethermalization using near-term quantum computers},\
  }\href {https://doi.org/10.1103/PRXQuantum.4.030320} {\bibfield  {journal}
  {\bibinfo  {journal} {PRX Quantum}\ }\textbf {\bibinfo {volume} {4}},\
  \bibinfo {pages} {030320} (\bibinfo {year} {2023})}\BibitemShut {NoStop}%
\bibitem [{\citenamefont {Granet}\ and\ \citenamefont
  {Dreyer}(2025)}]{PRXQuantum.6.010333}%
  \BibitemOpen
  \bibfield  {author} {\bibinfo {author} {\bibfnamefont {E.}~\bibnamefont
  {Granet}}\ and\ \bibinfo {author} {\bibfnamefont {H.}~\bibnamefont
  {Dreyer}},\ }\bibfield  {title} {\bibinfo {title} {{Dilution of Error in
  Digital Hamiltonian Simulation}},\ }\href
  {https://doi.org/10.1103/PRXQuantum.6.010333} {\bibfield  {journal} {\bibinfo
   {journal} {PRX Quantum}\ }\textbf {\bibinfo {volume} {6}},\ \bibinfo {pages}
  {010333} (\bibinfo {year} {2025})}\BibitemShut {NoStop}%
\bibitem [{\citenamefont {Chertkov}\ \emph {et~al.}(2026)\citenamefont
  {Chertkov}, \citenamefont {Chen}, \citenamefont {Lubasch}, \citenamefont
  {Hayes},\ and\ \citenamefont {Foss-Feig}}]{chertkov2024robustness}%
  \BibitemOpen
  \bibfield  {author} {\bibinfo {author} {\bibfnamefont {E.}~\bibnamefont
  {Chertkov}}, \bibinfo {author} {\bibfnamefont {Y.-H.}\ \bibnamefont {Chen}},
  \bibinfo {author} {\bibfnamefont {M.}~\bibnamefont {Lubasch}}, \bibinfo
  {author} {\bibfnamefont {D.}~\bibnamefont {Hayes}},\ and\ \bibinfo {author}
  {\bibfnamefont {M.}~\bibnamefont {Foss-Feig}},\ }\bibfield  {title} {\bibinfo
  {title} {Robustness of near-thermal dynamics on digital quantum computers},\
  }\href {https://doi.org/10.1103/nn2w-jxpf} {\bibfield  {journal} {\bibinfo
  {journal} {Phys. Rev. Res.}\ }\textbf {\bibinfo {volume} {8}},\ \bibinfo
  {pages} {013255} (\bibinfo {year} {2026})}\BibitemShut {NoStop}%
\bibitem [{\citenamefont {Carleo}\ \emph {et~al.}(2012)\citenamefont {Carleo},
  \citenamefont {Becca}, \citenamefont {Schir{\'o}},\ and\ \citenamefont
  {Fabrizio}}]{carleo2012localization}%
  \BibitemOpen
  \bibfield  {author} {\bibinfo {author} {\bibfnamefont {G.}~\bibnamefont
  {Carleo}}, \bibinfo {author} {\bibfnamefont {F.}~\bibnamefont {Becca}},
  \bibinfo {author} {\bibfnamefont {M.}~\bibnamefont {Schir{\'o}}},\ and\
  \bibinfo {author} {\bibfnamefont {M.}~\bibnamefont {Fabrizio}},\ }\bibfield
  {title} {\bibinfo {title} {Localization and glassy dynamics of many-body
  quantum systems},\ }\href {https://doi.org/10.1038/srep00243} {\bibfield
  {journal} {\bibinfo  {journal} {Scientific reports}\ }\textbf {\bibinfo
  {volume} {2}},\ \bibinfo {pages} {243} (\bibinfo {year} {2012})}\BibitemShut
  {NoStop}%
\bibitem [{\citenamefont {Vicentini}\ \emph {et~al.}(2022)\citenamefont
  {Vicentini}, \citenamefont {Hofmann}, \citenamefont {Szabó}, \citenamefont
  {Wu}, \citenamefont {Roth}, \citenamefont {Giuliani}, \citenamefont {Pescia},
  \citenamefont {Nys}, \citenamefont {Vargas-Calderón}, \citenamefont
  {Astrakhantsev},\ and\ \citenamefont {Carleo}}]{netket3:2022}%
  \BibitemOpen
  \bibfield  {author} {\bibinfo {author} {\bibfnamefont {F.}~\bibnamefont
  {Vicentini}}, \bibinfo {author} {\bibfnamefont {D.}~\bibnamefont {Hofmann}},
  \bibinfo {author} {\bibfnamefont {A.}~\bibnamefont {Szabó}}, \bibinfo
  {author} {\bibfnamefont {D.}~\bibnamefont {Wu}}, \bibinfo {author}
  {\bibfnamefont {C.}~\bibnamefont {Roth}}, \bibinfo {author} {\bibfnamefont
  {C.}~\bibnamefont {Giuliani}}, \bibinfo {author} {\bibfnamefont
  {G.}~\bibnamefont {Pescia}}, \bibinfo {author} {\bibfnamefont
  {J.}~\bibnamefont {Nys}}, \bibinfo {author} {\bibfnamefont {V.}~\bibnamefont
  {Vargas-Calderón}}, \bibinfo {author} {\bibfnamefont {N.}~\bibnamefont
  {Astrakhantsev}},\ and\ \bibinfo {author} {\bibfnamefont {G.}~\bibnamefont
  {Carleo}},\ }\bibfield  {title} {\bibinfo {title} {{NetKet 3: Machine
  Learning Toolbox for Many-Body Quantum Systems}},\ }\href
  {https://doi.org/10.21468/SciPostPhysCodeb.7} {\bibfield  {journal} {\bibinfo
   {journal} {SciPost Phys. Codebases}\ ,\ \bibinfo {pages} {7}} (\bibinfo
  {year} {2022})}\BibitemShut {NoStop}%
\bibitem [{\citenamefont {Verstraete}\ \emph {et~al.}(2004)\citenamefont
  {Verstraete}, \citenamefont {Garc\'{\i}a-Ripoll},\ and\ \citenamefont
  {Cirac}}]{Verstraete2004_mpdo}%
  \BibitemOpen
  \bibfield  {author} {\bibinfo {author} {\bibfnamefont {F.}~\bibnamefont
  {Verstraete}}, \bibinfo {author} {\bibfnamefont {J.~J.}\ \bibnamefont
  {Garc\'{\i}a-Ripoll}},\ and\ \bibinfo {author} {\bibfnamefont {J.~I.}\
  \bibnamefont {Cirac}},\ }\bibfield  {title} {\bibinfo {title} {{Matrix}
  {Product} {Density} {Operators}: {Simulation} of {Finite-Temperature} and
  {Dissipative} {Systems}},\ }\href
  {https://doi.org/10.1103/PhysRevLett.93.207204} {\bibfield  {journal}
  {\bibinfo  {journal} {Phys. Rev. Lett.}\ }\textbf {\bibinfo {volume} {93}},\
  \bibinfo {pages} {207204} (\bibinfo {year} {2004})}\BibitemShut {NoStop}%
\bibitem [{\citenamefont {Barthel}\ \emph {et~al.}(2009)\citenamefont
  {Barthel}, \citenamefont {Schollw\"ock},\ and\ \citenamefont
  {White}}]{Barthel2009}%
  \BibitemOpen
  \bibfield  {author} {\bibinfo {author} {\bibfnamefont {T.}~\bibnamefont
  {Barthel}}, \bibinfo {author} {\bibfnamefont {U.}~\bibnamefont
  {Schollw\"ock}},\ and\ \bibinfo {author} {\bibfnamefont {S.~R.}\ \bibnamefont
  {White}},\ }\bibfield  {title} {\bibinfo {title} {Spectral functions in
  one-dimensional quantum systems at finite temperature using the density
  matrix renormalization group},\ }\href
  {https://doi.org/10.1103/PhysRevB.79.245101} {\bibfield  {journal} {\bibinfo
  {journal} {Phys. Rev. B}\ }\textbf {\bibinfo {volume} {79}},\ \bibinfo
  {pages} {245101} (\bibinfo {year} {2009})}\BibitemShut {NoStop}%
\bibitem [{\citenamefont {Zu}\ \emph {et~al.}(2021)\citenamefont {Zu},
  \citenamefont {Machado}, \citenamefont {Ye}, \citenamefont {Choi},
  \citenamefont {Kobrin}, \citenamefont {Mittiga}, \citenamefont {Hsieh},
  \citenamefont {Bhattacharyya}, \citenamefont {Markham}, \citenamefont
  {Twitchen}, \citenamefont {Jarmola}, \citenamefont {Budker}, \citenamefont
  {Laumann}, \citenamefont {Moore},\ and\ \citenamefont {Yao}}]{Zu_2021_hydro}%
  \BibitemOpen
  \bibfield  {author} {\bibinfo {author} {\bibfnamefont {C.}~\bibnamefont
  {Zu}}, \bibinfo {author} {\bibfnamefont {F.}~\bibnamefont {Machado}},
  \bibinfo {author} {\bibfnamefont {B.}~\bibnamefont {Ye}}, \bibinfo {author}
  {\bibfnamefont {S.}~\bibnamefont {Choi}}, \bibinfo {author} {\bibfnamefont
  {B.}~\bibnamefont {Kobrin}}, \bibinfo {author} {\bibfnamefont
  {T.}~\bibnamefont {Mittiga}}, \bibinfo {author} {\bibfnamefont
  {S.}~\bibnamefont {Hsieh}}, \bibinfo {author} {\bibfnamefont
  {P.}~\bibnamefont {Bhattacharyya}}, \bibinfo {author} {\bibfnamefont
  {M.}~\bibnamefont {Markham}}, \bibinfo {author} {\bibfnamefont
  {D.}~\bibnamefont {Twitchen}}, \bibinfo {author} {\bibfnamefont
  {A.}~\bibnamefont {Jarmola}}, \bibinfo {author} {\bibfnamefont
  {D.}~\bibnamefont {Budker}}, \bibinfo {author} {\bibfnamefont {C.~R.}\
  \bibnamefont {Laumann}}, \bibinfo {author} {\bibfnamefont {J.~E.}\
  \bibnamefont {Moore}},\ and\ \bibinfo {author} {\bibfnamefont {N.~Y.}\
  \bibnamefont {Yao}},\ }\bibfield  {title} {\bibinfo {title} {Emergent
  hydrodynamics in a strongly interacting dipolar spin ensemble},\ }\href
  {https://doi.org/10.1038/s41586-021-03763-1} {\bibfield  {journal} {\bibinfo
  {journal} {Nature}\ }\textbf {\bibinfo {volume} {597}},\ \bibinfo {pages}
  {45–50} (\bibinfo {year} {2021})}\BibitemShut {NoStop}%
\bibitem [{\citenamefont {Wannier}(1950)}]{PhysRev.79.357}%
  \BibitemOpen
  \bibfield  {author} {\bibinfo {author} {\bibfnamefont {G.~H.}\ \bibnamefont
  {Wannier}},\ }\bibfield  {title} {\bibinfo {title} {Antiferromagnetism. the
  triangular ising net},\ }\href {https://doi.org/10.1103/PhysRev.79.357}
  {\bibfield  {journal} {\bibinfo  {journal} {Phys. Rev.}\ }\textbf {\bibinfo
  {volume} {79}},\ \bibinfo {pages} {357} (\bibinfo {year} {1950})}\BibitemShut
  {NoStop}%
\bibitem [{\citenamefont {Anderson}(1987)}]{anderson1987resonating}%
  \BibitemOpen
  \bibfield  {author} {\bibinfo {author} {\bibfnamefont {P.~W.}\ \bibnamefont
  {Anderson}},\ }\bibfield  {title} {\bibinfo {title} {The resonating valence
  bond state in la2cuo4 and superconductivity},\ }\href@noop {} {\bibfield
  {journal} {\bibinfo  {journal} {science}\ }\textbf {\bibinfo {volume}
  {235}},\ \bibinfo {pages} {1196} (\bibinfo {year} {1987})}\BibitemShut
  {NoStop}%
\bibitem [{\citenamefont {Granet}\ \emph {et~al.}(2025)\citenamefont {Granet},
  \citenamefont {Lin}, \citenamefont {Hémery}, \citenamefont {Hagshenas},
  \citenamefont {Andres-Martinez}, \citenamefont {Stephen}, \citenamefont
  {Ransford}, \citenamefont {Arkinstall}, \citenamefont {Allman}, \citenamefont
  {Campora}, \citenamefont {Cooper}, \citenamefont {Delaney}, \citenamefont
  {Dreiling}, \citenamefont {Estey}, \citenamefont {Figgatt}, \citenamefont
  {Foltz}, \citenamefont {Gaebler}, \citenamefont {Hall}, \citenamefont
  {Husain}, \citenamefont {Isanaka}, \citenamefont {Kennedy}, \citenamefont
  {Kotibhaskar}, \citenamefont {Madjarov}, \citenamefont {Mills}, \citenamefont
  {Milne}, \citenamefont {Park}, \citenamefont {Reed}, \citenamefont
  {Neyenhuis}, \citenamefont {Bohnet}, \citenamefont {Foss-Feig}, \citenamefont
  {Potter}, \citenamefont {Nigmatullin}, \citenamefont {Iqbal},\ and\
  \citenamefont {Dreyer}}]{granet2025}%
  \BibitemOpen
  \bibfield  {author} {\bibinfo {author} {\bibfnamefont {E.}~\bibnamefont
  {Granet}}, \bibinfo {author} {\bibfnamefont {S.-H.}\ \bibnamefont {Lin}},
  \bibinfo {author} {\bibfnamefont {K.}~\bibnamefont {Hémery}}, \bibinfo
  {author} {\bibfnamefont {R.}~\bibnamefont {Hagshenas}}, \bibinfo {author}
  {\bibfnamefont {P.}~\bibnamefont {Andres-Martinez}}, \bibinfo {author}
  {\bibfnamefont {D.~T.}\ \bibnamefont {Stephen}}, \bibinfo {author}
  {\bibfnamefont {A.}~\bibnamefont {Ransford}}, \bibinfo {author}
  {\bibfnamefont {J.}~\bibnamefont {Arkinstall}}, \bibinfo {author}
  {\bibfnamefont {M.~S.}\ \bibnamefont {Allman}}, \bibinfo {author}
  {\bibfnamefont {P.}~\bibnamefont {Campora}}, \bibinfo {author} {\bibfnamefont
  {S.~F.}\ \bibnamefont {Cooper}}, \bibinfo {author} {\bibfnamefont {R.~D.}\
  \bibnamefont {Delaney}}, \bibinfo {author} {\bibfnamefont {J.~M.}\
  \bibnamefont {Dreiling}}, \bibinfo {author} {\bibfnamefont {B.}~\bibnamefont
  {Estey}}, \bibinfo {author} {\bibfnamefont {C.}~\bibnamefont {Figgatt}},
  \bibinfo {author} {\bibfnamefont {C.}~\bibnamefont {Foltz}}, \bibinfo
  {author} {\bibfnamefont {J.~P.}\ \bibnamefont {Gaebler}}, \bibinfo {author}
  {\bibfnamefont {A.}~\bibnamefont {Hall}}, \bibinfo {author} {\bibfnamefont
  {A.}~\bibnamefont {Husain}}, \bibinfo {author} {\bibfnamefont
  {A.}~\bibnamefont {Isanaka}}, \bibinfo {author} {\bibfnamefont {C.~J.}\
  \bibnamefont {Kennedy}}, \bibinfo {author} {\bibfnamefont {N.}~\bibnamefont
  {Kotibhaskar}}, \bibinfo {author} {\bibfnamefont {I.~S.}\ \bibnamefont
  {Madjarov}}, \bibinfo {author} {\bibfnamefont {M.}~\bibnamefont {Mills}},
  \bibinfo {author} {\bibfnamefont {A.~R.}\ \bibnamefont {Milne}}, \bibinfo
  {author} {\bibfnamefont {A.~J.}\ \bibnamefont {Park}}, \bibinfo {author}
  {\bibfnamefont {A.~P.}\ \bibnamefont {Reed}}, \bibinfo {author}
  {\bibfnamefont {B.}~\bibnamefont {Neyenhuis}}, \bibinfo {author}
  {\bibfnamefont {J.~G.}\ \bibnamefont {Bohnet}}, \bibinfo {author}
  {\bibfnamefont {M.}~\bibnamefont {Foss-Feig}}, \bibinfo {author}
  {\bibfnamefont {A.~C.}\ \bibnamefont {Potter}}, \bibinfo {author}
  {\bibfnamefont {R.}~\bibnamefont {Nigmatullin}}, \bibinfo {author}
  {\bibfnamefont {M.}~\bibnamefont {Iqbal}},\ and\ \bibinfo {author}
  {\bibfnamefont {H.}~\bibnamefont {Dreyer}},\ }\href
  {https://arxiv.org/abs/2511.02125} {\bibinfo {title} {Superconducting pairing
  correlations on a trapped-ion quantum computer}} (\bibinfo {year} {2025}),\
  \Eprint {https://arxiv.org/abs/2511.02125} {arXiv:2511.02125 [quant-ph]}
  \BibitemShut {NoStop}%
\bibitem [{\citenamefont {Alam}\ \emph
  {et~al.}(2025{\natexlab{a}})\citenamefont {Alam}, \citenamefont {Bosse},
  \citenamefont {{\v{C}}epait{\.e}}, \citenamefont {Chapman}, \citenamefont
  {Clinton}, \citenamefont {Crichigno}, \citenamefont {Crosson}, \citenamefont
  {Cubitt}, \citenamefont {Derby}, \citenamefont {Dowinton} \emph
  {et~al.}}]{alam2025fermionic}%
  \BibitemOpen
  \bibfield  {author} {\bibinfo {author} {\bibfnamefont {F.}~\bibnamefont
  {Alam}}, \bibinfo {author} {\bibfnamefont {J.~L.}\ \bibnamefont {Bosse}},
  \bibinfo {author} {\bibfnamefont {I.}~\bibnamefont {{\v{C}}epait{\.e}}},
  \bibinfo {author} {\bibfnamefont {A.}~\bibnamefont {Chapman}}, \bibinfo
  {author} {\bibfnamefont {L.}~\bibnamefont {Clinton}}, \bibinfo {author}
  {\bibfnamefont {M.}~\bibnamefont {Crichigno}}, \bibinfo {author}
  {\bibfnamefont {E.}~\bibnamefont {Crosson}}, \bibinfo {author} {\bibfnamefont
  {T.}~\bibnamefont {Cubitt}}, \bibinfo {author} {\bibfnamefont
  {C.}~\bibnamefont {Derby}}, \bibinfo {author} {\bibfnamefont
  {O.}~\bibnamefont {Dowinton}}, \emph {et~al.},\ }\bibfield  {title} {\bibinfo
  {title} {Fermionic dynamics on a trapped-ion quantum computer beyond exact
  classical simulation},\ }\href {https://arxiv.org/abs/2510.26300} {\bibfield
  {journal} {\bibinfo  {journal} {arXiv preprint arXiv:2510.26300}\ } (\bibinfo
  {year} {2025}{\natexlab{a}})}\BibitemShut {NoStop}%
\bibitem [{\citenamefont {Alam}\ \emph
  {et~al.}(2025{\natexlab{b}})\citenamefont {Alam}, \citenamefont {Bosse},
  \citenamefont {{\v{C}}epait{\.e}}, \citenamefont {Chapman}, \citenamefont
  {Clinton}, \citenamefont {Crichigno}, \citenamefont {Crosson}, \citenamefont
  {Cubitt}, \citenamefont {Derby}, \citenamefont {Dowinton} \emph
  {et~al.}}]{alam2025programmable}%
  \BibitemOpen
  \bibfield  {author} {\bibinfo {author} {\bibfnamefont {F.}~\bibnamefont
  {Alam}}, \bibinfo {author} {\bibfnamefont {J.~L.}\ \bibnamefont {Bosse}},
  \bibinfo {author} {\bibfnamefont {I.}~\bibnamefont {{\v{C}}epait{\.e}}},
  \bibinfo {author} {\bibfnamefont {A.}~\bibnamefont {Chapman}}, \bibinfo
  {author} {\bibfnamefont {L.}~\bibnamefont {Clinton}}, \bibinfo {author}
  {\bibfnamefont {M.}~\bibnamefont {Crichigno}}, \bibinfo {author}
  {\bibfnamefont {E.}~\bibnamefont {Crosson}}, \bibinfo {author} {\bibfnamefont
  {T.}~\bibnamefont {Cubitt}}, \bibinfo {author} {\bibfnamefont
  {C.}~\bibnamefont {Derby}}, \bibinfo {author} {\bibfnamefont
  {O.}~\bibnamefont {Dowinton}}, \emph {et~al.},\ }\bibfield  {title} {\bibinfo
  {title} {Programmable digital quantum simulation of 2d fermi-hubbard dynamics
  using 72 superconducting qubits},\ }\href {https://arxiv.org/abs/2510.26845}
  {\bibfield  {journal} {\bibinfo  {journal} {arXiv preprint arXiv:2510.26845}\
  } (\bibinfo {year} {2025}{\natexlab{b}})}\BibitemShut {NoStop}%
\bibitem [{\citenamefont {Viola}\ \emph {et~al.}(1999)\citenamefont {Viola},
  \citenamefont {Knill},\ and\ \citenamefont {Lloyd}}]{viola1999}%
  \BibitemOpen
  \bibfield  {author} {\bibinfo {author} {\bibfnamefont {L.}~\bibnamefont
  {Viola}}, \bibinfo {author} {\bibfnamefont {E.}~\bibnamefont {Knill}},\ and\
  \bibinfo {author} {\bibfnamefont {S.}~\bibnamefont {Lloyd}},\ }\bibfield
  {title} {\bibinfo {title} {{Dynamical Decoupling of Open Quantum Systems}},\
  }\href {https://doi.org/10.1103/PhysRevLett.82.2417} {\bibfield  {journal}
  {\bibinfo  {journal} {Phys. Rev. Lett.}\ }\textbf {\bibinfo {volume} {82}},\
  \bibinfo {pages} {2417} (\bibinfo {year} {1999})}\BibitemShut {NoStop}%
\bibitem [{\citenamefont {Wallman}\ and\ \citenamefont
  {Emerson}(2016)}]{Wallman2016}%
  \BibitemOpen
  \bibfield  {author} {\bibinfo {author} {\bibfnamefont {J.~J.}\ \bibnamefont
  {Wallman}}\ and\ \bibinfo {author} {\bibfnamefont {J.}~\bibnamefont
  {Emerson}},\ }\bibfield  {title} {\bibinfo {title} {Noise tailoring for
  scalable quantum computation via randomized compiling},\ }\href
  {https://doi.org/10.1103/PhysRevA.94.052325} {\bibfield  {journal} {\bibinfo
  {journal} {Phys. Rev. A}\ }\textbf {\bibinfo {volume} {94}},\ \bibinfo
  {pages} {052325} (\bibinfo {year} {2016})}\BibitemShut {NoStop}%
\bibitem [{\citenamefont {P.~Santos}\ \emph {et~al.}(2024)\citenamefont
  {P.~Santos}, \citenamefont {Bar},\ and\ \citenamefont
  {Uzdin}}]{P.Santos2024}%
  \BibitemOpen
  \bibfield  {author} {\bibinfo {author} {\bibfnamefont {J.}~\bibnamefont
  {P.~Santos}}, \bibinfo {author} {\bibfnamefont {B.}~\bibnamefont {Bar}},\
  and\ \bibinfo {author} {\bibfnamefont {R.}~\bibnamefont {Uzdin}},\ }\bibfield
   {title} {\bibinfo {title} {Pseudo twirling mitigation of coherent errors in
  non-clifford gates},\ }\href {https://doi.org/10.1038/s41534-024-00889-8}
  {\bibfield  {journal} {\bibinfo  {journal} {npj Quantum Information}\
  }\textbf {\bibinfo {volume} {10}},\ \bibinfo {pages} {100} (\bibinfo {year}
  {2024})}\BibitemShut {NoStop}%
\bibitem [{\citenamefont {Stricker}\ \emph {et~al.}(2020)\citenamefont
  {Stricker}, \citenamefont {Vodola}, \citenamefont {Erhard}, \citenamefont
  {Postler}, \citenamefont {Meth}, \citenamefont {Ringbauer}, \citenamefont
  {Schindler}, \citenamefont {Monz}, \citenamefont {M{\"u}ller},\ and\
  \citenamefont {Blatt}}]{Stricker_2020}%
  \BibitemOpen
  \bibfield  {author} {\bibinfo {author} {\bibfnamefont {R.}~\bibnamefont
  {Stricker}}, \bibinfo {author} {\bibfnamefont {D.}~\bibnamefont {Vodola}},
  \bibinfo {author} {\bibfnamefont {A.}~\bibnamefont {Erhard}}, \bibinfo
  {author} {\bibfnamefont {L.}~\bibnamefont {Postler}}, \bibinfo {author}
  {\bibfnamefont {M.}~\bibnamefont {Meth}}, \bibinfo {author} {\bibfnamefont
  {M.}~\bibnamefont {Ringbauer}}, \bibinfo {author} {\bibfnamefont
  {P.}~\bibnamefont {Schindler}}, \bibinfo {author} {\bibfnamefont
  {T.}~\bibnamefont {Monz}}, \bibinfo {author} {\bibfnamefont {M.}~\bibnamefont
  {M{\"u}ller}},\ and\ \bibinfo {author} {\bibfnamefont {R.}~\bibnamefont
  {Blatt}},\ }\bibfield  {title} {\bibinfo {title} {Experimental deterministic
  correction of qubit loss},\ }\href
  {https://doi.org/10.1038/s41586-020-2667-0} {\bibfield  {journal} {\bibinfo
  {journal} {Nature}\ }\textbf {\bibinfo {volume} {585}},\ \bibinfo {pages}
  {207} (\bibinfo {year} {2020})}\BibitemShut {NoStop}%
\bibitem [{Note2()}]{Note2}%
  \BibitemOpen
  \bibinfo {note} {After the low-temperature and hydrodynamics quench
  experiments concluded, we found that the optimal way to arrange the phases is
  to instead follow the rule: after each $X$ and $Y$ Pauli, the phases of
  subsequent $X$ and $Y$ gates flips. This is based on the fact that $\pm
  Xe^{+i\delta \theta _x X}Ye^{+i\delta \theta _y Y}= \mp Ye^{-i\delta \theta
  _y Y}Xe^{-i\delta \theta _x X} + O(\delta \theta _x \delta \theta _y)$. The
  intermediate-temperature and triangular lattice quenches used the optimal
  approach. The original approach still heuristically cancels the coherent
  error, though slightly suboptimally.}\BibitemShut {Stop}%
\bibitem [{\citenamefont {Erhard}\ \emph {et~al.}(2019)\citenamefont {Erhard},
  \citenamefont {Wallman}, \citenamefont {Postler}, \citenamefont {Meth},
  \citenamefont {Stricker}, \citenamefont {Martinez}, \citenamefont
  {Schindler}, \citenamefont {Monz}, \citenamefont {Emerson},\ and\
  \citenamefont {Blatt}}]{Erhard2019}%
  \BibitemOpen
  \bibfield  {author} {\bibinfo {author} {\bibfnamefont {A.}~\bibnamefont
  {Erhard}}, \bibinfo {author} {\bibfnamefont {J.~J.}\ \bibnamefont {Wallman}},
  \bibinfo {author} {\bibfnamefont {L.}~\bibnamefont {Postler}}, \bibinfo
  {author} {\bibfnamefont {M.}~\bibnamefont {Meth}}, \bibinfo {author}
  {\bibfnamefont {R.}~\bibnamefont {Stricker}}, \bibinfo {author}
  {\bibfnamefont {E.~A.}\ \bibnamefont {Martinez}}, \bibinfo {author}
  {\bibfnamefont {P.}~\bibnamefont {Schindler}}, \bibinfo {author}
  {\bibfnamefont {T.}~\bibnamefont {Monz}}, \bibinfo {author} {\bibfnamefont
  {J.}~\bibnamefont {Emerson}},\ and\ \bibinfo {author} {\bibfnamefont
  {R.}~\bibnamefont {Blatt}},\ }\bibfield  {title} {\bibinfo {title}
  {Characterizing large-scale quantum computers via cycle benchmarking},\
  }\href {https://doi.org/10.1038/s41467-019-13068-7} {\bibfield  {journal}
  {\bibinfo  {journal} {Nature Communications}\ }\textbf {\bibinfo {volume}
  {10}},\ \bibinfo {pages} {5347} (\bibinfo {year} {2019})}\BibitemShut
  {NoStop}%
\bibitem [{\citenamefont {Layden}\ \emph {et~al.}(2024)\citenamefont {Layden},
  \citenamefont {Mitchell},\ and\ \citenamefont {Siva}}]{Layden2024}%
  \BibitemOpen
  \bibfield  {author} {\bibinfo {author} {\bibfnamefont {D.}~\bibnamefont
  {Layden}}, \bibinfo {author} {\bibfnamefont {B.}~\bibnamefont {Mitchell}},\
  and\ \bibinfo {author} {\bibfnamefont {K.}~\bibnamefont {Siva}},\ }\bibfield
  {title} {\bibinfo {title} {Theory of quantum error mitigation for
  non-clifford gates},\ }\href {https://arxiv.org/abs/2403.18793} {\bibfield
  {journal} {\bibinfo  {journal} {arxiv preprint quant-ph 2403.18793}\ }
  (\bibinfo {year} {2024})}\BibitemShut {NoStop}%
\bibitem [{\citenamefont {Chen}\ \emph {et~al.}(2023)\citenamefont {Chen},
  \citenamefont {Liu}, \citenamefont {Otten}, \citenamefont {Seif},
  \citenamefont {Fefferman},\ and\ \citenamefont {Jiang}}]{Chen2023}%
  \BibitemOpen
  \bibfield  {author} {\bibinfo {author} {\bibfnamefont {S.}~\bibnamefont
  {Chen}}, \bibinfo {author} {\bibfnamefont {Y.}~\bibnamefont {Liu}}, \bibinfo
  {author} {\bibfnamefont {M.}~\bibnamefont {Otten}}, \bibinfo {author}
  {\bibfnamefont {A.}~\bibnamefont {Seif}}, \bibinfo {author} {\bibfnamefont
  {B.}~\bibnamefont {Fefferman}},\ and\ \bibinfo {author} {\bibfnamefont
  {L.}~\bibnamefont {Jiang}},\ }\bibfield  {title} {\bibinfo {title} {The
  learnability of pauli noise},\ }\href
  {https://doi.org/10.1038/s41467-022-35759-4} {\bibfield  {journal} {\bibinfo
  {journal} {Nature Communications}\ }\textbf {\bibinfo {volume} {14}},\
  \bibinfo {pages} {52} (\bibinfo {year} {2023})}\BibitemShut {NoStop}%
\bibitem [{\citenamefont {Proctor}\ \emph {et~al.}(2019)\citenamefont
  {Proctor}, \citenamefont {Carignan-Dugas}, \citenamefont {Rudinger},
  \citenamefont {Nielsen}, \citenamefont {Blume-Kohout},\ and\ \citenamefont
  {Young}}]{proctor2019}%
  \BibitemOpen
  \bibfield  {author} {\bibinfo {author} {\bibfnamefont {T.~J.}\ \bibnamefont
  {Proctor}}, \bibinfo {author} {\bibfnamefont {A.}~\bibnamefont
  {Carignan-Dugas}}, \bibinfo {author} {\bibfnamefont {K.}~\bibnamefont
  {Rudinger}}, \bibinfo {author} {\bibfnamefont {E.}~\bibnamefont {Nielsen}},
  \bibinfo {author} {\bibfnamefont {R.}~\bibnamefont {Blume-Kohout}},\ and\
  \bibinfo {author} {\bibfnamefont {K.}~\bibnamefont {Young}},\ }\bibfield
  {title} {\bibinfo {title} {Direct randomized benchmarking for multiqubit
  devices},\ }\href {https://doi.org/10.1103/PhysRevLett.123.030503} {\bibfield
   {journal} {\bibinfo  {journal} {Phys. Rev. Lett.}\ }\textbf {\bibinfo
  {volume} {123}},\ \bibinfo {pages} {030503} (\bibinfo {year}
  {2019})}\BibitemShut {NoStop}%
\bibitem [{\citenamefont {Bukov}\ \emph
  {et~al.}(2015{\natexlab{b}})\citenamefont {Bukov}, \citenamefont
  {D’Alessio},\ and\ \citenamefont {Polkovnikov}}]{Bukov_2015}%
  \BibitemOpen
  \bibfield  {author} {\bibinfo {author} {\bibfnamefont {M.}~\bibnamefont
  {Bukov}}, \bibinfo {author} {\bibfnamefont {L.}~\bibnamefont {D’Alessio}},\
  and\ \bibinfo {author} {\bibfnamefont {A.}~\bibnamefont {Polkovnikov}},\
  }\bibfield  {title} {\bibinfo {title} {Universal high-frequency behavior of
  periodically driven systems: from dynamical stabilization to {F}loquet
  engineering},\ }\href {https://doi.org/10.1080/00018732.2015.1055918}
  {\bibfield  {journal} {\bibinfo  {journal} {Advances in Physics}\ }\textbf
  {\bibinfo {volume} {64}},\ \bibinfo {pages} {139–226} (\bibinfo {year}
  {2015}{\natexlab{b}})}\BibitemShut {NoStop}%
\bibitem [{\citenamefont {Blanes}\ \emph {et~al.}(2009)\citenamefont {Blanes},
  \citenamefont {Casas}, \citenamefont {Oteo},\ and\ \citenamefont
  {Ros}}]{Blanes_2009}%
  \BibitemOpen
  \bibfield  {author} {\bibinfo {author} {\bibfnamefont {S.}~\bibnamefont
  {Blanes}}, \bibinfo {author} {\bibfnamefont {F.}~\bibnamefont {Casas}},
  \bibinfo {author} {\bibfnamefont {J.}~\bibnamefont {Oteo}},\ and\ \bibinfo
  {author} {\bibfnamefont {J.}~\bibnamefont {Ros}},\ }\bibfield  {title}
  {\bibinfo {title} {The {M}agnus expansion and some of its applications},\
  }\href {https://doi.org/10.1016/j.physrep.2008.11.001} {\bibfield  {journal}
  {\bibinfo  {journal} {Physics Reports}\ }\textbf {\bibinfo {volume} {470}},\
  \bibinfo {pages} {151–238} (\bibinfo {year} {2009})}\BibitemShut {NoStop}%
\bibitem [{\citenamefont {Zaletel}\ \emph {et~al.}(2015)\citenamefont
  {Zaletel}, \citenamefont {Mong}, \citenamefont {Karrasch}, \citenamefont
  {Moore},\ and\ \citenamefont {Pollmann}}]{Zaletel2014}%
  \BibitemOpen
  \bibfield  {author} {\bibinfo {author} {\bibfnamefont {M.~P.}\ \bibnamefont
  {Zaletel}}, \bibinfo {author} {\bibfnamefont {R.~S.~K.}\ \bibnamefont
  {Mong}}, \bibinfo {author} {\bibfnamefont {C.}~\bibnamefont {Karrasch}},
  \bibinfo {author} {\bibfnamefont {J.~E.}\ \bibnamefont {Moore}},\ and\
  \bibinfo {author} {\bibfnamefont {F.}~\bibnamefont {Pollmann}},\ }\bibfield
  {title} {\bibinfo {title} {Time-evolving a matrix product state with
  long-ranged interactions},\ }\href
  {https://doi.org/10.1103/PhysRevB.91.165112} {\bibfield  {journal} {\bibinfo
  {journal} {Physical Review B}\ }\textbf {\bibinfo {volume} {91}},\ \bibinfo
  {pages} {165112} (\bibinfo {year} {2015})}\BibitemShut {NoStop}%
\bibitem [{\citenamefont {Fannes}\ \emph {et~al.}(1992)\citenamefont {Fannes},
  \citenamefont {Nachtergaele},\ and\ \citenamefont
  {Werner}}]{fannes1992finitely}%
  \BibitemOpen
  \bibfield  {author} {\bibinfo {author} {\bibfnamefont {M.}~\bibnamefont
  {Fannes}}, \bibinfo {author} {\bibfnamefont {B.}~\bibnamefont
  {Nachtergaele}},\ and\ \bibinfo {author} {\bibfnamefont {R.~F.}\ \bibnamefont
  {Werner}},\ }\bibfield  {title} {\bibinfo {title} {Finitely correlated states
  on quantum spin chains},\ }\href@noop {} {\bibfield  {journal} {\bibinfo
  {journal} {Communications in mathematical physics}\ }\textbf {\bibinfo
  {volume} {144}},\ \bibinfo {pages} {443} (\bibinfo {year}
  {1992})}\BibitemShut {NoStop}%
\bibitem [{\citenamefont {{\"O}stlund}\ and\ \citenamefont
  {Rommer}(1995)}]{ostlund1995thermodynamic}%
  \BibitemOpen
  \bibfield  {author} {\bibinfo {author} {\bibfnamefont {S.}~\bibnamefont
  {{\"O}stlund}}\ and\ \bibinfo {author} {\bibfnamefont {S.}~\bibnamefont
  {Rommer}},\ }\bibfield  {title} {\bibinfo {title} {Thermodynamic limit of
  density matrix renormalization},\ }\href@noop {} {\bibfield  {journal}
  {\bibinfo  {journal} {Physical review letters}\ }\textbf {\bibinfo {volume}
  {75}},\ \bibinfo {pages} {3537} (\bibinfo {year} {1995})}\BibitemShut
  {NoStop}%
\bibitem [{\citenamefont {Vidal}(2003)}]{vidal2003efficient}%
  \BibitemOpen
  \bibfield  {author} {\bibinfo {author} {\bibfnamefont {G.}~\bibnamefont
  {Vidal}},\ }\bibfield  {title} {\bibinfo {title} {Efficient classical
  simulation of slightly entangled quantum computations},\ }\href@noop {}
  {\bibfield  {journal} {\bibinfo  {journal} {Physical review letters}\
  }\textbf {\bibinfo {volume} {91}},\ \bibinfo {pages} {147902} (\bibinfo
  {year} {2003})}\BibitemShut {NoStop}%
\bibitem [{\citenamefont {White}(1992)}]{white1992}%
  \BibitemOpen
  \bibfield  {author} {\bibinfo {author} {\bibfnamefont {S.~R.}\ \bibnamefont
  {White}},\ }\bibfield  {title} {\bibinfo {title} {Density matrix formulation
  for quantum renormalization groups},\ }\href@noop {} {\bibfield  {journal}
  {\bibinfo  {journal} {Physical review letters}\ }\textbf {\bibinfo {volume}
  {69}},\ \bibinfo {pages} {2863} (\bibinfo {year} {1992})}\BibitemShut
  {NoStop}%
\bibitem [{\citenamefont {Schollw{\"o}ck}(2011)}]{schollwock2011}%
  \BibitemOpen
  \bibfield  {author} {\bibinfo {author} {\bibfnamefont {U.}~\bibnamefont
  {Schollw{\"o}ck}},\ }\bibfield  {title} {\bibinfo {title} {The density-matrix
  renormalization group in the age of matrix product states},\ }\href@noop {}
  {\bibfield  {journal} {\bibinfo  {journal} {Annals of physics}\ }\textbf
  {\bibinfo {volume} {326}},\ \bibinfo {pages} {96} (\bibinfo {year}
  {2011})}\BibitemShut {NoStop}%
\bibitem [{Note3()}]{Note3}%
  \BibitemOpen
  \bibinfo {note} {In the sense that for every Trotter time $s$, there is a
  $k\in \{1, \protect \ldots , K\}$ such that $U_k U_{k-1} \protect \cdots U_1$
  implements the first $s$ Trotter steps.}\BibitemShut {Stop}%
\bibitem [{\citenamefont {Zhou}\ \emph {et~al.}(2020)\citenamefont {Zhou},
  \citenamefont {Stoudenmire},\ and\ \citenamefont {Waintal}}]{zhou2020}%
  \BibitemOpen
  \bibfield  {author} {\bibinfo {author} {\bibfnamefont {Y.}~\bibnamefont
  {Zhou}}, \bibinfo {author} {\bibfnamefont {E.~M.}\ \bibnamefont
  {Stoudenmire}},\ and\ \bibinfo {author} {\bibfnamefont {X.}~\bibnamefont
  {Waintal}},\ }\bibfield  {title} {\bibinfo {title} {What limits the
  simulation of quantum computers?},\ }\href@noop {} {\bibfield  {journal}
  {\bibinfo  {journal} {Physical Review X}\ }\textbf {\bibinfo {volume} {10}},\
  \bibinfo {pages} {041038} (\bibinfo {year} {2020})}\BibitemShut {NoStop}%
\bibitem [{\citenamefont {Thompson}\ \emph {et~al.}(2025)\citenamefont
  {Thompson}, \citenamefont {Soeteman}, \citenamefont {Cade},\ and\
  \citenamefont {Niesen}}]{thompson2025}%
  \BibitemOpen
  \bibfield  {author} {\bibinfo {author} {\bibfnamefont {A.~P.}\ \bibnamefont
  {Thompson}}, \bibinfo {author} {\bibfnamefont {A.}~\bibnamefont {Soeteman}},
  \bibinfo {author} {\bibfnamefont {C.}~\bibnamefont {Cade}},\ and\ \bibinfo
  {author} {\bibfnamefont {I.}~\bibnamefont {Niesen}},\ }\bibfield  {title}
  {\bibinfo {title} {Non-zero noise extrapolation: accurately simulating noisy
  quantum circuits with tensor networks},\ }\href@noop {} {\bibfield  {journal}
  {\bibinfo  {journal} {arXiv preprint arXiv:2501.13237}\ } (\bibinfo {year}
  {2025})}\BibitemShut {NoStop}%
\bibitem [{Note4()}]{Note4}%
  \BibitemOpen
  \bibinfo {note} {The fact that fidelity allows for more accurate
  extrapolations than bond dimension does make intuitive sense: the fidelity
  contains information on the accuracy of the simulation, whereas the bond
  dimension merely specifies the amount of classical resources
  used.}\BibitemShut {Stop}%
\bibitem [{\citenamefont {Mandr{\`a}}\ \emph {et~al.}(2025)\citenamefont
  {Mandr{\`a}}, \citenamefont {Astrakhantsev}, \citenamefont {Isakov},
  \citenamefont {Villalonga}, \citenamefont {Ware}, \citenamefont
  {Westerhout},\ and\ \citenamefont {Kechedzhi}}]{mandra2025heuristic}%
  \BibitemOpen
  \bibfield  {author} {\bibinfo {author} {\bibfnamefont {S.}~\bibnamefont
  {Mandr{\`a}}}, \bibinfo {author} {\bibfnamefont {N.}~\bibnamefont
  {Astrakhantsev}}, \bibinfo {author} {\bibfnamefont {S.}~\bibnamefont
  {Isakov}}, \bibinfo {author} {\bibfnamefont {B.}~\bibnamefont {Villalonga}},
  \bibinfo {author} {\bibfnamefont {B.}~\bibnamefont {Ware}}, \bibinfo {author}
  {\bibfnamefont {T.}~\bibnamefont {Westerhout}},\ and\ \bibinfo {author}
  {\bibfnamefont {K.}~\bibnamefont {Kechedzhi}},\ }\bibfield  {title} {\bibinfo
  {title} {A heuristic for matrix product state simulation of
  out-of-equilibrium dynamics of two-dimensional transverse-field ising
  models},\ }\href@noop {} {\bibfield  {journal} {\bibinfo  {journal} {arXiv
  preprint arXiv:2511.23438}\ } (\bibinfo {year} {2025})}\BibitemShut {NoStop}%
\bibitem [{\citenamefont {Niggemann}\ \emph {et~al.}(1997)\citenamefont
  {Niggemann}, \citenamefont {Kl{\"u}mper},\ and\ \citenamefont
  {Zittartz}}]{niggemann1997quantum}%
  \BibitemOpen
  \bibfield  {author} {\bibinfo {author} {\bibfnamefont {H.}~\bibnamefont
  {Niggemann}}, \bibinfo {author} {\bibfnamefont {A.}~\bibnamefont
  {Kl{\"u}mper}},\ and\ \bibinfo {author} {\bibfnamefont {J.}~\bibnamefont
  {Zittartz}},\ }\bibfield  {title} {\bibinfo {title} {Quantum phase transition
  in spin-3/2 systems on the hexagonal lattice—optimum ground state
  approach},\ }\href@noop {} {\bibfield  {journal} {\bibinfo  {journal}
  {Zeitschrift f{\"u}r Physik B Condensed Matter}\ }\textbf {\bibinfo {volume}
  {104}},\ \bibinfo {pages} {103} (\bibinfo {year} {1997})}\BibitemShut
  {NoStop}%
\bibitem [{\citenamefont {Nishino}\ and\ \citenamefont
  {Okunishi}(1998)}]{nishino1998density}%
  \BibitemOpen
  \bibfield  {author} {\bibinfo {author} {\bibfnamefont {T.}~\bibnamefont
  {Nishino}}\ and\ \bibinfo {author} {\bibfnamefont {K.}~\bibnamefont
  {Okunishi}},\ }\bibfield  {title} {\bibinfo {title} {A density matrix
  algorithm for 3d classical models},\ }\href
  {https://doi.org/10.1143/JPSJ.67.3066} {\bibfield  {journal} {\bibinfo
  {journal} {Journal of the Physical Society of Japan}\ }\textbf {\bibinfo
  {volume} {67}},\ \bibinfo {pages} {3066} (\bibinfo {year}
  {1998})}\BibitemShut {NoStop}%
\bibitem [{\citenamefont {Sierra}\ and\ \citenamefont
  {Martin-Delgado}(1998)}]{sierra1998density}%
  \BibitemOpen
  \bibfield  {author} {\bibinfo {author} {\bibfnamefont {G.}~\bibnamefont
  {Sierra}}\ and\ \bibinfo {author} {\bibfnamefont {M.}~\bibnamefont
  {Martin-Delgado}},\ }\bibfield  {title} {\bibinfo {title} {The density matrix
  renormalization group, quantum groups and conformal field theory},\
  }\href@noop {} {\bibfield  {journal} {\bibinfo  {journal} {arXiv preprint
  cond-mat/9811170}\ } (\bibinfo {year} {1998})}\BibitemShut {NoStop}%
\bibitem [{\citenamefont {Verstraete}\ and\ \citenamefont
  {Cirac}(2004)}]{verstraete2004renormalization}%
  \BibitemOpen
  \bibfield  {author} {\bibinfo {author} {\bibfnamefont {F.}~\bibnamefont
  {Verstraete}}\ and\ \bibinfo {author} {\bibfnamefont {J.~I.}\ \bibnamefont
  {Cirac}},\ }\bibfield  {title} {\bibinfo {title} {Renormalization algorithms
  for quantum-many body systems in two and higher dimensions},\ }\href@noop {}
  {\bibfield  {journal} {\bibinfo  {journal} {arXiv preprint cond-mat/0407066}\
  } (\bibinfo {year} {2004})}\BibitemShut {NoStop}%
\bibitem [{\citenamefont {Liao}\ \emph {et~al.}(2023)\citenamefont {Liao},
  \citenamefont {Wang}, \citenamefont {Zhou}, \citenamefont {Zhang},\ and\
  \citenamefont {Xiang}}]{liao2023simulation}%
  \BibitemOpen
  \bibfield  {author} {\bibinfo {author} {\bibfnamefont {H.-J.}\ \bibnamefont
  {Liao}}, \bibinfo {author} {\bibfnamefont {K.}~\bibnamefont {Wang}}, \bibinfo
  {author} {\bibfnamefont {Z.-S.}\ \bibnamefont {Zhou}}, \bibinfo {author}
  {\bibfnamefont {P.}~\bibnamefont {Zhang}},\ and\ \bibinfo {author}
  {\bibfnamefont {T.}~\bibnamefont {Xiang}},\ }\bibfield  {title} {\bibinfo
  {title} {Simulation of ibm's kicked ising experiment with projected entangled
  pair operator},\ }\href@noop {} {\bibfield  {journal} {\bibinfo  {journal}
  {arXiv preprint arXiv:2308.03082}\ } (\bibinfo {year} {2023})}\BibitemShut
  {NoStop}%
\bibitem [{\citenamefont {Alkabetz}\ and\ \citenamefont {Arad}(2021)}]{AlAr21}%
  \BibitemOpen
  \bibfield  {author} {\bibinfo {author} {\bibfnamefont {R.}~\bibnamefont
  {Alkabetz}}\ and\ \bibinfo {author} {\bibfnamefont {I.}~\bibnamefont
  {Arad}},\ }\bibfield  {title} {\bibinfo {title} {Tensor networks contraction
  and the belief propagation algorithm},\ }\href
  {https://doi.org/10.1103/PhysRevResearch.3.023073} {\bibfield  {journal}
  {\bibinfo  {journal} {Phys. Rev. Res.}\ }\textbf {\bibinfo {volume} {3}},\
  \bibinfo {pages} {023073} (\bibinfo {year} {2021})}\BibitemShut {NoStop}%
\bibitem [{\citenamefont {Jordan}\ \emph {et~al.}(2008)\citenamefont {Jordan},
  \citenamefont {Or\'us}, \citenamefont {Vidal}, \citenamefont {Verstraete},\
  and\ \citenamefont {Cirac}}]{JoOrViVeCi08}%
  \BibitemOpen
  \bibfield  {author} {\bibinfo {author} {\bibfnamefont {J.}~\bibnamefont
  {Jordan}}, \bibinfo {author} {\bibfnamefont {R.}~\bibnamefont {Or\'us}},
  \bibinfo {author} {\bibfnamefont {G.}~\bibnamefont {Vidal}}, \bibinfo
  {author} {\bibfnamefont {F.}~\bibnamefont {Verstraete}},\ and\ \bibinfo
  {author} {\bibfnamefont {J.~I.}\ \bibnamefont {Cirac}},\ }\bibfield  {title}
  {\bibinfo {title} {{Classical Simulation of Infinite-Size Quantum Lattice
  Systems in Two Spatial Dimensions}},\ }\href
  {https://doi.org/10.1103/PhysRevLett.101.250602} {\bibfield  {journal}
  {\bibinfo  {journal} {Phys. Rev. Lett.}\ }\textbf {\bibinfo {volume} {101}},\
  \bibinfo {pages} {250602} (\bibinfo {year} {2008})}\BibitemShut {NoStop}%
\bibitem [{\citenamefont {Jiang}\ \emph {et~al.}(2008)\citenamefont {Jiang},
  \citenamefont {Weng},\ and\ \citenamefont {Xiang}}]{JiWeXi08}%
  \BibitemOpen
  \bibfield  {author} {\bibinfo {author} {\bibfnamefont {H.~C.}\ \bibnamefont
  {Jiang}}, \bibinfo {author} {\bibfnamefont {Z.~Y.}\ \bibnamefont {Weng}},\
  and\ \bibinfo {author} {\bibfnamefont {T.}~\bibnamefont {Xiang}},\ }\bibfield
   {title} {\bibinfo {title} {{Accurate Determination of Tensor Network State
  of Quantum Lattice Models in Two Dimensions}},\ }\href
  {https://doi.org/10.1103/PhysRevLett.101.090603} {\bibfield  {journal}
  {\bibinfo  {journal} {Phys. Rev. Lett.}\ }\textbf {\bibinfo {volume} {101}},\
  \bibinfo {pages} {090603} (\bibinfo {year} {2008})}\BibitemShut {NoStop}%
\bibitem [{\citenamefont {Baxter}(1968)}]{baxter1968dimers}%
  \BibitemOpen
  \bibfield  {author} {\bibinfo {author} {\bibfnamefont {R.~J.}\ \bibnamefont
  {Baxter}},\ }\bibfield  {title} {\bibinfo {title} {Dimers on a rectangular
  lattice},\ }\href@noop {} {\bibfield  {journal} {\bibinfo  {journal} {Journal
  of Mathematical Physics}\ }\textbf {\bibinfo {volume} {9}},\ \bibinfo {pages}
  {650} (\bibinfo {year} {1968})}\BibitemShut {NoStop}%
\bibitem [{\citenamefont {Nishino}\ and\ \citenamefont
  {Okunishi}(1996)}]{nishino1996corner}%
  \BibitemOpen
  \bibfield  {author} {\bibinfo {author} {\bibfnamefont {T.}~\bibnamefont
  {Nishino}}\ and\ \bibinfo {author} {\bibfnamefont {K.}~\bibnamefont
  {Okunishi}},\ }\bibfield  {title} {\bibinfo {title} {Corner transfer matrix
  renormalization group method},\ }\href@noop {} {\bibfield  {journal}
  {\bibinfo  {journal} {Journal of the Physical Society of Japan}\ }\textbf
  {\bibinfo {volume} {65}},\ \bibinfo {pages} {891} (\bibinfo {year}
  {1996})}\BibitemShut {NoStop}%
\bibitem [{\citenamefont {Fishman}\ \emph {et~al.}(2018)\citenamefont
  {Fishman}, \citenamefont {Vanderstraeten}, \citenamefont {Zauner-Stauber},
  \citenamefont {Haegeman},\ and\ \citenamefont
  {Verstraete}}]{fishman2018faster}%
  \BibitemOpen
  \bibfield  {author} {\bibinfo {author} {\bibfnamefont {M.}~\bibnamefont
  {Fishman}}, \bibinfo {author} {\bibfnamefont {L.}~\bibnamefont
  {Vanderstraeten}}, \bibinfo {author} {\bibfnamefont {V.}~\bibnamefont
  {Zauner-Stauber}}, \bibinfo {author} {\bibfnamefont {J.}~\bibnamefont
  {Haegeman}},\ and\ \bibinfo {author} {\bibfnamefont {F.}~\bibnamefont
  {Verstraete}},\ }\bibfield  {title} {\bibinfo {title} {Faster methods for
  contracting infinite two-dimensional tensor networks},\ }\href@noop {}
  {\bibfield  {journal} {\bibinfo  {journal} {Physical Review B}\ }\textbf
  {\bibinfo {volume} {98}},\ \bibinfo {pages} {235148} (\bibinfo {year}
  {2018})}\BibitemShut {NoStop}%
\bibitem [{\citenamefont {Lubasch}\ \emph
  {et~al.}(2014{\natexlab{b}})\citenamefont {Lubasch}, \citenamefont {Cirac},\
  and\ \citenamefont {Ba\~nuls}}]{LuCiBa14b}%
  \BibitemOpen
  \bibfield  {author} {\bibinfo {author} {\bibfnamefont {M.}~\bibnamefont
  {Lubasch}}, \bibinfo {author} {\bibfnamefont {J.~I.}\ \bibnamefont {Cirac}},\
  and\ \bibinfo {author} {\bibfnamefont {M.-C.}\ \bibnamefont {Ba\~nuls}},\
  }\bibfield  {title} {\bibinfo {title} {Algorithms for finite projected
  entangled pair states},\ }\href {https://doi.org/10.1103/PhysRevB.90.064425}
  {\bibfield  {journal} {\bibinfo  {journal} {Phys. Rev. B}\ }\textbf {\bibinfo
  {volume} {90}},\ \bibinfo {pages} {064425} (\bibinfo {year}
  {2014}{\natexlab{b}})}\BibitemShut {NoStop}%
\bibitem [{\citenamefont {Tindall}\ and\ \citenamefont
  {Fishman}(2023)}]{tindall2023gauging}%
  \BibitemOpen
  \bibfield  {author} {\bibinfo {author} {\bibfnamefont {J.}~\bibnamefont
  {Tindall}}\ and\ \bibinfo {author} {\bibfnamefont {M.}~\bibnamefont
  {Fishman}},\ }\bibfield  {title} {\bibinfo {title} {Gauging tensor networks
  with belief propagation},\ }\href@noop {} {\bibfield  {journal} {\bibinfo
  {journal} {SciPost Physics}\ }\textbf {\bibinfo {volume} {15}},\ \bibinfo
  {pages} {222} (\bibinfo {year} {2023})}\BibitemShut {NoStop}%
\bibitem [{\citenamefont {Tindall}\ \emph {et~al.}(2024)\citenamefont
  {Tindall}, \citenamefont {Fishman}, \citenamefont {Stoudenmire},\ and\
  \citenamefont {Sels}}]{tindall2024efficient}%
  \BibitemOpen
  \bibfield  {author} {\bibinfo {author} {\bibfnamefont {J.}~\bibnamefont
  {Tindall}}, \bibinfo {author} {\bibfnamefont {M.}~\bibnamefont {Fishman}},
  \bibinfo {author} {\bibfnamefont {E.~M.}\ \bibnamefont {Stoudenmire}},\ and\
  \bibinfo {author} {\bibfnamefont {D.}~\bibnamefont {Sels}},\ }\bibfield
  {title} {\bibinfo {title} {Efficient tensor network simulation of ibm's eagle
  kicked ising experiment},\ }\href
  {https://doi.org/10.1103/PRXQuantum.5.010308} {\bibfield  {journal} {\bibinfo
   {journal} {PRX Quantum}\ }\textbf {\bibinfo {volume} {5}},\ \bibinfo {pages}
  {010308} (\bibinfo {year} {2024})}\BibitemShut {NoStop}%
\bibitem [{\citenamefont {Tindall}\ \emph {et~al.}(2025)\citenamefont
  {Tindall}, \citenamefont {Mello}, \citenamefont {Fishman}, \citenamefont
  {Stoudenmire},\ and\ \citenamefont {Sels}}]{tindall2025dynamics}%
  \BibitemOpen
  \bibfield  {author} {\bibinfo {author} {\bibfnamefont {J.}~\bibnamefont
  {Tindall}}, \bibinfo {author} {\bibfnamefont {A.}~\bibnamefont {Mello}},
  \bibinfo {author} {\bibfnamefont {M.}~\bibnamefont {Fishman}}, \bibinfo
  {author} {\bibfnamefont {M.}~\bibnamefont {Stoudenmire}},\ and\ \bibinfo
  {author} {\bibfnamefont {D.}~\bibnamefont {Sels}},\ }\bibfield  {title}
  {\bibinfo {title} {Dynamics of disordered quantum systems with two-and
  three-dimensional tensor networks},\ }\href@noop {} {\bibfield  {journal}
  {\bibinfo  {journal} {arXiv preprint arXiv:2503.05693}\ } (\bibinfo {year}
  {2025})}\BibitemShut {NoStop}%
\bibitem [{\citenamefont {Evenbly}\ \emph {et~al.}(2024)\citenamefont
  {Evenbly}, \citenamefont {Pancotti}, \citenamefont {Milsted}, \citenamefont
  {Gray},\ and\ \citenamefont {Chan}}]{EvEtAl24}%
  \BibitemOpen
  \bibfield  {author} {\bibinfo {author} {\bibfnamefont {G.}~\bibnamefont
  {Evenbly}}, \bibinfo {author} {\bibfnamefont {N.}~\bibnamefont {Pancotti}},
  \bibinfo {author} {\bibfnamefont {A.}~\bibnamefont {Milsted}}, \bibinfo
  {author} {\bibfnamefont {J.}~\bibnamefont {Gray}},\ and\ \bibinfo {author}
  {\bibfnamefont {G.~K.-L.}\ \bibnamefont {Chan}},\ }\href
  {https://arxiv.org/abs/2409.03108} {\bibinfo {title} {{Loop Series Expansions
  for Tensor Networks}}} (\bibinfo {year} {2024}),\ \Eprint
  {https://arxiv.org/abs/2409.03108} {arXiv:2409.03108 [quant-ph]} \BibitemShut
  {NoStop}%
\bibitem [{\citenamefont {Lubasch}\ \emph
  {et~al.}(2014{\natexlab{c}})\citenamefont {Lubasch}, \citenamefont {Cirac},\
  and\ \citenamefont {Bañuls}}]{Lubasch:2014}%
  \BibitemOpen
  \bibfield  {author} {\bibinfo {author} {\bibfnamefont {M.}~\bibnamefont
  {Lubasch}}, \bibinfo {author} {\bibfnamefont {J.~I.}\ \bibnamefont {Cirac}},\
  and\ \bibinfo {author} {\bibfnamefont {M.-C.}\ \bibnamefont {Bañuls}},\
  }\bibfield  {title} {\bibinfo {title} {Unifying projected entangled pair
  state contractions},\ }\href {https://doi.org/10.1088/1367-2630/16/3/033014}
  {\bibfield  {journal} {\bibinfo  {journal} {New Journal of Physics}\ }\textbf
  {\bibinfo {volume} {16}},\ \bibinfo {pages} {033014} (\bibinfo {year}
  {2014}{\natexlab{c}})}\BibitemShut {NoStop}%
\bibitem [{\citenamefont {Ferris}\ and\ \citenamefont
  {Vidal}(2012)}]{ferris2012perfect}%
  \BibitemOpen
  \bibfield  {author} {\bibinfo {author} {\bibfnamefont {A.~J.}\ \bibnamefont
  {Ferris}}\ and\ \bibinfo {author} {\bibfnamefont {G.}~\bibnamefont {Vidal}},\
  }\bibfield  {title} {\bibinfo {title} {Perfect sampling with unitary tensor
  networks},\ }\href@noop {} {\bibfield  {journal} {\bibinfo  {journal}
  {Physical Review B—Condensed Matter and Materials Physics}\ }\textbf
  {\bibinfo {volume} {85}},\ \bibinfo {pages} {165146} (\bibinfo {year}
  {2012})}\BibitemShut {NoStop}%
\bibitem [{\citenamefont {Gray}\ and\ \citenamefont {Chan}(2024)}]{gray:2024}%
  \BibitemOpen
  \bibfield  {author} {\bibinfo {author} {\bibfnamefont {J.}~\bibnamefont
  {Gray}}\ and\ \bibinfo {author} {\bibfnamefont {G.~K.-L.}\ \bibnamefont
  {Chan}},\ }\bibfield  {title} {\bibinfo {title} {Hyperoptimized approximate
  contraction of tensor networks with arbitrary geometry},\ }\href
  {https://doi.org/10.1103/PhysRevX.14.011009} {\bibfield  {journal} {\bibinfo
  {journal} {Phys. Rev. X}\ }\textbf {\bibinfo {volume} {14}},\ \bibinfo
  {pages} {011009} (\bibinfo {year} {2024})}\BibitemShut {NoStop}%
\bibitem [{\citenamefont {Gray}\ and\ \citenamefont
  {Kourtis}(2021)}]{Gray:2021}%
  \BibitemOpen
  \bibfield  {author} {\bibinfo {author} {\bibfnamefont {J.}~\bibnamefont
  {Gray}}\ and\ \bibinfo {author} {\bibfnamefont {S.}~\bibnamefont {Kourtis}},\
  }\bibfield  {title} {\bibinfo {title} {Hyper-optimized tensor network
  contraction},\ }\href {https://doi.org/10.22331/q-2021-03-15-410} {\bibfield
  {journal} {\bibinfo  {journal} {{Quantum}}\ }\textbf {\bibinfo {volume}
  {5}},\ \bibinfo {pages} {410} (\bibinfo {year} {2021})}\BibitemShut {NoStop}%
\bibitem [{\citenamefont {Gray}(2018)}]{gray2018quimb}%
  \BibitemOpen
  \bibfield  {author} {\bibinfo {author} {\bibfnamefont {J.}~\bibnamefont
  {Gray}},\ }\bibfield  {title} {\bibinfo {title} {quimb: a python library for
  quantum information and many-body calculations},\ }\href
  {https://doi.org/10.21105/joss.00819} {\bibfield  {journal} {\bibinfo
  {journal} {Journal of Open Source Software}\ }\textbf {\bibinfo {volume}
  {3}},\ \bibinfo {pages} {819} (\bibinfo {year} {2018})}\BibitemShut {NoStop}%
\bibitem [{\citenamefont {Begu{\v{s}}i{\'c}}\ \emph {et~al.}(2023)\citenamefont
  {Begu{\v{s}}i{\'c}}, \citenamefont {Hejazi},\ and\ \citenamefont
  {Chan}}]{beguvsic2023simulating}%
  \BibitemOpen
  \bibfield  {author} {\bibinfo {author} {\bibfnamefont {T.}~\bibnamefont
  {Begu{\v{s}}i{\'c}}}, \bibinfo {author} {\bibfnamefont {K.}~\bibnamefont
  {Hejazi}},\ and\ \bibinfo {author} {\bibfnamefont {G.~K.}\ \bibnamefont
  {Chan}},\ }\bibfield  {title} {\bibinfo {title} {Simulating quantum circuit
  expectation values by {C}lifford perturbation theory},\ }\href
  {https://doi.org/10.48550/arXiv.2306.04797} {\bibfield  {journal} {\bibinfo
  {journal} {arXiv preprint arXiv:2306.04797}\ } (\bibinfo {year}
  {2023})}\BibitemShut {NoStop}%
\bibitem [{\citenamefont {Lerch}\ \emph {et~al.}(2024)\citenamefont {Lerch},
  \citenamefont {Puig}, \citenamefont {Rudolph}, \citenamefont {Angrisani},
  \citenamefont {Jones}, \citenamefont {Cerezo}, \citenamefont {Thanasilp},\
  and\ \citenamefont {Holmes}}]{lerch2024efficient}%
  \BibitemOpen
  \bibfield  {author} {\bibinfo {author} {\bibfnamefont {S.}~\bibnamefont
  {Lerch}}, \bibinfo {author} {\bibfnamefont {R.}~\bibnamefont {Puig}},
  \bibinfo {author} {\bibfnamefont {M.}~\bibnamefont {Rudolph}}, \bibinfo
  {author} {\bibfnamefont {A.}~\bibnamefont {Angrisani}}, \bibinfo {author}
  {\bibfnamefont {T.}~\bibnamefont {Jones}}, \bibinfo {author} {\bibfnamefont
  {M.}~\bibnamefont {Cerezo}}, \bibinfo {author} {\bibfnamefont
  {S.}~\bibnamefont {Thanasilp}},\ and\ \bibinfo {author} {\bibfnamefont
  {Z.}~\bibnamefont {Holmes}},\ }\bibfield  {title} {\bibinfo {title}
  {Efficient quantum-enhanced classical simulation for patches of quantum
  landscapes},\ }\bibfield  {journal} {\bibinfo  {journal} {arXiv preprint
  arXiv:2411.19896}\ }\href {https://doi.org/10.48550/arXiv.2411.19896}
  {10.48550/arXiv.2411.19896} (\bibinfo {year} {2024})\BibitemShut {NoStop}%
\bibitem [{\citenamefont {Fontana}\ \emph {et~al.}(2023)\citenamefont
  {Fontana}, \citenamefont {Rudolph}, \citenamefont {Duncan}, \citenamefont
  {Rungger},\ and\ \citenamefont {C{\^\i}rstoiu}}]{fontana2023classical}%
  \BibitemOpen
  \bibfield  {author} {\bibinfo {author} {\bibfnamefont {E.}~\bibnamefont
  {Fontana}}, \bibinfo {author} {\bibfnamefont {M.~S.}\ \bibnamefont
  {Rudolph}}, \bibinfo {author} {\bibfnamefont {R.}~\bibnamefont {Duncan}},
  \bibinfo {author} {\bibfnamefont {I.}~\bibnamefont {Rungger}},\ and\ \bibinfo
  {author} {\bibfnamefont {C.}~\bibnamefont {C{\^\i}rstoiu}},\ }\bibfield
  {title} {\bibinfo {title} {Classical simulations of noisy variational quantum
  circuits},\ }\href@noop {} {\bibfield  {journal} {\bibinfo  {journal} {arXiv
  preprint arXiv:2306.05400}\ } (\bibinfo {year} {2023})}\BibitemShut {NoStop}%
\bibitem [{\citenamefont {Schmitt}\ \emph {et~al.}(2022)\citenamefont
  {Schmitt}, \citenamefont {Rams}, \citenamefont {Dziarmaga}, \citenamefont
  {Heyl},\ and\ \citenamefont {Zurek}}]{schmitt2022quantum}%
  \BibitemOpen
  \bibfield  {author} {\bibinfo {author} {\bibfnamefont {M.}~\bibnamefont
  {Schmitt}}, \bibinfo {author} {\bibfnamefont {M.~M.}\ \bibnamefont {Rams}},
  \bibinfo {author} {\bibfnamefont {J.}~\bibnamefont {Dziarmaga}}, \bibinfo
  {author} {\bibfnamefont {M.}~\bibnamefont {Heyl}},\ and\ \bibinfo {author}
  {\bibfnamefont {W.~H.}\ \bibnamefont {Zurek}},\ }\bibfield  {title} {\bibinfo
  {title} {Quantum phase transition dynamics in the two-dimensional
  transverse-field ising model},\ }\href
  {https://doi.org/10.1126/sciadv.abl6850} {\bibfield  {journal} {\bibinfo
  {journal} {Science Advances}\ }\textbf {\bibinfo {volume} {8}},\ \bibinfo
  {pages} {eabl6850} (\bibinfo {year} {2022})}\BibitemShut {NoStop}%
\bibitem [{\citenamefont {Carleo}\ and\ \citenamefont
  {Troyer}(2017)}]{carleo2017solving}%
  \BibitemOpen
  \bibfield  {author} {\bibinfo {author} {\bibfnamefont {G.}~\bibnamefont
  {Carleo}}\ and\ \bibinfo {author} {\bibfnamefont {M.}~\bibnamefont
  {Troyer}},\ }\bibfield  {title} {\bibinfo {title} {Solving the quantum
  many-body problem with artificial neural networks},\ }\href
  {https://doi.org/10.1126/science.aag2302} {\bibfield  {journal} {\bibinfo
  {journal} {Science}\ }\textbf {\bibinfo {volume} {355}},\ \bibinfo {pages}
  {602} (\bibinfo {year} {2017})}\BibitemShut {NoStop}%
\bibitem [{\citenamefont {Czarnik}\ \emph {et~al.}(2019)\citenamefont
  {Czarnik}, \citenamefont {Dziarmaga},\ and\ \citenamefont
  {Corboz}}]{czarnik2019time}%
  \BibitemOpen
  \bibfield  {author} {\bibinfo {author} {\bibfnamefont {P.}~\bibnamefont
  {Czarnik}}, \bibinfo {author} {\bibfnamefont {J.}~\bibnamefont {Dziarmaga}},\
  and\ \bibinfo {author} {\bibfnamefont {P.}~\bibnamefont {Corboz}},\
  }\bibfield  {title} {\bibinfo {title} {Time evolution of an infinite
  projected entangled pair state: An efficient algorithm},\ }\href
  {https://doi.org/10.1103/PhysRevB.99.035115} {\bibfield  {journal} {\bibinfo
  {journal} {Physical Review B}\ }\textbf {\bibinfo {volume} {99}},\ \bibinfo
  {pages} {035115} (\bibinfo {year} {2019})}\BibitemShut {NoStop}%
\bibitem [{\citenamefont {Dziarmaga}(2022)}]{dziarmaga2022simulation}%
  \BibitemOpen
  \bibfield  {author} {\bibinfo {author} {\bibfnamefont {J.}~\bibnamefont
  {Dziarmaga}},\ }\bibfield  {title} {\bibinfo {title} {Simulation of many-body
  localization and time crystals in two dimensions with the neighborhood tensor
  update},\ }\href {https://doi.org/10.1103/PhysRevB.105.054203} {\bibfield
  {journal} {\bibinfo  {journal} {Physical Review B}\ }\textbf {\bibinfo
  {volume} {105}},\ \bibinfo {pages} {054203} (\bibinfo {year}
  {2022})}\BibitemShut {NoStop}%
\bibitem [{\citenamefont {Kochkov}\ and\ \citenamefont
  {Clark}(2018)}]{kochkov2018variational}%
  \BibitemOpen
  \bibfield  {author} {\bibinfo {author} {\bibfnamefont {D.}~\bibnamefont
  {Kochkov}}\ and\ \bibinfo {author} {\bibfnamefont {B.~K.}\ \bibnamefont
  {Clark}},\ }\bibfield  {title} {\bibinfo {title} {Variational optimization in
  the ai era: Computational graph states and supervised wave-function
  optimization},\ }\href@noop {} {\bibfield  {journal} {\bibinfo  {journal}
  {arXiv preprint arXiv:1811.12423}\ } (\bibinfo {year} {2018})}\BibitemShut
  {NoStop}%
\bibitem [{\citenamefont {Medvidovi{\'c}}\ and\ \citenamefont
  {Carleo}(2021)}]{medvidovic2021classical}%
  \BibitemOpen
  \bibfield  {author} {\bibinfo {author} {\bibfnamefont {M.}~\bibnamefont
  {Medvidovi{\'c}}}\ and\ \bibinfo {author} {\bibfnamefont {G.}~\bibnamefont
  {Carleo}},\ }\bibfield  {title} {\bibinfo {title} {Classical variational
  simulation of the quantum approximate optimization algorithm},\ }\href
  {https://doi.org/10.1038/s41534-021-00440-z} {\bibfield  {journal} {\bibinfo
  {journal} {npj Quantum Information}\ }\textbf {\bibinfo {volume} {7}},\
  \bibinfo {pages} {101} (\bibinfo {year} {2021})}\BibitemShut {NoStop}%
\bibitem [{\citenamefont {Guti{\'e}rrez}\ and\ \citenamefont
  {Mendl}(2022)}]{gutierrez2022real}%
  \BibitemOpen
  \bibfield  {author} {\bibinfo {author} {\bibfnamefont {I.~L.}\ \bibnamefont
  {Guti{\'e}rrez}}\ and\ \bibinfo {author} {\bibfnamefont {C.~B.}\ \bibnamefont
  {Mendl}},\ }\bibfield  {title} {\bibinfo {title} {Real time evolution with
  neural-network quantum states},\ }\href
  {https://doi.org/10.22331/q-2022-01-20-627} {\bibfield  {journal} {\bibinfo
  {journal} {Quantum}\ }\textbf {\bibinfo {volume} {6}},\ \bibinfo {pages}
  {627} (\bibinfo {year} {2022})}\BibitemShut {NoStop}%
\bibitem [{\citenamefont {Sinibaldi}\ \emph {et~al.}(2023)\citenamefont
  {Sinibaldi}, \citenamefont {Giuliani}, \citenamefont {Carleo},\ and\
  \citenamefont {Vicentini}}]{sinibaldi2023unbiasing}%
  \BibitemOpen
  \bibfield  {author} {\bibinfo {author} {\bibfnamefont {A.}~\bibnamefont
  {Sinibaldi}}, \bibinfo {author} {\bibfnamefont {C.}~\bibnamefont {Giuliani}},
  \bibinfo {author} {\bibfnamefont {G.}~\bibnamefont {Carleo}},\ and\ \bibinfo
  {author} {\bibfnamefont {F.}~\bibnamefont {Vicentini}},\ }\bibfield  {title}
  {\bibinfo {title} {Unbiasing time-dependent variational monte carlo by
  projected quantum evolution},\ }\href
  {https://doi.org/10.22331/q-2023-10-10-1131} {\bibfield  {journal} {\bibinfo
  {journal} {Quantum}\ }\textbf {\bibinfo {volume} {7}},\ \bibinfo {pages}
  {1131} (\bibinfo {year} {2023})}\BibitemShut {NoStop}%
\bibitem [{\citenamefont {Barison}\ \emph {et~al.}(2021)\citenamefont
  {Barison}, \citenamefont {Vicentini},\ and\ \citenamefont
  {Carleo}}]{barison2021efficient}%
  \BibitemOpen
  \bibfield  {author} {\bibinfo {author} {\bibfnamefont {S.}~\bibnamefont
  {Barison}}, \bibinfo {author} {\bibfnamefont {F.}~\bibnamefont {Vicentini}},\
  and\ \bibinfo {author} {\bibfnamefont {G.}~\bibnamefont {Carleo}},\
  }\bibfield  {title} {\bibinfo {title} {An efficient quantum algorithm for the
  time evolution of parameterized circuits},\ }\href
  {https://doi.org/10.22331/q-2021-07-28-512} {\bibfield  {journal} {\bibinfo
  {journal} {Quantum}\ }\textbf {\bibinfo {volume} {5}},\ \bibinfo {pages}
  {512} (\bibinfo {year} {2021})}\BibitemShut {NoStop}%
\bibitem [{\citenamefont {Donatella}\ \emph {et~al.}(2023)\citenamefont
  {Donatella}, \citenamefont {Denis}, \citenamefont {Le~Boit{\'e}},\ and\
  \citenamefont {Ciuti}}]{donatella2023dynamics}%
  \BibitemOpen
  \bibfield  {author} {\bibinfo {author} {\bibfnamefont {K.}~\bibnamefont
  {Donatella}}, \bibinfo {author} {\bibfnamefont {Z.}~\bibnamefont {Denis}},
  \bibinfo {author} {\bibfnamefont {A.}~\bibnamefont {Le~Boit{\'e}}},\ and\
  \bibinfo {author} {\bibfnamefont {C.}~\bibnamefont {Ciuti}},\ }\bibfield
  {title} {\bibinfo {title} {Dynamics with autoregressive neural quantum
  states: Application to critical quench dynamics},\ }\href
  {https://doi.org/10.1103/PhysRevA.108.022210} {\bibfield  {journal} {\bibinfo
   {journal} {Physical Review A}\ }\textbf {\bibinfo {volume} {108}},\ \bibinfo
  {pages} {022210} (\bibinfo {year} {2023})}\BibitemShut {NoStop}%
\bibitem [{\citenamefont {{Sinibaldi}}\ \emph {et~al.}(2024)\citenamefont
  {{Sinibaldi}}, \citenamefont {{Hendry}}, \citenamefont {{Vicentini}},\ and\
  \citenamefont {{Carleo}}}]{Sinibaldi2024}%
  \BibitemOpen
  \bibfield  {author} {\bibinfo {author} {\bibfnamefont {A.}~\bibnamefont
  {{Sinibaldi}}}, \bibinfo {author} {\bibfnamefont {D.}~\bibnamefont
  {{Hendry}}}, \bibinfo {author} {\bibfnamefont {F.}~\bibnamefont
  {{Vicentini}}},\ and\ \bibinfo {author} {\bibfnamefont {G.}~\bibnamefont
  {{Carleo}}},\ }\bibfield  {title} {\bibinfo {title} {{Time-dependent Neural
  Galerkin Method for Quantum Dynamics}},\ }\href
  {https://ui.adsabs.harvard.edu/abs/2024arXiv241211778S} {\bibfield  {journal}
  {\bibinfo  {journal} {arXiv:2412.11778}\ } (\bibinfo {year}
  {2024})}\BibitemShut {NoStop}%
\bibitem [{\citenamefont {Van~de Walle}\ \emph {et~al.}(2025)\citenamefont
  {Van~de Walle}, \citenamefont {Schmitt},\ and\ \citenamefont
  {Bohrdt}}]{VandeWalle2024}%
  \BibitemOpen
  \bibfield  {author} {\bibinfo {author} {\bibfnamefont {A.}~\bibnamefont
  {Van~de Walle}}, \bibinfo {author} {\bibfnamefont {M.}~\bibnamefont
  {Schmitt}},\ and\ \bibinfo {author} {\bibfnamefont {A.}~\bibnamefont
  {Bohrdt}},\ }\bibfield  {title} {\bibinfo {title} {Many-body dynamics with
  explicitly time-dependent neural quantum states},\ }\href@noop {} {\bibfield
  {journal} {\bibinfo  {journal} {Machine Learning: Science and Technology}\ ,\
  \bibinfo {pages} {045011}} (\bibinfo {year} {2025})}\BibitemShut {NoStop}%
\bibitem [{\citenamefont {Roth}\ and\ \citenamefont
  {MacDonald}(2021)}]{roth2021group}%
  \BibitemOpen
  \bibfield  {author} {\bibinfo {author} {\bibfnamefont {C.}~\bibnamefont
  {Roth}}\ and\ \bibinfo {author} {\bibfnamefont {A.~H.}\ \bibnamefont
  {MacDonald}},\ }\bibfield  {title} {\bibinfo {title} {Group convolutional
  neural networks improve quantum state accuracy},\ }\href@noop {} {\bibfield
  {journal} {\bibinfo  {journal} {arXiv preprint arXiv:2104.05085}\ } (\bibinfo
  {year} {2021})}\BibitemShut {NoStop}%
\bibitem [{Note5()}]{Note5}%
  \BibitemOpen
  \bibinfo {note} {A single hidden-layer CNN with an even activation function,
  e.g., {poly6} or $\ln (\cosh ( z))$, and zero bias leads to an $\protect
  \mathbb {Z}_2$ symmetrized ansatz with respect to flipping all the spins.
  Since the $\protect \mathbb {Z}_2$ symmetry of the Ising model is broken by
  our initial state, we use the {poly5} activation function, which is an odd
  function, in all our calculations.}\BibitemShut {Stop}%
\bibitem [{Note6()}]{Note6}%
  \BibitemOpen
  \bibinfo {note} {A single hidden-layer CNN with the $\ln (\cosh ( z))$
  activation function is exactly equivalent to the symmetrized RBM. A single
  hidden-layer CNN with the {poly6} activation function serves as an
  holomorphic ansatz closed to the symmetrized RBM and has been shown to give
  competitive results to the state of the art tensor network simulation~\cite
  {Schmitt2020}.}\BibitemShut {Stop}%
\bibitem [{\citenamefont {{Gravina}}\ \emph {et~al.}(2024)\citenamefont
  {{Gravina}}, \citenamefont {{Savona}},\ and\ \citenamefont
  {{Vicentini}}}]{Gravina2024}%
  \BibitemOpen
  \bibfield  {author} {\bibinfo {author} {\bibfnamefont {L.}~\bibnamefont
  {{Gravina}}}, \bibinfo {author} {\bibfnamefont {V.}~\bibnamefont
  {{Savona}}},\ and\ \bibinfo {author} {\bibfnamefont {F.}~\bibnamefont
  {{Vicentini}}},\ }\bibfield  {title} {\bibinfo {title} {{Neural Projected
  Quantum Dynamics: a systematic study}},\ }\bibfield  {journal} {\bibinfo
  {journal} {arXiv:2410.10720}\ }\href
  {https://doi.org/10.48550/arXiv.2410.10720} {10.48550/arXiv.2410.10720}
  (\bibinfo {year} {2024})\BibitemShut {NoStop}%
\end{thebibliography}
\end{document}